\documentclass[11pt]{article}
\usepackage{jheppub}
\usepackage[utf8]{inputenc}
\usepackage{calligra}
\DeclareMathAlphabet{\mathpzc}{OT1}{pzc}{m}{it}
\usepackage[T1]{fontenc}
\usepackage[normalem]{ulem}
\usepackage{braket,slashed,bm}
\usepackage{amsmath}
\usepackage{mathrsfs}
\usepackage{graphicx}
\usepackage{floatrow}
\newfloatcommand{capbtabbox}{table}[][\FBwidth]
\usepackage{caption}
\usepackage{subcaption}
\usepackage{enumitem}
\usepackage{array}
\usepackage{tikz}
\usetikzlibrary{arrows,decorations.pathmorphing,backgrounds,positioning,fit,petri,automata,shadows,calendar,mindmap,
decorations.markings,calc}

\def\beq{\begin{equation}}
\def\eeq{\end{equation}}
\def\bea{\begin{eqnarray}}
\def\eea{\end{eqnarray}}
\def\nn{\nonumber \\}
\def\hyp{\mathsf{y}}
\renewcommand{\a}{\alpha}

\renewcommand{\d}{\delta}
\newcommand{\g}{\gamma}
\newcommand{\s}{\sigma}
\newcommand{\de}{\partial}
\newcommand{\Q}{\mathcal{Q}}
\newcommand{\CHlt}{C_{Hl}^{(3)}}
\newcommand{\CHqt}{C_{Hq}^{(3)}}
\newcommand{\CHls}{C_{Hl}^{(1)}}
\newcommand{\CHqs}{C_{Hq}^{(1)}}
\newcommand{\haew}{\hat\alpha_{ew}}
\newcommand{\ct}{c_\theta}
\newcommand{\st}{s_\theta}
\newcommand{\hst}{s_{\hat\theta}}
\newcommand{\hct}{c_{\hat\theta}}
\newcommand{\hsdt}{s_{2\hat\theta}}
\newcommand{\hcdt}{c_{2\hat\theta}}

\newcommand{\U}{\mathbf{U}}
\newcommand{\T}{\mathbf{T}}
\newcommand{\V}{\mathbf{V}}
\newcommand{\F}{\mathcal{F}}
\DeclareMathOperator{\Tr}{Tr}
\newcommand{\hc}{\text{h.c.}}
\newcommand{\that}{{\hat{\theta}}}
\newcommand{\gsm}[2]{g^{#1,SM}_{#2}}
\renewcommand{\Re}{{\rm Re}}
\newcommand{\titlemath}{\texorpdfstring}
\newcommand{\abs}[1]{\left|#1\right|}
\title{The Standard Model as an Effective Field Theory}
\author[a]{Ilaria Brivio and Michael Trott}

\affiliation[a]{Niels Bohr International Academy \& Discovery Center,
Niels Bohr Institute, University of Copenhagen,
Blegdamsvej 17, DK-2100, Copenhagen, Denmark}

\abstract{Projecting measurements of the interactions of the known Standard Model (SM) states
into an effective field theory (EFT) framework is an important goal of the LHC physics program.
The interpretation of measurements
of the properties of the Higgs-like boson in an EFT allows one to consistently study the properties of this state,
while the SM is allowed to eventually break down at higher energies. In this review, basic concepts relevant to the construction of such EFTs are reviewed pedagogically.
Electroweak precision data is discussed as a historical example of some importance
to illustrate critical consistency issues in interpreting experimental data in EFTs.
A future precision Higgs phenomenology program can benefit from the projection of raw experimental results into
consistent field theories such as the SM, the SM supplemented with
higher dimensional operators (the SMEFT) or an Electroweak chiral Lagrangian with a dominantly $J^P = 0^+$ scalar (the HEFT).
We discuss the developing SMEFT and HEFT approaches, that are consistent versions of such EFTs, systematically
improvable with higher order corrections, and comment on the
pseudo-observable approach.
We review the challenges that have been overcome in developing EFT methods for LHC studies, and the challenges that remain.}

\begin{document}
\maketitle

\section{Introduction}
Our understanding of the nature of fundamental interactions
can advance through a direct discovery of a new particle, or indirectly.
Knowledge gathered through indirect
methods has historically been the leading indication of a new particle or theoretical framework.
It has also been the case that such indirect knowledge is usually ambiguous in that it can be an indication of several possible
models. This is essentially due to the decoupling theorem \cite{Appelquist:1974tg}
which formalizes how the non-analytic structure of correlation functions due to heavy states
are projected out when matching onto a low energy effective field theory (EFT).
The discovery of a new particle is a clarifying event, as it usually removes such ambiguities.

The Large Hadron Collider (LHC) was constructed based on the expectation that the
functional mechanism by which $\rm SU_L(2) \times U_Y(1) \rightarrow U_{em}(1)$ below the unitarity violation
scale(s) dictated by the massive $W^\pm, Z$ vector bosons \cite{Cornwall:1974km,Vayonakis:1976vz,Lee:1977yc,Lee:1977eg,Chanowitz:1978uj,Chanowitz:1978mv}
would be revealed by probing the $\rm TeV$ energy range, with strong theoretical prejudice in favour of the Brout-Englert-Higgs mechanism influencing design choices \cite{Eichten:1984eu}.
In addition, it was expected that other beyond the Standard Model (SM) states involved in this
mechanism could also be discovered in the characteristic energy range that LHC is exploring. The first expectation for LHC has been met to date with
the discovery of a dominantly
$J^P = 0^+$ boson \cite{Chatrchyan:2012jja,Aad:2013xqa,Chatrchyan:2013mxa}
consistent with the SM Higgs boson \cite{Aad:2012tfa,Chatrchyan:2012xdj}.

The lack of additional new state
discoveries at LHC (to date) is perhaps unsurprising considering the large global data set consistent with the SM.
In recent years, the direct discovery of new\footnote{At least arguably fundamental.}
states has become less frequent; the last three such discoveries being the
top quark in 1995 \cite{Abe:1995hr,Abachi:1995iq}, the reporting of direct evidence of the tau neutrino
in 2000 \cite{Kodama:2000mp}, and the Higgs-like boson discovered
in 2012 \cite{Aad:2012tfa,Chatrchyan:2012xdj}.
The possibility that the next direct discovery of a new particle
is a prospect for the far experimental future is manifest. This expectation is
supported by the lack of
statistically significant deviations from SM predictions in the global data set, which
can be largely a result of at least a moderate degree of decoupling of physics beyond the SM to higher
energy scales ($\gg m_{Z,W,h}$). Avoiding unproductive melancholy,
this is motivation for increasing our understanding of all manner of SM
physics to improve our theoretical predictions of experimental results.
Thereby we sharpen the theoretical tools that allow us to indirectly search for
physics beyond the SM.

This motivation is supported by the fact that all of these latest discoveries
of new states $\{t, \nu_\tau, h\}$ were preceded by decades of indirect evidence gathered using EFT techniques, requiring precise
SM predictions. Further, this argument also supports developing EFT methods to capture the low energy effect of physics beyond the SM, as
only focusing efforts on improving SM predictions
is insufficient the moment a real deviation is discovered.
In addition, the unavoidable theoretical ambiguity associated with indirect knowledge
of physics beyond the SM means that a singular theoretical explanation of such a deviation from the SM is unlikely to
be epistemologically assured. It is important to be able to systematically understand such an anomalous measurement
in a well defined field theory framework, that also dictates correlated deviations in other processes
to distinguish between competing explanations. After all, any
successful explanation of such a deviation must be consistent with the global data set, not just the observable
deviating from the SM.

It can be remarkably efficient to approach such tests of consistency by projecting a particular model
into an EFT framework in the presence of some degree of decoupling. This is the case so long as the EFT is
well developed, so that properly interfacing with the global data set
can be done in a one time matching calculation.
To this end, it is essential to systematically improve
our understanding of the EFTs that can accommodate SM deviations in advance of any such discovery of the SM breaking down in describing the data.
It is also critical to encode the current data set into a form that maximizes its future utility
when the SM can no longer successfully describe higher energy measurements.

This review is focused on these tasks. We discuss
the recent developments in using indirect methods to study the Higgs-like scalar, related signals,
and the development of two EFT frameworks. Both of these frameworks describe the known
SM particles that lead to non-analytic structure in the correlation functions measured
in particle physics experiments to date, in some region of phase space.  These theories are distinguished by the
nature of the low energy (infrared -IR) limit of physics beyond the SM being assumed.
When the SM Higgs doublet is present in the EFT construction, the EFT
is known as the Standard Model Effective Field Theory (the SMEFT). Conversely,
when the SM Higgs doublet is not present, the
Higgs Effective Field Theory (the HEFT) is constructed. Due to the lack of any
clear experimental indication to choose between these approaches at this time, it is important
to minimize theoretical bias when reporting LHC data.
For this reason, it can also be advantageous to project raw experimental data in terms of cross section
measurements into gauge invariant pseudo-observables in some cases, that are constructed by expanding around the poles
of the SM states. These pseudo-observable decompositions can be related to multiple theoretical frameworks,
such as these two EFTs. We also discuss this developing paradigm and the relationship between these various approaches.

The outline of this review is as follows. In Section \ref{sec:history} we briefly review historical, and more traditional uses of EFT.
In Section \ref{precision-motivation} we then discuss the current pressing motivation for a precision
Higgs phenomenology program using EFT methods. In Section \ref{basics} we review pedagogically the
key points leading to the structure of EFTs. We then turn to discussing
the candidate field theories to use to interpret the global data set. First the SM
is reviewed in Section \ref{SMsection} to fix notation. We then discuss the SMEFT in Section \ref{SMEFTsec}
and the HEFT in Section \ref{HEFTsec}. Some issues that are being currently debated in the literature
are reviewed in Section \ref{bloodbath}.
In Section \ref{preliminarymeasurements} we discuss and review pedagogically
the distinction between $S$ matrix elements, Lagrangian parameters and pseudo-observables
with an emphasis on the differences between these concepts that
are accentuated in the presence of an EFT such as the SMEFT or HEFT.
We apply this understanding to LEPI-II pseudo-observable measurements and interpretations in Section \ref{LEPgoodandbad}.
In Section \ref{Higgsfun} we discuss how many of these concepts and subtleties appear again in the interpretation
of the measurements of the Higgs-like scalar at LHC, and review the $\kappa$ formalism and proposals to go beyond
this formalism in the long term LHC program.
In Section \ref{topsection} we discuss the application of EFTs to top quark measurements at LHC.
Finally, in Section \ref{stateoftheart} we summarize the state of affairs early in
LHC Run II and sketch out some expected future developments. The Appendix presents a series of LO results in a unified notation
for SMEFT shifts to $\bar{\psi} \psi \rightarrow V \rightarrow \bar{\psi} \psi$ scattering through $V = \{Z,W\}$
gauge boson scattering, Higgs production and decay, vector boson scattering and $h V$ production using a $\{\haew, \hat{G}_F, \hat{m}_Z \}$
input parameter set.

The intended audience for this review is a mixture of experts and novices and both theorists and experimentalists. The presentation is geared to aid
a new Ph.D. student with a solid quantum field theory background to jump into this area of research.
We hope that experts in the field will also benefit from some of the discussion
on the conceptual and technical aspects of these interesting examples of EFTs
incorporating the presence of the Higgs-like boson.\footnote{We apologize in advance for overlooking
any references in this literature.}

\section{Features of the EFT landscape}\label{sec:history}
EFT is now a common tool used in many areas of particle physics, and increasingly in related areas of physics.
The main reason EFT has become a standard theoretical tool is that it allows one to study large sets of experimental data in a systematically improvable
field theory approach. This is the case without  having to assume the theory used is valid to arbitrarily high energies.
We discuss some of the physics underlying this view in Section \ref{basics}.
Considering that the SM will eventually break down at higher energies/shorter distance scales, this makes EFT extensions of the SM key tools to develop
in the modern "data rich" era.
EFT methods generally come to the fore when large amounts of data are at hand to constrain the many parameters that usually arise in such a construction.
This is now occurring for studies of the Higgs boson and the top quark, due to the successful operation of LHC. This is the reason that this review is more focused
on studying these particles using EFT methods.

These efforts are beginning in earnest using an approach to EFT that has a long history. An early influential example of EFT is given in Fermi's theory of $\beta$ decay. In retrospect, Fermi theory is an effective operator approach to
describe $\mu^- \rightarrow e^- + \bar{\nu}_e+ \nu_\mu$ via an assumed Lagrangian \cite{Fermi:1934hr}
\bea
\mathcal{L}_{GF} = - \tilde G_F (\bar{\psi}_i \gamma^\mu P_L \psi_j)(\bar{\psi}_k \gamma^\mu P_L \psi_l).
\eea
The left handed structure of the interactions was only fixed in due time with experimental input, and Fermi's approach was even more general in its initial formulation.
The postulated interaction was introduced with a free coupling fit to data -- $ \tilde G_F$.\footnote{We turn our attention back to Fermi theory in Section \ref{sec:renorm-nonrenorm}.}
This EFT description of a decay was advanced and found to be manifestly useful to study experimental results,
before any solid experimental evidence of the existence of a $W$ boson, or the SM itself, was at hand.
This illustrates a key point of underlying the power of EFT: it is not required to know the underlying UV completion to use EFT methods.
For more discussion on this point for the case of Fermi theory, see Refs.~\cite{Gavela:2016bzc,Manohar:2018aog}.

The amount of data available in high energy collisions $\sqrt{s} \gg m_{W,Z}$ has historically been limited compared to
lower energy collisions or decays, such as $\beta$ decay. The historical use of EFT has been focused on using
the relatively larger data sets gathered on lower energy phenomena, such as in flavour physics and/or studies of bound states of Quantum Chromodynamics (QCD) as a result.
Another important example of an early application of EFT was the study of bound states of QCD in the 1960's, and Weinberg's calculation of
pion scattering lengths. This occurred before any clear experimental evidence of the existence of quarks or any understanding of QCD was at hand.
These calculations were first performed using assumed symmetry principles (partially conserved axial currents) in Ref.~\cite{Weinberg:1966kf} and otherwise free parameters
were again fit to the data. It was soon understood that a non-linear realization of
$\rm SU(2) \times SU(2)$ chiral symmetry allowed a more elegant and general understanding of the physics at work in Ref.~\cite{PhysRev.166.1568}. A clear discussion on this point is presented in Ref.~\cite{Weinberg:1978kz}. The understanding of non-linear realizations of symmetries, commonly used in EFT applications, was simultaneously advanced and generalized in the classic works of
Coleman et al. in Refs.~\cite{Coleman:1969sm,Callan:1969sn}.

Once supplied with the solid theoretical hammer that is EFT, and a clear conceptual foundation of this approach developed coincident with the resurgent interest in field theory in the 1970's, the theoretical community found many nails.
An explosion of applications and well defined EFT's has emerged in the past few decades. An incomplete summary of some important applications of EFT include the following.
\begin{itemize}
\item{ChPT. Following the pioneering studies of the 1960's the study of the $\pi$, K and $\eta$  mesons using chiral perturbation theory (ChPT) methods has been systematically developed for decades.
Some of the key papers of this development are Refs.~\cite{Gasser:1982ap,Gasser:1983yg,Gasser:1984gg,Ecker:1988te}. For reviews of this approach see Ref.~\cite{Pich:1995bw,Pich:2018ltt}}.
\item{LEFT. Fermi's theory has been systematically extended into a complete description of a low energy phenomena where higher dimensional operators are used to describe
flavour conserving and violating contact interactions. The operators of this EFT are generated when the $W,Z,h,t$ particles of the SM are integrated out, and can also have beyond the SM matching contributions.
This EFT is used extensively in studies of flavour transitions of QCD bound states at low energies for decades.
Recently, this approach has been further systematized in Ref.~\cite{Jenkins:2017jig,Jenkins:2017dyc}, which determined the complete one loop anomalous dimension
of the theory and the matching onto the framework of the SMEFT. Some of the results in the Appendix are defined in this EFT.}
\item{HQET. A systematic expansion in the ratio $\Lambda_{QCD}/m_b \ll 1$ underlies the Heavy Quark Effective Theory. This theory describes
the interactions of a heavy quark with soft partons and has been applied to describe $B$ meson decays and oscillations. An excellent resource to learn and calculate in HQET is
Ref.~\cite{Manohar:2000dt}.}
\item{SCET. Building on the idea of the large energy effective field theory (LEET) \cite{Dugan:1990de}, and initially motivated out of the study of
summing Sudakov logarithms in $B \rightarrow X_s \gamma$ decay that had failed in the LEET formalism, the Soft Collinear Effective Field Theory was developed
in Refs.~\cite{Bauer:2000ew,Bauer:2000yr,Bauer:2001ct,Bauer:2001yt}. This EFT describes the interactions of particles of relatively different energies $Q$ and the small
expansion parameter is $\Lambda_{QCD}/Q$ in most applications. A good introduction to SCET is given in Ref.~\cite{Becher:2014oda}.}
\end{itemize}

EFT's continue to be developed and added to this incomplete list. In this review, we mostly focus on the SMEFT and HEFT effective field theories that are currently experiencing an intense development related to LHC
experimental results. First we set the stage by discussing the strong motivation for precision Higgs studies and lay out the basic ideas underlying EFT.

\section{The need for a precision Higgs phenomenology program}\label{precision-motivation}

Further developing the theoretical methods discussed in this review is
not an idle pursuit. It is reasonable to expect that the properties of the dominantly $J^P = 0^+$ scalar boson
could be perturbed by physics beyond the SM, and a precision Higgs phenomenology program could uncover such
perturbations.

The reason for this expectation is the idea that the SM Higgs mechanism describing
$\rm SU_L(2) \times U_Y(1) \rightarrow U_{em}(1)$ is likely to be only an
effective description. This belief is deeply rooted in the historical origin
of the Higgs mechanism itself. The
Higgsed phase of the SM (see discussion in Refs.~\cite{Weinberg:1996kr,witten})
can be understood to be analogous to the ideas that first emerged in the
Landau-Ginzburg effective model of superconductivity \cite{Ginzburg:1950sr}.
The Landau-Ginzburg action functional is given by \cite{witten}
\bea
LG(s) = \int_{\Re^3} d x^3 \left[\frac{1}{2}|(d - 2 \, i \, e \, A)s|^2
+ \frac{\gamma}{2}\left(|s|^2 - a^2 \right)\right],
\eea
where $A$ is the vector potential of Electromagnetism, $d$ is a derivative defined on $\Re^3$, $e$ is the electric charge
and $s$ is a section of the (squared) unitary complex line bundle of
Electromagnetism. This action has an energetically favoured minimum for $|s| = a$
and $(d - 2 \, i \, e \, A)s = 0$  when $\gamma>0$, leading to a topologically flat
line bundle in the superconducting phase, and the exclusion of the magnetic field
from the superconducting material.\footnote{Interestingly, the topology of the scalar
manifold defined by the Higgs doublet $H$ will form a crucial discriminant between the SMEFT and HEFT theories in the discussion that follows.}

The (partial) action of the SM Higgs \cite{Higgs:1964pj,Englert:1964et,Guralnik:1964eu} is directly analogous
\bea\label{Haction}
S_H = \int \, d^4 x \,  \left(|D_\mu H|^2 -\lambda \left(H^\dagger H -\frac12 v^2\right)^2 \right),
\eea
with $H$ is the Higgs doublet and $D$ is the covariant derivative of the $\rm SU_L(2) \times U_Y(1)$ theory.
This theory has an energetically favoured minimum at $\langle H^\dagger H\rangle = v^2/2$. Expanding
around the minimum of the potential (that defines the EW vacuum background field) leads to the massive $W^\pm_\mu,Z_\mu$
vector bosons, due to the Higgsing of $\rm SU_L(2) \times U_Y(1) \rightarrow U_{em}(1)$.
Massless $\rm SU_L(2) \times U_Y(1)$ vector boson field configurations are then energetically excluded.

Landau-Ginzburg theory is not a fundamental theory.
It is a functional effective description
of superconductivity that can be related
to a theory of Cooper pairs, such as BCS \cite{Bardeen:1957mv} theory.
The connection drawn between Landau-Ginzburg theory and the Higgsed phase in Yang-Mills theory by Anderson \cite{Anderson:1963pc},
leads to an expectation of a shorter distance (or higher energy i.e ultraviolet -- UV) completion/origin of the Higgs mechanism.\footnote{
A direct analogy would have the Higgs as a composite field, similar to the Cooper pair of BCS theory.}
In this manner,  the curious appearance of a classical Higgs potential with
a chosen ``Mexican hat'' form and an explicit scale $v$ in the SM Lagrangian
is understandable as a general low energy parameterization of underlying physics leading to
an effective $\rm SU_L(2) \times U_Y(1) \rightarrow U(1)_{em}$.
It is possible that this parameterization is not appropriate as an IR limit
of a UV sector leading to the observed massive gauge bosons. This is a way to understand the difference between HEFT and SMEFT that will be discussed
in more detail below.

\subsection{Quantum corrections to Higgs properties and potential}
\begin{figure}[h!]
\includegraphics[width=1\textwidth]{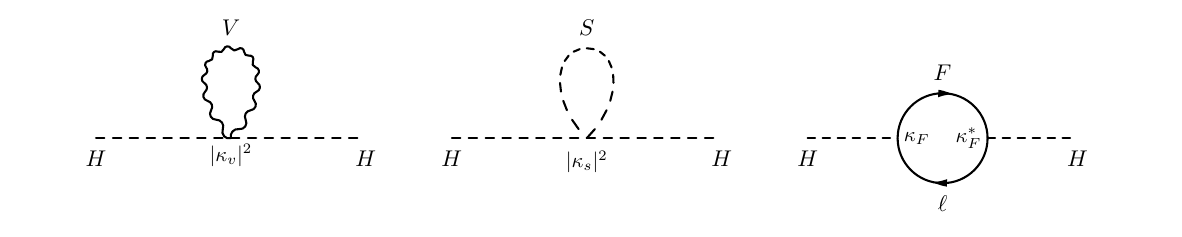}
\caption{\label{fig:threshold} One loop corrections giving threshold matching contributions to $H^\dagger H$.
We have used a common notation to simplify presentation in Eq.~\ref{crosssectionpert}, but
note that $|\kappa_v|^2,|\kappa_s|^2$ can be negative.}
\end{figure}

The previous section advanced an argument in support of a more fundamental description of
Electroweak symmetry breaking (EWSB) in that the Higgs potential
has no quantum or dynamical origin in the SM; it is a parameterization.
Functionally, it is a directly assumed classical potential --
that is extremely sensitive to UV physics including quantum corrections.
The reason for this is that the dimension of the $(H^\dagger H)$ operator is two.
Dimensional analysis indicates that this operator can receive dimensionful corrections
due to heavy states that extend the SM. Such heavy states generally couple to the field $H$, or composite operators
involving $H$, to perturb the properties of the Higgs when integrated out,\footnote{In some cases low energy effects can
be present modifying Higgs properties that do not satisfy this requirement, due to reducing the field theory
with the Equations of Motion (EOM) to a minimal basis
reshuffling the appearance of IR physics effects.}
see Fig.~\ref{fig:threshold}.
The Higgs field transforms as an $\rm SU_L(2)$ doublet and has hypercharge $\hyp_H = 1/2$, so
composite operators involving $H$ that allow dimension $\leq 4$ couplings
to singlet fermion ($F$), scalar bi-linear ($S^\dagger S$), and (non-gauged) vector fields $V_\mu$
are of the form\footnote{$H^\dagger D^\mu H V_\mu + h.c$ can be directly shown to not lead
to a threshold correction, by integrating by parts.}
\bea
\Delta \mathcal{L} =  |\kappa_v|^2 \, (H^\dagger H) \, V^\dagger_\mu \, V^{\mu} - |\kappa_s|^2 \, (H^\dagger H) \, (S^\dagger S) + \kappa_F \, \overline{F}\,  \tilde{H}^\dagger \ell_{L}+ h.c.
\eea
where $\ell$ is a lepton $\rm SU_L(2)$ doublet with hypercharge $\hyp_\ell = -1/2$.
These interaction terms lead to threshold matching contributions in the Higgs potential\footnote{The threshold matching contributions
are obtained by calculating in dimensional regularization (DR) in $d = 4 - 2 \, \epsilon$ dimensions using $\rm \overline{MS}$ subtraction, and Taylor
expanding the resulting amplitudes in the limit $v^2/m_i^2 < 1$.}
\bea
\Delta V(H^\dagger H) \simeq H^\dagger H \left(\frac{3 \, |\kappa_v|^2 \, m_v^2 \ N_v}{16 \, \pi^2} + \frac{|\kappa_s|^2 \, m_s^2 \, N_s}{16 \, \pi^2}
- \frac{|\kappa_F|^2 \, m_F^2 \, N_F}{16 \, \pi^2} \right) + \cdots,
\eea
where $m_i$ is the mass of the corresponding field, and $N_i$ can result from the sum over
the degrees of freedom in an internal (flavour) symmetry group of the field $i$.
When avoiding tuning the bare Higgs mass in the classical Lagrangian against quantum corrections,\footnote{Tuning parameters at the Lagrangian level to avoid these conclusions can be understood to be
best avoided in the following way. As Lagrangian parameters can always be related to measured quantities through
$S$ matrix elements, and eliminated in relationships between $S$ matrix elements, parameter tuning can be understood to be
equivalent to assuming precise relationships between a series of independent measured quantities, in order to
have a further $S$ matrix element take on a value not expected by naive dimensional analysis.} then it follows that $m_h \sim |\kappa_i| \, m_i \sqrt{N_i}/4 \,\pi$.
This is the reason the mass of the SM Higgs is expected to be proximate (up to a loop factor)
to beyond the SM mass scales.

One can turn the relation between $m_h$ and $m_i$ around. Then corrections to cross sections that are probed through
a measurement exploiting a propagating SM state (that goes on-shell) scale as
\bea\label{crosssectionpert}
\frac{\sigma_{SM + i}}{\sigma_{SM}}\simeq \frac{1}{16 \, \pi^2}\, \left(\frac{N_i^2 \, |\kappa_i|^2 \kappa'}{g_{SM} \, \lambda} \right) + \cdots
\eea
Here we have introduced another coupling between the new physics state and the SM ($\kappa'$) and Taylor expanded out the
non-analytic structure of the tree level propagating state $i$. The non-analytic structure of the propagating SM state is essentially unchanged in this limit
and cancels out in the ratio. $g_{SM}$ corresponds to
a generic SM coupling. This result argues that $\sim \%$ level deviations in Higgs properties can occur
in scenarios that avoid parameter tuning. This estimate is subject to the following qualifications:
\begin{itemize}
\item{Differences between coupling constants can lead to further enhancements/suppressions.}
\item{This estimate implicitly assumed one local contact operator was introduced (at tree level)
interfering with the SM. It has been proven \cite{Jiang:2016czg} that when considering tree level
effects, subject to the condition that
a flavour symmetry is not explicitly broken and the UV scales have a dynamical origin, multiple operators
are always present.}
\item{If the state $i$ does not lead to any corrections to $\sigma_{SM + i}$ at tree level, then a further
$\sim 1/16 \pi^2$ suppression occurs. On the other hand, when considering one loop effects,
the multiplicity of operators present is generically very large due to one loop mixing, see Section \ref{onelooprunning} for more detail.\footnote{In some exceptional cases,
$\mathcal{L}_6$ operators do not mix. This can be understood at an operator level using operator weights and related helicity and unitarity
arguments \cite{Jenkins:2013sda,Alonso:2014rga,Cheung:2015aba}.}}
\end{itemize}
 This rough and schematic understanding is nevertheless validated in
exact calculations in some popular new physics models, see Refs.~\cite{Gupta:2012mi,Dawson:2013bba,deFlorian:2016spz}
for more discussion.

\subsection{Higgs substructure}\label{substructure}
EFTs capture the Taylor expanded effects of particle exchange
at tree and loop level, but
also encode multi-pole expansions of underlying structure \cite{weinberglectures,Georgi:1994qn,Jenkins:2013fya,trottmichigan}
that are generic in field configurations set up
by charge distributions that are spatially separated.
The classic example is the multi-pole expansion of an electrostatic charge distribution
\cite{multipole,Jackson:1998nia}. This physics is not directly or trivially identified
in general with tree or loop level particle exchange diagrams
in the presence of non-perturbative bound states (such as a composite Higgs), and offers further hopes for perturbations of
Higgs properties that could be experimentally measured.

When an EFT is capturing the consistent low energy limit of a
strongly interacting light composite Higgs, the multi-pole
expansion should be considered \cite{Jenkins:2013fya,trottmichigan}.\footnote{This point has recently been re-emphasized in Ref.~\cite{Liu:2016idz}.}
The possibility that the Higgs field is composite
underlies a significant fraction of the interest in the EFT methods discussed
in this review.  Composite Higgs models are still of experimental interest. These ideas emerged in the 80's
in Refs.~\cite{Dimopoulos:1981xc,Kaplan:1983fs,Kaplan:1983sm,Georgi:1984ef,Georgi:1984af,Dugan:1984hq}
with early studies also exploring the possibility of dynamical mass generation
in extra dimensional scenarios \cite{Hosotani:1983xw,Manton:1979kb}.
These ideas re-emerged and were extended in the late 90's
in the context of Little Higgs constructions \cite{ArkaniHamed:2002qx,ArkaniHamed:2002qy},
and extra dimensional models aimed at dynamical mass generation
\cite{Antoniadis:2001cv,Hall:2001zb,Scrucca:2003ra,Scrucca:2003ut}
with the Holographic composite Higgs models consolidating many
of these developments in Refs.~\cite{Contino:2003ve,Agashe:2003zs}.

In analyzing scattering off of non-perturbative bound states, leading to a multi-pole expansion, perturbative methods fail by definition. One can gain some intuition
on how this scattering is represented in an EFT by considering
the solutions to the time independent Schr\"{o}dinger equation of a non-local potential of a fixed target,
represented as $V(\bf{r},\bf{r'})$. Such a potential
can mimic the extended nature of the composite particle.
With appropriate boundary conditions this scattering is described by the Lippmann–Schwinger equation \cite{PhysRev.79.469};
outgoing wavefunctions are related to those incoming by a partial wave transition matrix that
satisfies the integral equation
\bea
T_\ell({\bf{k}},{\bf{k}}';E) = V_\ell({\bf{k}},{\bf{k}}') + \frac{2}{\pi} \int_0^\infty d |{\bf{q}}| \, q^2 \frac{V_\ell({\bf{k}}',{\bf{q}}) T_\ell({\bf{q}},{\bf{k}};E)}{E - q^2/\mu + i \epsilon}.
\eea
Here ${\bf{k}},{\bf{k}}'$ are the Fourier momentum of the radial coordinate $\bf{r},\bf{r'}$ and $\mu$ is the reduced mass.
The asymptotic scattering is described by a partial wave scattering matrix: $S_\ell(k) = 1 + 2 i k f_\ell(k)$,
which can be parameterized by a partial wave phase shift $S_\ell(k) = e^{2 i \delta_\ell(k)}$, for spherically symmetric potentials.
The multi-pole expansion in this case is the fact that the phase shift parameter
characterizing the scattering matrix has a power series expansion in $k^2$.
For the partial wave $\ell = 0$, this expansion is given as
\bea
k \cot \delta_0(k) = -\frac{1}{a_0} + \frac{1}{2}\, r_0 \, k^2 - C_2 r_0^3 \, k^4 + \cdots
\eea
There is an expansion in derivatives acting on a field $F$ of the state associated with
asymptotic wavefunction scattering
off of the fixed target field $\mathcal{T}$ as a series of interactions of the form
$\sim \{\mathcal{T} \,  F \, F$, $\mathcal{T} \,  F \nabla^2 F $,
$\mathcal{T} \,  F \nabla^4 F\}$. The effective range expansion in the parameters
$\{a_0,r_0,C_2 r_0^3\}$ are analogous to Wilson coefficients in the relativistic EFT.
The bound state substructure generates a series of
scales that characterize the multi-pole expansion. This is generic
in EFTs describing the scattering off of bound states. See Refs.~\cite{Caswell:1985ui,Luke:1996hj} for
related discussion on non-relativistic bound states in EFTs.

\begin{figure}[t]
    \centering
\parbox{0.495\textwidth}{\includegraphics[width=0.55\textwidth]{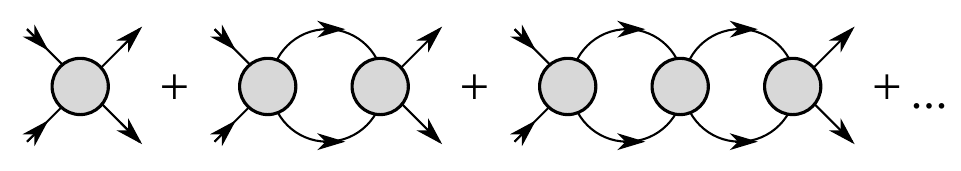}}
\parbox{0.3\textwidth}{\includegraphics[width=0.25\textwidth]{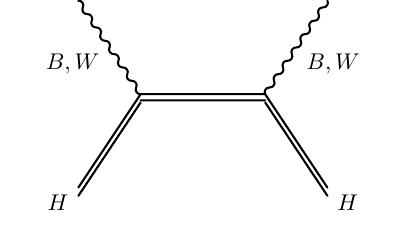}}
\caption{\label{fig:multi-pole} Diagrams relevant for the multi-pole expansion. The left figure illustrates the infinite
sum of bubble diagrams in a ``ladder approximation'' in the field theory equivalent of the non-relativistic Schr\"{o}dinger equation, while
the right diagram illustrates the scattering of the $\rm SU_L(2) \times U_Y(1)$ gauge fields off of the constituent
particles of a composite Higgs.}
\end{figure}

In nucleon-nucleon ($NN$) EFT \cite{Weinberg:1990rz,Kaplan:1998tg,Kaplan:1998we} an analogy
to the non-relativistic scattering case is extensive.
The time independent Schr\"{o}dinger equation corresponds to a
summation of an infinite set of Feynman diagrams in the EFT, defining a scattering amplitude $\mathcal{A}$
as shown in Fig.~\ref{fig:multi-pole}.
The Born series expansion of the
related Lippmann–Schwinger equation can be mapped to an infinite sum of ladder diagrams
that depends on the particles exchanged in the EFT and the kinematics of the poles dominating the
convolution integrals between the nucleon potential and the non-relativistic propagators.
The amplitude defined by this Born series is related to the phase shift \cite{Kaplan:1998tg}
\bea
|{\bf{p}}| \, \cot \delta({\bf{p}}) = i \, |{\bf{p}}| + \frac{4 \, \pi}{M} \frac{1}{\mathcal{A}}.
\eea
Here ${\bf{p}}$ is the three momentum of the $NN$ system and $M$ is the mass scale of the nucleon.
$\mathcal{A}$ is a scattering amplitude that corresponds to the infinite sum of ladder diagrams.
Again a power series expansion of the phase shift leads to the multi-pole expansion. The parameters
characterizing the effective range expansion of nucleon scattering take on values that differ by an order of magnitude \cite{Kaplan:1998tg}
and depends in a non-trivial manner on the spectrum of states retained in the EFT as the bound state is near threshold.

For a composite Higgs,
one can consider the Higgs to be analogous to the nucleon, and the convolution integral of the Higgs self interaction potential
to be made with Higgs propagators and beyond the SM states involved in EW symmetry breaking,
such as vector resonances analogous to the $\rho$ meson in chiral perturbation theory.
Considering current LHC data, there is little motivation to assume that such a $\rho$ state is accidentally lighter
than the remaining states in the strong sector. Scattering results developed in analogy to
$NN$ scattering EFT then lead to a multi-pole expansion in derivative operators involving the Higgs field.

In addition, when considering the multi-pole expansion in terms of the $\rm SU_L(2) \times U_Y(1)$ gauge fields, scattering off
the bound constituents that make up the composite Higgs can occur. This is the case
if the constituents are charged
under the SM gauge groups, or a larger group that contains the SM as a subgroup as illustrated in
Fig.\ref{fig:multi-pole} (right).\footnote{We restrict our attention
to the EW interactions due to the expectation that new physics underlying Higgs compositeness would be associated
with EW symmetry breaking.}

A composite Higgs has an associated multi-pole expansion.
Unfortunately, using crossing symmetry (i.e. rotating Fig.\ref{fig:multi-pole} (right) $90^\circ$ counterclockwise)
to consider the interaction potential as only describing
the composite Higgs state is inconsistent with a non-relativistic EFT approach related to the
Schr\"{o}dinger equation.\footnote{As crossing symmetry relations are between Mandelstam variables constructed out of full four vectors.} Furthermore, the summation of subsets of diagrams in ``ladder approximations''
to the convolution integrals is not valid in general. Noting all of these concerns, the multi-pole expansion
can be associated with
the suppression scale and Wilson coefficients of the $\rm U(3)^5$ symmetric operators\footnote{See Table ~\ref{op59} for the operator definitions corresponding to these Wilson coefficients.}
\bea
\lambda_{Mul}^2 \simeq \{\frac{C_{H \Box}}{\Lambda^2}, \frac{C_{HD}}{\Lambda^2},
\frac{C_{HWB}}{\Lambda^2},\frac{C_{HW}}{\Lambda^2},\frac{C_{HB}}{\Lambda^2} \}.
\eea
When these operators are all constrained so that $\lambda_{Mul} \ll \lambda_h$,
where the Compton wavelength of the Higgs $\lambda_h = \hbar/m_h \, c$, the Higgs boson is effectively interacting as a point-like particle,
when considering these dimension six operators. As we have scaled the operators by $\Lambda$
 associated with particles integrated out of the spectrum, the Wilson coefficients
of the operators involved in the multi-pole expansion can be expected to differ from order one values --
if the scales characterizing the effective range expansion are distinct from the mass scale of the states
integrated out of the theory constructing the EFT.

To summarize: considering the possibility of compositeness and the related multi-pole expansion, the UV sensitivity of Higgs properties, and
the classical nature of the assumed SM EW symmetry breaking potential, a precision Higgs phenomenology program to probe for
indirect hints of physics beyond the SM is well motivated.

\section{Basics of EFT}\label{basics}
A Taylor expansion in dimensionless ratios was used in the previous sections
to simplify the results. That such a simplification can occur
is consistent with the intuitive understanding that IR physics can be calculated without reference to the details of all UV physics.
This is generic in
observables calculated in a Quantum Field Theory (QFT)
{ so long as limited theoretical precision is all that is demanded.}
Manifestly this is true for the SM;
which despite being a QFT that is not well defined to arbitrarily high energies,\footnote{Due to the presence of Landau poles
in the $\rm SU_L(2) \times U_Y(1)$ theory.}
has still been validated to be an adequate description of LHC data considering current experimental precision.

\subsection{Separation of scales, renormalization and local/analytic expansions}\label{scales}
EFT is a set of ideas that justifies why this systematic separation of the physics of different scales
can be true in field theory.\footnote{For excellent reviews on EFT
see Refs.~\cite{Georgi:1994qn,Kaplan:1995uv,Manohar:1996cq,Polchinski:1992ed,Pich:1998xt,Rothstein:2003mp,Skiba:2010xn,Burgess:2007pt}.
The pioneering works developing the modern understanding of EFT include
Refs.~\cite{Coleman:1969sm,Callan:1969sn,Appelquist:1974tg,Witten:1976kx,Weinberg:1978kz,
Gilman:1979bc,Weinberg:1980wa}.}
Renormalization also separates IR and UV physics, but
EFT is more than a statement that QFTs are systematically renormalizable. Furthermore,
the success of renormalization programs in QFTs can be understood as an EFT consequence in an intuitive way.\footnote{The old field theory approach of stressing of the distinction between bare and renormalized parameters is
drawn when correlation functions involving the parameters are considered to be measured or predicted to arbitrary precision.
In this sense, the EFT understanding of renormalization is consistent with such lore.}
When renormalizing a QFT the short distance physics in the theory one calculates in is modified, and the effects of
regularizing such physics is absorbed into the low energy parameters of the effective theory.
How this modification takes place is illustrative of scale separation in EFTs.

Consider calculating an amplitude at one loop
using dimensional regularization in $d = 4 - 2 \, \epsilon$ dimensions \cite{'tHooft:1972fi}.\footnote{We use $\rm \overline{MS}$ subtraction,
by introducing $n$ powers of $\hat{\mu}^{(4-d)/2} = (\mu \, \sqrt{e^{\gamma}/4 \pi})^{(4-d)/2}$ for each power of the coupling
present defining the amplitude, so that the renormalized coupling remains dimensionless. Here $\gamma$ is the Euler-Mascheroni
constant. The arguments in this section are formulated for a one
loop example but they generalize to higher loop orders directly.}
The amplitude can be expressed by using the master formula for Minkowski space momentum integrals
\bea
M_I = \int \frac{d^d q}{(2 \, \pi)^d (\hat{\mu}^2)^{n \,(d-4)/4}} \, \frac{(q^2)^\alpha}{(q^2 - \Delta^2)^\beta}
= \frac{i \,(-1)^{\alpha-\beta}}{(\Delta^2)^{\beta - \alpha - d/2}} \,
\frac{\Gamma(\alpha + d/2)\Gamma(\beta - \alpha - d/2)}{\Gamma(d/2)\Gamma(\beta)},
\eea
with a four momentum $q^\mu$ and a factor $\Delta$ introduced for the characteristic scales (masses, kinematic invariants) in the amplitude.
Following the discussion of Georgi \cite{Georgi:1994qn}, consider integrating over the $\epsilon$ momentum space of such an amplitude
after Wick rotating to Euclidean momentum space and factorizing the loop momentum as $q^2 = q^2_\epsilon + q^2_E$. Restricting one's attention to
divergent terms one finds \cite{Georgi:1994qn}
\bea\label{regularizationresult}
M_I \propto  \int \frac{d^4 q_E}{(2 \, \pi)^4}
\frac{(q_E^2)^\alpha}{(q_E^2 + \Delta^2)^\beta}
\left[\frac{\Gamma(\beta +\epsilon)}{\Gamma(\beta)}\right]
\left[\frac{4 \, \pi \mu^2}{q_E^2 + \Delta^2}\right]^{\epsilon}.
\eea
The last two factors in square brackets are both finite as $\epsilon \rightarrow 0$, but there is an important difference between them.
The final term does not significantly change the amplitude so long as all the scales are similar $\mu^2 \sim \Delta^2 \sim q^2_E$.
On the other hand, this term leads to a modification (an introduced regularization) of the amplitude that can become significant for small $\epsilon$
if $\mu^2 \gg  \Delta^2 + q_E^2$ or $\mu^2 \ll \Delta^2 + q_E^2$,
i.e. when highly separated scales are present in the amplitude. In this manner, the universal subtractions present in systematically renormalizing a
QFT are understood to correspond to UV physics effects that have been systematically removed out of the lower
energy theory by such a regularization for $\mu^2 \ll \Delta^2 + q_E^2$. That such a separation of IR and UV physics can occur
is the key idea of EFT and this can be understood to be an underlying reason for renormalization to work. The case
$\mu^2 \gg  \Delta^2 + q_E^2$ has a different meaning, it corresponds to an IR divergence, and we discuss this divergence below.

Counterterms are of a simple universal analytic form when using DR.
This is also the case in other regularization schemes, such as schemes with dimensionful regulators
that directly satisfy the decoupling theorem \cite{Appelquist:1974tg}.
One might doubt if the regularization of divergences due to arbitrary UV physics sectors can be subtracted out
of a prediction of a lower energy observable
in this simple manner. Formally, this can be understood to follow from a proof supplied by Bogoliubov
and Parasiuk\footnote{We thank F. Herzog for this reference.} on the analytic nature of counterterms in 1957
\cite{Bogoliubov:1957gp}. Renormalization Group (RG) based arguments also support this understanding,
as demonstrated by Polchinski in Ref.~\cite{Polchinski:1983gv}, as do the diagrammatic
arguments of Weinberg's power counting theorem \cite{Weinberg:1959nj}.
Recently the
formal proof of the renormalizability in effective field theories has also been studied with
increased mathematical rigor in Ref.~\cite{costellorenormalization}.

A less formal and more intuitive understanding of the universal nature of the subtractions follows from considering
the constraints of the global symmetries in the EFT, Lorentz invariance,
and the fact that the non-analytic structure of correlation functions\footnote{Here we refer to the poles and cuts in
the momentum space of the spectral function defined in analogy to the K\"{a}ll\'{e}n-Lehmann \cite{Kallen:1952zz,Lehmann1954} two point spectral function.}
is only generated when intermediate states propagate on-shell. When subtracting
the effects of UV physics acting to regularize divergences in the full theory systematically out of a lower energy amplitude, far below the characteristic mass scales of such UV states,
these states are off-shell. The correlation functions can be simplified by Taylor expanding in
the ratio of the separated scales, and are well approximated by the first few  analytic terms in the expansion.
Any divergence thereby has an analytic form. The locality of the subtractions is because
off-shell exchange of the virtual particles (of mass $\sim M$) occurs, but it is
local as it is limited
to short times and distances by the uncertainty principle \cite{Burgess:2007pt}
\bea\label{uncertainty}
\Delta t \,  \Delta E  \sim  \Delta t \, M  > 1 \rightarrow  \Delta t \sim \frac{1}{M},  \quad \quad
\Delta |x| \, \Delta |p|   \sim \Delta |x| \,  M >1 \rightarrow \Delta |x| \sim \frac{1}{M}.
\eea
Here we are using units where $\hbar = 1 = c$.\footnote{Unless otherwise noted we use such ``God-given" units
in this review.}
Renormalization understood in this manner does not draw any fundamental distinction
between theories with only interaction terms limited to mass dimension $d \leq 4$ and
EFTs with a tower of higher dimensional operators. The renormalizability is understood to be
possible due to the fact that the only way that high energy physics integrated out of the low
energy theory can modify the lower energy construction is through a tower of local analytic operators.
In both cases the renormalizability of the theories follows from the separation of scales
that allows the Taylor expansion.

This reasoning also holds for the non-divergent
contributions of UV physics approximated in an EFT by expanding in a ratio of scales.
As a result, an EFT is a field theory with a tower
of local analytic operators of dimension $d$ divided by $d-4$ powers of
a suppression scale characteristic of the UV physics removed from the EFT construction.\footnote{In some
exceptional cases EFTs can be constructed with non-local operators. This is usually due to distinguishing
field excitations as retained or removed from the EFT by assigning a four momentum of a particle excitation of a field (not the $p^2$ Lorentz invariant
used to distinguish on or off-shell) some scaling rules.
See Refs.~\cite{Isgur:1988gb,Isgur:1989vq,Mannel:1994pm,Brambilla:1999xf,Bauer:2000ew,Bauer:2000yr} for famous examples.}
A well constructed EFT is designed to capture the relevant low energy physics to predict a set of experimental measurements,
while exploiting the simplifications that result from such expansions as soon as possible.
The essential and key idea is to
separate the description of the processes under study
into IR (i.e. infrared or long distance) propagating states and their interactions, captured by
the local and analytic operator expansion, and the UV dependent
short distance Wilson coefficients, i.e. construct
\bea
\mathcal{L}_{EFT} \simeq \sum_i C_i^{UV}(\mu) \, O_i^{IR}(\mu).
\eea
Taking this reasoning to its logical conclusion
gives a commonly held set of ``prime directives'' of effective field theorists:\footnote{We acknowledge M. Luke for
this nomenclature.}
\begin{itemize}
\item{Isolate and separate a series of characteristic scales in observables.}
\item{Construct the Lagrangian of the EFT only out of the degrees of freedom that
lead to non-trivial structure
in the correlation functions. These are propagating on-shell states
(i.e. with $p^2 \sim m^2$), ideally with only one scale defining the EFT closely related to the scales
identified in the previous step. In short, expand ASAP, at the Lagrangian level.}
\item{Calculate in the EFT without unnecessary reference to the UV physics that is  is decoupled.
Some UV dependence is present, but it is sequestered in the EFT into the short distance Wilson coefficients in the matching procedure.
In contrast, in the EFT the operators encode the IR physics describing long distance propagating
states.}
\item{Use a mass independent renormalization scheme, such as dimensional regularization.}
\end{itemize}

The last two points are related to the requirement of matching and some technical consequences that
result from a renormalization and subtraction scheme choice that we discuss in more detail in the following sections.

\begin{figure}[t]\centering
 \includegraphics[width=.9\textwidth]{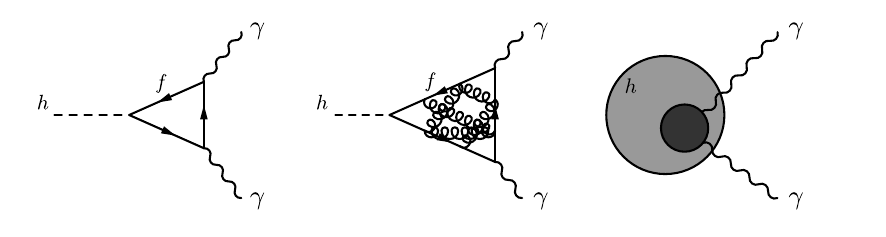}
 \caption{Illustration of various UV physics scenarios captured in the SMEFT for $h \rightarrow \gamma \gamma$.
 The leftmost figure illustrates perturbative mediators (a fermion $f$) leading to the decay $h \rightarrow \gamma \gamma$.
 The middle figure illustrates this decay being mediated by UV states in a strongly interacting field theory
 where the diagram sum does not converge.
 The rightmost figure illustrates the possibility of Higgs substructure leading to the SMEFT multi-pole expansion
 with the outer circle indicating the Compton wavelength of the Higgs $\lambda_h$ and the shaded interior region a
 substructure scale characterized by $\lambda_{Mul}$.}\label{substructure_1}
\end{figure}

The physics that can be captured in the SMEFT in this manner as a consistent IR limit is
only limited by the assumptions that $ \Lambda > v$, and the existence of a Higgs doublet in the EFT construction.
Various cases of beyond the SM physics are illustrated in Fig.~\ref{substructure_1}.
When this strict separation of scales is maintained, i.e. the SMEFT is treated as a general EFT retaining all of the operators that are
allowed by the assumed symmetries, powerful model independent conclusions can result.
All of the cases in Fig.~\ref{substructure_1}, and combinations of such cases in possible UV physics sectors,
have to project onto a series of local and analytic operators with various Wilson coefficients
in the EFT expansion. This point holds {even when the UV physics cannot be calculated with known field theory techniques}.

If this separation of scales is violated, then
the resulting statements and analysis, even if constructed and framed in EFT language,
are not EFT conclusions. Such conclusions can be model dependent or simply ill-defined.
This issue is very well known in research areas where EFT techniques have been dominant for decades,
but is less appreciated when applying EFT techniques to characterize and constrain new physics at LHC,
which is a continual source of debates in the literature. The key problem that can be introduced when
going beyond EFT is the introduction of an assumption heavy hypothetical UV physics sector,
which can result in the lack of a consistent IR limit, rendering the EFT framework used inconsistent
and without meaning. To avoid this problem the key requirements that UV assumptions must address are
\begin{itemize}
\item{The IR limit of a UV theory must be well defined,
which requires that the UV theory is written down.
In particular, the origin of the scales in a UV theory should be specified
to have a possibility of a meaningful IR limit.}
\item{If a strong interaction is present in a UV completion, and the mass
scale characterizing bound states is $\Lambda \gg v$, then non-perturbative contributions can exist (see Fig.\ref{substructure_1} right)
and should not be assumed to vanish without a precise justification. Assuming
that such non-perturbative effects are absent, or negligible, in the EFT projects into a strong, and at times undefined, condition on
UV completions that can generate the EFT. Again we emphasize that one of the core strengths of the standard approach to EFT
is the ability to characterize and constrain such physics rendering UV assumptions
on strongly coupled physics completely avoidable.}
\end{itemize}

\subsection{The decoupling theorem}\label{decouplingtheorem}
The previous section outlines the basic intuition underlying
EFT methods that is formalized in the Appelquist-Carazzone decoupling theorem \cite{Appelquist:1974tg}
(see also Symanzik \cite{Symanzik:1973vg}).
This theorem played an important role in the emergence of EFT methods in the 1970s.
Examining this result in detail shows how renormalization scheme choice is also a technical issue
of some importance when calculating in EFTs, as in the SM.
The decoupling theorem is developed studying a set of massless gauge fields, denoted $A_\mu$
(and referred to as ``vector mesons'' at times in Ref.~\cite{Appelquist:1974tg}),
that are coupled to a set of massive fermions, denoted $\Psi$. The Lagrangian is
\bea\label{nondc}
\mathcal{L}_{dc} = - \frac{1}{4} F_{\mu \nu}^a F_a^{\mu \nu} + \bar{\Psi} i \, \slashed{D} \Psi - \bar{\Psi} m \Psi - \delta m \bar{\Psi}\Psi,
\eea
where
\bea
F^a_{\mu \nu} &=& \partial_\mu \, A_\nu^a - \partial_\nu \, A_\mu^a - g f^{abc} A_{b, \mu} A_{c, \nu}, \\
(D_\mu \Psi)_n &=& \partial_\mu \Psi_n + i [T_{a}\, A_\mu^a]_n,
\eea
and finally $\left[T_a,T_b\right] = i f_{abc} T^a$ defines the Lie algebra of the gauge group, with coupling $g$.\footnote{
We have modified some notational conventions compared to Ref.~\cite{Appelquist:1974tg} to maintain a common notation throughout the review.}
Here $\delta m$ explicitly denotes the mass counterterm. The decoupling theorem is stated as \cite{Appelquist:1974tg}:
\\
\\
{\it{For any 1PI Feynman graph with external vector mesons only but containing internal fermions,
when all external momenta (i.e. $p^2$) are small relative to $m^2$, then apart from coupling constant and field strength renormalization the graph
will be suppressed by some power of $m$ relative to a graph with the same number of external vector mesons
but no internal fermions.}\footnote{Here the exact wording of  Ref.~\cite{Appelquist:1974tg} is edited for clarity.}}
\\
\\
Removing the field whose quantum is a heavy particle from the Lagrangian used to calculate experimental observables, based on the
decoupling theorem, is known as ``integrating out'' a particle from the theory.
The decoupling theorem is stated and proven
for the specific field theory in Eq.~\ref{nondc}, but the arguments used to prove it generalize directly to other theories.\footnote{
For example, for a detailed discussion and proof on decoupling for scalar field theory with two fields
when $d=6$ see Collins \cite{collins_1984}.}
The generalization of this result to arbitrary field theories can be given in terms of all n-point Green's functions $G^{n}$ as
\bea
\prod_i^N Z_i G_{full}^{n}(p_1,p_2 \cdots p_n; \mu) = \prod_j^M Z_j \, G_{EFT}^{n}(p_1,p_2 \cdots p_n; \mu) + \frac{1}{m^2}
\prod_i^k Z_i \, G_{EFT}^{' n}(p_1,p_2 \cdots p_n; \mu)+ \cdots
\eea
where $Z_{i..N}$ is the set of renormalizations required to render the full theory finite,
$Z_{i..M}$ is the set of renormalizations required to render the leading $d \le 4$ terms in the effective theory finite
in an on-shell scheme. $Z_{i..M..k}$ includes these renormalizations and the additional renormalizations required to also render the local
contact operators suppressed by $1/m^2$ finite. This theorem is formally establishing that if the intermediate heavy fields
do not go on-shell (i.e never satisfy $p^2 \simeq m^2$) they modify the leading local analytic operator structures through renormalization
in the lower energy Lagrangian, or add additional interactions suppressed
by powers of $1/m$. This is expected considering the schematic arguments of the previous section.

The proof of the decoupling theorem is non-trivial. The statement is for {\it any 1PI graph},
i.e. can be an arbitrarily high order in perturbation theory. Due to this, the renormalization
scheme chosen has an important impact on the arguments required to establish the proof.
In Ref.~\cite{Appelquist:1974tg} the scheme used defines $\delta m$ to fix the fermion self energy to vanish
at $\slashed{k} = m$. The remaining counterterms are subtractions defined at off-shell Euclidean momentum subtraction points ($p^2 = - \mu^2$).
Wavefunction renormalization conditions fix the propagator to have its tree level form.
Using this scheme Ref.~\cite{Appelquist:1974tg} considered divergent and finite terms in arbitrary
1PI Feynman graphs and established the decoupling theorem exhaustively. Careful attention is paid
in the proof to ensure that a well defined IR limit ($p^2/m^2 <1$) is under consideration, by examining
IR safe observables consistent with the KLN theorem \cite{Kinoshita:1962ur,Lee:1964is}. Equally important
is a careful consideration of sub-divergences (that are sensitive to the
regularization scheme used) in establishing the theorem.
An implicit dependence on an off-shell subtraction scheme is present in the decoupling theorem.

One of the
``prime directives'' of EFT is a direct consequence of the decoupling theorem: calculate in the EFT without unnecessary reference to the UV physics.
This is required as the UV physics is decoupled and simply removed from an EFT in a controlled fashion.
Its IR effects are reproduced to a limited precision and encoded in the matching procedure
in the Wilson coefficients of the EFT.

\subsection{Non-decoupling physics}

The decoupling theorem has some exceptions. This should be surprising considering the generality of the arguments that have been advanced in the previous sections.
Calculating in field theories to an approximate precision,
in the presence of separated scales, is usefully thought of using the techniques of EFT. Such EFTs
are constructed based on the separation of scales that underlies decoupling.
However, no theorem can escape the constraints of its exact wording and assumptions, and this is also true for the decoupling theorem.
Several examples of ``non-decoupling" effects are discussed in the literature.
Heavy physics of this form does not imply that an
EFT is impossible to construct to capture an IR limit of some UV physics. It just enforces the construction of the EFT to take on a particular form,
usually by requiring that a non-linear representation of a symmetry be used.\footnote{Again the existence of the SMEFT and the HEFT can be understood to be related
to this fact, as non-decoupling effects in the scalar sector are a possibility.}

\subsubsection{The \titlemath{$\rho$}{rho} parameter}

Non-decoupling effects can occur when heavy states and the light states are
embedded in the same representations of a symmetry group in the full theory.
Divergences can be forbidden by the linearly realized symmetry,
 due to cancellation between the particles of different masses embedded in such a (softly broken) representation
of a symmetry group. When
 some of the states are no longer in the spectrum in the EFT, the counterterms are no longer forbidden
 by the linearly realized symmetry.
 Then perturbative corrections can grow with the mass of the state removed from the theory.

\begin{figure}[t]\centering
 \includegraphics[width=.9\textwidth]{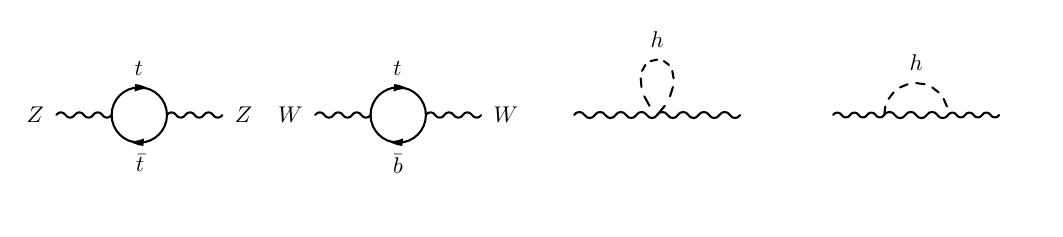}
 \caption{SM one loop corrections to the $\rho$ parameter.}\label{rhocorrections}
\end{figure}

 The practical signal of this physics can be the appearance of numerically
 larger perturbative corrections when the heavy state is still retained in the theory,
 and at times an additional mass dependence outside of logarithms in such corrections. This can be the case as in the loop corrections the masses
sometimes act to regulate the divergences when the symmetry is linearly realized.
Several historical examples of this form of non-decoupling are present in the literature, in $\nu e$ scattering \cite{Adler:1969gk},
 in large $\mathcal{O}(\alpha_s)$ corrections (due to quark doublet mass splittings)
 to the axial neutral current \cite{Collins:1978wz} and in the behavior of one loop corrections
 \cite{Marciano:1974vp,Toussaint:1978zm,Veltman:1976rt,Veltman:1977kh}
 to the ratio of charged and neutral currents in the SM, due to the diagrams shown in Fig.~\ref{rhocorrections}.

We discuss this latter case of the $\rho$ parameter \cite{Ross:1975fq}, defined as the ratio of charged and neutral currents at low energies,
as an example. The $\rho$ parameter has the perturbative expansion, with one loop contributions shown in Fig.\ref{rhocorrections}
(in $\rm \overline{MS}$) which give
\bea
\rho \simeq \frac{\bar{g}_Z^2 \, \bar{m}_W^2}{\bar{g}_2^2 \, \bar{m}_Z^2}  + \frac{N_c \, \hat{G}_F}{8 \, \pi^2 \sqrt{2}}
\left(m_t^2 + m_b^2 - 2 \frac{m_t^2 \, m_b^2}{m_t^2 - m_b^2} \log \left(\frac{m_t^2}{m_b^2} \right) \right) - \frac{11 \hat{G}_F \, \hat{M}_Z^2
s_{\bar \theta}^2}{24 \sqrt{2} \pi^2} \log \left(\frac{m_h^2}{m_Z^2} \right).
\eea
The Higgs mass dependent correction is not exceedingly large and it is not related in mass to
another particle in the spectrum by a linear realization of a symmetry. The limit $m_h \rightarrow \infty$ can be taken,
which still leads to a non-linear realization of $\rm SU_L(2) \times U_Y(1)$ as the Higgs field and its vacuum expectation
value are related when this symmetry is linearly realized. The effective theory construction
of Refs~\cite{Appelquist:1980vg,Longhitano:1980iz,Longhitano:1980tm,Appelquist:1993ka,Feruglio:1992wf} results when
the limit $m_h \rightarrow \infty$ is taken.
The corrections due to the heavy Higgs matched onto this EFT are not suppressed by explicit powers of $m_h^2$ in their leading contributions.
The full results of this form are given in Refs.~\cite{Herrero:1994iu,Dittmaier:1995ee}.
This is an example of non-decoupling effects that deviate from a naive expectation formed from the decoupling theorem.

Even larger corrections come about due to splitting the quark masses in the limit $m_t \gg m_b$. Note that
\bea\label{limiteqn}
m_t^2 + m_b^2 - 2 \frac{m_t^2 \, m_b^2}{m_t^2 - m_b^2} \log \left(\frac{m_t^2}{m_b^2}\right) \rightarrow 0,
\eea
in the limit $m_t\rightarrow m_b$. Integrating out the top, while leaving the $b$ quark in the spectrum, leads to a theory
without a linearly realized $\rm SU_L(2)$ symmetry. Furthermore, $m_t$ is acting to regulate an integral,
which is a reason that it appears as a polynomial outside of the logarithm.
As $m_t = y_t \,v/\sqrt{2}$ the limit $m_t \rightarrow \infty$ must correspond to
$v \rightarrow \infty$, $y_t \rightarrow \infty$, or both. The former limit is interesting, as
the corrections given Eq.~\ref{limiteqn}
vanish if $m_t/m_b \rightarrow 1$ as $v \rightarrow \infty$.
Then
$\rm SU_L(2)$ can again be linearly realized. The limit $y_t \gg 1$
is a strong coupling limit, leading to a breakdown of perturbation theory. Then the
decoupling theorem's implicit assumption of a valid perturbation theory no longer holds.
In the case of the $\rho$ parameter, the non-decoupling effects come about due to this strong coupling limit
in addition to the differences in the realization of the symmetries of the full theory
and the low energy EFT.\footnote{In the case of the $\rho$ parameter
the corrections shown are also the leading violations of custodial symmetry, as an additional subtlety.}

\subsubsection{Weak interactions}
When
exact symmetries are present in a subset of interactions in the EFT, such symmetries can first be broken
explicitly by the heavy fields integrated out. Then the leading operator mediating a process can be due to the local contact
operator correction to the EFT suppressed by $m^2$ (in the case of weak interactions a suppression by $m_W^2$),
but with no {\it relative} suppression compared to any leading order effect, which is absent. This is another way
in which non-decoupling effects can come about.

The weak interactions are an important example of this form of non-decoupling. The SM is defined in Section \ref{SMsection}. Flavour violating effects
in the SM due to the weak interactions having an intricate pattern that encodes non-decoupling physics of this form.
In the limit that the Yukawa interactions of the SM vanish, $Y_{u,d,e} \rightarrow 0$
a ${\rm U}(3)^5$ global flavour symmetry group of the SM is present.
We define this group through the relation between the weak (unprimed) basis and the mass (primed) basis as
\begin{align}\label{weakmassrotations}
u_L &= \mathcal{U}(u,L)\, u_L^\prime, & u_R &= \mathcal{U}(u,R)\, u_R^\prime,
& \nu_L &= \mathcal{U}(\nu,L)\, \nu_L^\prime, \\
d_L &= \mathcal{U}(d,L)\, d_L^\prime, & d_R &= \mathcal{U}(d,R)\, d_R^\prime, &
e_L &= \mathcal{U}(e,L)\, e_L^\prime, & e_R &= \mathcal{U}(e,R)\, e_R^\prime.
\end{align}
Each $\mathcal{U}$ rotation defines a ${\rm U}(3)$ flavour group. The ${\rm U}(3)^5$ group of the SM is defined as
\bea
{\rm U}(3)^5 = \mathcal{U}(u,R) \times \mathcal{U}(d,R) \times \mathcal{U}(Q,L)\times \mathcal{U}(\ell,L) \times \mathcal{U}(e,R).
\eea
The relative $\mathcal{U}$ rotations between components of the lepton and quark ${\rm SU}_L(2)$ doublet fields define the PMNS
and CKM matrices as
\bea
V_{\rm CKM} = \mathcal{U}(u,L)^\dagger \, \mathcal{U}(d,L), \quad \quad U_{\rm PMNS} = \mathcal{U}(e,L)^\dagger \, \mathcal{U}(\nu,L).
\eea
Consistent with the discussion in the previous section, the unbroken ${\rm U}(3)^5$ flavour symmetry of the SM forbids
divergences corresponding to flavour violating interactions.

The $\mathcal{U}(u,R) \times \mathcal{U}(d,R) \times \mathcal{U}(e,R)$ rotations commute
with the weak interaction generators. At tree level the neutral current interactions to the left handed doublet fields ($\psi_L$)
couple to the diagonal generators
\bea
{\bf \hyp_i}\psi_L = y_i \, \left( \begin{array}{cc}
1 & 0 \\
0 & 1
\end{array} \right)\psi_L , \quad
\tau^3 \psi_L = \left( \begin{array}{cc}
1 & 0 \\
0 & -1
\end{array} \right) \psi_L.
\eea
which also commutes with the $\mathcal{U}(Q,L)\times \mathcal{U}(\ell,L)$ rotations between the weak and mass eigenstates. No tree level flavour changing neutral currents
follows.

This symmetry is broken when the
charged currents that interact in the weak eigenbasis of the SM propagate, and quark mass differences are retained
in the resulting amplitudes. This distinguishes the mass and weak eigenstates of the SM.
Flavour violating effects come about due to the relative rotation of the SM states in $\psi_L$
proportional to $V_{\rm CKM},U_{\rm PMNS}$, that distinguishes the components of the
$\psi_L$ doublets, and appear in interactions proportional
to $\tau^{1,2}$ through which the charged currents couple. This leads to flavour changing charged currents at tree level in the SM.
Nevertheless, if all weak or mass eigenstates are summed over, and flavour violating spurions are
neglected, the flavour symmetry is again restored due to the rotation between the eigenbases being unitary.

\begin{figure}[t]\centering
 \includegraphics[width=.9\textwidth]{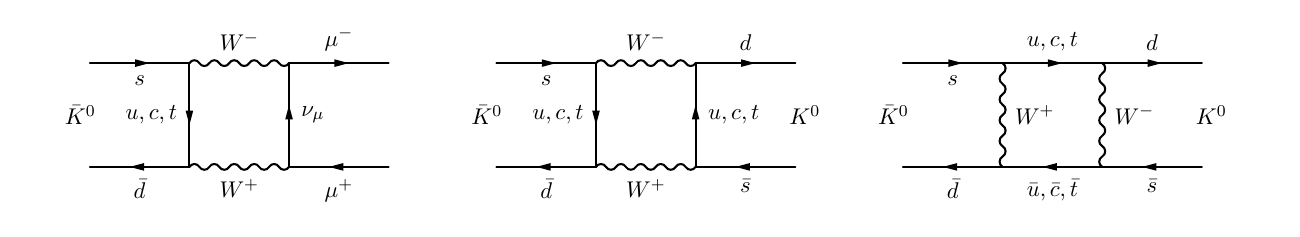}
 \caption{SM one loop corrections to kaon decay and mixing.}\label{kaoncorrections}
\end{figure}

Consider the diagrams in Fig.~\ref{kaoncorrections} as an illustrative example. The leftmost diagram gives a contribution to the decay
$K_L \rightarrow \mu^+ \, \mu^-$.\footnote{The Kaon mesons are defined by their quark content as $K^0 = d \,\bar{s}, \bar{K^0} = \bar{d} \, s$
and $K_L = (d \,\bar{s} - s \bar{d})/\sqrt{2}$, $K_S = (d \,\bar{s} + s \bar{d})/\sqrt{2}$.}
Constructing the EFT useful for the measurement scale $\mu^2 \simeq m_K^2$, both the top quark
 and the $W$ boson are not on-shell fields propagating for longer distances (compared to the measurement scale),
and hence integrated out. The EFT so constructed is the LEFT, briefly introduced in Section \ref{sec:history}.
 The leading operator mediating the decay is given by \cite{Inami:1980fz}
\bea\label{kaon1}
\mathcal{L}_{K_L \rightarrow \mu^+ \, \mu^-} = \frac{\hat{G}_F}{ \sqrt{2}} \frac{\haew}{2 \, \pi \hst^2} \, V_{ts}^\star \, V_{td} Y(x_t) \, \left(\bar{s} \, \gamma^\mu P_L \, d \right)
\left(\bar{\mu} \, \gamma^\mu P_L \, \mu \right) + h.c + \cdots
\eea
Here $x_t = m_t^2/m_W^2$ and retained is the contribution from the top quark in the loop
that breaks flavour symmetry. We neglect higher order (and penguin diagram) contributions to the decay.
See Refs.~\cite{Buchalla:1992zm,Buchalla:1993bv,Buchalla:1993wq,Buchalla:1995vs} for more discussion
on such corrections.

The ``non-decoupling" effects are indicated by the presence of polynomial powers of $m_t^2$ in the numerator, similar to the case of the
$\rho$ parameter.
Again, integrating out the top quark will lead to a non-linearly realized $\rm SU_L(2)$ symmetry,
but the situation is different than in the case of the $\rho$ parameter, where the top mass scale also regulates the integral.
Fig.~\ref{kaoncorrections} (left) is a naively convergent integral (in an appropriately chosen gauge).
The $m_{t,c,u}$  dependence as a polynomial mass contribution outside of a logarithm follows from the flavour breaking pattern of the SM matched onto the lower scale EFT.

The result reflects the Glashow-Iliopoulos-Maiani \cite{Glashow:1970gm}
(GIM) mechanism describing the relevant phenomenological
suppression by powers of the quark masses in addition to the weak couplings as
a result of the $\rm U(3)^5$  symmetry breaking pattern of the SM. The GIM mechanism is
also a statement
that in the limit of
vanishing quark masses, Eq.~\ref{kaon1} (and similar flavour changing amplitudes for other processes)
exactly vanish as due to unitarity
\bea
\sum_i  \, V^\star_{is} \, V_{id} = 0.
\eea

One loop contributions to Kaon mixing also respect the GIM mechanism,
and include the contributions shown in Fig.~\ref{kaoncorrections} (right two diagrams).
Reproducing the SM amplitudes following from the weak interactions at
low energies in an EFT is non-trivial. A proper treatment
summing all logarithms again uses the LEFT by integrating out a series of SM fields $\{h,W,Z,t,c\}$ in sequence,
in the renormalization group evolution down to the experimental scale
of $K^0 -\bar{K^0}$ mixing. For a discussion on the reproduction of the SM result in the LEFT see Ref.~\cite{Manohar:1996cq}.

The sum of these graphs (combined with Goldstone boson diagrams for gauge independence)
gives the leading result \cite{Gilman:1978wm,Gilman:1979bc,Shifman:1975tn,Vainshtein:1975sv}
\bea\label{kaon2}
\mathcal{L}_{K^0 -\bar{K^0}} = \frac{\hat{G}_F}{ \sqrt{2}} \frac{\haew}{4 \, \pi \hst^2} \, \sum_i V_{is}^\star \, V_{id} \,
\sum_j V_{js} \, V_{jd}^\star \,  \bar{E}(x_i,x_j) \, \left(\bar{s} \, \gamma^\mu P_L \, d \right)
\left(\bar{d} \, \gamma^\mu P_L \, s \right) + h.c + \cdots
\eea
where
\bea
\bar{E}(x_i,x_j) &=& - x_i \, x_j \, \left(\frac{1}{x_i - x_j}\left[\frac{1}{4} - \frac{3}{2}\frac{1}{x_i-1}
- \frac{3}{4}\frac{1}{(x_i-1)^2}\right]\log x_i, \right. \nn
&+& \left.\frac{1}{x_j - x_i}\left[\frac{1}{4} - \frac{3}{2}\frac{1}{x_j-1}
- \frac{3}{4}\frac{1}{(x_j-1)^2}\right]\log x_j - \frac{3}{4}\frac{1}{(x_i-1)(x_j-1)}\right)
\eea
Again $x_{i,j} = m_{i,j}^2/m_W^2$ and these indices sum over up quark flavours.
We neglect here higher order corrections, see
Refs.~\cite{Gilman:1978wm,Gilman:1979bc,Shifman:1975tn,Vainshtein:1975sv,Buchalla:1995vs}
for further discussion. The ``non-decoupling"
flavour breaking structure of the SM interactions is present in that the result is proportional to
four powers of quark masses in Eq.~\ref{kaon2}.

The many instances of non-decoupling effects
in the SM should generate caution when choosing between the SMEFT and HEFT formalisms to capture the low energy limit of
physics beyond the SM. The possibility of non-decoupling effects related to the discovered $0^+$ boson
with mass $\sim 125 \,  {\rm GeV}$ is pressing and well motivated. One of the key distinctions between the SMEFT and HEFT constructions
is the latter is arguably more appropriate to capture the IR limit of such non-decoupling UV physics coupled to the $0^+$ boson.

\subsubsection{Renormalization scheme dependence and decoupling}
A renormalization scheme is composed of a method to regulate divergent integrals, and a subtraction
scheme choice. When relationships between physically measured $S$ matrix elements are determined in
perturbation theory, the regularization and subtraction scheme choice has no physical effect.
When discussing renormalized Lagrangian parameters per se and intermediate results for observables in terms of these parameters, scheme dependence is present.
Ref.~\cite{Appelquist:1974tg} used a renormalization scheme which performs subtractions
at an off-shell Euclidean momentum point.
This approach to renormalization has the benefit of making decoupling manifest, which is not the case in dimensional regularization (DR)
when $\rm \overline{MS}$ is used as a subtraction scheme.

\begin{figure}[t]\centering
 \includegraphics[width=.6\textwidth]{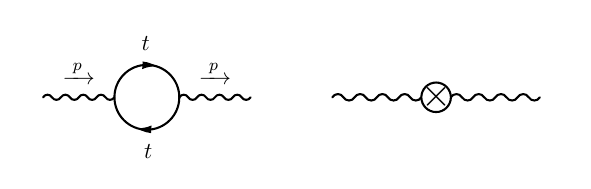}
 \caption{One loop contribution to the running of the QED coupling constant due to a fermion; shown is the case when the fermion is
 the top.}\label{diagrams_2pf2}
\end{figure}

Consider the Lagrangian for quantum electrodynamics (QED), and the running of the QED coupling $e$ due to fermions $\psi_f$.
The Lagrangian is given by
\bea
\mathcal{L}^{0}_{QED} = -\frac{1}{4} \, F^{\mu \, \nu}_{0} \, F_{\mu \, \nu}^{0} + \bar{\psi}_f^{0}\gamma_\mu (i \, \partial^\mu - e_0 \, Q_f \, A^\mu_0)\, \psi_f^0
- m_f^0\, \bar{\psi}_f^{0} \, \psi_f^0,
\eea
with $F^{\mu \, \nu}_{0} =\partial^\mu \, A^\nu_{0}-\partial^\nu \, A^\mu_{0}$ the QED field strength tensor composed of bare fields,
indicated with $0$ labels.
The one loop diagram shown in Fig.~\ref{diagrams_2pf2} (left) gives
\bea\label{dimregerunning}
i \, \mathcal{A} = - e_0^2 \, Q^2 \,\mu^{4-d} \, \int \frac{d^d  q}{(2 \pi)^d} \frac{{\rm Tr}\left(\gamma^\mu (\slashed{q}+ \slashed{p} + m_f)\gamma^{\nu}
(\slashed{q}+ m_f)\right)}{((q+p)^2- m_f^2)(q^2-m_f^2)},
\eea
which is divergent.
The bare fields are related to the renormalized fields (denoted with r labels)
by introducing the renormalization constants $Z_i$
\begin{align}
A^\nu_0 &= \sqrt{Z_A} A^\nu_r, & \psi_f^0 &= \sqrt{Z_{\psi_f}} \psi_f^r, \\
e^0 &= Z_e  \, \mu^\epsilon \,  e^r, & m^0_f &= Z_{m_f}\, m^r_f.
\end{align}
$\mu^\epsilon$ is introduced as dimensional regularization with $d= 4 - 2 \epsilon$ is used to regulate the divergent integrals.
The counterterm that performs the subtraction for $\Pi_{AA}(p^2)$ is indicated in  Fig.~\ref{diagrams_2pf2}  (right) and gives
\bea
-\frac{i}{4} \, Z_A \, (p^\mu \, p^\nu - p^2 g^{\mu \, \nu}).
\eea
Choosing the manner in which the divergence is subtracted fixes $Z_A$
and defines the subtraction scheme.
Defining a renormalization condition for $\Pi_{AA}(p^2)$ where the diagram is subtracted at the Euclidean momentum
point $p^2 = - M_f^2$ removes the divergence, and gives the sum for Fig.~\ref{diagrams_2pf2}
\bea
i \, \mathcal{A}_{M_f} = - i \, \frac{e_0^2 \, Q^2 \, \mu^{2 \epsilon}}{2 \, \pi^2} \, (p^\mu \, p^\nu - p^2 g^{\mu \, \nu}) \int_0^1 dx
\log \frac{m_f^2 - p^2 \, x \, (1-x)}{m_f^2 + M_f^2 \, x \, (1-x)}.
\eea
Alternatively, the $\rm \overline{MS}$ scheme subtracts the $\epsilon$ poles and a set of constant terms
due to the rescaling $ \mu \rightarrow \mu \,(e^\gamma/4 \pi)^{1/2}$.
Using $\rm \overline{MS}$ the sum for Fig.~\ref{diagrams_2pf2} is given by
\bea
i \, \mathcal{A}_{\rm \overline{MS}} = - i \, \frac{e_0^2 \, Q^2}{2 \, \pi^2} \, (p^\mu \, p^\nu - p^2 g^{\mu \, \nu}) \int_0^1 dx
\log \frac{m_f^2 - p^2 \, x \, (1-x)}{\mu^2}.
\eea
In either case, the Ward identities of the theory due to unbroken $\rm U(1)_{em}$ fix
\bea
Z_e = 1/\sqrt{Z_A},
\eea
to order $e^2_0$. The bare coupling is independent of the renormalization scheme choice
so expanding the derivative of the bare coupling with respect to $M_f$ gives
\bea
\beta(e) = \frac{e_0^3 \, Q^2}{2 \, \pi^2} \, \int_0^1 \, dx \, \frac{M_f^2 \, x^2 \, (1-x)^2}{m_f^2 + M_f^2 \, x \, (1-x)}.
\eea
The running of $\beta(e)$ calculated in this manner exhibits manifest decoupling. For $m_f < M_f$
one finds
\bea\label{msrunninge}
\beta(e) \simeq \frac{e_0^3 \, Q^2}{12 \, \pi^2},
\eea
while for $M_f < m_f$ one has
\bea
\beta(e) \simeq \frac{e_0^3 \, Q^2}{60 \, \pi^2} \, \frac{M_f^2}{m_f^2}.
\eea
When renormalizing the theory for measurements
made at scales $\sim - M^2_f < m^2_f$, decoupling of the effects of the heavy fermion is manifest.
For $\rm \overline{MS}$ one finds the result in Eq.~\ref{msrunninge}
for all $\mu$. In this case, to implement decoupling appropriately one must set $\beta(e) \simeq 0$ for $\mu \lesssim m_f$
by hand.\footnote{This point holds for $\beta$ functions, and also for other theoretical quantities
such as the effective potential \cite{Bando:1992np,Bando:1992wy,Casas:1998cf}.}

The effects of heavy particles contributing to an experimental measurement
that do not propagate
on shell are still encoded in local contact operators in the EFT,
no matter what subtraction scheme is chosen.
Connecting back to the initial regularization result in Eq.~\ref{regularizationresult}, large logarithms can be
present expanding this equation when using DR and $\rm \overline{MS}$, if $\mu^2 \ll m_f^2$ (or $\mu^2 \ll \Delta^2$ in the
notation of Section \ref{scales}), indicating the regularization of an amplitude. A poorly behaved perturbative
expansion results if decoupling is not imposed by hand in this
scheme.

Considering the requirement of modifying the beta functions by hand,
it could be surprising that using $\rm \overline{MS}$ and DR is strongly preferred in modern EFT calculations.
This renormalization scheme makes the power counting of the EFT manifest and directly preserved
in loop calculations. This is an important technical simplification that overwhelms the drawback of having to impose decoupling
by hand. See Section \ref{section:dimensionfulreg} for further discussion
on this point.

\subsection{Matching}\label{matchingexamples}
An EFTs dynamics is defined without the need to extensively reference the details of any UV completion. This is fortunate, as
the EFT and the UV completion are quite different.
They do not have the same high energy behavior and each theory is renormalized separately, with a
different set of counterterms.

An EFT is a self-consistent field theory capable of predicting $S$ matrix elements for a range of energies
where the expansion leading to the EFT is convergent.
At times an EFT can be constructed to faithfully reproduce the predictions of
a UV completion in a low energy limit. As the correspondence
between the EFT and the UV theory is limited,
when this is done, the theories must be
matched to ensure the predictions agree. This procedure fixes the
free parameters (the Wilson coefficients) of the tower
of higher dimensional operators that make up $\mathcal{L}_{EFT}$.
When the UV completion is weakly coupled, the matching procedure
can be directly carried out in perturbation theory.
To fix a set of $n$ free parameters in the EFT, a set of $n$ linearly independent
$S$ matrix elements\footnote{We define the $S$ matrix precisely below in Section \ref{Smatrix}.} are calculated in both the EFT and in the UV completion.
The results are equated in the IR limit that defines the EFT. Denote this limit
as $p^2 \ll M^2$ with $M^2$ some heavy mass scale of a state in the UV completion and not in the EFT.
Fixing
\bea\label{matching}
\langle p_1 \cdots p_a|S_{1..n}|k_1 \cdots k_b \rangle^{\rm UV}_{p^2 \ll M^2} \equiv
\langle p_1 \cdots p_a|S_{1..n}|k_1 \cdots k_b \rangle^{\rm EFT},
\eea
so defines the matching conditions that fixes the Wilson coefficients in terms of the
parameters of the UV completion, and the couplings of the perturbative
expansion.
This procedure works at tree level, and order by order in perturbation theory where both UV and IR divergences
can occur.

The divergences can be neglected in practice and the matching condition
is defined by the finite parts of Eq.~\ref{matching}.
This follows from the UV divergences being subtracted by the corresponding counterterms on each side
of Eq.~\ref{matching}.
The IR divergences correspond to the case $\mu^2 \gg  \Delta^2 + q_E^2$
in Section~\ref{scales}. These divergences are also regulated in dimensional regularization
but a key defining condition of the EFT construction is that {the IR physics of the EFT and the UV completion
is the same.} Matching fixes the Wilson coefficients to reflect the short distance UV physics integrated out of
the EFT. The reason is simple and intuitive, the IR physics that is not modified
in transitioning from the UV theory to an EFT description cancels in the matching. This includes
the IR divergences themselves. The correction that remains
to match onto the Wilson coefficients is then only due to the UV physics integrated out of the EFT.
When the matching procedure is carried out in DR and $\rm \overline{MS}$, the following
technical simplifications also occur:
\begin{itemize}
\item{Scaleless integrals vanish and the IR and UV divergences in such integrals cancel.
A simple example of this is given by the wavefunction renormalization factor of a massless fermion in QCD
with the $\psi$ two point function having divergences
\bea
\frac{\alpha_s \, C_F}{4 \, \pi} \, i \slashed{p} \left[\frac{1}{\epsilon_{IR}} - \frac{1}{\epsilon_{UV}}\right].
\eea
This can simplify matching calculations dramatically, as diagrams that are scaleless in the EFT can be
neglected when using DR.}
\item{In calculating one loop matchings, expanding intermediate results in small $\epsilon$,
or dimensionless ratios, before Feynman parameter integrals are carried out, can be justified
so long as the modifications in the results cancel in the matching condition.}
\item{Quadratic divergences are represented as $\epsilon$ poles.
Dimensionful threshold corrections in the matching conditions occur at tree level (in the EFT),
and also as one loop running corrections to EFT parameters \cite{Jenkins:2013zja}.}
\item{Usually the matching conditions are evaluated at the scale of particles integrated out of the theory
to define the EFT. This is not required, but is advantageous as it acts to minimize potentially large logs
in the perturbatively expanded matching equations,
when such matching is combined with Renormalization Group Evolution (RGE) running.}
\end{itemize}

\subsubsection{Matching examples}\label{sec:matchingexamples}
``Integrating out a field" as nomenclature follows from the path integral approach
to defining an effective action as developed by Wilson \cite{Wilson:1971bg,Wilson:1971dh}.
The effective action $S_{\rm eff}[\phi]$  retaining a light field $\phi$, and removing the heavy field $\Phi$
is defined such that
\begin{equation}
e^{iS_{\rm eff}[\phi]} =\int d\Phi\, e^{iS[\phi,\Phi]}/\int d\Phi\, e^{iS[\phi,0]} \,,
\end{equation}
where the integral is over all field values $\Phi$. Hence the heavy field is ``integrated out".
At times this procedure can be carried out formally in the path integral while using the Equations of Motion (EOM) for the theory.
Such manipulations can require the interactions to be of a limited form to be formally justified.

Integrating out a heavy field and determining the matching conditions for the Wilson coefficients
at tree level can always be done using Feynman diagram techniques directly. Again the EOM are used
and this approach can be easier in some cases than directly determining $S$ matrix elements in the full and effective
theories as in Eq.~\ref{matching}, and solving the resulting system of equations.

As a set of examples, consider the case of a heavy SM singlet scalar field ($S$) and a
heavy SM singlet (Weyl) fermion ($N$). The Lagrangian for the former case for $d \leq 4$ interactions is
\bea
\mathcal{L}_{SM+S} = \mathcal{L}_{SM} + \frac{1}{2}\left((\partial_\mu S)(\partial^\mu S) - m_S^2 S^2 \right) - \frac{\kappa_1}{2} \, S^2 H^\dagger H - \Lambda_1 \, S \, H^\dagger \, H
- \Lambda_2 \, S^3 - \kappa_2 \, S^4.
\eea
The SM is defined in Section \ref{SMsection}, $\kappa_{1,2}$ are dimensionless couplings while
$\Lambda_{1,2}$ have mass dimension one. For field values $\langle H^\dagger \, H \rangle < \langle S^2 \rangle < m_S^2$
and $p^2 < m_S^2$ one can solve the equation of motion for $S$ and Taylor expand around the classical solution
finding
\bea
S \simeq - \frac{\Lambda_1 \, H^\dagger H}{m_S^2} + \cdots,
\eea
which substituted back into the initial Lagrangian gives
\bea
\mathcal{L}_{SMEFT} = \mathcal{L}_{SM} + \frac{\Lambda_1^2}{2 \, m_S^2} \, (H^\dagger \, H)^2 + \mathcal{L}^{(6)} + \cdots.
\eea
The leading correction term shown can be absorbed into a finite shift of the SM Higgs self coupling, consistent with the
decoupling theorem. The terms in $\mathcal{L}_6$ are given by
\bea
\mathcal{L}^{(6)} = - \frac{\Lambda_1^2}{m_S^4} \mathcal{Q}_{H \Box} + \left(\frac{\Lambda_2 \, \Lambda_1}{m_S^2} -\frac{\kappa_1}{2}\right) \,\frac{\Lambda_1^2}{m_S^4} \, \mathcal{Q}_H
\eea
using the Warsaw basis for $\mathcal{L}_6$.\footnote{See Section \ref{SMEFTbasis} for details on basis choice and $\mathcal{L}_6$.}
The matching contributions are organized due to the IR operator forms that result, not naive scalings in $m_S$.
The coefficients of terms in $\mathcal{L}_6$ are expected to be overall $\propto 1/m_S^2$.
The naive expectation of the ordering of the operator forms in powers of $m_S$ is generically upset due to the presence of
dimensionful couplings, as purposefully illustrated here.
Naturalness considerations imply that $\Lambda_{1,2}^2 \lesssim m_S^2 \, 16 \, \pi^2$.
When this bound is saturated, large Wilson coefficients result. Even when the bound is not saturated,
the presence of such dimensionful couplings can lead to $\mathcal{O}(1)$ Wilson coefficients  ($ \times 1/m_S^2$).
This is a particular concern when considering matching to strongly interacting UV physics sectors, below the scale
present in the confining phase of such a sector. Such dimensionful couplings can then be present
in the interactions of the resulting bound states, unless forbidden by a symmetry, and can
directly upset any intuition based on perturbative matching in a UV coupling $g^\star$ and taking a
limit $g^\star \rightarrow 4 \, \pi $.\footnote{See Section \ref{badidea} for more discussion on such an approach.}

As another example, consider the case of a SM singlet Weyl fermion integrated out in the UV.
This scenario corresponds to the Seesaw model \cite{Minkowski:1977sc,GellMann:1980vs,Yanagida:1979as,Mohapatra:1979ia} for generating massive
Neutrino's, and has been studied in an EFT context in
Refs.\cite{Broncano:2002rw,Gavela:2009cd,Gavela:2008ra,Abada:2007ux,Broncano:2003fq,Bonnet:2012kz,Bonnet:2009ej,delAguila:2008ir,delAguila:2012nu,Bhattacharya:2015vja,Angel:2012ug}.
The Lagrangian can be defined as $\mathcal{L}_{SM} + \mathcal{L}_{N_p}$
where
\bea\label{basicL}
2 \, \mathcal{L}_{N_p}=  \overline{N_p} (i\slashed{\partial} - m_{p})N_p - \overline{\ell_{L}^\beta} \tilde{H} \omega^{p,\dagger}_\beta  N_p -  \overline{\ell_{L}^{c \beta}} \tilde{H}^* \, \omega^{p,T}_\beta N_p - \overline{N_p} \, \omega^{p,*}_\beta \tilde{H}^T \ell_{L}^{c \beta}  - \overline{N_p}\, \omega^p_\beta \tilde{H}^\dagger \ell_{L}^\beta.  \label{LNa}
\eea
The couplings $\omega^p_\beta = \{x_\beta,y_\beta,z_\beta\}$ are complex vectors in flavour space that absorbed the Majorana phases.
$p=\{1,2,3\}$ is summed over. Integrating out the $N_p$ at tree level
by taking the $p^2 < m_p^2$ limit of the tree level exchange diagram
gives the result $\mathcal{L}_{SMEFT} = \mathcal{L}_{SM} + \mathcal{L}^{(5)} + \cdots$ where
\bea\label{eqn:l5}
\mathcal{L}^{(5)} = \frac{c_{\beta \, \kappa}}{2} \,  \left(\overline{\ell^{c, \beta}_{L}} \, \tilde{H}^\star\right) \left(\tilde{H}^\dagger \, \ell_{L}^{\kappa}\right) \,  + h.c.
\eea
and $c_{\beta \, \kappa} =  (\omega^p_\beta)^T \, \omega^p_\kappa/m_p$. The matching is onto the
leading correction to the SM dimension four Lagrangian \cite{Weinberg:1979sa,Wilczek:1979hc}.
The notation used here is that the $c$ superscript in Eq.~\ref{eqn:l5} corresponds to a charge conjugated Dirac four component
spinor defined as $\psi^c  = C \overline{\psi}^T$ with $C= - i \gamma_2 \, \gamma_0$ in the chiral basis.
$\ell_L^c$ denotes the doublet lepton field that is chirally projected and subsequently charge conjugated.
See Ref.~\cite{Elgaard-Clausen:2017xkq} for further notational details. Expanding the result
around the vacuum expectation value for the Higgs field gives experimentally required Neutrino masses.

Calculating higher order corrections to the matching results can also be directly
determined using standard Feynman diagram techniques. Conversely, a naive path integral approach to integrating out a field
can be practically limited to leading order calculations.
Determining higher order perturbative matching corrections is straightforward in the case of SM singlet fields in a UV sector.
The required loop corrections are only present on the left or the right hand side of Eq.~\ref{matching}
due to the different symmetry groups of the SM and the UV sector in this case.
Alternatively, when UV field content integrated out is charged under the SM gauge groups,
loop corrections in the EFT and in the full theory (on both sides of Eq.~\ref{matching}) are required at each order in perturbation theory.
The $S$ matrix elements are calculated to higher orders in perturbation theory in the full theory and the EFT.
UV divergences are canceled by counterterms dictated by the subtraction and regularization scheme chosen.
IR divergences and constant terms cancel in the matching calculations, and the Wilson coefficients are
then determined to the desired order by the UV physics removed from the EFT. Higher order terms in the operator expansion (which is usually referred to as the non-perturbative expansion of the EFT
in the literature) can also be determined using these techniques, and mixed perturbative and non-perturbative
contributions.
For a sample of excellent examples of matching see
Ref.~\cite{Manohar:2003vb,Dawson:2012di,Henning:2014wua,Gorbahn:2015gxa,Brehmer:2015rna,Freitas:2016iwx,Dawson:2015oha,Chen:2014xwa,Chen:2017hak,Dawson:2017vgm}.

\subsubsection{Covariant Derivative Expansion matching}
Matching typically requires the computation of a large number of diagrams in the full theory
and the EFT, which is done choosing a convenient gauge, and subsequently recombining the results into
gauge invariant effective operators. This can be cumbersome when determining matching calculations to higher orders
in the expansions present in the EFT. A recently developed technique, that goes under the name of the
Covariant Derivative Expansion (CDE), is aimed at simplifying this computation by resorting to more advanced functional methods.
This technique has two main advantages: it does not require the evaluation of Feynman diagrams
because the matching is done at the action level and, at the same time, it seeks to preserve manifest
gauge invariance at all the stages of the calculation. The CDE method has been introduced
in the modern EFT context in Ref.~\cite{Henning:2014wua} reviving an approach previously
explored in the 80's~\cite{Gaillard:1985uh,Cheyette:1987qz} for other applications. In the following we
summarize the main argument of~\cite{Henning:2014wua}. At the action level, the matching of
a theory containing both heavy fields $\Phi$ and light fields $\phi$ onto an EFT that
 describes only the $\phi$ degrees of freedom again amounts to constructing an effective action $S_{\rm eff}[\phi]$ such that
\begin{equation}
e^{iS_{\rm eff}[\phi]} =\int d\Phi\, e^{iS[\phi,\Phi]}/\int d\Phi\, e^{iS[\phi,0]}\,.
\end{equation}
Using a saddle-point approximation, which is valid for a perturbative expansion up
to one loop, and expanding the heavy fields around their background values $\Phi=\Phi_c+\eta$ one obtains
\begin{equation}\label{Seff_tree+loop}
 S_{\rm eff.}[\phi] \simeq S(\Phi_c) + \frac{i}{2}\Tr\log\left(-\left.\frac{\d^2 S}{\d\Phi^2}\right|_{\Phi_c}\right),
\end{equation}
where the first term contains the structures obtained integrating out the heavy field in
tree level diagrams, while the second encodes the contributions generated at one-loop.
Eq.~\ref{Seff_tree+loop} can be evaluated explicitly assuming a generic (universal)
structure for the Lagrangian of the UV model. For example, if the heavy field is a complex scalar, one has
\begin{equation}
 \mathcal{L}_{UV} \supseteq -\Phi^\dag(D^2+M^2+U(x))\Phi + \left(\Phi^\dag B(x)+\hc\right)+\mathcal{O}(\Phi^3)
\end{equation}
where $D_\mu = \de_\mu-i A_\mu$ is a covariant derivative and $U(x)$, $B(x)$ are arbitrary
model-dependent expressions containing light fields. The tree-level matching
contribution $S(\Phi_c)$ is derived in a standard fashion replacing $\Phi\to\Phi_c$ in $\mathcal{L}_{UV}$,
where $\Phi_c$ is the solution of the EOM for $\Phi$, and expanding the resulting
Lagrangian in inverse powers of $M$. The final result is
\begin{equation}
\begin{aligned}
\Delta \mathcal{L}_{\rm eff, tree} =& -B^\dag\left[-D^2-M^2-U\right]^{-1}B+\mathcal{O}(\Phi_c^3) \\
 \simeq& \, B^\dag M^{-2} B + B^\dag M^{-2}[-D^2-U] M^{-2} B + \dots
 \end{aligned}
\end{equation}
where we have dropped the $x$ dependence. Note that the field $\Phi$ is
generally a multiplet, so that the mass term $M$ is a matrix, which does not necessarily commute with $(D^2+U)$.

The evaluation of the one loop piece is slightly more involved.
The most general result can be found in Ref.~\cite{Henning:2014wua} together with a detailed
derivation. The final expression obtained expanding up to dimension 6 is
\begin{equation}
\begin{aligned}
\Delta &\mathcal{L}_{\text{eff,1-loop}} =\frac{c_s}{(4\pi)^2} \, \Tr \, \Bigg\{
 +M^4\bigg[-\frac{1}{2} \Big(\log \frac{M^2}{\mu^2} -\frac{3}{2}\Big) \bigg]   +M^2\bigg[-\Big(\log \frac{M^2}{\mu^2} - 1\Big) \, U\bigg]  \\
&\qquad +M^0\Bigg[-\frac{1}{12}\Big(\log \frac{M^2}{\mu^2} - 1\Big) \, G_{\mu\nu}'^2 - \frac{1}{2} \log \frac{M^2}{\mu^2} \, U^2 \Bigg] \\
&\qquad + \frac{1}{M^2}\Bigg[ \frac{1}{60} \, \big(D_{\mu}G_{\mu\nu}'\big)^2 - \frac{1}{90} \, G_{\mu\nu}'G_{\nu\s}'G_{\s\mu}' +\frac{1}{12} \, (D_{\mu}U)^2 - \frac{1}{6}\, U^3 - \frac{1}{12}\, U G_{\mu\nu}'G_{\mu\nu}' \Bigg]  \\
&\qquad +\frac{1}{M^4} \Bigg[\frac{1}{24} \, U^4 - \frac{1}{12}\, U \big(D_{\mu}U\big)^2 + \frac{1}{120}\, \big(D^2U\big)^2 +\frac{1}{24} \, \Big( U^2 G'_{\mu\nu}G'_{\mu\nu} \Big)  \\
&\qquad \qquad \qquad + \frac{1}{120} \, \big[(D_{\mu}U),(D_{\nu}U)\big] G'_{\mu\nu} - \frac{1}{120}\, \big[U[U,G'_{\mu\nu}]\big] G'_{\mu\nu} \Bigg]  \\
&\qquad +\frac{1}{M^6} \Bigg[-\frac{1}{60} \, U^5 + \frac{1}{20} \, U^2\big(D_{\mu}U\big)^2 + \frac{1}{30} \, \big(UD_{\mu}U\big)^2 \Bigg]
+ \frac{1}{M^8} \bigg[ \frac{1}{120} \, U^6 \bigg]
 \Bigg\} .
\label{eqn:CDE_universal_lag}
\end{aligned}
\end{equation}
Here $c_s=\{1/2,1\}$ for a real and complex scalar $\Phi$ respectively and $G'_{\mu\nu}=[D_\mu,D_\nu]$.
Finally, the trace is over internal indices, i.e. Lorentz, flavour, gauge indices etc.
Inserting into Eq.~\ref{eqn:CDE_universal_lag} the expressions of $U$ and $G_{\mu\nu}$ defined in a
specific model, one immediately obtains a sum of dimension six operators
whose coefficients are automatically matched with the UV model.
Eq.~\ref{eqn:CDE_universal_lag} is universal in the sense that it can be applied not only
to the case of scalar $\Phi$ but also when the heavy field is a fermion or vector boson,
as detailed in Ref.~\cite{Henning:2014wua}. Although extremely practical, this expression has two
main defects:
\begin{enumerate}
\item{It holds only in the case of degenerate heavy states, in which the mass matrix $M$ is diagonal and
commutes with the other structures;}
\item{The $\Delta \mathcal{L}_{\text{eff,1-loop}}$ computed in this way accounts only
for loop diagrams in which all the internal lines are heavy. Mixed heavy-light loops are missing
because the light field have been treated as background fields and therefore only enter as
external lines~\cite{delAguila:2016zcb,Boggia:2016asg}.}
\end{enumerate}
Point 1 was addressed in Ref.~\cite{Drozd:2015rsp}, that generalized Eq.~\ref{eqn:CDE_universal_lag} to the
case of a non-degenerate multiplet $\Phi$ obtaining an expression
for the effective action at one loop that was named the UOLEA (Universal One Loop Effective Action).
Point 2 represents a deeper problem in the basic CDE technique described above, which requires
a modification of the functional treatment. Different solutions have been proposed
in Refs.~\cite{Henning:2016lyp,Ellis:2016enq,Fuentes-Martin:2016uol}. Both Ref.~\cite{Henning:2016lyp} and Ref.~\cite{Ellis:2016enq} expand
the functional analysis of Ref.~\cite{Henning:2014wua} with the inclusion of the fluctuations around the background
fields for the light degrees of freedom $\phi$ and both suggest a method that requires the subtraction of
non-local terms from the functional determinant. This step is avoided with the alternative method
proposed in Refs.~\cite{Fuentes-Martin:2016uol}, that builds upon Refs.~\cite{Dittmaier:1995cr,Dittmaier:1995ee} and employs
the ``expansion-by-regions'' technique for the evaluation of the loop integrals ~\cite{Beneke:1997zp,Smirnov:2002pj,Jantzen:2011nz}.
One defines
the multiplet $\varphi=(\Phi,\phi)$ and
\begin{equation}
 \Delta \mathcal{L}_{\text{eff,1-loop}} = \frac{1}{2}\varphi^\dag \left.\frac{\d^2 \mathcal{L}_{UV}}{\d\varphi^* \d \varphi}\right|_{\varphi_c} =
 \varphi^\dagger \begin{pmatrix} \Delta_H& X_{HL}^\dag\\ X_{HL}& \Delta_L\end{pmatrix}\varphi,
\end{equation}
where the last term contains a block matrix so that the heavy fields $\Phi$ are contracted
by $\Delta_H$, the light ones by $\Delta_L$ and $X_{HL}$ is a mixed term.
The key idea of Ref.~\cite{Fuentes-Martin:2016uol} is to perform a field transformation that brings the
matrix to a block diagonal form ${\rm diag}(\tilde\Delta_H,\Delta_L)$, shifting the effect of $X_{HL}$ into
the heavy-particle contribution while leaving the $\Delta_L$ unchanged. This procedure
gives $\tilde\Delta_H = \Delta_H-X_{HL}^\dag\Delta_L^{-1} X_{HL}$ which, in the scalar case, can be
expressed in the notation of Ref.~\cite{Henning:2014wua} as
\begin{equation}
\tilde\Delta_H = -(D^2+M^2+\tilde U), \qquad\qquad \tilde U = U(x)+U_{HL}(x,p)\,,
\end{equation}
where $U_{HL}$ comes from the field redefinition and carries a dependence on the loop momentum $p$.
The effective action takes the form $S_{\rm eff}[\phi] = i c_s \Tr\log\tilde\Delta_H$ and it generates all
the loop diagrams with at least one heavy internal propagator. The desired result for the mixed heavy-light one
loop contributions to the EFT matching are obtained performing the loop momentum integrals in $\tilde\Delta_H$ only in the ``hard'' region,
i.e. first expanding out all the low-energy scales, that are small in the limit $p\sim M$, and then
integrating over the full $d$-dimensional $p$ space. This method makes use of dimensional regularization and
is known as ``expansion-by-regions''~\cite{Beneke:1997zp,Smirnov:2002pj,Jantzen:2011nz}  discussed in more detail in Section \ref{sec:regions}.
As a result,
the contribution to the dimension six effective Lagrangian in Eq.~\ref{eqn:CDE_universal_lag} is extended by the
inclusion of \cite{Fuentes-Martin:2016uol}:
\begin{equation}
\begin{aligned}\label{eq:LHuniversal}
\Delta\mathcal{L}_{\text{eff,1-loop}}^{HL}&=-ic_s\,\int\frac{d^dp}{\left(2\pi\right)^d}\bigg\{
\frac{1}{p^2-M^2}\,\mbox{tr}_{\rm s}\left(\tilde U\right)+\frac{1}{2}\frac{1}{\left(p^2-M^2\right)^2}\,\mbox{tr}_{\rm s}\left(\tilde U^2\right)\\
&+\frac{1}{3}\frac{1}{\left(p^2-M^2\right)^3}\,\left[\mbox{tr}_{\rm s}\left(\tilde U^3\right)+\mbox{tr}_{\rm s}\left(\tilde UD^2\tilde U\right)+2ip^\mu\,\mbox{tr}_{\rm s}\left(\tilde UD_\mu \tilde U\right)\right]\\
&+\frac{1}{4}\frac{1}{\left(p^2-M^2\right)^4}\,\Big[\mbox{tr}_{\rm s}\left(\tilde U^4\right)+2ip^\mu\,\mbox{tr}_{\rm s}\left(\tilde U^2 D_\mu \tilde U\right)+2ip^\mu\,\mbox{tr}_{\rm s}\left(\tilde U D_\mu \tilde U^2\right)\\
&\hspace{3cm}+\,\mbox{tr}_{\rm s}\left(\tilde U^2 D^2 \tilde U\right)+\,\mbox{tr}_{\rm s}\left(\tilde U D^2 \tilde U^2\right)-4\,p^\mu p^\nu\,\mbox{tr}_{\rm s}\left(\tilde U D_\mu D_\nu \tilde U\right)\\
&\hspace{3cm}+2ip^\mu\,\mbox{tr}_{\rm s}\left(\tilde U D^2D_\mu \tilde U\right)+2ip^\mu\,\mbox{tr}_{\rm s}\left(\tilde U D_\mu D^2 \tilde U\right)
+\mbox{tr}_{\rm s}\left(\tilde U (D^2)^2\, \tilde U\right)\Big]\\
&+\frac{1}{5}\frac{1}{\left(p^2-M^2\right)^5}\,\Big[\mbox{tr}_{\rm s}\left(\tilde U^5\right)+2ip^\mu\,\mbox{tr}_{\rm s}\left(\tilde U^3D_\mu \tilde U\right)+2ip^\mu\,\mbox{tr}_{\rm s}\left(\tilde U^2D_\mu \tilde U^2\right)\\
&\hspace{3cm}+2ip^\mu\,\mbox{tr}_{\rm s}\left(\tilde UD_\mu \tilde U^3\right)-4p^\mu p^\nu\,\mbox{tr}_{\rm s}\left(\tilde U^2 D_\mu D_\nu \tilde U\right)\\
&\hspace{3cm}-4p^\mu p^\nu\,\mbox{tr}_{\rm s}\left(\tilde U D_\mu \tilde U D_\nu \tilde U\right)-4p^\mu p^\nu\,\mbox{tr}_{\rm s}\left(\tilde U D_\mu D_\nu \tilde U^2\right)\\
&\hspace{3cm}-8i\,p^\mu p^\nu p^\rho\,\mbox{tr}_{\rm s}\left(\tilde U D_\mu D_\nu D_\rho \tilde U\right\}\Big]\bigg\}
+\mathcal{L}_{\rm EFT}^F+\mathcal{O}\left(M^{-3}\right)\,.
\end{aligned}
\end{equation}
Here $\mbox{tr}_{\rm s}$ is a subtracted trace defined as
\begin{align}
\mbox{tr}_{\rm s}f(\tilde U,D_\mu)\equiv\mbox{tr}\left(f(\tilde U,D_\mu)-f(U,D_\mu)-\Theta_f\right)\,,
\end{align}
where $f$ is an arbitrary function of $\tilde U$ and covariant derivatives, while $\Theta_f$ generically denotes all the terms
with covariant derivatives at the rightmost of the original trace. The subtraction of the terms containing
only $U$ avoids a double counting, as these contributions are already contained in Eq.~\ref{eqn:CDE_universal_lag}.
Finally, $\mathcal{L}_{\rm EFT}^F$ contains all the terms containing open derivatives, that eventually combine into operators with
field-strength tensors.
The evaluation of this piece truncated at dimension-six terms is quite complex: a universal expression, although achievable in principle, is not yet available to date.

Although not definitive, the results presented in Ref.~\cite{Fuentes-Martin:2016uol} represented a key step in the development of CDE techniques.
In particular, they highlighted that the heavy-light structures could be directly inferred from the heavy-only ones and they paved the way for the development of a covariant diagrammatic formalism~\cite{Zhang:2016pja}. The latter allows an immediate and efficient evaluation of functional quantities in terms of gauge-invariant operators, while providing a graphical representation that keeps track of the CDE expansion\footnote{Analogously, Feynman diagrams allow to compute correlation functions and make manifest the organization of different contributions in a perturbative expansion.}. This powerful formalism has been adopted recently in Ref.~\cite{Ellis:2017jns} to compute explicitly all the universal terms (operators and coefficients) in the UOLEA functional in~\cite{Drozd:2015rsp}, with the addition of the heavy-light ones and for both the degenerate and non-degenerate cases. These results extend the construction in Ref.~\cite{Drozd:2015rsp} and supersede those of Ref.~\cite{Fuentes-Martin:2016uol}, providing universal expressions for CDE matching that can be applied to a given model without the need of reiterating the functional procedure.

As mentioned above, the evaluation of open derivative terms is still missing at this stage, but it is expected to become available in the near future, again thanks to covariant diagrams techniques. The same holds for mixed bosonic-fermionic loops. The calculation of these two categories of effects will complete the ``UOLEA program'' to supply a completely universal, gauge-invariant result for EFT matching.

The power of the CDE approach has recently been illustrated in many examples in the literature. A stand out example is the demonstration of how the
known one loop matching MSSM results of Ref.~\cite{Pierce:1996zz} can be elegantly determined using the modern CDE
approach \cite{Wells:2017vla}.

\subsubsection{Method of regions}\label{sec:regions}
Since the first version of this review was written, a rather comprehensive tree level matching dictionary for integrating out UV field content
(when the underlying theory is perturbative) has been reported in Ref.~\cite{deBlas:2017xtg}. This set of matching results is for the SMEFT up to $\mathcal{L}^{(6)}$.

Building further on this result is theoretically well motivated. The power of EFT is most apparent when going beyond leading order in the EFT expansion parameter in matching, and in determining
perturbative corrections to observable processes. For studies to advance beyond leading order in perturbation theory consistently  requires that matching calculations
be performed beyond leading order. Recent advances to this end include the one loop CDE matching
results discussed in the previous section, where the technique of ``expansion-by-regions'' (or method of regions) was also used. In this section we summarize some of the physics underlying 
this latter approach.

The method of regions \cite{Beneke:1997zp,Smirnov:2002pj,Jantzen:2011nz} is an elegant way to determine matching coefficients.
The utility of the method of regions relies on the point that non-analytic structure of a full theory is projected out in matching.
Using this fact actively simplifies a matching calculation. This simplification can be important to enable the determination of a matching coefficient at one loop, or higher orders in the EFT expansion.
The method of regions allows one to evaluate Eqn.~\ref{matching}
without performing the full theory calculation first with all mass scales retained. This can be done by expanding a loop result in the EFT expansion {\it before integrating}.

Consider matching the seesaw model given in Eqn.~\ref{basicL} at one loop to the SMEFT. A one loop calculation to determine the matching to the Higgs two point function is defined 
by expanding a result in the ratio of scales $v_T^2/m_r^2 <1$, where $m_r$ is the mass of the $N_r$ Majorana particle.
The mass of the charged lepton field is denoted $m_{\ell}$ and we retain both mass scales for illustrative purposes initially. 

\begin{figure}[t]
\includegraphics[width=6cm]{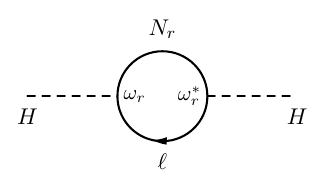}
\caption{Diagram contributing to the Higgs two-point-function in the type-I seesaw model.}\label{diagram_seesawHH}
\end{figure}

The only diagram contributing is drawn in Fig.~\ref{diagram_seesawHH} and it gives
\bea
i \Pi^{full} _{H H^\dagger}(p^2)&=& 
- 2\,  |\omega_r^2| \,  \int \frac{d^4 \ell}{(2 \pi)^4} \int_0^1 dx \frac{\ell^2 + x (x-1)p^2}{[\ell^2 - \Delta]^2}, \nn
&=& 
- \frac{i |\omega_r^2|}{16 \, \pi^2 \, \epsilon} (2 m_\ell^2 + 2 m_r^2 - p^2)  - \frac{i |\omega_r^2|}{48\, \pi^2} (3 m_\ell^2 + 3 m_r^2 - p^2), \\
&+& \frac{i|\omega_r^2|}{8 \, \pi^2}  \int_0^1 dx (2m_r^2 (x-1) - x (2m_\ell^2 + 3 p^2 (x-1)) \log[\frac{\mu^2}{m_\ell^2 x + (x-1) (p^2 x - m_r^2)}]\,, \nonumber
\eea
where $\Delta = m_\ell^2 x + (x-1) (p^2 x - m_r^2)$.
Recall that the UV theory is renormalized directly with its own set of counterterms. This effectively removes the first term in the expression.
The second term in the expression leads to a threshold matching correction.  This physics is the origin of what is called the Hierarchy problem, and we will return to this physics in the following sections.
The difficulties of loop calculations with multiple scales are illustrated in the last term of this expression. At higher loop orders, this complication grows and becomes a serious technical hurdle.
On the other hand, expanding in ratios of the scales present in the problem simplifies the result directly. This expansion can be done
when matching onto the EFT even before integrating is the key point.

First consider the dependence on the scale $m_\ell$ in this last term. This is an IR scale
that is present in the SMEFT and in the full theory. The full theory in this case is $\mathcal{L}_{SM} + \mathcal{L}_{N}$  and has non-analytic dependence on this scale in
predicted $S$ matrix elements. This occurs when the intermediate $\ell$ state goes on-shell. This non-analytic behavior is common to the EFT and the UV theory and so cancels out in the matching.
By definition, the EFT (in this case the SMEFT) reproduces the IR of the full theory and this statement holds at arbitrary orders in perturbation theory, including
for the non-analytic behavior of the full theory loop
diagrams dependence on the scale $m_\ell$. For this reason one can expand in $m_\ell/(p,m_r)$ {\it before integrating} and neglect $m_\ell$ in determining the matching coefficient.

Now, consider expanding in the remaining scales of the problem, $p^2$ and $m_r^2$. The propagator of the heavy Majorana field can be expanded as
\bea\label{eftsimp}
\frac{1}{k^2 - m_r^2} = - \frac{1}{m_r^2} \left[ 1+ \frac{k^2}{m_r^2} +  \cdots \right]
\eea
where $k^2$ is the loop momentum when considering the loop integral in the SMEFT for this matrix element. Using this result one can simplify the expression for the two point function in the SMEFT to
\bea
i \Pi^{EFT} _{H H^\dagger}(p^2)&=& \frac{|\omega_r^2|}{2 \, m_r^2} \int \frac{d^4 k}{(2 \pi)^4} \frac{k^2}{k^2} + \cdots.
\eea
Here the loop momentum has been shifted after the propagator expansion in Eqn.~\ref{eftsimp} has been performed.
All of the terms that are present are scaleless integrals that vanish in dimensional regularization when all IR scales are expanded out. For this reason, the matching result in Eqn.~\ref{matching} can be directly determined  by performing the simpler calculation in the full theory, where all IR mass scales such as $m_\ell$ was expanded out from the start, and the one loop contribution from the SMEFT in the matching is dropped. One only needs to perform the much simpler loop integral that remains, and expand in $p^2/m_r^2$ the result. This determines the matching coefficient. For the two point function
one finds the contribution
\bea
i \Pi^{full,exp} _{H H^\dagger}(p^2)&=& \frac{-i |\omega_r^2| \, m_r^2}{8 \pi^2} \left(1 + \log \frac{\mu^2}{m_r^2} \right)  +  \frac{i |\omega_r^2| \, p^2}{32 \pi^2} \left(1 + 2\log \frac{\mu^2}{m_r^2} \right). \nonumber
\eea
A finite field redefinition is used to cancel the last term in this expression, and the final matching result to the Higgs two point function is the first term.

This procedure illustrates the utility of expanding in the scales of the problem before integrating when performing matching calculations. The wide use of the method of regions is essentially due to it using this physics
systematically. For more discussion, see Refs.~\cite{Beneke:1997zp,Smirnov:2002pj,Jantzen:2011nz,Manohar:2006nz,Manohar:2018aog}.

\subsection{Choose any scheme, so long as it is dimensional regularization and \titlemath{$\rm \overline{MS}$}{MS}}
In matching calculations, divergences can be dropped as the UV divergences in each theory are canceled by UV counterterms,
and the IR divergences in the full theory and the EFT coincide by definition. If such divergences
are retained in the calculation, they must be regulated.
Any physical conclusion is independent of a regulator choice, in the limit the regulator
is taken to infinity.\footnote{For a recent discussion on this point in more detail see Ref.\cite{tHooft:2007hyi}.} Nevertheless, the ease of obtaining physical conclusions,
and performing loop calculations depends on the regulator choice.
The discussions in the previous sections were made using  DR.
Using a dimensionless regulator is now standard in most EFT studies, and can be essential
in EFT studies of the Higgs boson. The reason for this is that the SM is classically scaleless in the limit $v \rightarrow 0$ ($m_h \rightarrow 0$ as a result)
and a dimensionful regulator explicitly breaks this (anomalous) symmetry by introducing a dimensionful cut off scale.

It took decades for the benefits of DR to be fully appreciated in
the EFT community. This was due to some history and some misunderstanding.
The history is due to the key initial ideas of
the RG emerging into broad use\footnote{A precursor of these ideas were
presented by St\"uckelberg and Petermann in a prescient work \cite{Stueckelberg:1953dz}.} following the
pioneering condensed matter studies of Kadanoff \cite{Kadanoff:1966wm},
Wilson \cite{Wilson:1971bg,Wilson:1971dh} and Wilson and Fisher \cite{Wilson:1971dc}.
The partition function in the early cases studied using the RG was constructed out of a reduced number of degrees of freedom
by integrating out a series of high energy modes as
\bea
Z = \int^{\Lambda_1} \mathcal{D} \, \phi^1_i \, e^{- S_1(\phi^2_i)} = \int^{\Lambda_2} \mathcal{D} \, \phi^2_i \, e^{- S_2(\phi^2_i)}
 \cdots = \int^{\Lambda_n} \mathcal{D} \, \phi^n_i \, e^{- S_n(\phi^n_i)},
\eea
where $\Lambda^n > \Lambda^{n-1}$. At each step in integrating out a momentum shell, the action
is matched by fixing its parameters and redefining the fields, to reproduce the low energy physics of the action
with the higher energy modes removed. The differential version of this procedure gives the RG
Equations which are local and analytic, as can be understood from the general arguments of
the previous sections. That the initial applications \cite{Kadanoff:1966wm,Wilson:1971bg,Wilson:1971dh} of the RG was formulated with a dimensionful regulator makes
perfect sense as the modes integrated out were discretizing physically separated spin configurations, with a
corresponding dimensionful Fourier transformed momentum. The misunderstanding that slowed the adoption of DR was the perception
that as it integrates over all momenta, it does not faithfully limit the EFT to momenta where it is defined,
introducing an inconsistency. This is incorrect, as discussed in Section \ref{basics}.

A key step forward in the development
of EFTs was the realization that such a dimensionful regularization is not necessary, and best
avoided. The main reasons for this are that such regulators
make power counting suspect,
and calculations technically challenging. For example, the method of regions approach in the previous section relies
on using dimensional regularization and the fact that scaleless integrals vanish when using this regulator. 

\subsubsection{\titlemath{$Z$}{Z} decay and dimensionful regulators}\label{section:dimensionfulreg}
\begin{figure}[t]
\begin{tikzpicture}
[
decoration={
	markings,
	mark=at position 0.55 with {\arrow[scale=1.5]{stealth'}};
}]

\draw (0,0) circle (0.75);
\filldraw (0.6,-0.1) rectangle (0.8,0.1);

% top
\draw  [postaction=decorate] (0:0.75) -- + (30:1.30) ;
\draw [postaction=decorate] (-20:2.0) -- +(150:1.40);
\draw  [->][ultra thick]  (90:0.75)  -- + (0:-0.1) ;
\draw  [->][ultra thick]  (270:0.75)  -- + (0:0.01) ;

% 1

\draw [decorate,decoration=snake] (-1.7,0) -- (-0.75,0);

\node [left][ultra thick] at (-1.7,0) {$Z$};
\node [above][ultra thick] at (0,0.8) {$t$};
\node [below][ultra thick] at (0,-0.8) {$t$};
\node [right][ultra thick] at (1.80,0.7) {$\bar{\psi}$};
\node [right][ultra thick] at (1.80,-0.7) {$\psi$};

\end{tikzpicture}
\caption{\label{ffdiagam}Four fermion operator contribution to $Z \rightarrow e^+ \, e^-$ at one loop.}
\end{figure}
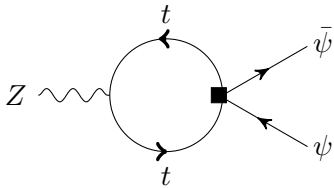

The problems of dimensionful regulators are well illustrated by the example of four fermion operators generating the decay
$Z \rightarrow \overline{\ell} \, \ell$ at one loop. The results dependent on $y_t$
are known \cite{Hartmann:2016pil} in the $\rm U(3)^5$ limit.
The Effective Lagrangian generated
has the terms
\bea
\mathcal{L}_{Z,eff} = - \, 2 \, 2^{1/4} \, \sqrt{\hat{G}_F} \, \hat{m}_Z \,
\bar{\ell}_s \, \gamma_\mu \left[(\bar{g}^{\ell}_L)_{ss} \, P_L +  (\bar{g}^{\ell}_R)_{ss} \, P_R \right] \ell_s,
\eea
where at one loop the four fermion operators give a correction \cite{Hartmann:2016pil}
\bea\label{leptonloopshift}
\Delta (g^{\ell}_L)_{ss}&=& \frac{N_c \, \hat{m}_t^2}{16 \, \pi^2 \, \Lambda^2} \,  \log \left[\frac{\mu^2}{\hat{m}_t^2} \right] \,
 \left[C_{\substack{\ell q\\ ss33}}^{(1)} - C_{\substack{\ell q\\ ss33}}^{(3)} - C_{\substack{\ell u\\ ss33}} \right],
 \\
%%%%%%%%%%%%%%%%%%%%%%%%%%%%%%%%%%%%%%%%%%%%%%%%%%%%%%%%%%%%%%%%%
\Delta(g^{\ell}_R)_{ss}&=&  \frac{N_c \, \hat{m}_t^2}{16 \, \pi^2 \, \Lambda^2} \,  \log \left[\frac{\mu^2}{\hat{m}_t^2} \right], \,
\left[-C_{\substack{eu\\ ss33}}+ C_{\substack{q e \\ 33ss}} \right].
\eea
Here we are using the notation of Ref.~\cite{Hartmann:2016pil} where hat superscripts correspond to measured quantities at tree level.
These results come about in DR and $\rm \overline{MS}$ in the following manner. The anomalous dimensions
required were determined in Ref.~\cite{Jenkins:2013wua} using DR and exploiting the Background Field method \cite{DeWitt:1967ub,tHooft:1973bhk,Abbott:1981ke}. This method
preserves the symmetries of the SMEFT when gauge fixing as fields in the action are split into classical ($\phi$) and quantum ($Q$) components
\bea
S(Q) \rightarrow S(Q+ \phi).
\eea
A gauge fixing term then breaks the gauge invariance of the
quantum fields while maintaining the gauge invariance of the classical background fields,
this makes gauge independent counterterms easier to determine.
The Background Field method can be directly implemented in DR which also preserves the symmetries of the Lagrangian \cite{tHooft:1973bhk}.
The direct calculation of Fig.~\ref{ffdiagam} leads to contributions of the form
\bea\label{ffermionactualloop}
i \, \mathcal{A} &\propto& 4 \, N_c \, (\bar{g}^{\ell}_{L,SM} \, \bar{g}_{L,Q} +
\bar{g}^{\ell}_{R,SM} \, \bar{g}_{R,Q}) \int \frac{d^D l}{(2 \pi)^D} \frac{(D-2)}{D} \, \frac{l^2}{(l^2 - \hat{m}_t^2)^2}, \nn
&-& 4 \, N_c \,  (\bar{g}^{\ell}_{L,SM} \, \bar{g}_{R,Q} +
\bar{g}^{\ell}_{R,SM} \, \bar{g}_{L,Q}) \, \int \frac{d^D l}{(2 \pi)^D}  \, \frac{\hat{m}_t^2}{(l^2 - \hat{m}_t^2)^2}.
\eea
The notation $\bar{g}_{L/R,Q}$ corresponds to $P_{L/R}$ in the four fermion operators $Q_i$
inserted in the loop diagrams. In DR and $\rm \overline{MS}$ the only scale in the loops to make up the dimensions of the $1/\Lambda^2$ suppression
is the top quark mass, and the results combine in a non-trivial manner to cancel the pole
of Ref.~\cite{Jenkins:2013wua}, leaving the finite terms in Eq.~\ref{leptonloopshift}.

In the case of a dimensionful regulator where $D=4$ a cut off is introduced to the loop momentum $\sim \Lambda$.
Generally such regulators violate gauge invariance, as they regulate $p^\mu$ and not $D^\mu$. Furthermore,
translation invariance of the momenta in the propagators can be broken, which stands in the way of using the Feynman/Schwinger
trick to combine propagators in loop calculations. This makes identifying gauge independent
counter-terms a challenge and the use of the Background Field method unfeasible. Assuming that the
gauge invariant anomalous dimensions were determined using a dimensionful regulator (somehow),
the first term in Eq.~\ref{ffermionactualloop} gives
\bea
\int^\Lambda \frac{d^4 l}{(2 \ \pi)^4}  \frac{l^2}{(l^2 - \hat{m}_t^2)^2} \simeq  \frac{\Lambda^2}{16 \, \pi^2}.
\eea
The $\Lambda^2$
dependence acts to cancel the $1/\Lambda^2$ suppression of the operators in  $\bar{g}_{L/R,Q}$.
This results in an $\mathcal{O}(1)$
shift of the amplitude, due to a violation of the power counting.\footnote{DR is not without its own challenges, in particular
the definition of $\gamma_5$ in $d$ dimensions requires a scheme choice. See the recent discussion in Ref.~\cite{Hartmann:2016pil}
on the appearance of this issue in Fig.~\ref{ffdiagam}.} Dimensionful regulators are a nuisance to be avoided.

\subsubsection{Avoiding regulator dependence}
In DR, power counting violations due to the regulator choice do not occur in the insertion of higher dimensional operators in loop diagrams.
The $\mu$ scale introduced in the regulation procedure only appears in the Logs as in
Eq.~\ref{leptonloopshift}. Large mass scales that are present in the UV theory
can appear outside of Logs, in threshold matching corrections to $(H^\dagger \, H)$,
as shown in Section~\ref{precision-motivation}. This is the appearance of the Hierarchy problem using DR.

The Hierarchy problem is the need
to stabilize the scale invariance violating coefficient of $(H^\dagger \, H)$
against perturbations proportional to scales $\Lambda \gg \hat{m}_h$. A concrete example of these perturbations was given in
Section \ref{sec:regions} where a correction to $\Pi_{H H^\dagger} \propto |\omega_r|^2 \, m_r^2$ was found. Here $m_r \gg \hat{m}_h$ is the heavy mass scale
of putative Majorana states leading to Neutrino masses. This result was developed in dimensional regularization.
Conversely, using a hard cut off regulator makes it challenging to disentangle the
perturbation to  $(H^\dagger \, H)$ due to UV physics from unphysical regulator dependence.
Explicit breaking of scale invariance by the regulator is
unfortunate, as the mass parameter of the Higgs is the only classical source of the violation of
scale invariance in the SM.\footnote{See Refs.~\cite{Bardeen:1995kv,Meissner:2006zh,Foot:2007iy,Shaposhnikov:2008xi,Lindner:2014oea,
Helmboldt:2016mpi,Davoudiasl:2014pya,Brivio:2017dfq} for related discussion on scale invariance and the Hierarchy problem.}
This can lead to a different point of view as to what the Hierarchy problem is,
and what can solve the Hierarchy problem.

A regulator dependent argument can be formulated focused on the effect of the top quark on the operator $(H^\dagger H)$ at one loop. The relevant diagram is the leftmost entry in
Fig.~\ref{fig:threshold} which gives a coefficient to this operator of the form
\bea
i \, \mathcal{A} &=& -\frac{|y_t|^2 \, N_c}{2} \int^\Lambda \frac{d^4 l}{(2 \pi)^4} \frac{l^2 + 4 \, \hat{m}_t^2}{(l^2 - \hat{m}_t^2)^2}, \\
&=& -\frac{|y_t|^2 \, N_c \, \Lambda^2}{32 \, \pi^2} + \cdots
\eea
Such a regulated quadratic divergence in the case of hard cut off can be compared to DR
where the mass scale in the loop is $\hat{m}_t^2$ and the $\epsilon$ pole is subtracted.
The appearance of the quadratic divergence can be interpreted as a signal of the
Hierarchy problem that is regulator independent,
when it is indicating a threshold correction in a manner consistent with a
dimensionless regulator such as DR.

The standard conclusion that $\rm TeV$ scale states are motivated to appear in multiplets that
stabilize $(H^\dagger H)$ is valid and regulator independent. In DR, this is the statement that
one loop threshold corrections to $(H^\dagger H)$ scale $\propto (y_a^2 \, m_a^2 \pm y_b^2 \, m_b^2)/(16 \, \pi^2)$
where $m_{a,b}$ are states in such a multiplet with couplings $y_{a,b}$ to $H$. Symmetries can be built into models to
suppress such threshold corrections, and stabilize the dimension two operator. For example, unbroken supersymmetry by construction fixes $y_a = y_b$
and $m_{a} =m_{b}$ with the difference in sign between terms being present due to the different spin states in supermultiplets.
In a regulator independent manner, a precision Higgs phenomenology program is well motivated to search for
the low energy signatures of physics beyond the SM that acts to
stabilize the Higgs mass.

\section{Candidate field theories: the SM, SMEFT and HEFT}
The main field theories  discussed in this review used to interpret LHC, LEP and other low energy data, are the SM, the SMEFT or the HEFT.
The choice of which field theory to use is distinguished by the assumption on the size of possible new physics effects compared
to the achievable experimental resolution ($\Delta E_r$) at current and future facilities, and an assessment of the propagating
states in the particle spectrum. By using the SM, one assumes
that it will always hold when interpreting the data that
\bea
\Delta E_r \gg \frac{C_i \, v^2}{g_{SM} \, \Lambda^2}, \frac{C_i \, p^2}{g_{SM} \, \Lambda^2}.
\eea
Here each $C_i$ is a Wilson coefficient in the SMEFT or HEFT that corresponds to the ``pole expansion''
ratio $v^2/\Lambda^2$ or the derivative expansion ratio $p^2/\Lambda^2$.
We define the SM to fix our notation in Section \ref{SMsection}. Both the HEFT and the SMEFT
follow from the expectation that it is possible that
\bea
\Delta E_r \lesssim \frac{C_i \, v^2}{g_{SM} \, \Lambda^2}, \frac{C_i \, p^2}{g_{SM} \, \Lambda^2},
\eea
will occur in the near future, or in the longer term. This assumption is reasonable to adopt.
These EFTs are further distinguished by the presence of a Higgs doublet (or not) in the construction.
In the SMEFT (see Section \ref{SMEFTsec}) the EFT is constructed with an explicit Higgs doublet,
while in the HEFT (see Section \ref{HEFTsec}) no such doublet is included.

\subsection{The Standard Model}\label{SMsection}
We define the SM Lagrangian \cite{Glashow:1961tr,Weinberg:1967tq,Salam:1968rm},
with conventions consistent with Refs.~\cite{Jenkins:2013zja,Jenkins:2013wua,Alonso:2013hga}, as
\bea\label{sm1}
\mathcal{L} _{\rm SM} &=& -\frac14 G_{\mu \nu}^A G^{A\mu \nu}-\frac14 W_{\mu \nu}^I W^{I \mu \nu} -\frac14 B_{\mu \nu} B^{\mu \nu}
+ \! \! \! \! \! \sum_{\psi=q,u,d,\ell,e} \overline \psi\, i \slashed{D} \, \psi \\
&+&(D_\mu H)^\dagger(D^\mu H) -\lambda \left(H^\dagger H -\frac12 v^2\right)^2- \biggl[ H^{\dagger j} \overline d\, Y_d\, q_{j}
+ \widetilde H^{\dagger j} \overline u\, Y_u\, q_{j} + H^{\dagger j} \overline e\, Y_e\,  \ell_{j} + \hbox{h.c.}\biggr], \nonumber
\eea
where $H$ is an $\rm SU_L(2)$ scalar doublet.\footnote{The alert reader will notice the lack of dual field strength terms of the
form ${\rm Tr} \left[F^{\mu \, \nu} \tilde{F}_{\mu \, \nu} \right]$ for the Yang-Mills fields $F = \{W,G\}$, which can be present
\cite{tHooft:1974kcl,Polyakov:1974ek}.
Here and below the dual fields are defined with the convention $\widetilde F_{\mu \nu} =(1/2) \epsilon_{\mu \nu \alpha \beta} F^{\alpha \beta}$
with $\epsilon_{0123}=+1$.
The measurements of the electric dipole moment of the neutron indicate that such topological terms \cite{Belavin:1975fg,tHooft:1976rip,Peccei:1977ur} are
strongly suppressed for QCD, and the accidental conservation of $B+L$ in the SM allows the neglect of such terms for electroweak theory in the SM
\cite{tHooft:1976rip,Callan:1976je}.}
 With this normalization convention,
the Higgs boson mass is $m_H^2=2 \, \lambda \, v^2$. The vacuum expectation value (vev) acts to break $\rm SU_L(2) \times U_Y(1) \rightarrow U_{em}(1)$ and
is defined as
$\langle H^\dagger \, H \rangle= v^2/2$ in the SM, with
$v \sim 246$~GeV.
The gauge covariant derivative is defined by the states its $\rm SU_c(3) \times SU_L(2) \times U_Y(1)$
generators act on, and we use the conventional form
\bea
D_\mu = \partial_\mu + i g_3 T^A A^A_\mu + i g_2  t^I W^I_\mu + i g_1 {\bf \hyp_i} B_\mu.
\eea
The ${\bf \hyp_i}$ is the $\rm U_Y(1)$ hypercharge generator. The $T^A$ are the
$\rm SU_c(3)$ generators, the Gell-Mann matrices, with normalization ${\rm Tr}(T^A T^B) = 2 \delta^{AB}$.
The $t^I=\tau^I/2$ are the $\rm SU_L(2)$ generators, the Pauli matrices, taken to be
\bea
\tau^1 = \left( \begin{array}{cc}
0 & 1 \\
1 & 0
\end{array} \right), \quad \quad
\tau^2 = \left( \begin{array}{cc}
0 & -i \\
i & 0
\end{array} \right), \quad \quad
\tau^3 = \left( \begin{array}{cc}
1 & 0 \\
0 & -1
\end{array} \right).
\eea
For example, $H$ has hypercharge $\hyp_H=1/2$, is a $\rm SU_c(3)$
singlet and a $\rm SU_L(2)$ doublet so that $D$ acting on $H$ is given by the matrix equation
$D_\mu H = (\partial_\mu +  i g_2  t^I W^I_\mu + i g_1 \, B_\mu/2) H$.
$\rm SU_L(2)$ indices are usually denoted as $\{i,j,k\}$ and $\{I,J,K\}$ in the fundamental and adjoint representations respectively. The $\rm SU_c(3)$ indices
in the adjoint representation are instead denoted as $\{A,B,C\}$, each of which runs from $\{1..8\}$.
$\widetilde H$ is defined by $\widetilde H_j = \epsilon_{jk} H^{\dagger\, k}$
where the $\rm SU_L(2)$ invariant tensor $\epsilon_{jk}$ is defined by $\epsilon_{12}=1$ and $\epsilon_{jk}=-\epsilon_{kj}$, $j,k=\{1,2\}$.

All fermion fields have a suppressed flavour index in Eq.~\ref{sm1}. We conventionally denote these indices by $\{p,r,s,t\}$
that each run over $\{1,2,3\}$ for the three generations.
The fermion mass matrices are $M_{u,d,e}=Y_{u,d,e}\, v /\sqrt 2$.
$Y_{u,d,e}$ and $M_{u,d,e}$ are complex Yukawa matrices in flavour space.
Explicitly reintroducing flavour indices one has
\begin{align}
H^{\dagger j} \, \overline d\, Y_d\, q_{j} &= H^{\dagger j} \, \overline d_p\, [Y_d]_{pr}\, q_{rj}.
\end{align}
The fermion fields $q$ and $\ell$ are left-handed fields, i.e. they transform as $(1/2,0)$ under
the restricted Lorentz group $\rm SO^+(3,1)$.\footnote{Formally the spinors are defined as finite dimensional representations of
the orthochronous Lorentz transformations. The label $(a,b)$ corresponds to the spin of the representation,
where the $\rm SU(2) \times SU(2)$ group that is locally isomorphic to $\rm SO^+(3,1)$ has been used to label the
2 component Weyl spinor subgroups of a four component Dirac spinor.
As the Lorentz group is connected it is related to the
universal cover group $\rm SL(2,{\bf C})$ which is sometimes itself referred to as the Lorentz group in other applications.} The $u$, $d$ and $e$ are right-handed fields and transform as
$(0,1/2)$. The chiral projectors have the convention $\psi_{L/R} = P_{L/R} \, \psi$ where
$P_{R/L} = \left(1 \pm \gamma_5 \right)/2$.
The matter fields of the SM have the charges and representations
\bea\label{SMfields}
\begin{array}{lcccccccccccc}
\hline \hline
{\rm Field} \, & & & \rm SU_c(3) \, & & & \rm SU_L(2) \, & & & \rm U_Y(1) \, & & & \rm SO^+(3,1) \\
\hline
q_i = (u_L^i,d_L^i)^T & & & {\bf 3} & & & {\bf 2} & & & {1/6} & & & (1/2,0) \\
u_i = \left\{u_R,c_R,t_R\right\} & & & {\bf 3} & & & {\bf 1} & & & {2/3} & & & (0,1/2) \\
d_i = \left\{d_R,s_R,b_R\right\} & & & {\bf 3} & & & {\bf 1} & & & {-1/3} & & & (0,1/2) \\
\ell_i = (\nu_L^i,e_L^i)^T & & & {\bf 1} & & & {\bf 2} & & & {-1/2} & & & (1/2,0) \\
e_i = \left\{e_R,\mu_R,\tau_R\right\} & & & {\bf 1} & & & {\bf 1} & & & {-1} & & & (0,1/2) \\
H  & & & {\bf 1} & & & {\bf 2} & & & {1/2} & & & (0,0) \\
\hline
\hline
\end{array}
\eea
The fields in Eq.~\ref{sm1} are in the weak eigenbasis, where
\begin{align}
q_1 &= \left[ \begin{array}{c} u_L \\ d_L^\prime \end{array}\right], &
q_2 &= \left[ \begin{array}{c} c_L \\ s_L^\prime \end{array}\right], &
q_3 &= \left[ \begin{array}{c} t_L \\ b_L^\prime \end{array}\right], &
\ell_1 &= \left[ \begin{array}{c} \nu^{e \, \prime}_L \\ e_L \end{array}\right], &
\ell_2 &= \left[ \begin{array}{c} \nu^{\mu \, \prime}_L \\ \mu_L \end{array}\right], &
\ell_3 &= \left[ \begin{array}{c} \nu^{\tau \, \prime}_L \\ \tau_L \end{array}\right], &
\end{align}

\begin{align}
\left[ \begin{array}{c} d_L^\prime \\ s_L^\prime \\ b_L^\prime \end{array}\right] &=
V_{\rm CKM} \, \left[ \begin{array}{c} d_L \\ s_L \\ b_L \end{array}\right], &
\left[ \begin{array}{c} \nu^{e \, \prime}_L \\ \nu^{\mu \, \prime}_L \\ \nu^{\tau \, \prime}_L \end{array}\right] &=
U_{\rm PMNS} \, \left[ \begin{array}{c} \nu^{e}_L \\ \nu^{\mu}_L \\ \nu^{\tau}_L\end{array}\right].
\end{align}

To date, the SM is a successful EFT for physics at and below the energies probed by LHC $\sqrt{s} \lesssim 3 \, {\rm TeV}$.
The SM does not explain
the experimental evidence for dark matter
\cite{Zwicky:1933gu,Rubin:1970zza,Ade:2015xua}, Baryogenesis \cite{Kuzmin:1985mm,Sakharov:1967dj}
and neutrino masses \cite{Bahcall:1964gx,Davis:1964hf,Fukuda:1998mi,Ahmad:2001an,Ahmad:2002jz}.
These experimental facts, and the theoretical arguments of Section \ref{precision-motivation}, argue for
embedding the SM in a more complete model of fundamental interactions and particles --
of some unknown form.

\subsubsection{The Standard Model EOM}\label{SMEOM}
The SM equations of motion (EOM) play a role in the choice of a SMEFT operator basis, and the removal of redundant operators.
We summarize the well known SM EOM here, which follow from demanding a stationary action $S_{SM} =\int \mathcal{L}_{SM} \, d^4 x$
with respect to a variation due to each SM field. We follow the notation and presentation of Ref.~\cite{Jenkins:2013zja}.
For the Higgs field one has
\begin{align}
D^2 H_k -\lambda v^2 H_k +2 \lambda (H^\dagger H) H_k + \overline q^j\, Y_u^\dagger\, u \epsilon_{jk} + \overline d\, Y_d\, q_k +\overline e\, Y_e\,  l_k
&=0 \,,
\label{eomH}
\end{align}
while the fermion field EOM are given by
\begin{align}
i\slashed{D}\, q_j &= Y_u^\dagger\, u\, \widetilde H_j + Y_d^\dagger\, d\, H_j \,, &
i\slashed{D}\,  d &= Y_d\,  q_j\, H^{\dagger\, j} \,, &
i\slashed{D}\, u &= Y_u\, q_j\, \widetilde H^{\dagger\, j}\,, \nn
i\slashed{D} \, l_j &= Y_e^\dagger\, e  H_j  \,, &
i\slashed{D} \, e &= Y_e\, l_j H^{\dagger\, j}\,,
\label{eompsi}
\end{align}
and the gauge field EOM are given as
\begin{align}
\left[D^\alpha , G_{\alpha \beta} \right]^A &= g_3  j_\beta^A, &
\left[D^\alpha , W_{\alpha \beta} \right]^I &= g_2  j_\beta^I, &
D^\alpha B_{\alpha \beta} &= g_1  j_\beta.
\label{eomX}
\end{align}
Note that $\left[D^\alpha , F_{\alpha \beta} \right]$ is the covariant derivative in the adjoint representation in the notation above.
Hermitian derivative notation is introduced as
\begin{align}
H^\dagger \, i\overleftrightarrow D_\beta H &= i H^\dagger (D_\beta H) - i (D_\beta H)^\dagger H \,, &
\hspace{0.1cm} H^\dagger \, i\overleftrightarrow D_\beta^I H &= i H^\dagger \tau^I (D_\beta H) - i (D_\beta H)^\dagger \tau^I H.
\end{align}
Using this notation, the gauge currents are
\begin{align}
  j_\beta &=\sum_{\psi=u,d,q,e,l}
  \overline \psi \, \hyp_i \gamma_\beta  \psi +
  \frac12 H^\dagger \, i\overleftrightarrow{D}_\beta H, &  \quad
j_\beta^I &= \frac 12 \overline q \, \tau^I \gamma_\beta  q + \frac12 \overline l \, \tau^I \gamma_\beta  l
+ \frac12 H^\dagger \, i\overleftrightarrow D_\beta^I H\,, \nn
j_\beta^A &=\sum_{\psi'=u,d,q} \overline \psi \, T^A \gamma_\beta  \psi.
\end{align}
We use the notation $\psi=\{u,d,q,e,l\}$ to sum over all SM fermions, and $V = \{B,W,G\}$
to sum over the SM gauge fields. Note that these EOM have corrections due to $\mathcal{L}^{(5)}+\mathcal{L}^{(6)} + \cdots$
in the SMEFT, that must be included for a consistent matching to higher orders in the non-perturbative expansion.
Such corrections are also $\mathcal{L}^{(n)}$ basis dependent.

\subsection{The Standard Model Effective Field Theory}\label{SMEFTsec}

The SMEFT is a consistent EFT generalization of the SM
constructed out of a series of $\rm SU_c(3) \times SU_L(2) \times U_Y(1)$ invariant higher dimensional operators, built out of SM fields
and including an $H$ field as defined in Table \ref{SMfields}.
The idea of the SMEFT is that extensions to the SM are assumed to involve massive particles heavier than the measured vev, which sets
the scale (up to coupling suppression) of the SM states. In addition, it is assumed that any
non-perturbative matching effects are characterized by a scale parametrically separated from the EW scale
and the observed Higgs-like boson is embedded in the $\rm SU_L(2)$ Higgs doublet.

The SMEFT follows from these assumptions and
is defined as
\bea
\mathcal{L}_{SMEFT} = \mathcal{L}_{SM} + \mathcal{L}^{(5)} + \mathcal{L}^{(6)} + \mathcal{L}^{(7)} + ...,
\quad \quad \mathcal{L}^{(d)}= \sum_{i = 1}^{n_d} \frac{C_i^{(d)}}{\Lambda^{d-4}} Q_i^{(d)} \hspace{0.25cm} \text{ for $d > 4$.}
\eea
The operators  $Q_i^{(d)}$ are suppressed by $d-4$ powers of the cutoff scale $\Lambda$,
and the $C_i^{(d)}$ are the Wilson coefficients.
The number of non-redundant operators in $\mathcal{L}^{(5)}$, $\mathcal{L}^{(6)}$, $\mathcal{L}^{(7)}$ and $\mathcal{L}^{(8)}$ is
known \cite{Buchmuller:1985jz,Grzadkowski:2010es,Weinberg:1979sa,Wilczek:1979hc,Abbott:1980zj,Lehman:2014jma,Lehman:2015coa,Henning:2015alf}.
Furthermore, the general algorithm to determine operator bases at higher orders developed
in Refs.~\cite{Lehman:2015via,Lehman:2015coa,Henning:2015daa,Henning:2015alf} makes the SMEFT defined to all orders in the expansion
in local operators. Note that when transitioning to the SMEFT,
symmetry arguments leading to a neglect of dual field strength terms in $\mathcal{L}_{SM}$ should be reformulated, as such terms multiplied
by $(H^\dagger H)$ appear in $\mathcal{L}_6$. The dual field strength terms should not be casually neglected.

In the SMEFT $\rm SU_L(2) \times U_Y(1) \rightarrow \rm U(1)_{em}$ is Higgsed as in the SM.
The minimum of the Higgs potential is now determined including the effect of
the operator $Q_H \equiv \left(H^\dagger H \right)^3 $, which modifies the scalar doublet potential to the form \cite{Alonso:2013hga}
\begin{align}
V(H^\dagger H) &= \lambda \left(H^\dagger H -\frac12 v^2\right)^2 - C_H \left( H^\dagger H \right)^3,
\label{pot}
\end{align}
yielding the new minimum
\begin{align}
\langle H^\dagger H \rangle &= \frac{v^2}{2} \left( 1+ \frac{3 C_H v^2}{4 \lambda} \right) \equiv \frac12 v_T^2,
\end{align}
on expanding the exact solution $(\lambda- \sqrt{\lambda^2-3 C_H \lambda v^2})/(3 C_H)$ to first order in $C_H$.
This expansion assumes a mass gap to the
scale(s) of new physics (referred to schematically as $\Lambda$) which leads to the expansion parameter $\sim C_H \, \bar{v}_T^2/\Lambda^2 < 1$.
The dependence on $\Lambda$ was suppressed in the previous equations. We absorb the cut off scale into the Wilson coefficients
as a notational
choice unless otherwise noted.

The SMEFT is an enormously powerful consistent field theory to use to characterize the low energy limit
of physics beyond the SM. Even if a full model extension of the SM becomes experimentally supported in the future,
the SMEFT can still be a useful and appropriate tool to use to interface with large swaths of experimental data
below the characteristic scale(s) $\Lambda$ of a new physics sector.  It cannot be emphasized too strongly that the systematic
development of this framework is expected to have an important return on investment of the time expended on it.
The payoff in terms of improved
scientific conclusions being enabled from the ever growing data set of measurements of SM states below
the scale $\Lambda$ is clear. This payoff can be starkly contrasted to the return on time invested when developing
the predictions of a particular model, or even a set of models, if the many assumptions of the model are
not experimentally validated. Considering the current global data set of particle physics,
adopting the IR assumptions that define the SMEFT seems to be a very reasonable compromise
between utility and generality of the theoretical framework assumed to accommodate the
certain fact that the SM is an incomplete description of reality, while the LHC data set is indicating some degree of decoupling
is present to the scales involved with extending the SM.

\subsubsection{Operator bases in the SMEFT}\label{SMEFTbasis}
In this work we use the first\footnote{This basis was built using the foundation laid down in Refs.~\cite{Buchmuller:1985jz,Leung:1984ni,AguilarSaavedra:2010zi}.}
non-redundant operator basis for $\mathcal{L}^{(6)}$ determined in the literature, as given in Ref.~\cite{Grzadkowski:2010es}.
This construction has come to be known as the ``Warsaw basis''. To fix notation we include the complete summary of the
baryon number conserving operators in this basis in Table \ref{op59}, defined using notational conventions consistent with the previous sections.

The development of the Warsaw basis \cite{Grzadkowski:2010es} underlies the systematic development of the SMEFT in recent years.
It is a surprising fact that the full reduction of an operator basis for $\mathcal{L}^{(6)}$ to a non-redundant form took to 2010,
a full twenty four years after Ref.~\cite{Buchmuller:1985jz} was published.
%--------------------------------------------------------------------------------------------
\begin{table}[!h]
\begin{center}
\small
\begin{minipage}[t]{4.45cm}
\renewcommand{\arraystretch}{1.5}
\begin{tabular}[t]{c|c}
\multicolumn{2}{c}{$1:X^3$} \\
\hline
$Q_G$                & $f^{ABC} G_\mu^{A\nu} G_\nu^{B\rho} G_\rho^{C\mu} $ \\
$Q_{\widetilde G}$          & $f^{ABC} \widetilde G_\mu^{A\nu} G_\nu^{B\rho} G_\rho^{C\mu} $ \\
$Q_W$                & $\epsilon^{IJK} W_\mu^{I\nu} W_\nu^{J\rho} W_\rho^{K\mu}$ \\
$Q_{\widetilde W}$          & $\epsilon^{IJK} \widetilde W_\mu^{I\nu} W_\nu^{J\rho} W_\rho^{K\mu}$ \\
\end{tabular}
\end{minipage}
\begin{minipage}[t]{2.7cm}
\renewcommand{\arraystretch}{1.5}
\begin{tabular}[t]{c|c}
\multicolumn{2}{c}{$2:H^6$} \\
\hline
$Q_H$       & $(H^\dag H)^3$
\end{tabular}
\end{minipage}
\begin{minipage}[t]{5.1cm}
\renewcommand{\arraystretch}{1.5}
\begin{tabular}[t]{c|c}
\multicolumn{2}{c}{$3:H^4 D^2$} \\
\hline
$Q_{H\Box}$ & $(H^\dag H)\Box(H^\dag H)$ \\
$Q_{H D}$   & $\ \left(H^\dag D_\mu H\right)^* \left(H^\dag D_\mu H\right)$
\end{tabular}
\end{minipage}
\begin{minipage}[t]{2.7cm}

\renewcommand{\arraystretch}{1.5}
\begin{tabular}[t]{c|c}
\multicolumn{2}{c}{$5: \psi^2H^3 + \hbox{h.c.}$} \\
\hline
$Q_{eH}$           & $(H^\dag H)(\bar l_p e_r H)$ \\
$Q_{uH}$          & $(H^\dag H)(\bar q_p u_r \widetilde H )$ \\
$Q_{dH}$           & $(H^\dag H)(\bar q_p d_r H)$\\
\end{tabular}
\end{minipage}

\vspace{0.25cm}

\begin{minipage}[t]{4.7cm}
\renewcommand{\arraystretch}{1.5}
\begin{tabular}[t]{c|c}
\multicolumn{2}{c}{$4:X^2H^2$} \\
\hline
$Q_{H G}$     & $H^\dag H\, G^A_{\mu\nu} G^{A\mu\nu}$ \\
$Q_{H\widetilde G}$         & $H^\dag H\, \widetilde G^A_{\mu\nu} G^{A\mu\nu}$ \\
$Q_{H W}$     & $H^\dag H\, W^I_{\mu\nu} W^{I\mu\nu}$ \\
$Q_{H\widetilde W}$         & $H^\dag H\, \widetilde W^I_{\mu\nu} W^{I\mu\nu}$ \\
$Q_{H B}$     & $ H^\dag H\, B_{\mu\nu} B^{\mu\nu}$ \\
$Q_{H\widetilde B}$         & $H^\dag H\, \widetilde B_{\mu\nu} B^{\mu\nu}$ \\
$Q_{H WB}$     & $ H^\dag \tau^I H\, W^I_{\mu\nu} B^{\mu\nu}$ \\
$Q_{H\widetilde W B}$         & $H^\dag \tau^I H\, \widetilde W^I_{\mu\nu} B^{\mu\nu}$
\end{tabular}
\end{minipage}
\begin{minipage}[t]{5.2cm}
\renewcommand{\arraystretch}{1.5}
\begin{tabular}[t]{c|c}
\multicolumn{2}{c}{$6:\psi^2 XH+\hbox{h.c.}$} \\
\hline
$Q_{eW}$      & $(\bar l_p \sigma^{\mu\nu} e_r) \tau^I H W_{\mu\nu}^I$ \\
$Q_{eB}$        & $(\bar l_p \sigma^{\mu\nu} e_r) H B_{\mu\nu}$ \\
$Q_{uG}$        & $(\bar q_p \sigma^{\mu\nu} T^A u_r) \widetilde H \, G_{\mu\nu}^A$ \\
$Q_{uW}$        & $(\bar q_p \sigma^{\mu\nu} u_r) \tau^I \widetilde H \, W_{\mu\nu}^I$ \\
$Q_{uB}$        & $(\bar q_p \sigma^{\mu\nu} u_r) \widetilde H \, B_{\mu\nu}$ \\
$Q_{dG}$        & $(\bar q_p \sigma^{\mu\nu} T^A d_r) H\, G_{\mu\nu}^A$ \\
$Q_{dW}$         & $(\bar q_p \sigma^{\mu\nu} d_r) \tau^I H\, W_{\mu\nu}^I$ \\
$Q_{dB}$        & $(\bar q_p \sigma^{\mu\nu} d_r) H\, B_{\mu\nu}$
\end{tabular}
\end{minipage}
\begin{minipage}[t]{5.4cm}
\renewcommand{\arraystretch}{1.5}
\begin{tabular}[t]{c|c}
\multicolumn{2}{c}{$7:\psi^2H^2 D$} \\
\hline
$Q_{H l}^{(1)}$      & $(H^\dag i\overleftrightarrow{D}_\mu H)(\bar l_p \gamma^\mu l_r)$\\
$Q_{H l}^{(3)}$      & $(H^\dag i\overleftrightarrow{D}^I_\mu H)(\bar l_p \tau^I \gamma^\mu l_r)$\\
$Q_{H e}$            & $(H^\dag i\overleftrightarrow{D}_\mu H)(\bar e_p \gamma^\mu e_r)$\\
$Q_{H q}^{(1)}$      & $(H^\dag i\overleftrightarrow{D}_\mu H)(\bar q_p \gamma^\mu q_r)$\\
$Q_{H q}^{(3)}$      & $(H^\dag i\overleftrightarrow{D}^I_\mu H)(\bar q_p \tau^I \gamma^\mu q_r)$\\
$Q_{H u}$            & $(H^\dag i\overleftrightarrow{D}_\mu H)(\bar u_p \gamma^\mu u_r)$\\
$Q_{H d}$            & $(H^\dag i\overleftrightarrow{D}_\mu H)(\bar d_p \gamma^\mu d_r)$\\
$Q_{H u d}$ + h.c.   & $i(\widetilde H ^\dag D_\mu H)(\bar u_p \gamma^\mu d_r)$\\
\end{tabular}
\end{minipage}

\vspace{0.25cm}

\begin{minipage}[t]{4.75cm}
\renewcommand{\arraystretch}{1.5}
\begin{tabular}[t]{c|c}
\multicolumn{2}{c}{$8:(\bar LL)(\bar LL)$} \\
\hline
$Q_{\ell \ell}$        & $(\bar l_p \gamma_\mu l_r)(\bar l_s \gamma^\mu l_t)$ \\
$Q_{qq}^{(1)}$  & $(\bar q_p \gamma_\mu q_r)(\bar q_s \gamma^\mu q_t)$ \\
$Q_{qq}^{(3)}$  & $(\bar q_p \gamma_\mu \tau^I q_r)(\bar q_s \gamma^\mu \tau^I q_t)$ \\
$Q_{\ell q}^{(1)}$                & $(\bar l_p \gamma_\mu l_r)(\bar q_s \gamma^\mu q_t)$ \\
$Q_{\ell q}^{(3)}$                & $(\bar l_p \gamma_\mu \tau^I l_r)(\bar q_s \gamma^\mu \tau^I q_t)$
\end{tabular}
\end{minipage}
\begin{minipage}[t]{5.25cm}
\renewcommand{\arraystretch}{1.5}
\begin{tabular}[t]{c|c}
\multicolumn{2}{c}{$8:(\bar RR)(\bar RR)$} \\
\hline
$Q_{ee}$               & $(\bar e_p \gamma_\mu e_r)(\bar e_s \gamma^\mu e_t)$ \\
$Q_{uu}$        & $(\bar u_p \gamma_\mu u_r)(\bar u_s \gamma^\mu u_t)$ \\
$Q_{dd}$        & $(\bar d_p \gamma_\mu d_r)(\bar d_s \gamma^\mu d_t)$ \\
$Q_{eu}$                      & $(\bar e_p \gamma_\mu e_r)(\bar u_s \gamma^\mu u_t)$ \\
$Q_{ed}$                      & $(\bar e_p \gamma_\mu e_r)(\bar d_s\gamma^\mu d_t)$ \\
$Q_{ud}^{(1)}$                & $(\bar u_p \gamma_\mu u_r)(\bar d_s \gamma^\mu d_t)$ \\
$Q_{ud}^{(8)}$                & $(\bar u_p \gamma_\mu T^A u_r)(\bar d_s \gamma^\mu T^A d_t)$ \\
\end{tabular}
\end{minipage}
\begin{minipage}[t]{4.75cm}
\renewcommand{\arraystretch}{1.5}
\begin{tabular}[t]{c|c}
\multicolumn{2}{c}{$8:(\bar LL)(\bar RR)$} \\
\hline
$Q_{le}$               & $(\bar l_p \gamma_\mu l_r)(\bar e_s \gamma^\mu e_t)$ \\
$Q_{lu}$               & $(\bar l_p \gamma_\mu l_r)(\bar u_s \gamma^\mu u_t)$ \\
$Q_{ld}$               & $(\bar l_p \gamma_\mu l_r)(\bar d_s \gamma^\mu d_t)$ \\
$Q_{qe}$               & $(\bar q_p \gamma_\mu q_r)(\bar e_s \gamma^\mu e_t)$ \\
$Q_{qu}^{(1)}$         & $(\bar q_p \gamma_\mu q_r)(\bar u_s \gamma^\mu u_t)$ \\
$Q_{qu}^{(8)}$         & $(\bar q_p \gamma_\mu T^A q_r)(\bar u_s \gamma^\mu T^A u_t)$ \\
$Q_{qd}^{(1)}$ & $(\bar q_p \gamma_\mu q_r)(\bar d_s \gamma^\mu d_t)$ \\
$Q_{qd}^{(8)}$ & $(\bar q_p \gamma_\mu T^A q_r)(\bar d_s \gamma^\mu T^A d_t)$\\
\end{tabular}
\end{minipage}

\vspace{0.25cm}

\begin{minipage}[t]{4.7cm}
\renewcommand{\arraystretch}{1.5}
\begin{tabular}[t]{c|c}
\multicolumn{2}{c}{$8:(\bar LR)(\bar RL)+\hbox{h.c.}$} \\
\hline
$Q_{ledq}$ & $(\bar l_p^j e_r)(\bar d_s q_{tj})$
\end{tabular}
\end{minipage}
\begin{minipage}[t]{5.2cm}
\renewcommand{\arraystretch}{1.5}
\begin{tabular}[t]{c|c}
\multicolumn{2}{c}{$8:(\bar LR)(\bar L R)+\hbox{h.c.}$} \\
\hline
$Q_{quqd}^{(1)}$ & $(\bar q_p^j u_r) \epsilon_{jk} (\bar q_s^k d_t)$ \\
$Q_{quqd}^{(8)}$ & $(\bar q_p^j T^A u_r) \epsilon_{jk} (\bar q_s^k T^A d_t)$
\end{tabular}
\end{minipage}
\begin{minipage}[t]{5.4cm}
\renewcommand{\arraystretch}{1.5}
\begin{tabular}[t]{c|c}
\multicolumn{2}{c}{$8:(\bar LR)(\bar L R)+\hbox{h.c.}$} \\
\hline
$Q_{lequ}^{(1)}$ & $(\bar l_p^j e_r) \epsilon_{jk} (\bar q_s^k u_t)$ \\
$Q_{lequ}^{(3)}$ & $(\bar l_p^j \sigma_{\mu\nu} e_r) \epsilon_{jk} (\bar q_s^k \sigma^{\mu\nu} u_t)$
\caption{\label{op59}$\mathcal{L}_6$ of Refs.~\cite{Grzadkowski:2010es} as given in Ref.~\cite{Alonso:2013hga}. The flavour labels $p,r,s,t$ on the $Q$ operators are suppressed on the left hand side of
the tables.}
\end{tabular}
\end{minipage}
\end{center}
\end{table}
%--------------------------------------------------------------------------------------------
Preceding the development of
the Warsaw basis, innumerable works employed
subsets of operators to perform phenomenological studies of the low energy effects of
various models. Examples of this form of analysis include
Refs.~\cite{Hagiwara:1993ck,Arzt:1994gp,Gounaris:1998ni,Manohar:2006gz,Manohar:2006ga,Barger:2003rs,Giudice:2007fh,Grinstein:2007iv}.
In the literature, the Buchm\"uller and Wyler result \cite{Buchmuller:1985jz} is frequently referred to as an
operator basis, we note that it is fully specified and well defined, but overcomplete, unlike the Warsaw basis.
In addition the SILH subset of operators \cite{Giudice:2007fh} is sometimes referred to as an operator basis in some literature,
as is the HISZ subset of operators \cite{Hagiwara:1993ck}. This is perhaps due to the
influential nature of these works. To avoid confusion we stress that neither of these works
contains a complete set of $\mathcal{L}^{(6)}$ operators, and certainly not a well defined minimal non-redundant basis.
Subtleties involving flavour indices are present in extending the subset of
operators in Ref.~\cite{Giudice:2007fh} to a full basis, see Ref.~\cite{Alonso:2013hga} for a discussion.
As a result, the emergence of a full basis including these ``SILH operators" required some
further efforts to resolve these issues ~\cite{Contino:2013kra,Elias-Miro:2013eta}.
A full basis was defined in Ref.~\cite{Elias-Miro:2013eta} for the first time incorporating these flavour index subtleties.

\subsubsection{Removing redundant operators}\label{redundant}
Ref.~\cite{Leung:1984ni} reported a complete set of operators that satisfies $\rm SU_c(3) \times SU_L(2) \times U_Y(1)$ symmetry. This
was the first step in constructing an $\mathcal{L}_6$ operator basis, but
such a construction is overcomplete. Combinations of operators are present, whose Wilson coefficients
vanish when observables are calculated. This is due to the EOM
relating the field variables when external states go on-shell. Making small field redefinitions on the SM fields $\mathcal{O}(1/\Lambda^2)$
one can remove
the combination of $\mathcal{L}_6$ operators that will vanish in this manner directly in the Lagrangian, instead of
having the cancellation occur when the $S$ matrix element is constructed.\footnote{
This procedure can be understood heuristically as aligning the field variables in the SMEFT more directly with
classical (asymptotic) poles and the residues of the propagators at these poles, when the external particles go on-shell.}
When this is done, the SM fields have a meaning that is contextual in the SMEFT and are fixed in a particular operator basis with a chosen set of
$\mathcal{L}_6$ operators.
The set of $\mathcal{O}(1/\Lambda^2)$ bosonic field redefinitions that preserve $\rm SU_c(3) \times SU_L(2) \times U_Y(1)$ are\footnote{Here we restrict ourselves
to field redefinitions that have only dynamical field content, neglecting explicit $v^2/\Lambda^2$ corrections.}
\bea\label{set1redefine}
H_j' &\rightarrow& H_j + h_1 \frac{D^2 H_j}{\Lambda^2} + h_2 \frac{\bar{e} \, \ell_j \, Y_e}{\Lambda^2}
+ h_3 \frac{\bar{d} \, q_j \, Y_d}{\Lambda^2} + h_4 \frac{(\bar{u}  \epsilon \, q_j)^\star \, Y_u^\star}{\Lambda^2}
+ h_5 \frac{H^\dagger H \, H_j}{\Lambda^2}, \\
B'_\mu &\rightarrow& B_\mu + b_1 \frac{\bar{\psi} \gamma_\mu \psi}{\Lambda^2} + b_2 \frac{H^\dagger \, i\overleftrightarrow D_\mu H}{\Lambda^2}
+ b_3 \frac{D^\alpha B_{\alpha \mu}}{\Lambda^2} + b_4 \,\frac{H^\dagger H \, B_\mu}{\Lambda^2},\\
W^{I '}_\mu &\rightarrow& W^I_\mu + w_1 \frac{\bar{q} \sigma^I \gamma_\mu q}{\Lambda^2}
+ w_2 \frac{\bar{\ell} \sigma^I \gamma_\mu \ell}{\Lambda^2}
+ w_3 \frac{H^\dagger \overleftrightarrow D^I_\mu H}{\Lambda^2}
+ w_4 \frac{\left[D^\alpha , W_{\alpha \mu} \right]^I}{\Lambda^2} + w_5 \frac{H^\dagger H W^I_\mu}{\Lambda^2}, \\
G^{A \, '}_\mu &\rightarrow& G^A_\mu + \mathpzc{g}_1 \frac{\bar{q} T^A \gamma_\mu q}{\Lambda^2}
+ \mathpzc{g}_2 \frac{\bar{d} T^A \gamma_\mu d}{\Lambda^2}
+ \mathpzc{g}_3 \frac{\bar{u} T^A \gamma_\mu u}{\Lambda^2}
+ \mathpzc{g}_4 \frac{\left[D^\alpha , G_{\alpha \mu} \right]^A}{\Lambda^2} + \mathpzc{g}_5  \,\frac{H^\dagger H \, G^A_\mu}{\Lambda^2},
\eea
while the corresponding transformations on the right handed fermion fields are
\bea\label{set2redefine}
e' &\rightarrow& e + e_1 \frac{\bar{\ell} i\slashed{D} H Y_e^\dagger}{\Lambda^2}+ e_2 \frac{\bar{\ell} i \overleftarrow{\slashed{D}} H Y_e^\dagger}{\Lambda^2}
+ e_3 \frac{H^\dagger H e}{\Lambda^2} + e_4 \frac{D^2 e}{\Lambda^2},\\
d' &\rightarrow& d + d_1 \frac{\bar{q} i\slashed{D} H Y_d^\dagger}{\Lambda^2}+ d_2 \frac{\bar{q} i \overleftarrow{\slashed{D}} H Y_d^\dagger}{\Lambda^2}
+ d_3 \frac{H^\dagger H d}{\Lambda^2} + d_4 \frac{D^2 d}{\Lambda^2},\\
u' &\rightarrow& u + u_1 \frac{\bar{q} i\slashed{D} \tilde{H} Y_u^\dagger}{\Lambda^2}
+ u_2 \frac{\bar{q} i \overleftarrow{\slashed{D}} \tilde{H} Y_u^\dagger}{\Lambda^2}
+ u_3 \frac{H^\dagger H u}{\Lambda^2} + u_4 \frac{D^2 u}{\Lambda^2},
\eea
and finally the redefinitions of the left handed fermion fields are
\bea\label{set3redefine}
q'_j &\rightarrow& q_j + \mathpzc{q}_1 \frac{u i\slashed{D} \tilde{H}_j Y_u^\dagger}{\Lambda^2}
+ \mathpzc{q}_2 \frac{u i \overleftarrow{\slashed{D}} \tilde{H}_j Y_u^\dagger}{\Lambda^2}
+ \mathpzc{q}_3 \frac{d i\slashed{D} H _j Y_d^\dagger}{\Lambda^2}
+ \mathpzc{q}_4 \frac{d i \overleftarrow{\slashed{D}} H_j Y_d^\dagger}{\Lambda^2}
+ \mathpzc{q}_5 \frac{H^\dagger H q_j}{\Lambda^2} + \mathpzc{q}_4 \frac{D^2 q_j}{\Lambda^2}, \nn \\
\ell'_j &\rightarrow& \ell + l_1 \frac{e i\slashed{D} H_j Y_e^\dagger}{\Lambda^2}
+ l_2 \frac{e i \overleftarrow{\slashed{D}} H_j Y_e^\dagger}{\Lambda^2}
+ l_3 \frac{H^\dagger H \ell_j}{\Lambda^2} + l_4 \frac{D^2 \ell_j}{\Lambda^2}.
\eea
Here $\{h_a,b_a,w_a,e_a,d_a,u_a,\mathpzc{q}_a,l_a\}$ are free variables.
Performing field redefinitions with only a single $\mathcal{O}(1/\Lambda^2)$
term on the right hand side of each equation one can choose to cancel an operator out of a full set
of operators reported in Ref.~\cite{Leung:1984ni}. This
is known as removing redundant operators.
An example of this procedure is as follows.
The $B^\mu$ dependent, flavour symmetric terms in an overcomplete $\mathcal{L}_{SMEFT}$ are
\bea\label{Bprimeequation}
\mathcal{L}_{B'} &=& -\frac14 B'_{\mu \nu} B^{'\mu \nu}
- g_1 \,  \hyp_{\psi} \, \overline \psi\, \slashed{B'} \, \psi +
(D^\mu H)^\dagger (D_\mu H) + \mathcal{C}_{B}(H^\dagger \, \overleftrightarrow{D}^\mu H) (D^\nu B_{\mu \, \nu}),\nn
&+& \mathcal{C}_{BH} (D^\mu H)^\dagger \,  (D^\nu H) \,  B'_{\mu \, \nu}
+ C_{\substack{H l \\ tt}}^{(1)} Q_{\substack{H l \\ tt}}^{(1)} +
C_{\substack{H e \\ tt}} \, Q_{\substack{H e \\ tt}} + C_{\substack{H q \\ tt}}^{(1)} Q_{\substack{H q \\ tt}}^{(1)}
+ C_{\substack{H u \\ tt}} \, Q_{\substack{H u \\ tt}}, \nn
&+& C_{\substack{H d \\ tt}} \, Q_{\substack{H d \\ tt}}
+ C_{HB} \, Q_{HB} + C_{T} \, (H^\dagger \, \overleftrightarrow{D}^\mu H) \,  (H^\dagger \, \overleftrightarrow{D}^\mu H).
\eea
Performing the small field redefinition
\bea
B'_\mu &\rightarrow& B_\mu + b_2 \frac{H^\dagger \, i\overleftrightarrow D_\mu H}{\Lambda^2},
\eea
yields the result $\mathcal{L} _{B} -g_1 \, b_2 \Delta B$ where
\bea\label{EOMBshift}
\Delta B &=&
\hyp_l Q_{\substack{H l \\ tt}}^{(1)} +\hyp_e Q_{\substack{H e \\ tt}}
+ \hyp_q Q_{\substack{H q \\ tt}}^{(1)}+\hyp_u Q_{\substack{H u \\ tt}} + \hyp_d Q_{\substack{H d \\ tt}}, \nn
&+& \hyp_H \left(Q_{H \Box} +4 \, Q_{H D} \right)
+ \frac{1}{g_1}  B^{\mu \nu} \partial_\mu (H^\dagger i \overleftrightarrow{D}_\nu H).
\eea
Choosing $b_2$ to cancel one of the $\mathcal{L}_{B'}$ operators introduces
a shift in the Wilson coefficients of the remaining operators.
When the full set of such field redefinitions has been performed,
this corresponds to choosing a non-redundant basis.
The removal of redundant operators is always done in a gauge independent manner,
as this procedure is only justified by the invariance of observables (i.e. $S$ matrix elements) under
gauge independent field redefinitions.\footnote{See
Section \ref{Smatrix} for more details.}

Many field redefinitions are possible and
Eq.~\ref{set1redefine}-\ref{set3redefine} can introduce or remove the same operators.
It is essential to have a gauge independent algorithm to employ to systematically remove operator forms
to obtain a minimal non-redundant basis. In the Warsaw basis, the algorithm is the systematic removal of derivative operators
and an equally careful application of
Fierz identities to reduce out redundant four fermion operators (see also Ref.~\cite{AguilarSaavedra:2010zi}).
This reflects the approach of an on-shell EFT construction \cite{Gasser:1983yg,Georgi:1991ch}, so named because
the on-shell EOM are used to reduce out explicit
factors of $D^2 H$ and $\slashed{D} \psi$ in the higher order terms.
The derivative removing algorithm of the Warsaw basis was used to help develop
the results defining higher order corrections in the SMEFT operator expansion
reported in Refs.~\cite{Lehman:2015via,Lehman:2015coa,Henning:2015daa,Henning:2015alf}.
Building on past works enumerating flavour invariants in the SM \cite{Jenkins:2009dy,Hanany:2010vu}
and group invariants in SUSY theories \cite{Pouliot:1998yv} it has been found to be beneficial to
employ Hilbert series and a conformal algebra to systematize the counting of the number of
operators at even higher orders than $\mathcal{L}_6$ in the SMEFT, while systematically removing derivative operators
as in the Warsaw basis construction. Finally, the Warsaw basis algorithm was also essential to enabling
the one loop renormalization of the operators
in $\mathcal{L}_6$, that was developed in Refs.~\cite{Grojean:2013kd,Jenkins:2013zja,Jenkins:2013wua,Alonso:2013hga,Alonso:2014zka},
as discussed in Section~\ref{onelooprunning}.

Although the Warsaw basis removes derivative terms systematically,
not all of the derivative invariants acting on $H$ can be removed with such $\mathcal{O}(1/\Lambda^2)$
field redefinitions. For example, $\Box H$ does not appear in Eq.~\ref{set1redefine} as it does not satisfy
$\rm SU_L(2)_L \times U_Y(1)$ invariance as $H$ is not a singlet.
The distinction between $D^2 H$ and $\Box H$ is a
way to understand the relevance of the scalar manifold topology in determining what
derivative terms can be removed.\footnote{It is possible to misunderstand Ref.~\cite{Georgi:1991ch} on this point
as only a singlet scalar field is discussed in detail.}
Defining the doublet field $H$ to be decomposed into the real scalar fields $\vec{\phi}^T = \{\phi_1,\phi_2,\phi_3,\phi_4\}$ as
\bea
H = \left(\begin{array}{cc}
\phi_2 +i \phi_1 \\
\phi_4  - i \phi_3
\end{array}\right),
\eea
the Lagrangian derivative terms can be expressed as
\bea
\mathcal{L}_{derv} = \frac{1}{2} (\partial_\mu \vec{\phi}) \, \cdot \,  (\partial^\mu \vec{\phi})
+ \frac{C_{H\Box}}{\Lambda^2} \, \vec{\phi}^2 \Box \vec{\phi}^2 + \frac{C_{HD}}{\Lambda^2} (\vec{\phi} \cdot  (\partial^\mu \vec{\phi}))^2 + \cdots
\eea
This defines a target space metric for the scalar manifold that acts on $\partial^\mu \phi_i \partial_\mu \phi_j/2$ as
\bea
R_{ij} = \delta_{ij} + 2 \frac{\phi_i \, \phi_j}{\Lambda^2} (C_{HD} - 4 C_{H \Box}) + \cdots
\eea
The Riemann tensor $R^i_{jkl}$ associated with this manifold can be directly determined from $R_{ij}$,
and it does not vanish \cite{Burgess:2010zq,Alonso:2015fsp,Alonso:2016oah}.
Since the target space metric is not flat, there does not exist a field redefinition
which everywhere sets $R_{ij} = \delta_{ij}$. Not all of the $H$ self interaction
derivative terms can be removed with a gauge independent field redefinition as a result.

When expanding around the vev in unitary gauge, a distinction between $D^2 h$ and  $\Box h$ is absent.
As expected, the on-shell effective field theory approach of removing
all $p^2$ invariants for $h$ can be achieved with a {\it gauge dependent} field redefinition, and
$h$ can be canonically normalized for calculations with \cite{Alonso:2013hga,Hartmann:2015aia}
\bea
h  \rightarrow h\Big( 1 + (C_{H\Box}  - \frac{1}{4} C_{HD}) \bar{v}_T^2  \Big(1 + \frac{h}{\bar{v}_T} + \frac{h^2}{3 \bar{v}_T^2}  \Big) \Big).
\eea
This distinction between the ability to remove derivative terms on $h$ and not $H$
is problematic as $\rm SU_L(2) \times U_Y(1)$ is Higgsed,
and one uses $ \rm SU_L(2)$ global symmetry rotations to rotate the theory
to a form where only the $h$ field takes on a vev. This is also done when including corrections in the EFT. If this symmetry is broken
by assumption, or an ill defined Lagrangian convention choice, a mismatch between the vev and the fluctuations around the vev can be introduced
in a gauge dependent manner. This in turn can render a parameterization of new physics effects intrinsically
gauge dependent and unsuitable to use in the event that real (gauge independent) deviations from the
SM are found using EFT techniques. Such a parameterization can break structures in the EFT intrinsic to
its construction and well defined nature, leading to stronger
constraints that are misleading on the Wilson coefficient space. See Section \ref{subsec:reparam} for a discussion on
such a structure - that is known as a SMEFT reparameterization invariance, which when
not broken by assumption, or an inconsistent parameterization, requires that combination of data
sets be used to lift degeneracies between parameters. This issue can change bounds on
Wilson coefficients by orders of magnitude \cite{Berthier:2016tkq}, so great care should be taken to
avoid introducing inconsistencies of this form in SMEFT studies.

In addition to preforming the field redefinitions in Eqs.~\ref{set1redefine}-\ref{set3redefine}
one can directly translate between operator bases once a minimal non-redundant basis is found.
This is done by constructing $\mathcal{L}_6$ operator relations directly
out of the SM EOM's in Section~\ref{SMEOM} and then employing them to transition between bases.
As illustrated by Eq.~\ref{EOMBshift}
there is a gauge independent field redefinition that underlies the EOM relations
in Eq.~\ref{pops}.\footnote{We explicitly separate out the integration by parts identity $P_T = - Q_{H \Box} -4 Q_{H D}$.}
An example of these relationships is between the flavour singlet operator forms appearing in Ref.~\cite{Giudice:2007fh}
\begin{align}
\mathcal{P}_{HW} &=  -i \, g_2 \, (D^\mu H)^\dagger \, \tau^I \, (D^\nu H) \,  W^I_{\mu \, \nu}, & \hspace{1cm}
\mathcal{P}_{HB} &=  -i \, g_1 \, (D^\mu H)^\dagger \,  (D^\nu H) \,  B_{\mu \, \nu},\nn
\mathcal{P}_{W} &=  -\frac{i \, g_2}{2} \, (H^\dagger \,  \tau^I \, \overleftrightarrow{D}^\mu H) \,  (D^\nu W^I_{\mu \, \nu}), & \hspace{1cm}
\mathcal{P}_{B} &=  -\frac{i \, g_1}{2} \, (H^\dagger \, \overleftrightarrow{D}^\mu H) \,  (D^\nu B_{\mu \, \nu}), \nn
\mathcal{P}_{T} &=  (H^\dagger \, \overleftrightarrow{D}^\mu H) \,  (H^\dagger \, \overleftrightarrow{D}^\mu H),
\label{SILHY}
\end{align}
and the Warsaw basis operators in Table ~\ref{op59}, given by \cite{Alonso:2013hga}
\begin{align}
\mathcal{P}_{B}  &=
\frac12 \hyp_H  g_1^2 Q_{H \Box} +2g_1^2 \hyp_H Q_{H D}  + \frac12  g_1^2\left[\hyp_l Q_{\substack{H l \\ tt}}^{(1)} +\hyp_e Q_{\substack{H e \\ tt}}
+ \hyp_q Q_{\substack{H q \\ tt}}^{(1)}+\hyp_u Q_{\substack{H u \\ tt}}+ \hyp_d Q_{\substack{H d \\ tt}} \right],\nn
\mathcal{P}_{W} &= \frac34 g_2^2 Q_{H \Box}-\frac12 g_2^2 m_H^2 (H^\dagger H)^2 +2 g_2^2 \lambda Q_H +  \frac14  g_2^2\left[Q_{\substack{H l \\ tt}}^{(3)} + Q_{\substack{H q \\ tt}}^{(3)}\right] \nn
&+\frac12 g_2^2\left( [Y_u^\dagger]_{rs} Q_{\substack{ uH \\ rs}} + [Y_d^\dagger]_{rs} Q_{\substack{ dH \\ rs}}+ [Y_e^\dagger]_{rs} Q_{\substack{ eH \\ rs}}+h.c. \right),
\nn
\mathcal{P}_{HB}  &= \frac12 g_1^2 \hyp_H Q_{H \Box}  +2g_1^2 \hyp_H Q_{H D} - \frac12 \hyp_H g_1^2 Q_{H B} - \frac14 g_1g_2  Q_{H WB}, \nn
& +\frac12  g_1^2\left[\hyp_l Q_{\substack{H l \\ tt}}^{(1)} +\hyp_e Q_{\substack{H e \\ tt}} + \hyp_q Q_{\substack{H q \\ tt}}^{(1)}+\hyp_u Q_{\substack{H u \\ tt}}+ \hyp_d Q_{\substack{H d \\ tt}} \right] ,\nn
\mathcal{P}_{HW} &= \frac34 g_2^2 Q_{H \Box} -\frac12 g_2^2 m_H^2 (H^\dagger H)^2 +2 g_2^2 \lambda Q_H -\frac14 g_2^2 Q_{H W}-\frac12 \hyp_H g_1 g_2  Q_{H WB} + \frac14  g_2^2\left[Q_{H l}^{(3)} + Q_{H q}^{(3)} \right]
\nn
&+\frac12 g_2^2\left( [Y_u^\dagger]_{rs} Q_{\substack{ uH \\ rs}} + [Y_d^\dagger]_{rs} Q_{\substack{ dH \\ rs}}+ [Y_e^\dagger]_{rs} Q_{\substack{ eH \\ rs}}+h.c. \right).
\label{pops}
\end{align}
Note that the operators in Eq.~\ref{SILHY} do not carry flavour indices while the operators in
Eq.~\ref{pops} do carry flavour indices.
One needs to define the flavour indices of the operators
removed when changing basis in order to avoid ambiguities \cite{Alonso:2013hga}.
Respecting flavour symmetry is the reason Yukawa matrices appear
in Eqs.~\ref{set1redefine}-\ref{set3redefine} instead of arbitrary flavour matrices.

\subsubsection{Ad-hoc phenomenological Lagrangians}\label{adhoc}
There is a distinction between the concept of an operator basis in the SMEFT
and an incomplete ad-hoc phenomenological Lagrangian
used to characterize a subset of some SM deviations. Such an ad-hoc formalism can be constructed in a manner akin to a coordinate system basis choice for ${\bf \Re^3}$,
in a beyond the SM ``deviation space''. A core characteristic of ad-hoc constructions is
that the full Lagrangian is not specified, only a few terms, or a sector are defined.
An ad-hoc approach can have some uses as discussed in Refs.~\cite{Masso:2014xra,Gupta:2014rxa,Falkowski:2015fla},
but referring to such a construction
as a SMEFT operator basis has lead to enormous confusion in recent literature.
 The following problems can also occur when using ad-hoc phenomenological Lagrangians:
\begin{itemize}
\item{Utilizing unitary gauge to perform gauge dependent field redefinitions on only parts of a full SMEFT
Lagrangian does not satisfy the equivalence theorem \cite{Kallosh:1972ap}.\footnote{See Section \ref{Smatrix} for more discussion
on the equivalence theorem.} There is no formal expectation of gauge independent
results being obtained making such a transformation, even in LO results. An ad-hoc construction developed
in unitary gauge is very susceptible to gauge dependence for this reason. See
Section \ref{Smatrix} and Refs.~\cite{Passarino:2016pzb,Passarino:2016saj}
for more discussion.}
\item{Particular UV scenarios can lead to the expectation that only certain operators (times Wilson coefficients)
are present due to a vanishing of all other
Wilson coefficients at tree level, and can also lead
to the expectation that the Wilson coefficients of different operators differ by a loop factor. This cannot formally break the field redefinition
relations in Eqs.~\ref{set1redefine}-\ref{set3redefine} or the EOM relations Eq.~\ref{pops}
when the separation of scales present in the EFT construction is adhered to, as these are IR operator relations. At times such UV
assumptions are imposed removing operators, before a non-redundant basis is defined as the suppression is associated
with the operator, not the UV dependent Wilson coefficient. This is a mistake to avoid.}
\item{A clear sign of a phenomenological Lagrangian
is the inability of such an approach to accommodate and aid in determining loop corrections.}
\end{itemize}
To decide if a construction is truly an operator basis one can examine if the
field redefinitions in Eqs.~\ref{set1redefine}-\ref{set3redefine} can be used
to transform a complete set of $\rm SU_c(3) \times SU_L(2) \times U_Y(1)$ operators to the particular
chosen set or form. If this is not possible, then such a construction is
not an operator basis according to the definition above.
The ``Higgs Basis'' construction in Section II.2.1 of Ref.~\cite{deFlorian:2016spz} has not been shown to
satisfy this definition of a basis.

\subsubsection{One loop running of \titlemath{$\mathcal{L}_{SM} +\mathcal{L}_6$}{L6}}\label{onelooprunning}

Determining the closure of the full one loop anomalous dimension matrix
of $\mathcal{L}_6$ is an important check of the consistency of a basis.
The counterterm structure of $\mathcal{L}_6$ can be determined without
expanding around the vev of the Higgs boson, as the scales introduced when the Higgs takes
on a vev regulate the IR of the theory.
Using DR and $\rm \overline{MS}$ the full renormalization of the
$\mathcal{L}_6$ Warsaw basis
was reported in Refs.~\cite{Grojean:2013kd,Jenkins:2013zja,Jenkins:2013wua,Alonso:2013hga,Alonso:2014zka}
using this approach.
These results built upon the past results reported in
Refs.~\cite{Morozov:1985ef, Braaten:1990gq, Hagiwara:1993ck, Hagiwara:1993qt,
Elias-Miro:2013gya,Zhang:2013xya,Elias-Miro:2013mua,
Altarelli:1974exa,Shifman:1976ge,Gilman:1979bc,Floratos:1978ny,
Gilman:1979ud,Gilman:1982ap,Grinstein:1990tj,Bardeen:1978yd,
Buras:1990fn,Buchalla:1989we,Ciuchini:1993vr,Arzt:1992wz,Degrassi:2005zd,Gao:2012qpa,Mebane:2013cra,Mebane:2013zga}. It was found
that the Warsaw basis closes at one loop and the full $2499 \times 2499$
anomalous dimension matrix is now determined. This is the only basis
for which this has been demonstrated to date.\footnote{Interest in the renormalization of $\mathcal{L}_6$
continues. These results were distilled into the tool reported in Ref.~\cite{Celis:2017hod}.
Some results reported in an alternate scheme appeared in Ref.~\cite{Ghezzi:2015vva}
as did some partial results in an alternate basis in Ref.~\cite{Elias-Miro:2013eta}.
This continuing interest is also due to the curious structure of the anomalous dimension matrix \cite{Elias-Miro:2013gya,Elias-Miro:2013mua,Alonso:2013hga,Alonso:2014rga}.
This structure was vigorously misunderstood in the literature until it was explained in Ref.~\cite{Cheung:2015aba} as being due to helicity
and unitarity in the SMEFT. The explanation in Ref.~\cite{Cheung:2015aba} is UV independent,
it is a statement on the IR physics
captured in the operator forms in the SMEFT, so it is a valid EFT understanding of this structure.}

Several aspects of the results in Refs.~\cite{Grojean:2013kd,Jenkins:2013zja,Jenkins:2013wua,Alonso:2013hga,Alonso:2014zka}
can be understood to follow directly from the procedure
to construct an operator basis described in Section \ref{SMEFTbasis}.
The renormalization of operators in $\mathcal{L}_6$ mix down, modifying the running
of the $\mathcal{L}_{SM}$ parameters \cite{Jenkins:2013zja}. This is a scale dependent
result that indicates that the $\mathcal{L}^{(6)}$ operator bases are only defined when performing small field redefinitions
on the SM fields of $\mathcal{O}(1/\Lambda^2)$. Similarly, as is well known in past calculations in NRQCD
\cite{Bauer:1997gs,Pineda:2001ra}
the anomalous dimensions of redundant operators exhibit scheme and gauge dependence
that only cancel once an operator basis is reduced to its non-redundant form \cite{Jenkins:2013zja}.
This is another scale dependent sign that the field variables and operators
are not independent and  related by the EOM (as discussed in Section~\ref{redundant})
when a redundant basis is used. This is a reason to report results in a non-redundant basis.

For the same reason, one cannot use only the results of
Eq.~\ref{pops} to translate the anomalous dimensions determined in the Warsaw
basis to an alternate basis constructed out of the operators in Eq.~\ref{SILHY}.
The removal of operator forms
when defining the Warsaw basis used the full set of EOM results, not just the EOM
field redefinitions related to Eq.~\ref{pops}.
To map the anomalous dimension results of the Warsaw basis to another basis,
all of these EOM reductions must be undone, mapping the complete set of divergences to an overcomplete
$\mathcal{L}_{SMEFT}$. Subsequently, a gauge independent algorithm must be defined that maps
the full set of overcomplete divergences to a chosen non-redundant basis. If no such
algorithm exists, then this procedure cannot be carried out.
This is probably the reason that only the Warsaw basis has been completely renormalized to date.

Gauge independent field redefinitions have a central role in defining the SMEFT,
and this is manifest in the counterterm
structure of $\mathcal{L}_6$.
An ad-hoc phenomenological Lagrangian as discussed Section~\ref{adhoc}, that is
not obtained with such field redefinitions, is a challenge to ever renormalize
for this reason.\footnote{See also the discussion in Ref.~\cite{Passarino:2016pzb}.}

%%%%%%%%%%%%%%%%%%%%%%%%%%%%%%%%%%%%%%%%%%%%%%%%%%%%%
%%%%%%%%%%%%%%%%%%%%%%%%%%%%%%%%%%%%%%%%%%%%%%%%%%%%
\subsubsection{Functional redundancy/factorizability of the SMEFT}\label{functionalredundant}
A consequence of how operator bases are defined is that specifying an incomplete parameterization in only a few possible interaction terms,
or in a sector of interactions (like the Higgs sector) while simultaneously assuming ``SM-like"
interactions in other sectors, generally introduces inconsistencies into the SMEFT.

For example, the parameter $b_2$ can be chosen in Eq.~\ref{EOMBshift} so that the operator $Q_{\substack{H l}}^{(1)}$ is absent
in a basis.
This can only be done at the cost of shifting the Wilson coefficients of the remaining operators in
Eq.~\ref{EOMBshift}. Attempting to study the effects of the operator $Q_{H \Box}$ in
the process $h \rightarrow \bar{\psi} \, \psi$, that receives such a shift, must be done with care.
This operator's Wilson coefficient leads to a shift in the effective Yukawa coupling
$\mathcal{L} = - \mathcal{Y}_\psi  \, h \, \bar{\psi} \, \psi$ of the form
\bea
\mathcal{Y}_\psi = \frac{m_\psi}{v_T} \left[1 + C_{H \Box} \, v_T^2 \right] + \cdots.
\eea
Studying deviations in Higgs properties due to this operator, while simultaneously assuming a ``SM-like''
$Z$ boson interaction with $\bar{\psi} \, \psi$ to accommodate Electroweak precision data (EWPD) constraints in a global analysis
is inconsistent in general. The same shift in the Wilson coefficients of the operators
$Q_{\substack{H e}},Q_{\substack{H q}}^{(1)}, Q_{\substack{H u}},Q_{\substack{H d}}$ related to the shift
of $Q_{H \Box}$, that is introduced to remove the operator $Q_{\substack{H l}}^{(1)}$ from a basis,
could be uncovered in a study of $h \rightarrow \bar{\psi} \, \psi$ or $Z \rightarrow \bar{\psi} \, \psi$. This shift is required for consistency to be present
in anomalous $Z$ couplings, unless a series of other Wilson coefficients are tuned to cancel
out the expected correction.

If this cancellation is assumed one has stepped outside of a general SMEFT analysis.
If a parameterization is constructed that is designed to directly hide such correlations, enormous confusion
can be introduced into SMEFT studies. Considering such shifts as independent
one must impose the constraint of the chosen Wilson coefficients
canceling in all other processes in a global analysis. Even more arduous is to
impose this assumption on all other equivalent combinations of Wilson coefficients
(due to EOM relations) when using a non-redundant minimal basis. Not imposing this assumption
introduces an inconsistency which is referred to as a functional redundancy in Ref.~\cite{Trott:2014dma}.
Due to the large number of parameters present in the SMEFT, it is required to combine
a series of measurements to constrain the SMEFT Wilson coefficient space.
Functional redundancies can block
this combination of measurements being performed in a consistent manner and lead to spurious results.

Directly following from this subtlety is in what sense the SMEFT is factorizable into a subset of contributions when
predicting $S$ matrix elements. Approximations must be made in interfacing with experimental results.
The approximations that are safe to impose while maintaining a model independent
SMEFT analysis are {\it IR assumptions}. For more discussion on this point see Section \ref{EFTcolliderbasics}

Conversely, naively imposing UV assumptions can make the SMEFT
inconsistent as a field theory construction, and incapable of capturing the IR limit of a UV physics sector.
This can render such a SMEFT study ambiguous, with unclear implications for a UV physics sector, as it only uncovers
a distorted approximation of the true low energy constraints that a UV sector will face.
SMEFT studies walk a fine line between being powerful model independent conclusions,
and statements based on inconsistent field theory without any true UV implication or meaning. This fine
line is defined by consistency in the SMEFT analysis being enforced (or violated) to the precision and accuracy
demanded by the data.

\subsection{The Higgs Effective Field Theory}\label{HEFTsec}

\subsubsection{Minimal assumptions in the scalar sector}
Both the SM and the SMEFT constructions are field theories that
assume the existence of the scalar complex field $H$ defined in Table ~\ref{SMfields}
in the constructed Lagrangian.
The introduction of such a field is a consequence of requiring:
\begin{itemize}
  \item{(i) Three Goldstone bosons $\pi^I$, the longitudinal components of the EW gauge bosons.}
  \item{(ii) One singlet scalar $h$, corresponding to the physical Higgs boson, that ensures the \emph{exact} unitarity at
  all energies of scattering amplitudes with external $\pi^I$ fields. Pedagogical illustrations of this argument
  can be found in Refs.~\cite{Contino:2010rs,Contino:2010mh,Contino:2013gna}.}
\end{itemize}
Condition (i) is an IR assumption that is imperative for a correct description of the EW symmetry breaking.
Requirement (ii) can be relaxed and the EFT can remain self-consistent for lower energies scattering events
where the EFT is well defined. An EFT does not have to exactly preserve unitarity, it only has to be unitary up to the cut off scale
where the Taylor expansion used in its construction breaks down.
An EFT is dictated by describing the long distance propagating states that lead to non-analytic structure in the correlation functions of scattering amplitudes.
From this perspective, the ideal theoretical tool to describe scenarios without assumption (ii) being made
is the Higgs Effective Field Theory (HEFT).

In the HEFT the long distance propagating states are again the massive SM fermions and gauge bosons.
Instead of a $H$ field, a dominantly $J^P = 0^+$
singlet scalar state $h$ is included in the EFT with free couplings to the remaining SM states.  The HEFT is based on the Callan-Coleman-Wess-Zumino (CCWZ)
formalism~\cite{Coleman:1969sm,Callan:1969sn} and provides
a parameterization of the scalar sector with minimal IR assumptions.

This approach has been continually rediscovered over the years as it adheres to the EFT ``prime directive''
in the scalar sector.
Several ad-hoc parameterizations have independently emerged over the years along these lines \cite{Feruglio:1992wf,Burgess:1999ha,Grinstein:2007iv,
Barbieri:2007bh}
but none of these works developed a complete self-consistent EFT.\footnote{Ref.~\cite{Contino:2010mh} appears
after Refs.~\cite{Feruglio:1992wf,Burgess:1999ha,Grinstein:2007iv,Barbieri:2007bh} but
states that it introduces this parameterization to the literature, which has caused some confusion, as
Ref.~\cite{Contino:2010mh} cites Ref.~\cite{Barbieri:2007bh}.}
It is an important development that in recent years in Refs.~\cite{Feruglio:1992wf,Buchalla:2012qq,Alonso:2012px,Alonso:2012pz,Buchalla:2013rka,
Buchalla:2013mpa,Brivio:2013pma,Brivio:2014pfa,Gavela:2014vra,Gavela:2014uta,
Alonso:2014wta,Buchalla:2015wfa,Hierro:2015nna,Buchalla:2015qju,Brivio:2016fzo,Gavela:2016vte,
Merlo:2016prs,Hernandez-Leon:2017kea}
this theory has been advanced to a degree that it is now a consistent EFT description.

The size and pattern of any deviations from the SM discovered using EFT methods can indicate whether a HEFT or a SMEFT
description is appropriate. Any deviation from the SMEFT expectation that follows
from the exact $H$ doublet structure can carry significant information about the possible UV physics
matched onto the lower energy EFT description. Theoretical tools that can
consistently account for the possibility that
the Higgs boson may not be part of an exact $\rm SU_L(2)$ multiplet with the $\pi^I$,
and to probe experimentally this hypothesis as a part of a wider precision measurements program are essential.
The experimental collaborations have already started to address this issue, providing
independent measurements of several Higgs couplings and testing for the presence of anomalous
Lorentz structures (see e.g. Ref.~\cite{Khachatryan:2016vau}). At the present moment, all the observations
are compatible with the SM, and the SMEFT extension, but the evidence in favor of this option is still
not compelling. The presence of large uncertainties (roughly of order 10--20\% as discussed in Section \ref{Higgsfun}), together with the fact
that some decay and production channel are still not accessible, allows for significant deviations.

\subsubsection{Topology of the scalar manifold}

In parallel to the HEFT formalism being continually rediscovered over the years,
there has been a continual rediscovery of the resistance to this idea in the theoretical community.
This resistance is usually due to the field reparameterization
equivalence theorem \cite{Kallosh:1972ap} of $S$ matrix elements.
A consequence of this theorem is that a coordinate choice for a scalar
manifold does not matter for $S$ matrix elements. In the SM, the HEFT or the SMEFT,
the scalar manifold of interest is $\mathcal{M}(\pi^I,h)$.
A coordinate choice for this manifold has no physical effect.

The HEFT literature is not due to a misunderstanding of this point. This EFT
is designed to describe a set of possible low energy IR limits that are not consistent with
a predictive version of the SMEFT (i.e. when the operator expansion in the EFT converges). Specific examples
are given by the Dilaton constructions of Refs.~\cite{Halyo:1991pc,Goldberger:2008zz},
which have long been known to be represented by the HEFT. Nevertheless, a formulation of the HEFT/SMEFT discriminant
in terms of scalar manifold topology was not present in the literature until recently.

It is not surprising that this discriminant is topological in nature.
Manifold topology plays an important role in theories of symmetry
breaking such as in the SM, or in the Landau-Ginzburg effective theory of superconductivity (see Section \ref{precision-motivation}).
Topological distinctions of vacuum states are common in field theory, and have long been understood
to underlie the properties of the scattering of pions  \cite{Weinberg:1966kf,ArkaniHamed:2008gz}
based on the curvature of the scalar manifold describing the Goldstone bosons of chiral perturbation theory.
The heuristic embedding of possible deviations from SM expectations given by
$\rm SM \subset SMEFT \subset HEFT$ is also well known for many years, but the precise theoretical
statement is given as
\begin{itemize}
\item{The SM has a flat scalar manifold $\mathcal{M}$ and an $\rm O(4)$ fixed point
which is the unbroken global custodial group $\rm O(4) \sim SU_L(2) \times SU_R(2)$. This group is
respected by a subset of the SM Lagrangian \cite{Susskind:1978ms,Weinberg:1979bn,Sikivie:1980hm}.
When $\langle H^\dagger H \rangle = v^2/2$ a $\rm SU_{L+R}(2) = SU_c(2)$ subgroup (also generally referred to the custodial group)
is unbroken
that leads to the prediction $m_W = m_Z \cos \theta_W$.
The linearization lemma \cite{Coleman:1969sm} then allows for a local linear transformation of
the scalar manifold coordinates, which results in the embedding of the scalar fields into a linear multiplet $H$.
This is the group theory equivalent of assuming a $H$ field in the SM construction.}
\item{The SMEFT has a curved scalar manifold \cite{Burgess:2010zq,Alonso:2015fsp,Alonso:2016oah} due to the presence of two derivative Higgs operators.
In the Warsaw basis these operators are $Q_{HD},Q_{H \Box}$. The SMEFT also has a $\rm O(4)$ fixed point
which allows the EFT to be constructed with the linear multiplet $H$. The presence of this scalar curvature is a key point of
the SMEFT construction and is the reason that one cannot remove all the two derivative Higgs operators
from the SMEFT basis with gauge independent field redefinitions.}
\item{HEFT has a curved scalar manifold $\mathcal{M}$ and does not contain
a  $\rm O(4)$ fixed point \cite{Alonso:2015fsp,Alonso:2016oah}.}
\end{itemize}
This distinction between the $\rm SM$, $\rm SMEFT$ and $\rm HEFT$ is field redefinition invariant.\footnote{Note that in
Refs.~\cite{Espinosa:2012ir,Espinosa:2012vu,Espinosa:2012im,Grojean:2015ldw} and some other works this field redefinition invariant distinction between the HEFT and SMEFT is
obscured and Ref.~\cite{Giudice:2007fh} is incorrectly credited for the introduction of a nonlinear realization.}

\subsubsection{UV embeddings of HEFT}

The HEFT is a general field theory that can describe a wide variety of scenarios,
including composite Higgs models
\cite{Kaplan:1983fs,Kaplan:1983sm,Banks:1984gj,Georgi:1984ef,Georgi:1984af,
Dugan:1984hq,Agashe:2004rs,Gripaios:2009pe,Marzocca:2012zn,Feruglio:2016zvt}
and Dilaton constructions \cite{Halyo:1991pc,Goldberger:2008zz}
while reproducing the SM in a specific limit of parameter space.
The HEFT formalism is of most interest, if it captures the IR limit of
a UV completion of the SM.
Determining the necessary and sufficient conditions on UV dynamics that leads to the HEFT
construction at low energies is an unsolved problem. The naive
assumption that the distinction between SMEFT and HEFT is a statement of
the presence of a linear multiplet {in the UV sector integrated out of the theory}
is not correct. The assumption is an IR assumption on the nature of the EFT
that corresponds to an (unknown) set of criteria on UV dynamics.

This can be directly seen in the non-minimal coupling to gravity of the
Higgs doublet in a UV sector. In this case
the classical background field scalar manifold also plays a central role.
This Lagrangian is given as a tower of higher dimensional operators as
\bea\label{heft}
\mathcal{L}_{H,inf}= \sqrt{- \hat{g}} \left(\mathcal{L}_{SM} - \frac{M_p^2 \, \hat{R}}{2} - \xi H^\dagger H \,  \hat{R} \right).
\eea
This Lagrangian is known as the Higgs-Inflation Lagrangian, and was popularized in recent literature in Ref.~\cite{Bezrukov:2007ep}.
Here $\hat{g}$ is the determinant of the Jordan frame metric $\hat{g}_{\mu \nu}$, $M_p= 2.44 \times 10^{18} \, {\rm GeV}$ is the reduced Planck mass, $\hat{R}$ is the Ricci scalar
and $\xi$ is a dimensionless coupling. When $\langle H^\dagger H \rangle  \equiv \hat{v}^2/2 \gg v^2/2$
this Lagrangian leads to a flat Higgs potential due to mixing between the scalar state $h$ and a scalar component
of the graviton.\footnote{The scalar degree of freedom of gravity can only be removed in the usual manner with diffeomorphism
invariance in Minkowski space when this mixing is not present, which requires the theory is first canonically normalized.} This can be seen expanding the metric about Minkowski space
as $\hat{g}_{\mu \nu} = \eta_{\mu \nu} + h_{\mu \nu}/M_p$ giving
\bea
\delta \mathcal{L} = \frac{\xi}{M_p} (h + \hat{v})^2 \, \eta^{\mu \, \nu} \partial^2 h_{\mu \, \nu}.
\eea
This mixing was analyzed in a complete fashion in Ref.~\cite{Bezrukov:2010jz}
for the first time and becomes significant when $\sqrt{\langle H^\dagger H \rangle} \sim M_p/\xi$. At
this scale the canonical normalization of the $h$ field and the graviton leads to significant
non-linearities introduced into an EFT description of the propagating degrees of freedom
-- {resulting in the HEFT}. This explains the scattering results found related to the cut-off scale
behavior of this theory \cite{Burgess:2009ea,Barbon:2009ya,Burgess:2010zq,Hertzberg:2010dc,Burgess:2014lza}.
Although the UV dynamics includes a $H$ doublet, the IR EFT construction useful
for scattering calculations around background fields $\sqrt{\langle H^\dagger H \rangle} \gtrsim M_p/\xi$ is a version of the HEFT.
As this operator is induced renormalizing the SM in curved space, this physics is formally always present
and the (small) non-linearities due to gravity are present in EFT descriptions of Higgs physics.
Of course these effects are below the current (and most likely future!) experimental resolution.

The relevant point for LHC phenomenology is the possibility that dynamics present in a
UV sector that explains EWSB leads to similar IR effects, that can be accommodated in
the HEFT. In composite Higgs theories, some new physics states may mix with the $h$ field,
thus weakening the unitarity constraint as captured by the HEFT formalism, and permitting deviations from an exact doublet structure.
Due to our inability to calculate the low energy limit of all possible strongly interacting sectors to understand the precise conditions on dynamics that
can lead to such IR effects, this formalism should not be casually dismissed.
For recent discussions on IR limits associated with $\rm TeV$ scale physics
that require the HEFT, see Refs.~\cite{Alonso:2014wta,Alonso:2016btr,Buchalla:2016bse}.

\subsubsection{The HEFT Lagrangian, preliminaries}
In analogy with the formalism employed in chiral perturbation theory ($\chi$PT) for the pions of QCD,
in HEFT the three Goldstone bosons $\pi^I$ are embedded into a dimensionless unitary matrix
\begin{equation}\label{U.def}
 {\U = \exp\left(i \tau^I \pi^I/v\right)}\,,\qquad \U \mapsto L \U R^\dag\,,
\end{equation}
which transforms as a bi-doublet under global $\rm SU_L(2)\times SU_R(2)$ transformations.
Inside the exponential, the $\pi^I$ fields appear suppressed by the scale $v$ rather than by a heavier cutoff $\Lambda$.
This is a reasonable choice, as $v$ can be regarded as an order parameter of EWSB and the characteristic scale of the $\pi^I$.
This implies that, unlike in the SMEFT case, $\pi^I$ insertions are not suppressed in general.
Additional suppressions may be induced if the HEFT is matched to specific UV models
that associate these degrees of freedom with a heavier scale. This is the case of
Composite Higgs models, in which both the $\pi^I$ and $h$ fields arise weighted by a scale $f$
that satisfies the relation $4\pi f \geq \Lambda$, with $\Lambda$ being the cutoff of the EFT~\cite{Manohar:1983md}.
In this case, the scalar fields are always accompanied in the EFT by factors of $\xi = (v/f)^2<1$.

In the EW vacuum $\langle \U\rangle={\bf{1}}$ the global chiral symmetry is spontaneously broken down
to the custodial group $\rm SU_c(2)=SU_{L+R}(2)$. In addition, the $\rm SU_R(2)$ component is explicitly broken in
the EW sector by the Yukawa couplings and the gauged $U(1)$ subgroup with coupling $g_1$ generated by $\tau^3$.
It is convenient then to define objects that transform in the adjoint of $\rm SU_L(2)$,
to be employed as building blocks of a $\rm SU_c(3)\times SU_L(2)\times U_Y(1)$ invariant Lagrangian.
It is possible to construct a scalar and a vector as follows:
\begin{equation}
\begin{aligned}
 \T &= \U \tau^3 \U^\dag\qquad\qquad   & \T &\mapsto L \T L^\dag\\
 \V_\mu &= (D_\mu \U) \U^\dag& \V_\mu&\mapsto L \V_\mu L^\dag\,,
\end{aligned}
\end{equation}
where the covariant derivative of the $\U$ field is
\begin{equation}
 D_\mu\U=\de_\mu\U+\frac{ig_2}{2} W_\mu^I \tau^I\U- \frac{ig_1}{2} B_\mu \U\tau^3\,.
\end{equation}
Note that the field $\T$ is invariant under hypercharge transformations but not under the global $\rm SU_R(2)$ group
so it can be treated as a custodial symmetry breaking spurion.

The physical Higgs scalar is introduced as a gauge singlet $h$. This choice ensures the most general approach to the
physics of this state and makes the HEFT a versatile tool which includes the $\rm SU_L(2)$ doublet case
as a particular limit. Specifically, the SM Higgs doublet $H$ can be written as a fixed combination of the fields $h$ and $\U$ according to:
\begin{equation}
 \left(\tilde H \;\;H\right) = \frac{v+h}{\sqrt2}\,\U\,.
\end{equation}
Being a singlet, the $h$ field has completely arbitrary couplings, that are customarily encoded in generic functions~\cite{Feruglio:1992wf,Grinstein:2007iv,Contino:2010mh}
\begin{equation}\label{Fh}
 \F_i(h) = 1+ 2a_i\frac{h}{v}+b_i\frac{h^2}{v^2}+\dots
\end{equation}
that constitute another building block for the HEFT Lagrangian.
Note that here the polynomial structure should be interpreted as the result of a Taylor expansion in $(h/v)$, which may include an infinite number of terms.

\subsubsection{The HEFT Lagrangian}

The HEFT Lagrangian is composed of the gauge and fermion fields in Section \ref{SMsection}, and the scalar fields $\U$, $h$ defined in the previous section.
Its construction historically builds upon that of the EW
chiral Lagrangian~\cite{Appelquist:1980vg,Longhitano:1980iz,Longhitano:1980tm,Appelquist:1993ka,Feruglio:1992wf},
in which only the three $\pi^I$ fields were retained in the spectrum, while the physical Higgs was
assumed to be sufficiently heavy to be integrated out. Extensions with addition of extra
(pseudo-)scalar fields were also explored more recently~\cite{Brivio:2015kia,Buchalla:2016bse,Brivio:2017ije}.
The HEFT Lagrangian can be defined as
\begin{equation}
 \mathcal{L}_{HEFT} = \mathcal{L}_0+\Delta\mathcal{L}+\dots\,,
\end{equation}
where $\mathcal{L}_0$ contains the leading order terms and $\Delta\mathcal{L}$ includes first order deviations.
Unlike the SMEFT, it is not possible to classify the HEFT invariants based on just the canonical dimension. This is due to the fact that the HEFT is a fusion of $\chi$PT (in the scalar sector) with the SMEFT (in the fermions and gauge sector). Because these two theories follow different counting rules, the structure of the mixed expansion is complex.
A rigorous and self-contained method to determine the expected suppression of a given HEFT invariant represents a
non-trivial and subtle task that has been intensely debated in the literature. The discussion of this technical point is
postponed to Section~\ref{Section.countingHEFT}.

The leading Lagrangian $\mathcal{L}_0$
 can be written as\footnote{Due to the lack of a unique criterion
for the classification of the invariants, there is some freedom in the definition of the LO itself.
To define one universal rule that uniquely determines the
order a certain HEFT operator should be assigned, in a UV independent manner, is the essential challenge. This difficulty has been overcome in the literature by identifying a
set of heuristic criteria, which essentially follow from internal consistency requirements and some common sense considerations.
Unfortunately, some UV dependence is generally also present.
Nevertheless, there is substantial agreement in the community with respect to the organization of the HEFT expansion
which we stress here, despite disagreements on the definition of counting rules.
For definiteness, here we use the conventions of Ref.~\cite{Brivio:2016fzo}.
We stress this is not a value judgment, simply a convention. The same LO Lagrangian has
been previously adopted in Refs.~\cite{Alonso:2012px,Buchalla:2013rka}, although in a slightly different notation.
Different conventional choices were made in Refs.~\cite{Buchalla:2012qq,Gavela:2014vra,Brivio:2013pma}.}
\begin{equation}
\begin{aligned}
\mathcal{L}_0=& -\dfrac{1}{4} G_{\mu\nu}^A G^{A\,\mu\nu}
-\frac{1}{4} W_{\mu\nu}^I W^{I\,\mu\nu}-\dfrac{1}{4} B_{\mu\nu}B^{\mu\nu}+\sum_{\psi}\bar{\psi}i\slashed{D}\psi
+\\
&+\frac{1}{2}\de_\mu h \de^\mu h-\dfrac{v^2}{4}\Tr(\V_\mu \V^\mu)\F_C(h)-V(h)+\\
&-\frac{v}{\sqrt2}\left(\bar{q}\,\U\mathcal{Y}_Q(h) q_R+\hc\right)
-\frac{v}{\sqrt2}\left(\bar{\ell}\,\U\mathcal{Y}_Q(h) \ell_R +\hc\right)\,,
\end{aligned}
\label{Lag0}
\end{equation}
where the right-handed fermions have been collected into doublets of the global $\rm SU_R(2)$ symmetry
\begin{equation}
 q_R = \begin{bmatrix}u_R\\ d_R\end{bmatrix},\qquad \ell_R = \begin{bmatrix}0\\ e_R\end{bmatrix}\,
\end{equation}
and the Yukawa couplings include a dependence on $h$:
\begin{equation}
 \mathcal{Y}_{Q}(h) = \text{diag}\left(\sum_n Y_u^{(n)} \frac{h^n}{v^n},\; \sum_n Y_d^{n} \frac{h^n}{v^n}\right),
 \hspace*{5mm}
\mathcal{Y}_{L}(h) = \text{diag}\left(0,\; \sum_n Y_e^{(n)} \frac{h^n}{v^n}\right)\,.
\end{equation}
The first term in the sum ($n=0$) generates fermion masses, while higher orders describe the couplings with an arbitrary number of $h$ insertions.
The term $\Tr(\V_\mu \V^\mu)$ in the second line of Eq.~\ref{Lag0} contains the kinetic terms of the $\pi^I$ and the mass terms of the gauge bosons. Because the matrix $\U$ is adimensional, this invariant has canonical dimension 2 and therefore appears in the Lagrangian multiplied by a factor $v^2$. The reason why it is the scale $v$, and not the cutoff $\Lambda$, that multiplies this term is that this choice ensures a canonically normalized kinetic term for the $\pi^I$, given the definition of $\U$ in Eq.~\ref{U.def}. The same principle applies to the Yukawa couplings.

The Lagrangian $\mathcal{L}_0$ is equivalent to the SM Lagrangian up to the presence of an arbitrarily large number of Higgs and $\pi^I$ insertions and to the fact that the Higgs couplings (which includes the scalar potential) are parameterized by independent coefficients rather than being fixed by the doublet structure. Despite being allowed by symmetry, Higgs couplings to kinetic term structures are absent in $\mathcal{L}_0$. In the case of the fermion and $h$ kinetic terms, this is because any extra $\F(h)$ factor can be removed via field redefinitions~\cite{Buchalla:2013rka,Brivio:2016fzo}
and reabsorbed into the coefficients that parameterize the Higgs interactions. This can be understood to follow directly from the general arguments
on constructing on-shell EFTs given in Refs.~\cite{Gasser:1983yg,Georgi:1991ch}. In the case of gauge kinetic terms such couplings are chosen to be retained
in the basis,
but they are customarily classified as higher order effects due to a suppressed Wilson coefficient, under the assumption that the transverse components
of the gauge bosons do not couple strongly to the Higgs sector.
The same choice has been made for the suppression of the Wilson coefficient of the operator $\Tr(\T\V_\mu)^2$, which would belong to $\mathcal{L}_0$ according to
the derivative expansion of $\chi$PT, but that must carry an implicit suppression in its matching if the custodial symmetry
is assumed to be only weakly broken, as suggested by experimental observations.

The Lagrangian $\Delta\mathcal{L}$ contains the leading deviations from $\mathcal{L}_0$. Again, it is not possible to infer in a rigorous and
universal way which classes of operators belong to this order. The relative impact of two given invariants depends in general on
the kinematic regime considered (see Section~\ref{Section.countingHEFT} for further details).
It is possible to identify a set of invariants that \emph{can} be responsible for the largest deviations at least at energies up to a few TeV.
They amount to 148 independent invariants in total~\cite{Brivio:2016fzo}. A complete basis of
operators encompassing both bosonic and fermionic sector has been first
proposed in Ref.~\cite{Buchalla:2013rka} and an alternative set has been derived
independently in  Ref.~\cite{Brivio:2016fzo}. The two works agree on the classes of
operators, although there are differences in the choice of individual terms retained.
We describe the operators present sector by sector as follows.

{\bf Bosonic operators:} This must include chiral invariants with four derivatives, which are next-to-leading terms in the chiral expansion and can be reduced
 to an independent set with the structures\footnote{Not all of the Lorentz indices in these expressions match.
This indicates the possibility of multiple Lorentz contractions.}
 $$\begin{aligned}&(\V_\mu)^4, \quad(\V_\mu)^3 \de_\nu\F_i(h), \quad (\V_\mu)^2(\de_\nu\F_j(h))^2, \quad (\de_\mu\F_k(h))^4\\
 &X_{\mu\nu} (\V_\rho)^2,\quad X_{\mu\nu} \V_\rho \de_\sigma\F_l(h),\quad (X_{\mu\nu})^2\F_m(h)\,.
 \end{aligned}$$
 Note that operators in this category are not suppressed by any powers of $\Lambda$ in the Lagrangian, as they have formally canonical dimension 4.
The two-derivative operator $v^2 \Tr(\T\V_\mu)^2/4$ also belongs to $\Delta\mathcal{L}$ as explained above.
Finally, operators with the structure $(X_{\mu\nu})^3$ can be included at this order despite containing 6 derivatives. This can be justified assuming that, being composed of only transverse gauge boson,
these operators follow the ordering rules of the SMEFT (see Section~\ref{Section.countingHEFT}).

A complete set of interactions for the bosonic sector, constructed avoiding reduction via the EOM, was presented and studied phenomenologically in Ref.~\cite{Brivio:2013pma} (CP conserving terms) and Ref.~\cite{Gavela:2014vra} (CP odd terms).

{\bf Operators with fermions:} $\Delta\mathcal{L}$ contains operators with one fermionic current and up to two derivatives, as at least a subset of this category of invariants is required as counter-terms in the one-loop renormalization of $\mathcal{L}_0$. Schematically, they can be reduced to the set of independent structures
$$ (\bar\psi\gamma^\mu\psi) \V_\mu,\quad  (\bar\psi\psi)(\V_\mu)^2,\quad (\bar\psi\psi)\V_\mu \de^\mu\F_n(h),\quad (\bar\psi\sigma^{\mu\nu}\psi)(\V_\mu)\de_\nu\F_o(h), \quad (\bar\psi\sigma^{\mu\nu}\psi)(X_{\mu\nu})\,.
$$
Here operators with one derivative are unsuppressed, while operators with two derivatives are multiplied by $\Lambda^{-1}$, having canonical dimension 5.
Four-fermion operators appear in $\Delta\mathcal{L}$ as well, because a subset of terms in this class is required for the renormalization of $\mathcal{L}_0$.

These considerations lead to the construction of a consistent basis of independent operators in the HEFT,
reported by multiple groups, which is closed
under the EOM relations.

\subsection{SMEFT vs. HEFT}
$S$ matrix elements and relations between
$S$ matrix elements constructed in the SMEFT and the HEFT expansions are potentially distinguishable,
as the theories are distinct in a field redefinition invariant manner \cite{Alonso:2015fsp,Alonso:2016oah}.
The ordering of the theories in their IR assumptions is given by $\rm SM (H,\Lambda \rightarrow \infty) \subset SMEFT
(H,\Lambda \neq \infty) \subset HEFT (h,\Lambda \neq \infty)$.

Early statements on the need to experimentally check the assumption of a $H$ field in the
EFT used, appeared in
the literature before the turn on of LHC \cite{Burgess:1999ha,Grinstein:2007iv,Contino:2010mh}.
Developing a precise statement on what exact experimental result
would decide between these two frameworks conclusively is an ongoing challenge.
The differences between SMEFT and HEFT identified to date, and existing proposals
to take advantage of these differences are as follows.

\subsubsection{Higgs/Triple Gauge Couplings}
 While in SMEFT the Higgs couplings follow a polynomial dependence in $(v+h)^n$
 due to the $H$ doublet, in the HEFT they do not.
Current LHC results are compatible with $H$ in the spectrum, but with large uncertainties that allow for significant deviations.
A fundamental probe of this point would be the observation of double-Higgs production, which is
challenging at the LHC. A discussion of this process in relation to the HEFT can be found
in Refs.~\cite{Grinstein:2007iv,Contino:2010mh,Contino:2012xk,Grober:2010yv,Dolan:2012ac,Gouzevitch:2013qca,Azatov:2015oxa,Kling:2016lay,Bishara:2016kjn,deFlorian:2017qfk,Grober:2017gut}.

Another distinctive prediction of the HEFT is the decorrelation of triple gauge
couplings and Higgs-gauge bosons interactions, due to the fact that in this EFT the covariant
derivative $D_\mu H \sim \de_\mu h\U+(v+h)D_\mu\U$ is formally split into two independent pieces.
The possibility of isolating this effect experimentally has been studied in Refs.~\cite{Brivio:2013pma,Brivio:2016fzo}.

The appearance of Hermitian derivative terms acting on Higgs fields in the SMEFT of the form
$H^\dagger \, i\overleftrightarrow D_\beta H$ and $H^\dagger \, i\overleftrightarrow D_\beta^I H$ in $\mathcal{L}_6$
relates anomalous $Z$ couplings in $\bar{\psi} \psi \rightarrow Z \rightarrow \bar{\psi} \psi$ to
contact operator contributions to $p p \rightarrow h \rightarrow Z \, Z^\star \rightarrow \bar{\psi} \psi \, \bar{\psi} \psi$.
This correlation can be probed as a consistency test of the SMEFT accommodating any discovered deviations in the kinematic spectra
of $h \rightarrow Z \, \bar{\psi} \psi$ and $h \rightarrow Z \, Z^\star \rightarrow \bar{\psi} \psi \, \bar{\psi} \psi$ determined in the narrow width
approximation \cite{Isidori:2013cga,Brivio:2013pma}. For recent
analysis of this spectra in the SMEFT/HEFT, see
Refs.~\cite{Isidori:2013cga,Isidori:2013cla,Pomarol:2013zra,Buchalla:2013mpa,Grinstein:2013vsa,Gonzalez-Alonso:2015bha,Boselli:2017pef}.

\subsubsection{Organization of the expansion}
 As detailed in Section~\ref{Section.countingHEFT}, the HEFT Lagrangian is not organized as an expansion in canonical dimensions
 as in the SMEFT, but has a more complex structure. This is because the physical Higgs $h$ and the
$\pi^I$ containing matrix $\U$ are independent in HEFT.
A consequence is the re-shuffling the orders at which interactions appear in the expansion.

Insertions of longitudinal gauge bosons are less suppressed in HEFT.
Couplings of these fields or, equivalently, of the $\pi^I$
can be probed in high energy scattering. These processes are also potentially sensitive to the presence
of strong interactions in the EWSB sector. Several channels have been analyzed in HEFT,
including $V_L V_L \to V'_L V'_L$ ($V,V'=Z,W^\pm$)~\cite{Delgado:2013hxa},
$V_L V_L\to hh$~\cite{Contino:2010mh,Delgado:2015kxa}, $V_L V_L\to t\bar t$~\cite{Castillo:2016erh}
and  $\gamma\gamma\to V_LV_L$~\cite{Delgado:2014jda}. The latter is more promising,
having been already observed at the LHC~\cite{CMS-PAS-FSQ-13-008}.

Some interesting differences in the allowed interactions in HEFT compared
 to the SMEFT at fixed order in the power counting of each theory have been identified.
 For example, the operator $\epsilon_{\mu\nu\rho\s}\Tr(\T\V^\mu)\Tr(\V^\nu W^{\rho\s})\F(h)$,
introduces triple and quartic gauge couplings with an anomalous antisymmetric
 Lorentz structure in HEFT.\footnote{This operator violates explicitly the custodial symmetry group ($\rm SU_c(2)$)
 Therefore it may be suppressed in scenarios where this symmetry is broken only through standard effects (non-homogeneous Yukawas and $g'\neq0$).}.
 A phenomenological study of this term has been carried out in Ref.~\cite{Brivio:2013pma}.

\subsubsection{Number of operators/operator structure}
HEFT contains a larger number of invariants compared to the SMEFT order by order in the expansions of each theory.
In the flavour-blind limit, the complete HEFT basis $\Delta\mathcal{L}$ contains 148+h.c operators~\cite{Brivio:2016fzo},
compared to the 76 parameters when $n_f = 1$ in the SMEFT at $\mathcal{L}^{(6)}$ ~\cite{Grzadkowski:2010es,Alonso:2013hga}.

 Most of the operators appearing in HEFT contribute to the same interaction vertices in the SMEFT
 due to $\mathcal{L}^{(6)}$. When this is the case, the larger number of invariants
 is then present in the corrections to predicted $S$ matrix elements
 with the same interaction vertices.
 This does not lead to a measurable difference in a single observable as multiple observables are required
 to fix the free parameters.
Potentially, patterns of allowed deviations in multiple processes
 could be more easily accommodated in HEFT than the SMEFT.

\subsubsection{Tails vs poles}\label{tailspoles}
Differences between SMEFT and HEFT identified to date exist
in interactions involving at least three boson fields: $hhh$, $hhhh$, $hVV$,
 $VVV$. In each EFT, one can work in the canonically normalized basis of fields.
For pole observables, using the different EFTs label different unknown parameters in $V\bar{\psi}\psi$ and the mass terms $VV$.
Studying $\bar{\psi}\psi \rightarrow V \rightarrow \bar{\psi}\psi$ observables the
power counting of the corrections in each EFT scales as $C_i \, v^2/\Lambda^2$.
At present, no SM-resonant exchange (i.e. pole process)
measured at LEP, the Tevatron, or LHC
has deviated in a statistically significant manner from the SM expectation, see Section \ref{LEPgoodandbad}.

Future Run II data, the High Luminosity phase of the LHC, and future facilities
will offer more precise experimental results on pole observables, and $S$ matrix elements
that receive contributions from $hhh$, $hhhh$, $hVV$, $VVV$.
These off-shell vertices have a non-trivial relation to LHC observables,
with other possible deviations simultaneously present at each order in the
EFT expansions in vertex corrections.
When studying these processes, two expansions are present $\{v^2/\Lambda^2$, $p^2/\Lambda^2\}$
where $p^2$ stands for a general kinematic invariant, and in general $p^2/\Lambda^2 \rightarrow 1$
in the tail of a distribution.

It is an unsolved problem to develop a method for consistent
SMEFT/HEFT analysis of the tails of kinematic distributions
and to project the global constraints of each EFT into these distributions.\footnote{For recent results aimed at
this problem, see Refs.~\cite{Corbett:2012dm,Corbett:2012ja,Corbett:2013pja,Englert:2014cva,Liu:2016idz,Azatov:2016sqh,Falkowski:2016cxu}.}
Defining the interference of the SM for an observable $O$ with a tail expansion parameter factored out with HEFT or SMEFT as
$\langle \rangle_O$, to distinguish these theories and
conclude that the SMEFT cannot accommodate a discovered deviation from the SM, while the HEFT can accommodate such a deviation,
requires
\bea
\Delta\langle \mathcal{L}_6 -(\mathcal{L}_{0}+\Delta \mathcal{L})\rangle_O \frac{p^2}{\Lambda^2}> \langle \mathcal{L}^8 \rangle_O \,  \frac{(p^2)^2}{\Lambda^4}.
\eea
Due to the large multiplicity of operators appearing in the SMEFT expansion at $\mathcal{L}_8$
\cite{Lehman:2015via,Lehman:2015coa,Henning:2015daa,Henning:2015alf}, examining the distinguishability of these
theories in the limit $p^2/\Lambda^2 \lesssim 1$ (as opposed to $p^2/\Lambda^2 << 1$)
is difficult. An additional challenge is the
suppression of the interference terms due to $\mathcal{L}_6$ in several tail observables of this form following from
the Helicity non-interference arguments of Ref.~\cite{Dixon:1996wi,Mangano:1990by,Dixon:1993xd,Azatov:2016sqh,Falkowski:2016cxu}.
It remains to be conclusively shown if HEFT and SMEFT can be functionally distinguished in the tails of distributions.
Determining the global constraints on these theories from the LEP and LHC data sets is an essential step to examine this question quantitatively.

\section{Power counting}\label{bloodbath}
The SMEFT and the HEFT are constructed out of an infinite series of operators based on a separation of scales.
It is required that these theoretical
descriptions are consistent, and they are of interest if they are predictive. For this reason, it is important to establish:
\begin{itemize}
\item{How a Lagrangian term scales in a consistent manner with the dimensionful quantities in the theory,
i.e. the normalization of terms.}
\item{A prescription for ordering such invariants within the EFT expansion
to estimate the relative physical impact of any given Lagrangian term on a measurement, so that the {\it approximate theoretical precision}
of the EFT can be well defined.}
\end{itemize}
The procedure of establishing the consistent scaling/ordering of the terms in the Lagrangian is usually called ``power counting''.
There is a canonical interpretation of
this concept that runs through Refs.~\cite{Coleman:1969sm,Callan:1969sn,Weinberg:1978kz,Georgi:1991ch,Georgi:1994qn,Luke:1996hj,Cohen:1997rt,Luty:1997fk,
Jenkins:2013sda,Buchalla:2013eza,Gavela:2016bzc}
and many other works. We adopt this interpretation in this review.
A different usage of ``power counting'' also exists in the literature,
as we stress below.

\subsection{Counting rules for the SMEFT}

The SMEFT is built out of the SM fields and couplings that have the well defined canonical mass dimensions $\left[..\right]$.
Defining the total mass dimensions of the $\left[\mathcal{L}_{SM}\right] = d$, then
\bea\label{naivemass}
\left[\psi\right] = \frac{d-1}{2}, \quad \left[Y_i\right] = \frac{4-d}{2}, \quad \left[V\right] = \frac{d-2}{2},
\quad \left[g_i\right] = \frac{4-d}{2}, \quad \left[H\right] = \frac{d-2}{2},\quad \left[\lambda\right] = 4-d. \nn
\eea
A power counting in canonical mass dimension leads to an operator built out of field insertions
with total mass dimension $d$ multiplied by a factor $\Lambda^{4-d}$.
According to its canonical dimension, operators with $d \leq 4$ (the SM Lagrangian) are leading terms (LT),
operators with $d=6$ are next-to-leading terms (NLT) in the SMEFT and so on.
This power counting rule is an ordering and normalization statement on the EFT.

For the EFT to be predictive it is also required that the expansion parameters are perturbative
\bea
 C_i \, v^2/\Lambda^2 <1, \quad \quad C_i \, p^2/\Lambda^2 <1.
\eea
The accompanying Wilson coefficients are UV dependent. This condition translates into a condition
on the UV completion by insisting that the IR EFT construction is predictive.
A power counting in Naive Mass Dimension allows an estimate of the neglected higher order
terms in a calculation by varying the unknown $C^d_i v^{d-4}/\Lambda^{d-4},C^d_i |p|^{d-4}/\Lambda^{d-4}$ over a declared
range of values, allowing an interpretation of the data in the SMEFT.

\subsubsection{NDA counting}
One can determine a consistent set
of power counting rules that are a property of the SMEFT (and other field theories) that are distinct from just using the
canonical mass dimension estimate.
This alternative approach is known as Naive Dimensional
Analysis (NDA) \cite{Manohar:1983md}.
NDA normalization has a physical intuition underlying it.
One can determine the generation of operators $Q_i$ from operators $Q_j$ using
topological relations due to the full set of
connected diagrams in the EFT.
Applying this idea to the SMEFT, one can assume roughly homogeneous matching $C_i \sim C_j$ for operators in $\mathcal{L}_6$,
and demand that parameter tuning is avoided. Then one expects $|C_i| \gtrsim |\Delta C_i|$ where $|\Delta C_i|$ is the absolute value of the induced
Wilson coefficient for $C_i$ due to $\int \mathcal{L}_{SM} \times C_j \, Q_j$ from the allowed connected diagrams.

The generalized version of NDA~\cite{Gavela:2016bzc} of this form,
which builds upon the work in Refs.~\cite{Manohar:1983md,Cohen:1997rt,Luty:1997fk,Jenkins:2013sda,Buchalla:2013eza,Buchalla:2014eca}
gives a normalization of Lagrangian terms as
\begin{equation}\label{NDA}
 \frac{\Lambda^4}{16\pi^2} \left(\frac{\de_\mu}{\Lambda}\right)^D \left(\frac{4\pi V_\mu}{\Lambda}\right)^A
 \left(\frac{4\pi \psi}{\Lambda^{3/2}}\right)^F \left(\frac{4\pi H}{\Lambda}\right)^S\left(\frac{g}{4\pi}\right)^{N_g} \left(\frac{y}{4\pi}\right)^{N_y}
 \left(\frac{\lambda}{16\pi^2}\right)^{N_\lambda},
\end{equation}
for an interaction term with $D$ derivatives, $A$ gauge fields, $F$ fermion insertions and
$S$ scalar fields, accompanied by $N_g$ gauge coupling constants, $N_y$ Yukawas
and $N_\lambda$ quartic scalar couplings.

NDA was first determined in the context of the chiral quark \cite{Manohar:1983md}
model, but its approach has been successfully applied to many other EFT descriptions.
This is because it is a consistency condition on a EFT construction.
NDA does not change the Lagrangian. It
 makes manifest an ordering of the Wilson coefficients that
corresponds to a lack of tuning of parameters, consistent with its assumptions in the EFT.
This normalization is a property of the EFT
and the EOM generated by varying the action of the EFT (see Section \ref{SMEOM}) obeys this power counting as a result.

NDA supplies a reasonable {\it estimate} of the normalization of a Lagrangian term, which still multiplies an unknown
Wilson coefficient. Eq.~\ref{NDA} characterizes the scales in a UV sector as $\sim \Lambda$.
If a distinct set of scales exists in matching onto a UV sector, this can lead to the Wilson coefficients having to differ
significantly if a NDA normalization is used. Applying a NDA normalization to global fits in the SMEFT
the Wilson coefficients should be treated as free parameters for this reason.

\subsubsection{UV matching scenarios}\label{UVassumeaway}
A distinct approach to ordering parameters in the Lagrangian also goes under the name power counting
in the literature.
To clarify this difference we will  refer to this approach as ``Meta-matching''. The idea is to make more assumptions
on the UV sector, that leads to broad conditions on the Wilson coefficients.
The difficulty is in making UV assumptions in a well defined, precise, and consistent
manner. Meta-matching is closely related to power counting in that it defines a normalization
of Lagrangian terms in the EFT, this is why it is also referred to as power counting in some literature.
It is also a distinct UV assumption heavy approach
that blurs the strict separation of scales defining an EFT construction.
Despite this, it can still be a useful tool.

{\bf{Topological tree/loop Meta-matching:}} The most commonly used version of Meta-matching is the ``tree-loop''
classification scheme of Artz, Einhorn and Wudka \cite{Arzt:1994gp,Einhorn:2013kja}.
The idea is to study topologically all of the field content of weakly coupled renormalizable
models that can couple at tree level to the SM states exhaustively, and to determine which
$\mathcal{L}_6$ operators can be generated at tree level in matching to a basis.
For the Warsaw basis, the operators classified as ``tree'' in this manner for $\mathcal{L}_6$
are those in Classes $\{2,3,5,7,8\}$ in Table \ref{op59}. As the operators
in Classes $\{1,4,6\}$ contain field strengths, this classification scheme could have been
misunderstood as being related to a ``minimal coupling'' condition
in matching results. Artz, Einhorn and Wudka \cite{Arzt:1994gp} do not assert
this.\footnote{For further discussion on minimal coupling,
see Refs.~\cite{weinberglectures,Jenkins:2013fya}.}
The analysis in Ref.~\cite{Arzt:1994gp}
discussed the possibility of kinetic mixing of the form $V_{\mu \, \nu} F^{\mu \, \nu}$,
for $F$ a beyond the SM $\rm U(1)$ vector boson only briefly. This interaction is redundant,
so this does not limit the conclusions of Ref.~\cite{Arzt:1994gp}.

This scheme is an assertion of UV assumptions, not a consistency condition of the EFT, which is the point of NDA.
The results of Ref.~\cite{Arzt:1994gp} are limited to weakly coupled and renormalizable
UV scenarios, they do not apply if the UV contains an EFT or a strongly interacting theory \cite{Jenkins:2013fya}.
The classification scheme of Ref.~\cite{Arzt:1994gp} also cannot capture the low energy effects of the multi-pole expansion
discussed in Section \ref{substructure}.
Furthermore, $\mathcal{L}_6$ operators equated by the EOM have Wilson coefficients that differ in this scheme,
which can cause some confusion. An example is given in the EOM operator relation
\bea
\mathcal{P}_{HB}  &=& \frac12 g_1^2 \hyp_H Q_{H \Box}  +2g_1^2 \hyp_H Q_{H D} - \frac12 \hyp_H g_1^2 Q_{H B} - \frac14 g_1g_2  Q_{H WB}, \nn
 &+&\frac12  g_1^2\left[\hyp_l Q_{\substack{H l \\ tt}}^{(1)} +\hyp_e Q_{\substack{H e \\ tt}} + \hyp_q Q_{\substack{H q \\ tt}}^{(1)}
 +\hyp_u Q_{\substack{H u \\ tt}}+ \hyp_d Q_{\substack{H d \\ tt}} \right].
\eea
Here the Wilson coefficients of the operators $Q_{H B},Q_{H WB}$ are considered to be ``loop
level'', and the remaining operators are considered to have Wilson coefficients that are tree level in this scheme.
Recall that the operator forms are equated as a result of the freedom to redefine the SM fields by $\mathcal{O}(1/\Lambda^2)$
corrections without changing the $S$ matrix. It is important to understand the distinction between such IR relations
that are statements of consistency conditions in the SMEFT (in this case the freedom to change variables in a path integral
without physical effect) and the UV assumptions that can break the corresponding relations that follow between Wilson
coefficients.
Unfortunately, this distinction is frequently lost in the literature
where references to ``tree'' and ``loop'' operators abound, which blurs the
distinction between the IR operators and the UV matching coefficients
that follows from the separation of scales defining the EFT.

{\bf{One scale, one coupling Meta-matching:}}\label{badidea}
It was observed in Ref.~\cite{Georgi:1986kr} that the NDA results of
Ref.~\cite{Manohar:1983md} could follow from the assumption of NDA being
respected by QCD as an assumed high energy theory, and then considered to be matched
to $\chi$PT with $g_3 \simeq 4 \, \pi$. This point was not stressed, as it is well
known that when $\rm QCD$ is strongly interacting a useful EFT description
transitions to $\chi$PT treated as a bottom up EFT, as an essential singularity is present.
In the case of QCD with $m_q\rightarrow 0$, the beta function leads to the relation
\begin{align}
\left( \frac{\Lambda_{qcd}}{\mu} \right)^{b_0} &= e^{-8\pi^2/\left[ \hbar\, g_3^2(\mu)\right]}\, ,
\label{lqcd}
\end{align}
with the factors of $\hbar$ explicit ~\cite{Coleman:1975qj,Coleman:1978ae}.
$\Lambda_{qcd}$ is not due to summing the $g_3^2$ expansion to all orders with a choice $g_3 \simeq 4 \, \pi$
due to the essential singularity at $g_3^2\hbar=0$ in the $g_3^2$ and $\hbar$ expansions. No
useful perturbative matching can be performed in this limit.

Ref.~\cite{Georgi:1989xy} observed that if some of the
vector resonances of $\chi$PT were considered lighter than the remaining states,
an alternate normalization of the Lagrangian terms of $\chi$PT under this
assumption could be constructed.

A highly influential version of Meta-matching was put forth in Ref.~\cite{Giudice:2007fh}
based on arguing these two observations hold in a useful manner for general scenarios
of strongly interacting physics with a light pseudo-Goldstone Higgs. The modern version of the
Lagrangian normalization for a composite sector that results \cite{Panico:2015jxa} is given by
\bea\label{mistake}
\frac{m_\star^4}{g_\star^2} \left(\frac{\partial}{m_\star}\right)^A \left(\frac{g_\star \Pi}{m_\star}\right)^B \left(\frac{g_\star \sigma}{m_\star}\right)^C \left(\frac{g_\star \Psi}{m_\star^{3/2}}\right)^D.
\eea
Here the $\sigma$ model scale is given by $f = m_\star/g_\star$  where $m_\star$
and $g_\star$ are the corresponding one scale and one coupling assumed in the composite sector.
The $\Psi$, $\Pi$ and $\sigma$ are assumed states in the composite sector with the latter two
associated with the $\sigma$ model scale $f$. Further assuming that $g_\star < 4 \pi$
one can (at least hypothetically) hope to avoid the equivalent challenge of an essential singularity being present in perturbative
expansions in the strong coupling limit, and obtain a result consistent with the NDA
result of Ref.~\cite{Manohar:1983md} (with $4\pi$ replaced by $g_\star$), assuming a matching
onto the SM states making up $\mathcal{L}_6$. It is however necessary to stress that the presence of an essential singularity in the strong coupling limit
is the basic reason that any such approach taking the $g_\star \rightarrow 4 \pi$ (for example in Eq.~\ref{mistake})
in a naive analytic continuation is not expected to be useful or informative in general, and rather unwise on very basic grounds.
For a strongly coupled UV theory in general, there is no reason to expect such a simplistic counting to hold for this reason,
and also for the reasons discussed in Sections \ref{substructure}, \ref{matchingexamples}.

As initially formulated in Ref.~\cite{Giudice:2007fh}, this approach to Meta-matching
had several other essential inconsistencies, see discussion in Refs.~\cite{Manohar:1983md,weinberglectures,Jenkins:2013fya,Buchalla:2014eca,Liu:2016idz,Gavela:2016bzc}.
Alternative reasoning \cite{Buchalla:2014eca,Pomarol:2014dya,Panico:2015jxa,Contino:2016jqw} has since been advanced to
justify some of the results of Ref.~\cite{Giudice:2007fh} and other
assumptions of this work have been effectively abandoned \cite{Jenkins:2013fya,Buchalla:2014eca,Liu:2016idz}.

{\bf{Universal theories:}}
An oblique, or universal theory assumption \cite{Barbieri:2004qk,Elias-Miro:2013gya,Elias-Miro:2013mua,Wells:2015uba,Farina:2016rws,Alioli:2017jdo} is
another example of Meta-matching.
This idea asserts that the dominant effects of physics beyond the SM can be sequestered into
modifying the two point functions of the gauge bosons $\Pi_{WW,ZZ,\gamma Z}$.
The idea of oblique, or universal theories, is motivated out of past approaches to EWPD, where
the global $\rm S,T$ EWPD fit assumes that vertex corrections due to physics beyond the SM are neglected
- giving the ``oblique" qualifier \cite{Peskin:1991sw}.
It is problematic that the oblique, or universal, assumption is not field redefinition invariant,\footnote{A corollary to the point that
Lagrangian parameters are not physical (see Section \ref{preliminarymeasurements}) is that field redefinitions cannot be used
to satisfy defining physical conditions on the EFT, such as the oblique, or universal, assumption. For discussion
on this point see Ref.~\cite{Trott:2014dma}.} see
Section \ref{SMEOM}.

Oblique, or universal, assumptions were tolerable in the 1990's and early 2000's as the SM Higgs couples in a dominant fashion to $\Pi_{WW,ZZ}$
when generating the mass of the $W,Z$ bosons, and has small couplings to the light fermions, satisfying the assumption in a particular UV model
that was not directly experimentally supported.

LHC results now indicate the $W,Z$ bosons obtain their mass in a manner that is associated with the Higgs-like scalar.
Corrections to $\Pi_{WW,ZZ}$ can be included for the SM,
or more generally \cite{Barbieri:2007bh} due to the discovered scalar.
There is no strong theoretical support to maintain an oblique, or universal, assumption to constrain
the further perturbations due to new physics
scenarios in the SMEFT with EWPD. It is also unclear if this idea is
even a consistent theoretical concept.  The reason is that a fully defined mechanism of dynamical
mass generation in a UV sector has never been demonstrated to be
consistent with this assumption, to define a consistent IR limit. Furthermore, assumed
oblique, or universal, theories generate non-universal theories \cite{Trott:2014dma,Wells:2015cre} using the renormalization group to
run the operators matched onto from $\Lambda \rightarrow \hat{m}_z$. For these reasons,
this assumption has largely been abandoned in consistent SMEFT analyses in recent years.\footnote{For more discussion
on the limitations of a universal theories assumption, see Refs.~\cite{Cacciapaglia:2006pk,Trott:2017yhn}.}

\subsubsection{Meta-matching vs power counting}
EFT formalizes the separation of scales in a problem
so that Meta-matching assumptions are avoidable.
An analogy to the SM is useful. The SM predictions depend
on the numerical values of the gauge couplings $\{g_1,g_2,g_3\}$. Different values of the gauge
couplings in the SM lead to different predictions of $S$ matrix elements and relations between
$S$ matrix elements. The SM is studied
without invoking as necessary a reference to a grand unified theory (GUT) embedding, even though such an embedding would form a
UV boundary condition that dictates the values of $\{g_1,g_2,g_3\}$ at lower scales.
The same point holds in the SMEFT and HEFT, for different values of the Wilson coefficients.
It is largely a sterile and unproductive debate to dispute assumptions about the size of
$\{g_1,g_2,g_3\}$ or $C_i$. Measuring parameters is more productive.

Making Meta-matching assumptions in an EFT framework causes confusion and is the source of much
conflict in the literature. The separation of scales fundamental to the EFT construction
can be violated, the assumptions are frequently assigned to the IR operator forms, not the
UV dependent Wilson coefficients, and yet EFT language and concepts are used. The conclusions derived
are then not model independent EFT statements, although they are usually argued to
be ``broadly applicable" conclusions as a subjective assertion.

On the other hand, due to the large number of
parameters present in the HEFT and the SMEFT the reduction in the parameter
space seems required. We stress that it is universally accepted to use the {\it IR assumptions} of
experimentally motivated flavour symmetries
being present in the operator basis, which leads directly to a sufficient
reduction in Wilson coefficient parameters to enable a global constraint program.

At times extra assumptions are still
employed which can lead to much stronger conclusions studying the data. Such conclusions must be
defined in a consistent IR limit of a UV sector to be meaningful and,
even if this is the case, are limited by model dependence. Such results can be of great value
if they are properly qualified and defined.

\subsection{Counting rules for the chiral Lagrangians}\label{Section.countingHEFT}
Power counting in HEFT has also been intensely debated recently in
Refs.~\cite{Buchalla:2013eza,Buchalla:2014eca,Gavela:2016bzc,Buchalla:2016sop}.
In HEFT, it is not possible to define an ordering criterion among the
invariants that holds in any kinematic regime. Furthermore,
assumptions made on the size of the Wilson coefficients can have significant
phenomenological consequences. In this HEFT discussion, LT will indicate $\mathcal{L}_0$ while NLT is $\Delta\mathcal{L}$.

\subsubsection{\titlemath{$\chi$PT}{Chiral perturbation theory}}
The HEFT is a fusion of the SMEFT with $\chi$PT.
These two EFTs follow different power countings. One of the main applications of $\chi$PT is the
description of the octet of light QCD mesons ($\pi^{0,\pm},\,K^{0,\pm},\,\eta)$ at $E\ll\Lambda_{\rm QCD}$.
These particles are the Goldstone bosons of the spontaneous breaking of
the chiral symmetry $\rm SU_L(3)\times SU_R(3)$ into the diagonal $\rm SU_{L+R}(3)$ and they are collectively described by the field
\begin{equation}
 U = \exp\left(2i\frac{\mathbf{\Phi}}{f}\right),\quad \mathbf{\Phi} = \phi^A T^A\,,
\end{equation}
that transforms as $U\mapsto LUR^\dagger$ under the chiral group. Here $f$ is the meson decay constant\footnote{At this level all the eight mesons are
degenerate and share the same decay constant. The inclusion of higher-order terms breaks the $\rm SU_{L+R}(3)$ symmetry leading to the splitting $f_\pi\simeq93$~MeV, $f_K\simeq113$~MeV. }, that satisfies the relation $4\pi f \geq \Lambda$, where $\Lambda$ is the cutoff of the effective description~\cite{Manohar:1983md}.
Omitting the electromagnetic interactions, and in the limit of massless quarks, the leading order Lagrangian is
\begin{equation}\label{chiPT_kin}
 \mathcal{L}_2 =
\frac{f^2}{4}\Tr(\de_\mu U^\dagger \de^\mu U),
\end{equation}
where the presence of $f^2$ ensures canonical kinetic terms for the $\phi^A$ fields upon expanding the exponential.
Because the matrix $U$ is adimensional, this EFT is organized as an expansion in derivatives:
 \begin{equation}\label{chiPT_expansion}
 \mathcal{L}_{\chi{\rm PT}}=\mathcal{L}_2 + \mathcal{L}_4 + \mathcal{L}_6 +\dots\,,
\end{equation}
where $\mathcal{L}_d$ contains invariants with $d$ derivatives. This is Weinberg's power counting approach, which
is supported by a renormalization argument~\cite{Weinberg:1978kz}.
The Lagrangian $\mathcal{L}_2$ contains an infinite series in powers of the meson fields $\phi^A$.
The counterterms required to reabsorb the one-loop divergences appearing at this level correspond to interaction terms with
four derivatives. This can be seen with a topological analysis: an amplitude with $L$ loops and containing $N_d$ vertices with $d$ derivatives must scale with $D$ powers of momentum, where~\cite{Weinberg:1978kz}
\begin{equation}\label{Weinberg_counting}
 D = 2L+2 + \sum_d (d-2)N_d\,.
\end{equation}
Computing at one loop ($L=1$) with $d=2$ vertices gives only amplitudes with $D=4$, i.e. four derivatives. Computing at two loops with $L_2$ or at one loop with one insertion of a 4-derivative interaction plus an arbitrary number of $d=2$ vertices, gives instead 4-derivative terms and so on.
This is in contrast with the SMEFT case, where loops containing
one insertion of a $d$-dimensional operators generates divergences of order $\leq d$.\footnote{See Ref.~\cite{Jenkins:2013zja} for the complete set of these terms.}
Eq.~\ref{Weinberg_counting} implies that $\chi$PT is renormalizable at fixed order in the momentum expansion, which provides a solid iterative method for organizing the effective series: each order of the EFT must contain at least all the operators that are required as counterterms for the one-loop renormalization of the preceding order.

An alternative counting prescription for $\chi$PT is that of NDA~\cite{Manohar:1983md}. This provides a rule for establishing the normalization of a given operator and it is constructed so that the scaling assigned to a given invariant is independent of the loop order at which it is generated in the EFT. Restricted to our simple case, it states that the overall coefficient of a
  generic interaction with $D$ derivatives and $S$ pion fields is estimated by the formula
\begin{equation}\label{NDA_0}
 f^2\Lambda^2 \left(\frac{\de_\mu}{\Lambda}\right)^D  \left(\frac{\phi}{f}\right)^S \,.
\end{equation}
This rule assigns the correct coefficient $f^2$ to the Goldstones kinetic term (Eq.~\ref{chiPT_kin}).
The suppression in powers of $\Lambda$ assigned by the NDA rule is equivalent to the derivative counting. In the NDA notation, the $\chi$PT Lagrangian can be written
\begin{equation}\label{chiPT_expansion_NDA}
 \mathcal{L}_{\chi{\rm PT}}=\frac{f^2}{4}\left(\Tr(\de_\mu U^\dagger \de^\mu U) + \frac{1}{\Lambda^2}\mathcal{L}_4 + \frac{1}{\Lambda^4}\mathcal{L}_6 +\dots\right)\,.
\end{equation}
Requiring that operators with 4 derivatives must be weighted by a factor at least as large as a loop suppression one obtains the constraint $f^2/\Lambda^2\geq 1/(4\pi)^2$, i.e. $\Lambda\leq 4\pi f$.

$\chi$PT without electromagnetic interactions and in the limit of vanishing quark masses, is a simple example,
in which the three counting methods (derivative, loops, NDA) all lead to the same result. The scenario becomes more complicated
as soon as dimensionful quantities or fields, such as photons or mass terms,
are incorporated in the EFT. The explicit chiral symmetry breaking term
\begin{equation}
 \mathcal{L}_\chi =  \frac{f^2}{4}\Tr(U^\dagger \chi + \chi^\dagger U)\,,
\end{equation}
has $\chi$ as a quantity proportional to the quarks masses $\chi\sim{\rm diag}(M_u,M_d,M_s)$ that transforms as $\chi\to L\chi R^\dagger$. The derivative counting
can be extended to this object assuming that, being dimensionful, $\chi$ shall scale as $p$. In this way the expansion in $p/\Lambda$ is formally maintained, but upgraded to the nomenclature of ``chiral dimension'', which counts both the  number of derivatives
and $\chi$ insertions. The term $\mathcal{L}_\chi$ has chiral dimension 2 and therefore is a leading term, that
should be included in $\mathcal{L}_2$.

The scaling $\chi\sim p$ is formal and represents an approximation valid
only in the regime $p\simeq M_{\pi,K}$. The chiral dimension gives a correct estimate of the impact of a given operator only for processes with on-shell mesons, while it loses physical meaning as one moves outside this kinematic regime.
This is because the relative importance of two given operators (such as the kinetic term and $\mathcal{L}_\chi$) can be different
in $S$ matrix elements measured at distinct energies.

\subsubsection{HEFT counting}
HEFT is complex as an EFT due to the simultaneous presence of a scalar sector embedded in a $\chi$PT-like construction and of fermions and longitudinal gauge bosons, whose interactions are organized according to their canonical dimension.

Following the two-step procedure of power counting presented at the beginning of this section,
one first has to address the question of assigning weights to each Lagrangian term.
This can be done in a self-consistent EFT approach adopting the generalized version of
NDA~\cite{Gavela:2016bzc}, which builds upon the work in Refs.~\cite{Manohar:1983md,Cohen:1997rt,Luty:1997fk,Jenkins:2013sda,Buchalla:2013eza,Buchalla:2014eca}.
For the HEFT case, this modern NDA assigns the scaling
\begin{equation}\label{NDAh}
 \frac{\Lambda^4}{16\pi^2} \left(\frac{\de_\mu}{\Lambda}\right)^D \left(\frac{4\pi A_\mu}{\Lambda}\right)^A \left(\frac{4\pi \psi}{\Lambda^{3/2}}\right)^F \left(\frac{4\pi\phi}{\Lambda}\right)^S\left(\frac{g}{4\pi}\right)^{N_g} \left(\frac{y}{4\pi}\right)^{N_y},
\end{equation}
to an interaction with $D$ derivatives, $A$ gauge fields, $F$ fermion insertions and $S$ scalar fields (either the Goldstones or the physical Higgs), accompanied by $N_g$ gauge coupling constants and $N_y$ Yukawas.

Determining the scaling of a given term does not ensure the
existence of a unique ordering of effects in the EFT that holds at any energy regime
and that is also independent of any UV assumptions.
There are several different elements that coexist in the HEFT: one may choose to organize the expansion in loops, i.e. according to the requirement of an order by order renormalization, or rather to order the operators by their number of derivatives, chiral dimensions, inverse powers of $\Lambda$ or of $4\pi$. These rules are not generally consistent with each other
and they all operate simultaneously.
Such a choice can be made upon restricting to a particular kinematic regime or
assuming a particular class of UV completions.
It is not possible, to our knowledge, to select one of these elements as the unique dominating rationale
to order terms in importance for all $S$ matrix elements within the predictive
regime of validity of this EFT, and for all possible values of the Wilson coefficients.

To see the contradictions that may emerge between different countings, consider the operator $\epsilon_{IJK}W^I_{\mu\nu}W^{J\nu\rho}W^{K\mu}_\rho$.
This term is not required to reabsorb one-loop divergences of $\mathcal{L}_0$ and it has chiral dimension 6, so
it is a NNLT according to Weinberg's counting ($\mathcal{L}_6$ in the $\chi$PT series).\footnote{This is the case both with the classical notion of chiral dimension in $\chi$PT, defined in the previous paragraphs, and with the chiral dimension generalized to the HEFT case, defined in  Ref.~\cite{Buchalla:2013eza}.
Note that the latter requires the definition of the operator to include a factor $g_2^3$.}
It may be argued that, containing only transverse gauge fields,
this operator should not follow the chiral counting, but rather the SMEFT classification,
that sets it as a NLT. An alternative observation is that as this operator contains
two derivatives more than the gauge kinetic term, which belongs to $\mathcal{L}_0$,
this operator can be a NLT even in a chiral approach.
Finally, NDA assigns to this term a coefficient $4\pi/\Lambda^2$, which has
both a $(4\pi)$ enhancement and a $\Lambda^2$ suppression. The
different counting rules give contradictory estimates in this case because each rule
stems from a different assumption.  Depending on both the kinematic regime of interest
and particular values of Wilson coefficients considered, any
one of these arguments may drive the organization of the EFT:
for instance in the limit $p/\Lambda\ll v/\Lambda$ the derivative expansion
dominates over other criteria, but this is not the case in general.

It is necessary, whatever power counting is followed, that inconsistencies in the theory
are avoided. Classifying as NLT the operators appearing as counterterms
for the renormalization of $\mathcal{L}_0$ is required. Even within this category,
the physical impact of the invariants can be different and dependent on the
kinematic regime to which the HEFT is applied.
Two such structures are for instance $\bar Q_L\V_\mu\V^\mu\U Q_R$ and $(\bar Q_L\U Q_R)^2$.
The former has one fermionic current and two derivatives, that correspond to gauge bosons
insertions in unitary gauge, the latter is a four-fermion operator. The NDA weight
for the single-current term is $1/\Lambda$, and the one for the four-fermion term
is $(4\pi)^2/\Lambda^2$. So the two can have a comparable impact in a
specific kinematic region, but this isn't the case in general for all possible $S$ matrix elements.

There are many intrinsic challenges to a systematic classification of HEFT operators
into well defined orders. Some of those are present in $\mathcal{L}_0$ where one can select whether couplings
of the physical Higgs to the gauge kinetic terms should be included in $\mathcal{L}_0$ or not.
The choice made in Eq.~\ref{Lag0} to define these terms as NLT is phenomenologically sound,
but it requires a specific assumption that the transverse gauge fields not
being directly coupled to the scalar sector in the UV, leading to a suppressed Wilson coefficient.
Once the leading Lagrangian is fixed, the next order $\Delta\mathcal{L}$ can be identified
following:
\begin{itemize}
\item{All terms required for the one-loop renormalization on $\mathcal{L}_0$ must be included in $\Delta\mathcal{L}$.
 To this end, the one-loop renormalization of the scalar sector of the HEFT has been worked out
 in Refs.~\cite{Espriu:2013fia,Delgado:2013loa,Delgado:2014jda,Gavela:2014uta,Guo:2015isa,Alonso:2015fsp}.
 These results build upon previous results obtained in the absence of the $h$ degree of
 freedom~\cite{Herrero:1992zq,Herrero:1993nc,Herrero:1994iu,Dittmaier:1995cr,Dittmaier:1995ee}.}
\item{Structures that are not strictly required as counterterms, but that receive finite loop
contributions from $\mathcal{L}_0$ and/or have a similar field composition or NDA suppression
as the counterterms, should also be retained. An example corresponding to finite loops of $\mathcal{L}_0$ is
dipole operators, that have the structure $\bar\psi_L\s^{\mu\nu} \U\psi_R X_{\mu\nu}$.
An example of inclusion by similarity in the field composition are four-fermion operators with
vector/axial currents (schematically $\bar\psi\gamma_\mu\psi\,\bar\psi'\gamma^\mu\psi'$),
which are included by analogy with the four-fermion operators with scalar currents
($\bar\psi_L\U\psi_R\,\bar\psi'_L\U\psi'_R$) that are required for the renormalization of $\mathcal{L}_0$.}
 \item{Operator categories that are classified as NLT in at least one of the counting
 principles described above should be included, e.g. the term $\epsilon_{IJK}W^I_{\mu\nu}W^{J\nu\rho}W^{K\mu}_\rho$.
This point ensures that all the operators that can be relevant in at least one specific scenario are included,
preserving the generality of the HEFT description.}
 \item{The set of operators that form the $\Delta\mathcal{L}$ basis should close
 under the use of the EOMs.}
\end{itemize}
These rules lead to the construction of the bases of Refs.~\cite{Buchalla:2013rka,Brivio:2016fzo}.
The characterization of higher orders in the HEFT expansion poses further difficulties due to the mismatch among the expansions in different parameters.

\section{The \titlemath{$S$}{S}-matrix, Lagrangian parameters and measurements}\label{preliminarymeasurements}
Subtleties that can complicate the interpretation of experimental data in
EFTs are clarified by a clear definition of
$S$ matrix elements and their distinction from Lagrangian parameters. In this section we review
these two concepts for later use and emphasize the distinctions between them and their relation to measurements.
The aim of this section is to discuss the physical nature of $S$ matrix elements,
the unphysical nature of Lagrangian terms, and to then build up to how the pseudo-observable concept is introduced
to attempt to bridge between full fledged $S$ matrix elements and common gauge invariant quantities that appear in many measurements.

\subsection{\titlemath{$S$}{S}-matrix elements}\label{Smatrix}
Consider a set of interpolating spin-$\{0,1/2,1 \}$ fields, that are representations of  $\rm SO^+(3,1)$,
which we denote schematically as $\theta = \{\phi,\psi,V^\mu \}$. These fields are used to construct
asymptotic perturbative expansions that satisfy a set of global symmetries ($G$).\footnote{The global symmetry is emphasized as
the local gauge redundancy does not lead to relations between $S$ matrix elements. Gauging a symmetry does not give any additional
conserved charges (and corresponding $S$ matrix relationships) beyond those of the corresponding global symmetry, see Refs.~\cite{weinberglectures,Jenkins:2013fya}
for discussion.} We denote the Lagrangian
composed of $\theta$, constructed to manifestly
preserve $G$,
as $\mathcal{L}(\theta)$, and the sources of the fields as $J_\theta$. The
generating functional $W[J_\theta]$
of the connected Green's functions is then defined as
\bea\label{Eqn1.1generating}
e^{i \, W[J_\theta]} = \int \mathcal{D} \theta \, {\rm exp} \left[i \int d^4 x \left(\mathcal{L}(\theta) + \theta \, J_\theta \right) \right].
\eea
Here $\mathcal{D}$ is the measure of the functional integral.
Functional differentiation of $W[J_\theta]$ with respect to $n$ fields then defines an $n$-point time ordered correlation
function (or Green's function) between ground states ($\Omega$) of the interacting theory
\bea
\langle \Omega | T \theta(x_1) \, \theta(x_2) \cdots \theta(x_n)| \Omega \rangle
= (-i)^n \frac{\delta^n W}{\delta J_\theta(x_1) \cdots \delta J_\theta(x_n)}.
\eea
The LSZ reduction formula \cite{Lehmann:1954rq} defines the $n$-point $S$ matrix elements
related to these correlation functions as
\bea
\prod_1^a \int &d^4 x_i& \, e^{i p_i \cdot x_i} \prod_1^b \int d^4 y_i \, e^{i k_i \cdot y_i}
\langle \Omega | T \theta(x_1) \, \theta(x_2) \cdots \theta(x_a)\theta(y_1) \, \theta(y_2) \cdots \theta(y_b)| \Omega \rangle, \nn
&\simeq& \left(\prod_{i =1}^a \frac{i \sqrt{Z_i}}{p_i^2 - m^2_i+ i \epsilon}\right)
\left(\prod_{j =1}^b \frac{i \sqrt{Z_j}}{k_j^2 - m_j^2+ i \epsilon}\right) \langle p_1 \cdots p_a|S|k_1 \cdots k_b \rangle.
\eea
Here $n = a+b$, and $p_i$,$k_j$ are four momenta associated via the Fourier transform of each field,
which is also associated with a mass term $m_{i,j}$ and renormalization factor $Z_{i,j}$.
The $\simeq$ is due to the requirement to isolate each single particle pole experimentally
so that each $p_i^0 \rightarrow E_{p_i}$, $k_j^0 \rightarrow E_{k_j}$ when treating the particles as asymptotic states.

$S$ matrix elements, unlike Lagrangian parameters, directly define the measurable
quantities in the theory: the scattering and decay observables.\footnote{The $S$ matrix was first introduced in Ref.~\cite{PhysRev.52.1107}.} $S$ matrix elements
conserve overall four momentum and are unitary. The decomposition of an $S$ matrix element into an
Lorentz invariant amplitude $\mathcal{M}$ is given as
\bea
\langle p_1 \cdots p_a|S|k_1 \cdots k_b \rangle = I + i \,(2 \pi)^4  \, \delta^4(p_i - k_j) \, \frac{\mathcal{M}}{\sqrt{2 E_i}\sqrt{2 E_j}},
\eea
with implicit sum over $i,j$. For most scattering observables, $2 \rightarrow n$ processes are
sufficient to consider, which are given as
\bea
d \sigma = \frac{(2 \, \pi)^4 |\mathcal{M}|^2}{4 \sqrt{(p_1 \cdot p_2)^2 - m_1^2 \, m_2^2}} \, \delta^4 (p_1 + p_2 - \sum_j k_j) \prod_{j=1}^n
\frac{d^3 k_j}{(2 \pi)^3 \, E_j}.
\eea
Similarly, particle decays of an initial state with momentum ($P$) and mass ($M$) into $\sum_j k_j$ final states are given by
\bea
d \Gamma = \frac{(2 \, \pi)^4}{2 \, M} \, |\mathcal{M}|^2 \, \delta^4 (P - \sum_j k_j) \prod_{j=1}^n
\frac{d^3 k_j}{(2 \pi)^3 \, E_j}.
\eea
Measured event rates correspond to $d \sigma$ or $d \Gamma$ integrated over a measured phase space volume.
Such phase space integrations can be highly non-trivial, see Refs.~\cite{Byckling:1971vca,Olive:2016xmw} for excellent discussions
on overcoming this technical hurdle.
The relation between the differential scattering and decay observables and $S$ matrix elements is direct.
Intuitively, changing field variables on the path integral defining the generating functional in Eq.~\ref{Eqn1.1generating}
can be expected to have no physical effect on the $S$ matrix elements derived from it, as the field variables are essentially
dummy variables used to define the $G$ preserving perturbative expansions of the correlation functions in the path integral formulation
of the QFT. This intuition is correct, and not violated by quantum corrections.
Such variable changes do modify the
Lagrangian terms and the source terms in a correlated manner, which can be used to arrange the Lagrangian into a particular form. This
understanding of field redefinitions is formalized in what is known as the Equivalence theorem \cite{Kallosh:1972ap}, which is a precise formulation
of the invariance of $S$ matrix elements under
$G$ preserving field redefinitions for renormalized quantities. See Refs.~\cite{Chisholm:1961tha,Kamefuchi:1961sb,Coleman:1969sm,tHooft:1972qbu,
Bergere:1975tr, Politzer:1980me,Georgi:1991ch,Arzt:1993gz} for related discussion.
These formal developments are all focused on gauge
independent, and $G$ preserving, field redefinitions.

\subsection{Lagrangian parameters}\label{lagrangianparam}
Unlike $S$ matrix elements, individual Lagrangian terms are not invariant under field
redefinitions \cite{Kallosh:1972ap,Chisholm:1961tha,Kamefuchi:1961sb,Coleman:1969sm,tHooft:1972qbu,
Bergere:1975tr, Politzer:1980me,Georgi:1991ch,Arzt:1993gz,Passarino:2016pzb,Passarino:2016saj}.
Lagrangian parameters are neither directly, nor trivially related
to $S$ matrix elements, or measured observables. In modern times this understanding has
been advanced to a level where efforts are underway to avoid Lagrangian formulations
completely in highly symmetric field theories.\footnote{See Refs.~\cite{Elvang:2013cua,ArkaniHamed:2008gz} for discussion on this approach.}

Several historical examples of the equivalence of Lagrangians of a distinct interaction and field
variable form indicate that a particular interaction term in a Lagrangian
should not be mistaken for a physical $S$ matrix element. The famous equivalence of the Thirring \cite{Thirring:1958in} and sine-Gordon models under bosonization proven by
Coleman \cite{PhysRevD.11.2088} (see also Ref.~\cite{Mandelstam:1975hb})
illustrates the equivalence of a particular quantum theory with purely interacting fermions and bosons in each case. The Lagrangians are given by
\bea
\mathcal{L}_T = \bar{\psi} (i \slashed{d} - m) \psi - \frac{g}{2} (\bar{\psi}\gamma^\mu \psi)^2,
\quad \quad \mathcal{L}_{sG} = \frac{1}{2}\partial^\mu \phi \partial_\mu \phi + \frac{\alpha}{\beta^2} \cos (\beta \phi).
\eea
The couplings are identified as $\beta^2/4 \pi = 1/(1+ g/\pi)$. These Lagrangians describe the same physics despite
the drastically different field content and interaction terms \cite{PhysRevD.11.2088}.

The practical consequences of Lagrangian terms being distinct from physical $S$ matrix elements
is drastically different in the SM and in the SMEFT/HEFT field theories.
In the case of an operator basis choice in the SMEFT or HEFT, the general non-physical nature of individual Lagrangian terms manifests itself in the
freedom to pick different sets of Lagrangian terms to represent the same physics. This occurs via $\mathcal(1/\Lambda^2)$
small field redefinitions in the SMEFT\footnote{Again, only those that are gauge independent.} and the EOM play an essential
role in equating Lagrangian terms that are naively distinct.

In the SM, $\Lambda \rightarrow \infty$, so there is
no direct equivalent of these small field redefinitions.
The same point on the nature of Lagrangian terms is made by the use of auxiliary fields.
When performing gauge fixing, the BRST formalism \cite{Becchi:1975nq,Iofa:1976je}
introduces the non-propagating Lautrup-Nakanishi \cite{Lautrup, Nakanishi:1973fu}
auxiliary field $B^a$ into a non-abelian Lagrangian as
\bea
\mathcal{L}_{BRST} = -\frac{1}{4} F_a^{\mu \, \nu} \, F^a_{\mu \, \nu} + \bar{\psi} (i \slashed{D} - m) \psi
+ \frac{\xi}{2} (B^a)^2 + B^a \partial^\mu A_\mu^a + \bar{c} (- \partial^\mu D_\mu^{ac})c^c.
\eea
Performing the Gaussian functional integral over the $B^a$ field variable returns
\bea
\mathcal{L}_{FP} = -\frac{1}{4} F_a^{\mu \, \nu} \, F^a_{\mu \, \nu} + \bar{\psi} (i \slashed{D} - m) \psi
- \frac{1}{2 \, \xi}\left(\partial^\mu A_\mu^a\right)^2 + \bar{c} (- \partial^\mu D_\mu^{ac})c^c.
\eea
The two Lagrangians give equivalent $S$ matrix elements despite having a distinct form and naively differing field variables.
Another example is supplied by the Hubbard–Stratonovich transformation \cite{1957SPhD....2..416S,1959PhRvL...3...77H}
where a non-dynamical scalar field is ``integrated in''.
In this case the curvature $R^2$ action of a gravitational theory (i.e. Starobinsky inflation \cite{Starobinsky:1979ty})
\bea
\mathcal{L}_{inf} = \sqrt{\hat{g}} \left[-\frac{M_p^2}{2} \hat{R} + \zeta \hat{R}^2 \right],
\eea
is equivalent to the transformed Lagrangian with the auxiliary scalar field
\bea
\mathcal{L}'_{inf} = \sqrt{\hat{g}} \left[-\frac{M_p^2}{2} \hat{R} - 2 \, \alpha \, \Phi^2 \, \hat{R} - \Phi^4 \right].
\eea
This can be seen by performing a Gaussian functional integral with $\zeta = \alpha^2$. Performing the
Gaussian integrals modifies the field variables to a reduced (less redundant) set that still produce the
same $S$ matrix. In these cases the auxiliary field variables are non-propagating and do not have Fock spaces associated with them.
So the difference in the Lagrangians is a vanishing difference for the external states labeling $S$ matrix elements.
This is the analogy to the vanishing of the difference between operators equivalent by use of the on-shell EOM (as in Eq.~\ref{EOMBshift})
essential to choosing an operator basis in the EFT.\footnote{We stress that the point here is on the non-physical nature of Lagrangian parameters
in general not
that field redefinitions leave the Lagrangian alone invariant. In general this is not the case
as discussed in Section \ref{Smatrix}.}

The fact that Lagrangian parameters are not physical quantities is not changed when working with mass eigenstate fields,
or unitary gauge, in any way.

\subsection{Scattering measurements and renormalizable/non-renormalizable theories}\label{sec:renorm-nonrenorm}

The distinction between Lagrangian terms and $S$ matrix elements is crucial in the SMEFT and HEFT. It is also the case
that using naive physical intuition to assign SM Lagrangian terms
a naive physical meaning classically, although formally incorrect, is a reasonable rough approximation for a subset of parameters in the SM.
Essentially the naive physical intuition at work is accidentally supported
by the renormalizable nature of the SM Lagrangian, the small perturbative corrections in the EW sector of the SM,
and the related fact that the SM unstable states have small widths compared to their masses.
We discuss each of these points in turn and compare the differences that
appear when extending SM studies to the SMEFT and HEFT.

\subsubsection{The SM case}

Extracting the SM Lagrangian parameters $\{v, g_1, g_2, g_3, Y_i\}$
from measurements of $S$ matrix elements at tree level in the SM
is generally done in the presence of significant experimental correlations
and a degeneracy of parameters present in the mapping to the Lagrangian. This degeneracy is first an issue
at the level of perturbative and non-perturbative corrections to the Lagrangian
due to a quirk of renormalizable theories. In the renormalizable SM
on-shell vertices are unique in the interaction Lagrangian, and off-shell gauge coupling vertices
are related by global symmetries to on-shell coupling parameters. This is no longer
trivially the case when one transitions to EFT extensions of the SM.

An unusual case of a Lagrangian parameter extraction in the SM, is the extraction of $v$
from $\mu^- \rightarrow e^- + \bar{\nu}_e+ \nu_\mu$.
The physics at work is due to the
SM being a weakly coupled renormalizable theory describing a Higgsing of $\rm SU_L(2) \times U_Y(1) \rightarrow U(1)_{em}$, in conjunction
with neutrinos being weakly interacting particles.
Extracting $v$ in the SM is unusual as it can be extracted from a
$\bar{\psi}_L \, \psi_L \rightarrow \bar{\psi}_L \, \psi_L$ process at low energies where an effective
Lagrangian of the form
\bea
\mathcal{L}_{GF} = - \frac{4 G_F}{\sqrt{2}} V_{ij} \, V_{kl}^\star (\bar{\psi}_i \gamma^\mu P_L \psi_j)(\bar{\psi}_k \gamma^\mu P_L \psi_l),
\quad \quad \frac{4 G_F}{\sqrt{2}} = \frac{g_2^2}{2 \,M_w^2} = \frac{2}{v^2}.
\eea
is used.\footnote{Remarkably, interest in Fermi theory has been acute in recent years
in the literature, see Refs.~\cite{trottWG,Gavela:2016bzc,Contino:2016jqw}.
Our discussion is most consistent with Refs.~\cite{trottWG,Gavela:2016bzc}
due to the important role that the CKM and PMNS-matrix plays, and the standard understanding of EFT
adopted in this review.} Here $V_{ij}$ is the CKM or PMNS-matrix in the case of quarks
or leptons. The gauge coupling  $g_2$ cancels as a direct consequence that the SM has a Higgs mechanism
generating the gauge boson masses. This cancellation is not generic in EFTs but is a rather unique result of the SM.
Furthermore, as individual asymptotic eigenstates of neutrinos are not experimentally identified in this decay process,
practical measurements of $\mu^- \rightarrow e^- + \bar{\nu}_e+ \nu_\mu$ sum over all neutrino species.
Unitarity of the PMNS-matrix then leads to a direct extraction of $v$ from this decay, which is another unique feature
in extracting $v$ from $\mu^- \rightarrow e^- + \bar{\nu}_e+ \nu_\mu$.

Subsequently when using the input parameter set $\{\hat{G}_F,\hat{M}_{W},\hat{M}_Z\}$
mapping these measured quantities to the Lagrangian parameters $g_1,g_2$ can be done with a $m_Z$ pole scan
as performed at LEP \cite{ALEPH:2005ab}, and a study of transverse $W$ mass variables as performed at the Tevatron
\cite{Abazov:2012bv,Aaltonen:2012bp,Olive:2016xmw}
or similar studies that have begun at LHC \cite{Aaboud:2017svj}. This set of Lagrangian parameter extractions is afflicted with
significant experimental correlations.
\begin{itemize}
\item{In the case of the extraction of $\hat{m}_Z$ at LEP, despite the presence of
a resonance peak, the Lagrangian parameters are extracted from a fit
that simultaneously defines the pseudo-observables $\{\hat{m}_Z,\Gamma_Z, \sigma_{had}^0,R_e^0,R_\mu^0,R_\tau^0 \}$ \cite{ALEPH:2005ab}. See Section \ref{LEPgoodandbad}.}
\item{In the case of $\hat{m}_W$, due the calibration of the electromagnetic calorimeter
to $Z$ decays, the experimental extraction is effectively an extraction of the ratio $\hat{m}_W/\hat{m}_Z$ \cite{Abazov:2012bv,Aaltonen:2012bp,Olive:2016xmw}.}
\end{itemize}
These extractions of Lagrangian parameters from $S$ matrix elements take place in the
perturbative expansion of the EW theory. The precision of these measurements require that higher order
corrections be included. It is only because the EW interactions are
perturbative with typical leading loop corrections $ \lesssim \mathcal{O}(1\%)$
that the resulting parameter degeneracy introduced leads to small perturbations of the highly correlated central values
extracted in a naive tree level analysis of $\{v,g_1,g_2\}$.

If an input set of the form $\{\hat{G}_F,\haew,\hat{M}_Z\}$ is used
then the challenge of parameter degeneracy also appears due to the requirement to
run the extracted Lagrangian parameter $\haew$ through the hadronic
resonance region. Using an input parameter $\haew$ actually corresponds
to a simultaneous input set of $\{\haew,\nabla \alpha\}$
as extractions of $\haew$ are dominated by $q^2 \rightarrow 0$ measurements
determined by probing the Coulomb potential of a charged particle.
The low scale measurement extracts this parameter with the mapping to the two point functions $\Pi^{ab}$
\bea\label{alphaextract}
- i \, \left[\frac{4 \, \pi \, \hat{\alpha}(q^2)}{q^2}\right]_{q^2 \rightarrow 0}
\equiv \frac{- i \, \bar{e}_0}{q^2} \left[1 + {\rm Re} \frac{\Sigma^{AA}(m_Z^2)}{m_Z^2} - \nabla \alpha \right].
\eea
The finite terms in the low scale matching that are the largest effect are due to the vacuum polarization of the
photon in the $q^2 \rightarrow 0$ limit. This unknown term is given by the last two terms on the right hand side
of Eq.~\ref{alphaextract}, with the notation
\bea
\nabla \alpha = \left[\frac{{\rm Re} \Sigma^{AA}(m_Z^2)}{m_Z^2}  - \left[\frac{\Sigma^{AA}(q^2)}{q^2} \right]_{q^2 \rightarrow 0}\right].
\eea
This quantity parameterizes contributions to the two point function that have to be simultaneously
determined to define $\haew$. For further discussion see
Refs.~\cite{Hollik:1988ii,Kennedy:1988sn,Maksymyk:1993zm,Wells:2005vk,Olive:2016xmw,Freitas:2014hra}.
The uncertainty on $\nabla \alpha$ due to this parameter degeneracy completely dominates the
uncertainty on $\haew$ when it is used in LHC applications.

The requirement of simultaneous extractions and highly correlated Lagrangian parameters from $S$ matrix elements
in conjunction with other non-perturbative unknown parameters also returns in the cases of $\{g_3,Y_i\}$:
\begin{itemize}
\item{Extractions of $g_3$ from studies of $e^+ e^-$ event shapes simultaneously extract $\alpha_s$ and the leading non-perturbative
parameter from thrust distributions \cite{Davison:2008vx,Abbate:2012jh,Gehrmann:2012sc}.
Similarly the extraction of $g_3$ from the event shape $C$ parameter is a simultaneous extraction of
$\alpha_s$ and a leading non-perturbative parameter \cite{Parisi:1978eg,Donoghue:1979vi,Hoang:2015hka}.}
\item{The extraction of $m_b$ from inclusive $\bar{B} \rightarrow X_c \bar{\nu} \ell$ decays using HQET is a simultaneous extraction of
$\hat{m}_b$ and the leading non-perturbative corrections \cite{Bauer:2004ve}.}
\item{The extraction of light quark masses $\{\hat{m}_u,\hat{m}_d,\hat{m}_s,\hat{m}_c\}$ using Lattice QCD occurs with a simultaneous lattice cut-off parameter
and is related to $\rm \overline{MS}$ masses in perturbation theory \cite{Olive:2016xmw}. The same point holds for extractions
of $\hat{m}_c$ \cite{Allison:2008xk}.
Precise extractions of quark mass parameters $\{\hat{m}_u,\hat{m}_d,\hat{m}_s\}$ in $\chi$PT are extractions of
quark mass ratios \cite{Olive:2016xmw}, not single Lagrangian terms.}
\end{itemize}
Degeneracy and correlations in the space of Lagrangian parameters that are extracted from $S$ matrix elements is essentially unavoidable.
This is a reflection of the fact that Lagrangian parameters are not physical.
This remains the case despite a protection of the SM against parameter degeneracy due to the
nature of the Higgs mechanism in the SMEFT. As $\langle H^\dagger H \rangle = v^2/2$, interaction terms that differ by dimension
$d=2$ can lead to a degenerate interaction term of the remaining fields when expanding around the vacuum expectation value. As
$\mathcal{L}_{SM}$ has $d \leq 4$ this degeneracy is avoided in the SM, as an accidental simplification due to renormalizability.

\subsubsection{The EFT case}

The challenge of correlations and parameter degeneracy in relating Lagrangian terms
to $S$ matrix elements is serious in the SM, and this challenge
is even more acute in EFTs. The new parameters introduced in the Taylor
series defining the EFTs are local (see Eq.~\ref{uncertainty}), and several parameters appear simultaneously in the power counting expansions.
The presence of a Higgsed phase in an EFT now acts to increase parameter degeneracy.

Examining the field redefinition freedom that exists in defining
a $\mathcal{L}_6$ basis further reinforces the gap between the $S$ matrix
elements and the EFT parameters. Generally, when constructing a $\mathcal{L}_6$ basis, a
version of an on-shell effective field theory is constructed \cite{Georgi:1991ch,Gasser:1983yg}.
In this approach derivative terms are systematically removed (if possible)
and kinetic terms are canonically normalized. This shifts EFT corrections to vertices,
which increases parameter degeneracy in how $\mathcal{L}_6$ corrections modify SM predictions.
The natural expectation is a highly correlated fit space, and this is indeed frequently found \cite{Han:2004az,Skiba:2010xn,Falkowski:2014tna,Berthier:2015oma,Berthier:2015gja,Berthier:2016tkq,Brivio:2017bnu}.\footnote{See also Refs.~\cite{Ellis:2014dva, Ferreira:2016jea,Corbett:2015ksa, Corbett:2015mqf,Brehmer:2016nyr}.}

To expand on the example in the previous section in the context of an EFT extension of the SM,
the generic conclusion is that the fit spaces of Lagrangian parameters
are even more correlated. The accidental nature of the extraction of $v$ in the SM now projects onto a
three-fold Lagrangian parameter degeneracy
in the $\rm U(3)^5$ limit as
\begin{align}
\mathcal{L}_{G_F} &\equiv  -\frac{4\mathcal{G}_F}{\sqrt{2}} \, \left(\bar{\nu}_\mu \, \gamma^\mu P_L \mu \right) \left(\bar{e} \, \gamma_\mu P_L \nu_e\right),
\quad  -\frac{4\mathcal{G}_F}{\sqrt{2}} =  -\frac{2}{\bar{v}_T^2} - 4 \, \hat{G}_F \, \delta G_F,
\end{align}
where the leading order shift result is \cite{Alonso:2013hga,Berthier:2015oma}
\bea
\delta G_F &=& -\frac{1}{4 \, \hat{G}_F} \, \left(C_{\substack{ll \\ \mu ee \mu}} +  C_{\substack{ll \\ e \mu\mu e}}\right) + \frac{1}{2 \, \hat{G}_F} \left(C^{(3)}_{\substack{Hl \\ ee }} +  C^{(3)}_{\substack{Hl \\ \mu\mu }}\right).
\eea
Using a $\{\haew, \hat{m}_Z, \hat{G}_F\}$ input parameter set one finds the results
listed in the Appendix for $Z,W$ vertex terms in the SMEFT.
The Wilson coefficients that now appear $\it simultaneously$ in $\mathcal{L}_6$ in $\bar{\psi} \psi \rightarrow \bar{\psi} \psi$
scattering in the SMEFT (in addition to the SM couplings) are\footnote{The $\rm U(3)^5$ limit used here treats the two flavour contractions
$(C_{\substack{ll}} \delta_{mn} \, \delta_{op} + C'_{\substack{ll}}  \delta_{mp} \, \delta_{no})(\bar l_m \gamma_\mu l_n)(\bar l_o \gamma^\mu l_p)$ as independent \cite{Cirigliano:2009wk}.}
\begin{equation}
\tilde{C}_i \equiv \frac{\bar{v}_T^2}{\Lambda^2} \{
 C_{He}, C_{Hu}, C_{Hd}, \CHls, \CHlt, \CHqs, \CHqt, C_{HWB}, C_{HD}, C_{ll}, C_{ll}'
 \}.
\end{equation}

Furthermore, beginning at $\mathcal{L}_6$, an EOM degeneracy appears between these parameters in $\bar{\psi} \psi \rightarrow \bar{\psi} \psi$
and a subset of $\bar{\psi} \psi \rightarrow (\bar{\psi} \psi)^n$ amplitudes due to $\langle H^\dagger H \rangle$ being dimension two,
and the SMEFT having a Higgs mechanism.
See Section \ref{subsec:reparam} for further discussion on this resulting reparameterization invariance.\footnote{The first appearance of a bilinear
in $H$ outside the Higgs potential in the SM appears at dimension five, see Eqn.~\ref{eqn:l5}.
This Lagrangian term does not have a direct degeneracy with parameters already present up to $\mathcal{L}_4$ in the SM, due to the interplay of the global symmetry
structure of the SMEFT operator expansion with operator dimension \cite{deGouvea:2014lva,Kobach:2016ami}.}
Extracting $\{g_3, \hat{m}_b \}$ from lower energy data introduces an even larger set of $\mathcal{L}_6$ parameters.

Highly correlated Wilson coefficient fit spaces are a central feature of the SMEFT.
Even mild assumptions about parameter correlations (or lack thereof)
have a significant impact on the constrained Wilson coefficient space as a result.

\subsection{The idea of pseudo-observables}\label{POsectionintro}

The previous sections can be summarized as: $S$ matrix elements directly correspond to physical observables, and Lagrangian parameters do not.
Common Lagrangian parameters do feed into many observables in a manner that is sometimes consistent with
naive classical reasoning, despite classical intuition being misplaced in a QFT. A common reason that this
dichotomy between formally correct field theory and naive classical reasoning
persists for collider physics studies is frequently the fact
that the unstable states in the SM are {\it narrow}, i.e have the property that $\Gamma/m\ll 1$.

The ratios of the widths ($\Gamma_V$) of the unstable $V= \{W,Z\}$ bosons to their masses ($m_V$)
are \cite{Olive:2016xmw}
\bea
\frac{\hat{\Gamma}_Z}{\hat{m}_Z} = \frac{2.4952}{91.1876} \sim 0.03,  \quad \quad \quad \quad
\frac{\hat{\Gamma}_W}{\hat{m}_W} = \frac{2.085}{80.385} \sim 0.03.
\eea
The width of the Higgs is bounded (in the SM) to be \cite{Kauer:2012hd,Caola:2013yja,Campbell:2013una,Khachatryan:2014iha,Aad:2015xua}
\bea
\frac{\hat{\Gamma}_h}{\hat{m}_h}  \lesssim \frac{0.013}{125.09} \sim 10^{-4}.
\eea
The constraint on the Higgs width is model dependent \cite{Englert:2014aca,Cacciapaglia:2014rla,Azatov:2014jga}
and corresponds to a bound on $\hat{\Gamma}_h$ in conjunction with $\kappa_g$
and $\kappa_t$. Assuming that the width of the Higgs is only perturbed
from its SM value in the SMEFT, one can expand in $\hat{\Gamma}_h/\hat{m}_h$.

Narrow widths effectively factorize a full amplitude contributing to an $S$ matrix element up into sub-blocks. Frequently,
a naive classical intuition is also assigned to these sub-blocks, at least in a tree level
calculation.\footnote{See Ref.~\cite{Berdine:2007uv}
for a discussion on the explicit and implicit assumptions in the narrow width approximation.}
An unfortunate side effect of the highly correlated nature of the Wilson coefficient space in SMEFT studies
is that such factorizing up of observables with the narrowness of the SM states, can break flat directions in the Wilson coefficient space in an inconsistent manner,
and bias conclusions by orders of magnitude \cite{Berthier:2016tkq,Brivio:2017bnu}.

The phase space of the scattering events is populated in a manner that usually is dominated by
the impact of the narrow resonances, unless selection cuts are applied to remove all resonant contributions.
This remains true studying EFT extensions of the SM in colliders.
The narrowness of the SM states is actually fortunate for EFT studies at LHC,
as while simultaneously avoiding the large non-resonant QCD backgrounds, EFT approaches can exploit this
fact and gain a significant benefit -- {\it so long as the exploitation of the narrow width effect occurs in a well defined manner}.

\subsubsection{The naive narrow width expansion}
The most naive attempt to use the narrowness of the SM states to expand is the naive narrow width approximation.
Consider an $s$-channel scattering $\bar{\psi} \psi \rightarrow S(s)^\star \rightarrow \bar{\psi} \psi$
where $S =\{h,W,Z\}$.
Expanding around the limit $\Gamma_S/m \rightarrow 0$ is subtle.
It corresponds to a series around the kinematic point
in the amplitude where some propagators vanish due to an intermediate state going ``on-shell'' at $p^2 =m^2$.
It is not a trivial expansion in the $\Gamma_S/m$ ratio at the amplitude level, as this limit must be
taken in the sense of a distribution over the phase space.

Due to the Feynman propagator prescription \cite{Feynman:1949hz,Feynman:1950ir} defining the Green's function
when intermediate states go on-shell corresponds to a discontinuity in the imaginary part of the Feynman diagram.
Cutkosky \cite{Cutkosky:1960sp} developed formally the proof that the discontinuity across a branch cut when a stable intermediate state goes on-shell
in a Feynman diagram is given by the replacement of the propagator
\bea\label{narrowwidth}
\frac{dp^2}{p^2 - m^2 +i \, \epsilon} \rightarrow - 2 \, \pi \, i  \, \delta(p^2 - m^2)\,dp^2.
\eea
This approach can be extended to the case where the intermediate state is unstable,
leading to a generalization of the optical theorem. The most naive version of
the resulting reasoning is the narrow width approximation. Define the particle mass by the condition
\bea
m^2 - m_0^2 - {\rm Re} \, \Sigma(m_0^2) = 0,
\eea
where $m_0$ is the bare mass parameter.
The propagator pole is then shifted off of the real axis if the intermediate state decays,
in a manner that approximates a Breit-Wigner
distribution formula $\propto 1/(p^2 - m_0^2 + i \, m_0 \Gamma_S(p^2))$.
If $\Gamma_S/m \ll 1$, it can be approximated as a constant $\Gamma_S(p^2) \simeq \Gamma_S$ for the
phase space events populated by scattering through the resonance peak.
Then, in a distribution sense where the phase space is integrated over a sufficiently inclusive region, the appearance of a
propagator squared in a full cross section can be simplified.
Shifting the zero point of the symmetric $d p^2$ distribution, and performing the resulting integral gives
\bea
\int_{- \infty}^{\infty}\frac{d p^2}{(p^2 - m_0^2)^2 + m^2_0 \, \Gamma_S^2}
 = \int_{- \infty}^{\infty}\frac{d p^2}{(p^2)^2 + m^2_0 \, \Gamma_S^2} = \frac{1}{m_0 \, \Gamma_S} \int_{-\infty}^{\infty} \frac{dx}{1+ x^2}
= \frac{\pi}{m_0 \, \Gamma_S}.
\eea
To utilize this expansion for an on-shell intermediate state contributing to an $S$ matrix element we replace
\bea\label{unstablereplace}
\frac{d p^2}{(p^2 - m_0^2)^2 + m^2_0 \, \Gamma_S^2} \rightarrow  \frac{\pi}{m_0 \, \Gamma_S} \, \delta(p^2 - m^2)\, dp^2.
\eea
For a total $\bar{\psi}_k \psi_l \rightarrow S(s)^\star \rightarrow \bar{\psi}_i \psi_j$ cross section one obtains
\bea
\sigma &=& \frac{1}{32 \, \pi \,s} \, \int \frac{d^3 k_i}{(2 \pi)^3 \, E_j} \frac{d^3 k_j}{(2 \pi)^3 \, E_j}
|\mathcal{A}(\bar{\psi}_k \psi_l  \rightarrow S)|^2 |\mathcal{A}(S \rightarrow \bar{\psi}_i \psi_j)|^2 \frac{\pi}{m \, \Gamma_S} \, \delta(s - m^2), \nn
&=& \sigma(\bar{\psi}_k \psi_l  \rightarrow S) {\rm Br(S \rightarrow \bar{\psi}_i \psi_j)}.
\eea
This is the narrow width approximation. Using this approximation for on-shell production
corresponds to naive classical intuition and factorizes up phase space in a manner that is a
significant numerical benefit in evaluating cross sections.
The presence of narrow widths of the SM unstable states should be exploited to aid in studying the SMEFT and the HEFT
at LHC.
Attempting to utilize the naive narrow width approximation to achieve this end, is
problematic. Narrow width approximations are subject to severe limitations, as should be evident from the
derivation. The limitations include
\begin{itemize}
\item{The derivation is limited to on-shell kinematics and a tree level exchange, and not
performed in the presence of experimental cuts limiting the phase space.}
\item{Higher order corrections and renormalization is not obviously consistent with the derivation,
it is also not obvious that the approach is gauge invariant at higher orders.}
\item{The factorization of the cross section into sub-blocks for the limited
tree level exchange does not change the Hilbert space of the theory, and the excited unstable
$S$ state is still not an external particle. Technically the unstable particles do not allow
a plane wave expansion as asymptotic states; their energies are imaginary and the asymptotic
plane wave expansions either diverge or vanish.}
\item{It is unclear how to formally justify using replacements such as in Eq.~\ref{narrowwidth} for
other processes, even $\bar{\psi} \psi \rightarrow \bar{\psi} \psi \, \bar{\psi} \psi$ where one $S$ state
is on-shell while a second $S$ state is off-shell for the same region of phase space.
The formal developments of Cutkosky \cite{Cutkosky:1960sp} for stable intermediate states do not directly justify
replacing with Eq.~\ref{narrowwidth} in arbitrary Feynman diagrams for unstable intermediate states, and all points in phase space.}
\item{In the SMEFT or HEFT cases, the corrections to the narrow width approximations are
comparable to the order of the SMEFT corrections to the SM. One expects
in many cases $\Gamma_{W,Z}/M_{W/Z} \sim \bar{v}_T^2/\Lambda^2, p^2/\Lambda^2$. If a narrow width prescription adopted
is ambiguous or essentially an arbitrary scheme choice this can significantly impact the allowed parameter space
of the Wilson coefficients in the EFT using off-shell data, biasing global fit conclusions.}
\item{
Formally, the narrow width expansion and SMEFT expansion do not commute when the $\{\haew,\hat m_Z, \hat G_F\}$ input parameter scheme is used. To avoid an ambiguity, the ordering of the expansions should be defined, see Ref.~\cite{Berthier:2016tkq,Helset:2017mlf}. It is known that the ambiguity is experimentally constrained to be small, even in EFT extensions of the SM~\cite{Bjorn:2016zlr}.
This affects the results for the top width given in the Appendix (Eq.~\ref{top_width_smeft}).
}
\end{itemize}
The challenges of unstable states in field theory are well
known, see
Refs.~\cite{Jacob:1961zz,Veltman:1963th,Weldon:1975gu,Stuart:1991xk,Stuart:1991cc,Aeppli:1993rs,Beenakker:1994vn,Passarino:1999zh,Grunewald:2000ju,Grassi:2000dz,Actis:2006rc,Kniehl:2008cj,Kniehl:1998fn,Passarino:2010qk}
for formal developments. The modern approach of utilizing the narrowness of the SM unstable states is still challenged
by these issues, but some progress has been made.
Modern efforts to decompose a cross section into gauge invariant sub-blocks exploiting the narrowness
of the SM states avoid the most naive narrow width approximation.

\subsubsection{Double pole expansions}\label{doublepole}
Consider the process with the next level of complexity compared to $\bar{\psi}\, \psi \rightarrow S^\star \rightarrow \bar{\psi}\, \psi$.
When two propagators are present one has to determine an expansion exploiting narrow width enhancements for the process
\bea\label{doubleresonant}
\bar{\psi}\, \psi \rightarrow S^\star(s_{12}) \, S^\star(s_{34}) \rightarrow \bar{\psi}\, \psi \bar{\psi}\, \psi.
\eea
An extension of the naive narrow width approach can be developed directly by expanding the amplitude result
around the two $S$ poles, assuming the intermediate states are stable, giving the decomposition
\bea\label{PObreakdown}
\mathcal{A}(s_{12},s_{34}) &=& \frac{1}{s_{12} - \bar{m}_W^2} \frac{1}{s_{34} - \bar{m}_W^2} {\rm DR}[s_{12},s_{34}, d \Omega] + \frac{1}{s_{12} - \bar{m}_W^2}  {\rm SR_1}[s_{12},s_{34},d \Omega], \nonumber \\
&+&  \frac{1}{s_{34} - \bar{m}_W^2}  {\rm SR_2}[s_{12},s_{34},d \Omega] + {\rm NR}[s_{12},s_{34},d \Omega].
\eea
Here ${\rm DR}$, ${\rm SR_{1,2}}$ and $ {\rm NR}$ refer to the doubly resonant,
singly resonant and non-resonant contributions to the amplitude, respectively,
and $\Omega$ refers to all angular dependence defined in an $s_{12},s_{34}$ independent manner.
This expansion is defined as in Refs.~\cite{Beenakker:1994vn,Beenakker:1996kt,Bardin:1997gc,Denner:2000bj,Denner:2002cg}
and is not a trivial Feynman diagram decomposition for the off-shell phase space, but a reorganization of the full amplitude result
around the physical poles\footnote{One can understand that the situation is more subtle when
considering off-shell production as double resonant diagrams contributing to Eq.~\ref{doubleresonant}
are not gauge invariant as a subset of the full amplitude. The difference in axial and 't \!Hooft-Feynman gauge
expressions for the doubly resonant diagrams generates a single-resonant
diagram process  \cite{Beenakker:1994vn}.} in a Laurent expansion.
The residues of the poles are
gauge invariant as they can be experimentally measured (at least in principle).
This factorization of the process into gauge invariant sub-blocks can be considered an example
of a pseudo-observable decomposition, with the individual
terms in Eq.~\ref{PObreakdown} being pseudo-observables.

Adding the width of the unstable $S$ state into these pole expressions can be performed
after the residues are determined.
This approach, with perturbative corrections, underlies the SM prediction of the
$\bar{\psi}\, \psi \rightarrow W^\star(s_{12}) \, W^\star(s_{34}) \rightarrow \bar{\psi}\, \psi \bar{\psi}\, \psi$ process
in Refs.~\cite{Beenakker:1994vn,Beenakker:1996kt,Bardin:1997gc,Denner:2000bj,Denner:2002cg}.
The systematic perturbative improvement of a double-pole
decomposition with higher order radiative corrections and higher order terms in the $\Gamma/m$ expansion
is technically challenging. The challenge arises from the need to calculate soft photon radiative corrections that
are non-factorizable as well as factorizable in the sense of the pole decomposition.

In the case of
$\bar{\psi}\, \psi \rightarrow W^\star(s_{12}) \, W^\star(s_{34}) \rightarrow \bar{\psi}\, \psi \bar{\psi}\, \psi$
the non-factorizable corrections were small compared to the experimental uncertainty at LEP \cite{Denner:1998rh}.
This conclusion does not directly translate to the LHC experimental environment, or future colliders
where pseudo-observable approaches face significant challenges from the need to characterize radiative corrections.

\subsubsection{Complex mass scheme}
Probing for small SMEFT or HEFT corrections to the SM predictions
argues that improvements beyond a naive narrow width or tree level double pole approximation
could be required in the long term LHC program.
A systematically improvable theoretical framework is required to determine such corrections.
A preferred approach is to expand around unstable particle poles relying\footnote{This expansion is also known in some literature
as a multi-pole expansion. This expansion is distinct from the expansion discussed in Section \ref{substructure}.}
on the generalization of the idea of particle mass introduced in Refs.~\cite{moellerpole}
and developed for EW applications in Refs.~\cite{Weldon:1975gu,Stuart:1991xk,Stuart:1991cc,Aeppli:1993rs,Beenakker:1994vn,Passarino:2010qk,Passarino:2012cb,Denner:2014zga}.
The key observation is that an unstable particle corresponds to a pole on the second Riemann sheet of an
analytic continuation of the $S$ matrix.\footnote{For a detailed discussion on the formal development of this analytical continuation
for LHC processes, see Ref.~\cite{Passarino:2010qk}.} The complex mass of a state $S$ is the solution of
\bea
s - m_S^2 + \Sigma_{SS}(s) = 0,
\eea
with renormalized mass $m_S$ and self energy $\Sigma_{SS}$ and the negative imaginary solution is taken \cite{Passarino:2010qk}.\footnote{A drawback
when considering the SMEFT is
that this scheme is tied to on-shell renormalization schemes, while EFT studies generally use $\rm \overline{MS}$ subtraction.}
Use of the Nielsen identities \cite{Grassi:2001bz,Gambino:1999ai,Passarino:2010qk}
establishes that the position of the pole is gauge parameter independent in the SM and the SMEFT.
The decomposition of a propagator square in the cross section can be augmented from Eq.~\ref{narrowwidth} to be
\bea\label{unstablereplace2}
\frac{d p^2}{(p^2 - m_0^2)^2 + m^2_0 \, \Gamma^2} \rightarrow  \frac{\pi}{m_0 \, \Gamma} \, \delta(p^2 - m^2) + {\rm PV} \left[\frac{1}{(p^2 - m_0^2)^2} \right] dp^2,
\eea
where $\rm PV$ indicates a principal value. Simultaneously factorizing up the phase space in a manner consistent with this
pole decomposition allows a systematic analytical continuation of amplitudes into the complex mass scheme at tree level.
In the SM this approach was also pushed to report full one loop corrections to
$\bar{\psi}\, \psi  \rightarrow \bar{\psi}\, \psi \bar{\psi}\, \psi$, see Ref.~\cite{Denner:2005fg}.

The complex mass scheme treats resonant and non-resonant regions of phase space in a unified manner
and one loop calculations have a difficulty similar to one loop calculations in the SM. It is reasonable
to expect that this approach could be extended to one loop results in the SMEFT, but such full calculations have never
been carried out to date in any LHC process involving the Higgs boson.
As no explicit expansion in $\Gamma/m$ need be performed using the complex mass scheme, the application of this
approach to SMEFT studies is favored to systematically develop
pseudo-observables \cite{Passarino:2010qk,Passarino:2012cb,Ghezzi:2015vva,David:2015waa,deFlorian:2016spz}.

An alternative approach is to use unstable particle EFT that was developed in Refs.~\cite{Beneke:2003xh,Beneke:2004km,Beneke:2007zg}.
It is unclear if these results can be extended to a SMEFT study. It is also unclear if the comparative (one loop) benefits
of the complex mass scheme over the unstable particle EFT present in the SM  will persist in SMEFT studies. For more discussion comparing these approaches
in the SM, see Ref.~\cite{Beneke:2015vfa}.

\subsection{Basics of EFT studies at colliders}\label{EFTcolliderbasics}

Practically implementing EFT studies using data from modern colliders is challenging on several fronts. A large number of free parameters
is characteristic of the EFTs. A significant parameter degeneracy (until measurements are combined)
is also a fundamental feature. Utilizing a well defined naive narrow width approximation
is an essential step to factorizing up $S$ matrix elements and reducing the complexity of the analysis to a manageable level.
In addition, one can simultaneously exploit the relative scaling of leading corrections in the EFTs in how phase space is populated,
IR symmetry assumptions in the EFT, and the fact that parameters that violate symmetries approximately
preserved in the SM interfere in a numerically suppressed fashion. The resulting reduction in parameters enables
a systematic EFT program using LHC, and lower energy, data  to be practically carried out.

The relative population of phase space due to resonance enhancements or
suppressions is important at Hadron colliders, where an experimental measurement is always a combination of signal
and background processes. Exploiting this effect in EFT studies was
recently discussed more systematically in Ref.~\cite{Brivio:2017btx}, and we largely reproduce this discussion here.
\begin{figure}[t]
\begin{center}
\includegraphics[width=0.9\textwidth]{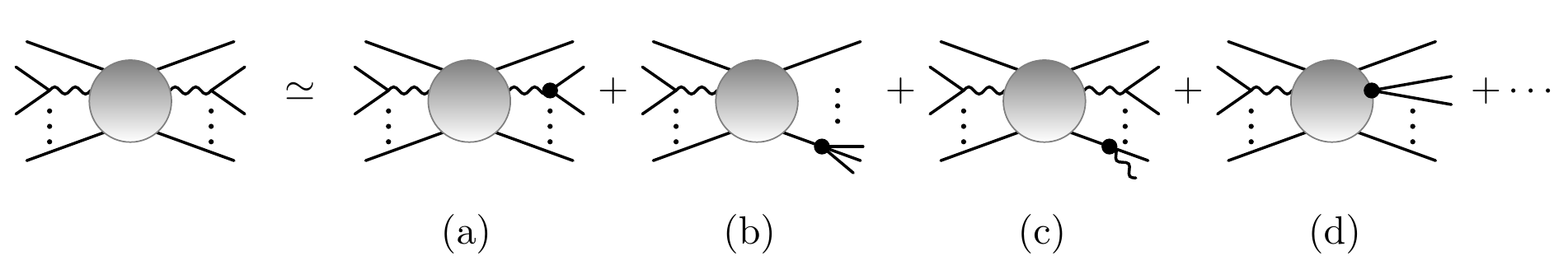}
\end{center}
\caption{A general scattering amplitude expanded in a series of EFT corrections. Here the black dot indicates
a possible insertion of $\mathcal{L}^{(6)}$
and shifts are only shown on the final states as an illustrative choice.}\label{poleparam}
\end{figure}
A general scattering amplitude is depicted in Fig.~\ref{poleparam}, and shows the decomposition
around the physical poles of the narrow propagating bosons $B$ of the SM
\bea
\mathcal{A} &=& \frac{\mathcal{A}_a(p_1^2, \cdots p_M^2)}{(p_1^2 - m_{B_1}^2 + i \Gamma_{B_1} m_{B_1}) \cdots (p_N^2 - m_{B_N}^2 + i \Gamma_{B_N} m_{B_N})}, \nn
 &+& \frac{\mathcal{A}_b(p_1^2, \cdots p_M^2)}{(p_1^2 - m_{B_1}^2 + i \Gamma_{B_1} m_{B_1}) \cdots (p_{N-1}^2 - m_{B_{N-1}}^2 + i \Gamma_{B_{N-1}} m_{B_{N-1}})},\nn
 &+& \cdots + \mathcal{A}_j(p_1^2, \cdots p_M^2).
\eea
If experimental selection cuts are made so that
the process is numerically dominated by a set of leading pole contributions of narrow bosons $B$, then
this phase space volume  $\Omega$  is
\bea
\left(\frac{d \sigma}{\d \Omega}\right)_{pole}  &\simeq& \left(\frac{d \sigma_{SM}}{\d \Omega}\right)^1 \, \left[1+ \mathcal{O}\left(\frac{C_i \, \bar{v}_T^2}{g_{SM}\Lambda^2}\right)
+ \mathcal{O}\left(\frac{C_j \, \bar{v}_T^2 \, m_B}{\Lambda^2 \, \Gamma_B}\right)\right], \\
&+& \left(\frac{d \sigma_{SM}}{\d \Omega}\right)^2 \,  \left[1+ \mathcal{O}\left(\frac{C_k \, p_i^2}{g_{SM}\Lambda^2}\right)\right]. \nonumber
\eea
The differential cross sections $\left(d \sigma_{SM}/\d \Omega\right)^{1,2}$ are distinct in each case, see
Ref.~\cite{Brivio:2017btx} for further details.
The interference with a complex Wilson coefficient denoted $C$,
that occurs when a resonance exchange is not present compared to the leading resonant SM signal result (shown in Fig.~\ref{poleparam} d))
scales as
\bea\label{intargument}
|\mathcal{A}|^2 &\propto& \left(\frac{g_{SM}^2}{(p_i^2 - m_B^2 + i \, \Gamma(p) m_B)} + \frac{C}{\Lambda^2} \right) \left(\frac{g_{SM}^2}{(p_i^2 - m_B^2 + i \, \Gamma(p) m_B)} + \frac{C}{\Lambda^2} \right)^\star \cdots \\
&\propto& \left[\frac{g_{SM}^2}{(p_i^2 - m_B^2)^2 + \Gamma_B^2 \, m_B^2} + \frac{(p_i^2 - m_B^2) (C/\Lambda^2 + C^\star/\Lambda^2) - i \Gamma_B \, m_B (C^\star/\Lambda^2 - C/\Lambda^2)}{(p_i^2 - m_B^2)^2 + \Gamma_B^2 \, m_B^2}\right] \cdots \nonumber
\eea
In the near on-shell region of phase space $(\sqrt{p_i^2} - m_B \sim \Gamma_B)$, the SMEFT then has the additional numerically subleading corrections
\bea
&\,&\left(\frac{d \sigma_{SM}}{\d \Omega}\right)^1 \,  \mathcal{O}\left(\frac{\Gamma_B \, m_B \, \{{\rm Re}(C),{\rm Im}(C)\}}{g_{SM}^2\Lambda^2}\right)
+ \left(\frac{d \sigma_{SM}}{\d \Omega}\right)^2 \, \mathcal{O}\left(\frac{\Gamma_B \, m_B \, \{{\rm Re}(C),{\rm Im}(C)\}}{g_{SM}^2\Lambda^2}\right)
\cdots
\eea
For this reason, the numerical effect of the parameters not resonantly enhanced are relatively suppressed by
\bea\label{scalingrule}
\left(\frac{\Gamma_B \, m_B}{\bar{v}_T^2}\right) \frac{\{{\rm Re}(C),{\rm Im}(C)\}}{g_{SM} \, C_i}, \quad \quad \left(\frac{\Gamma_B \, m_B}{p_i^2}\right) \frac{\{{\rm Re}(C),{\rm Im}(C)\}}{g_{SM} \,  C_k},
\eea
This relative numerical suppression occurs in addition to the power counting in the SMEFT and the combination of these
two suppressions is what is experimentally relevant.

In addition to maximally exploiting resonance enhancements or suppressions of parameters
other IR assumptions can be directly made.
Examples of such assumptions are
\begin{itemize}
\item{Symmetry assumptions made on the SMEFT or the HEFT operators. Typically these are global symmetry assumptions on Baryon or Lepton
number conservation, $\rm U(3)^5$ flavour symmetry or a flavour subgroup. Assumed IR symmetries lead to relations between
scattering amplitudes in the EFT, and hence lead to
constraints even if the symmetry is
spontaneously broken. Weinberg refers to such assumptions as ``algebraic
symmetries'' \cite{weinberglectures}, as they lead to algebraic relations between $S$ matrix elements.
Note that the IR limit of the full theory
by definition is reproduced in the EFT. For this reason, the assumption of an algebraic
symmetry in an EFT can directly enforce a UV class of theories which is required for matching.
The distinction of this assumption being an IR constraint on the EFT itself, is an important
conceptual point.}
\item{Taking the $Y_{u,d,c,s} \rightarrow 0$ limit in a process.\footnote{The dependence on these parameters referred to here is due
to $\mathcal{L}_{SM}$ and not a normalization choice on operator Wilson coefficients.}}
\item{Neglecting higher order SMEFT loops, i.e. those involving the SM fields and $\mathcal{L}_6$ that are perturbative
corrections determined in the IR EFT construction.}
\end{itemize}

Finally, we note that when a parameter is retained in the EFT that violates a symmetry approximately preserved in the SM,
the interference term is still numerically suppressed. This fact can also be used to neglect parameters that are numerically
small in contributions to measured observables. Such numerical suppressions affect a large number of parameters in Class
$5$, $6$, $7$ (in Table~\ref{op59}) for flavour changing neutral current contributions that interfere with small GIM suppressed processes
in the SM. Numerical suppressions of this form are also present for interference between the SM and the operators of Class $5$, $6$, $7$
that introduce CP violation, see Refs.~\cite{Chien:2015xha,Dekens:2013zca} for recent studies. Operators that introduce chirality
flips of the light SM fermions are also suppressed when interfering with the SM, leading to the possible neglect of some parameters in dipole
operators in Class $6$, and contributions from right handed currents induced bu $\mathcal{Q}_{Hud}$ in Class $7$ \cite{Alioli:2017ces}.
Taking all of this into account reduced sets of parameters can be well motivated in global fits in the SMEFT, see
Ref.~\cite{Brivio:2017btx} for more discussion.

IR assumptions, or simplifications of this form due to kinematics in scattering processes suppressing
dependence on the $C_i$, can be made so long as a
theoretical error for the SMEFT is introduced to accommodate the neglected higher order terms that violate the
assumption/simplification. In contrast, UV assumptions (setting $C_i = 0$, dropping operators and using an incomplete basis, or assuming $C_i/C_j \sim 16 \pi^2$ etc.) are dangerous.
The EOM makes it non-trivial and non-intuitive to determine how such assumptions modify the SMEFT framework into an
alternate consistent field theory formulation.

\section{The LEP example}\label{LEPgoodandbad}
We now turn to the interpretation of LEP data in EFT extensions of the SM, primarily focused on a SMEFT interpretation,
utilizing the results of the previous sections to frame this discussion.
Interpreting LEP data in EFT extensions of the SM
is illustrative of the challenges that the LHC experimental program will
face long term in enabling or disabling model independent re-interpretations of its
results. It is also informative as to how a program of decomposing
experimental data into pseudo-observables (PO) and mapping to EFT extensions of the SM
has comparative advantages and disadvantages if LHC data reporting attempts
to follow the same path as LEP, despite the very different collider environment.
Here we illustrate these challenges by comparing EFT and PO interpretations
of the legacy LEPI-II results.

\subsection{LEPI pseudo-observables and interpretation.}

The LEPI pseudo-observables are inferred from a data set that is a scan through the
$Z$ pole including $40 \, {\rm pb^{-1}}$ of off-peak data with $155 \, {\rm pb^{-1}}$ of on-peak data.
The narrowness of the $Z$ is directly exploited in the
LEPI PO set definitions, as summarized in the Appendix.

The LEPI PO results is an ideal case of model independent
reporting of experimental results with numerous consistency checks
on SM assumptions used to extract and define the PO. This fortunate result was enabled by the clean LEPI collision environment,
with known $e^+ e^-$ initial states scattering through a $Z$ resonance at relatively low
and fixed energies. The resulting legacy data reporting
still enables EFT interpretations to be revisited
and systematically improved over the years.
\begin{figure}[t]\centering
 \includegraphics[width=.9\textwidth]{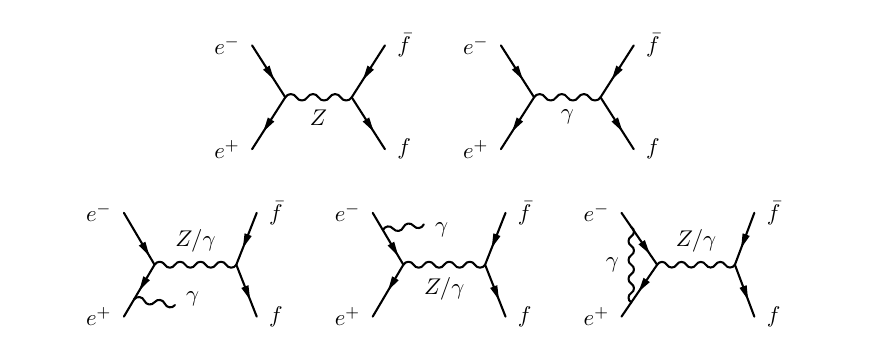}
 \caption{SM leading order contributions to $H^{tot}_{QED}$.}\label{LEPradiator}
\end{figure}

\subsubsection{Checking the SM-like QED radiation field assumption at LEP}
LEPI data is summarized in Ref.~\cite{ALEPH:2005ab}. The LEPI pseudo-observables sets
include flavour universal and flavour non-universal experimental results.
Leading order contributions to the resonant exchanges and the photon emissions are shown in Fig.~\ref{LEPradiator}.
We will restrict our detailed discussion to the flavour universal PO results.
The raw experimental results of the LEP pole scan through the $Z$ resonance are deconvolved with initial and final state
photon emissions being subtracted out, under a SM-like QED radiator function $H^{tot}_{QED}$ assumption.
The radiator function is calculated to third order in the SM in QED and the data is deconvolved
using this SM function. The measured pole scan is treated as a convolution of a electroweak kernel cross section $\sigma_{ew}(s)$
with $H^{tot}_{QED}$ as \cite{ALEPH:2005ab}
\bea
\sigma(s) = \int^{1}_{4 m_f^2/s} \, dz \, H^{tot}_{QED}(z,s) \, \sigma_{ew}(z \, s).
\eea
Radiator functions are also applied to
deconvolute the forward and backward cross sections $\sigma_{F/B}$ used to define the $A_{FB}$ pseudo-observable extractions.
The effect of this deconvolution is very dramatic. The peak is modified by $36 \%$ and its central value is shifted by $\rm 100 \, {\rm MeV}$,
which is fifty times larger than the quoted error on $\hat{m}_Z$ \cite{ALEPH:2005ab}.
The deconvolved cross section is then mapped to the PO set $\{\hat{m}_Z,\Gamma_Z, \sigma_{had}^0,R_e^0,R_\mu^0,R_\tau^0 \}$
given in Table \ref{EWtable-1},
and a subsequent mapping from PO to EFT parameters at tree or loop level is then made.
\begin{center}
\begin{table}[ht!]
\centering
\tabcolsep 8pt
\begin{tabular}{|c|c|c|c|c|}
\hline
Observable & Experimental Value & Ref. & SM Theoretical Value & Ref.  \\ \hline
$\hat{m}_Z $[GeV] & $91.1875 \pm 0.0021$ & \cite{Z-Pole} &-&-\\
$M_W $[GeV] & $80.385 \pm 0.015 $ & \cite{Group:2012gb} &$80.365 \pm 0.004$& \cite{Awramik:2003rn}\\
$\sigma_h^{0}$ [nb]& $41.540 \pm 0.037$ & \cite{Z-Pole}&$41.488 \pm 0.006$& \cite{Freitas:2014hra}\\
$\Gamma_{Z}$[GeV] &$2.4952 \pm 0.0023 $&\cite{Z-Pole} &$2.4943 \pm 0.0005$& \cite{Freitas:2014hra}\\
\hline
$R_{\ell}^{0}$ & $20.767 \pm 0.025$ &\cite{Z-Pole} &$20.752 \pm 0.005$& \cite{Freitas:2014hra}\\
$R_{b}^{0}$ & $0.21629 \pm 0.00066$ &\cite{Z-Pole} &$0.21580 \pm 0.00015$&\cite{Freitas:2014hra}\\
\hline
\hline
$R_{c}^{0}$ & $0.1721 \pm 0.0030$ &\cite{Z-Pole} &$0.17223 \pm 0.00005$& \cite{Freitas:2014hra}\\
$A_{FB}^{\ell}$ & $0.0171 \pm 0.0010$ & \cite{Z-Pole}&$0.01626 \pm 0.00008 $&\cite{Awramik:2006uz} \\
$A_{FB}^{c}$ & $0.0707 \pm 0.0035$ & \cite{Z-Pole}&$0.0738 \pm 0.0002$& \cite{Awramik:2006uz} \\
$A_{FB}^{b}$ & $0.0992 \pm 0.0016$ & \cite{Z-Pole} &$0.1033 \pm 0.0003$& \cite{Awramik:2006uz} \\
\hline\end{tabular}
\caption{Experimental and theoretical values of the LEPI pseudo-observables.
The results are grouped in terms of  the precision of the measurements made. The entries above the double line
are measured to better than percent accuracy, the entries below the double line are measured to an accuracy of a few percent.
Taken from Ref.~\cite{Berthier:2015gja}.
\label{EWtable-1}}
\end{table}
\end{center}

An immediate challenge to this procedure is the possibility that the QED radiator function $H^{tot}_{QED}$ is modified transitioning to the
SMEFT in a manner that dramatically biases the results. A modification of the current
\bea
\mathcal{L}_{A,eff}=\sqrt{4 \pi \hat{\alpha}}  \left[ Q_{x} \, J_{\mu}^{A, x}  \right] A^{\mu},
\eea
for $x = \{\ell,u,d\}$ due to SMEFT corrections at tree level is not present if $\haew$ is used as an input parameter in a global fit.
The modification of the dipole moment of the electron still leads to a potential bias to $H^{tot}_{QED}$.
The SMEFT electron dipole moment is given by \cite{Alonso:2013hga}
\begin{align}
\mathcal{L} &=\frac{ev}{\sqrt 2} \left(\frac{1}{g_1} C_{\substack{e B \\ rs }}  - \frac{1}{g_2} C_{\substack{e W \\ rs }}\right)\  \overline e_{r} \sigma^{\mu \nu} P_R e_{s}\, F_{\mu \nu}  +h.c.
\label{dipole2}
\end{align}
where $r$ and $s$ are flavour indices. Under a $\rm U(3)^5$ assumption dominantly broken by the SM Yukawas $C_{e B},C_{e W} \propto Y_e$,
yielding an effective suppression by a small fermion mass for the LEP events. Such dipole insertions contribute to the
$S$ matrix element of $e^+ e^- \rightarrow e^- e^+$ through a direct contribution to a $\gamma$ exchange between the initial and
final state, and also through modifying the external legs of the process in the LSZ formula in a disconnected contribution,
similar to the case of photon emissions in the effective Hamiltonian for inclusive $\bar{B} \rightarrow X_s \, \gamma$  \cite{Grinstein:1987vj}.
The SMEFT result of this form for has recently been calculated and is reported in Ref.~\cite{Boggia:2017hyq}.

Despite these facts supplying reasonable arguments that assuming $H^{tot}_{QED}$ is SM like in the
deconvolution procedure in the SMEFT introduces only a small theoretical bias, it was still checked at LEP what the impact is of
large anomalous $\gamma-Z$ interference effects on the reported PO. Using
a general $S$ matrix parameterization of this interference \cite{ALEPH:2005ab,Leike:1991pq,LEP-2}
the possibility of anomalous $\gamma-Z$ interference does change the inferred pseudo-observable results, primarily
the inferred value of $\hat{m}_Z$ by increasing the error by a factor of three compared to the quoted error in Table \ref{EWtable-1}.
The remaining pseudo-observables are modified by $20 \%$ of their quoted errors, when a $50 \%$ correction to the SM value
is introduced that is far in excess of a SMEFT correction expected to be $\sim {\rm few} \, \%$ \cite{ALEPH:2005ab,LEP-2}
if $\haew$ is not used as an input and $\Lambda \sim {\rm few} \, {\rm TeV}$.

A benefit of the $\gamma-Z$ interference cross check being performed, is that even when $\haew$ is not used as an input parameter,
deviations from the SM expectation in $\gamma-Z$ interference can be understood to only perturb EFT conclusions extracted
from the LEP PO. The simple addition of an extra theoretical error in interpreting the PO when a SMEFT interpretation is used
and $\haew$ is not an input is then justified.

A secondary benefit of the $S$ matrix parameterization check of LEP data is that it constrains another potential bias. Due to potential interference
with $\psi^4$ operators that are present in the SMEFT and feed into LEP data due to the presence of off the $Z$ pole
scattering events \cite{Berthier:2015oma,Han:2004az}, a potential bias in the LEP PO scales
as $\sim (m_Z \, \Gamma_Z/v^2)$ times a function of this ratio of off/on peak data \cite{Berthier:2015oma}.
The corresponding uncertainty does not disable using EWPD to obtain $\sim \%$ level constraints on the $C^6_i$,
as an anomalously large effect would also have shown up in the $S$ matrix cross check reported in Refs.~\cite{ALEPH:2005ab,LEP-2}.

LEPI results and cross checks establish that the assumption of a SM-like QED radiator function, and neglect of $\psi^4$
interference effects, is validated for the $Z$ pole scan data set. A slight increase in the errors
quoted on the pseudo-observable extractions is sufficient and appropriate if $\haew$ is not used as an input
to use the PO in the SMEFT. We stress that the LEP PO approach did not {simply assume only SM like radiative corrections,
it checked that a SM like QED radiation field
was present} for the processes of interest.
An assumption of a SM like radiation field for pseudo-observables, is essentially
an assumption that the multi-pole expansion that appears in the SMEFT derivative expansion directly does not
lead to a significant perturbation of the SM radiation field. This is a strong condition on UV dynamics
that should be avoided to maintain model independence.

An essential challenge for the PO program at LHC is to address the challenge of radiative corrections to Po's with suitable rigor to enable a precision
PO program to characterize the properties of the Higgs-like scalar in a model independent fashion. To date this challenge has not been met, but some initial
studies in the direction of characterizing such corrections have appeared, for example in Ref.~\cite{Bordone:2015nqa}.
We return to this point below.

\subsubsection{LEPI interpretations}
Most interpretations of EWPD that go beyond the PO level and make contact with specific models use the $\rm S,T$ formalism
(or related approaches
\cite{Kennedy:1988sn,Altarelli:1990zd,Golden:1990ig,Holdom:1990tc,Peskin:1990zt,Peskin:1991sw,Maksymyk:1993zm,Burgess:1993mg,Burgess:1993vc,Bamert:1996px}).
The $S,T$ oblique formalism parameterizes a few common corrections to the two point functions ($\Pi_{WW,ZZ,\gamma Z}$)
that feed into the extracted PO in the standard form \cite{Olive:2016xmw}
\bea\label{Sdefn}
\frac{\hat{\alpha}(m_Z)}{4 \, \hat{s}_Z^2 \, \hat{c}_Z^2} \, S &\equiv&
\frac{\Pi_{ZZ}^{new}(m_Z^2) - \Pi_{ZZ}^{new}(0)}{m_Z^2}
 - \frac{\hat{c}_Z^2-\hat{s}_Z^2}{\hat{c}_Z \, \hat{s}_Z} \frac{\Pi^{new}_{Z \, \gamma}(m_Z^2)}{m_Z^2} -  \frac{\Pi^{new}_{\gamma \, \gamma}(m_Z^2)}{m_Z^2}, \\
\hat{\alpha} T &\equiv& \frac{\Pi_{WW}^{new}(0)}{m_W^2}-\frac{\Pi_{ZZ}^{new}(0)}{m_Z^2}.
\eea
One calculates $\Pi_{WW,ZZ,\gamma Z}$ in a model and uses global fit results on EWPD with $\rm S,T$ corrections to constrain the model.
See Ref.~\cite{Baak:2014ora} for recent results of such fits. This can be done if
the conditions on the global $\rm S,T$ EWPD fit are satisfied; namely that vertex corrections due to
physics beyond the SM are neglected, which is the origin of the ``oblique" qualifier of EWPD \cite{Peskin:1991sw}.
The SM Higgs couples in a dominant fashion to $\Pi_{WW,ZZ}$
when generating the mass of the $W,Z$ bosons, and has small couplings to the light fermions due to the small Yukawa couplings,
so it satisfies the oblique assumption. This is the reason why this assumption was usually adopted before the Higgs-like scalar discovery.
In general, this assumption is very problematic as it is a UV condition, not an IR assumption in defining an EFT,
as the vertex correction operators are present in general. Furthermore, the oblique assumption is not
field redefinition invariant,\footnote{Attempts to translate the oblique condition into
a requirement to use
a particular operator basis by using these EOM relations are best ignored. As such attempts are based on a misconception of
the nature of field redefinitions (assuming that they can satisfy physical conditions), see Ref.~\cite{Trott:2014dma} for related discussion.
It has also been argued that the oblique requirement can be interpreted as a condition defining a class of UV theories
known as universal theories \cite{Barbieri:2004qk}. See Section \ref{UVassumeaway}
for more discussion on this idea.} as can be seen by inspecting the EOM that result
from SM field redefinitions, given in Section \ref{SMEOM}.

LHC results indicate the
$W,Z$ bosons obtain their mass in a manner that is associated with the Higgs-like scalar, see Section \ref{Higgsfun}.
Corrections to $\Pi_{WW,ZZ}$ can be included for the SM, or in extensions
due to this scalar, see Ref.~\cite{Barbieri:2007bh}. Once this is done,
there is no strong theoretical support for an oblique assumption to be invoked on the remaining perturbations to EWPD.
Dropping this problematic assumption leads to a SMEFT analysis which has several benefits.
For example, a SMEFT analysis permits the determination of higher order corrections
when interpreting EWPD, see Ref.~\cite{Hartmann:2016pil}.
\begin{figure}
 \centering
  \includegraphics[width=\textwidth]{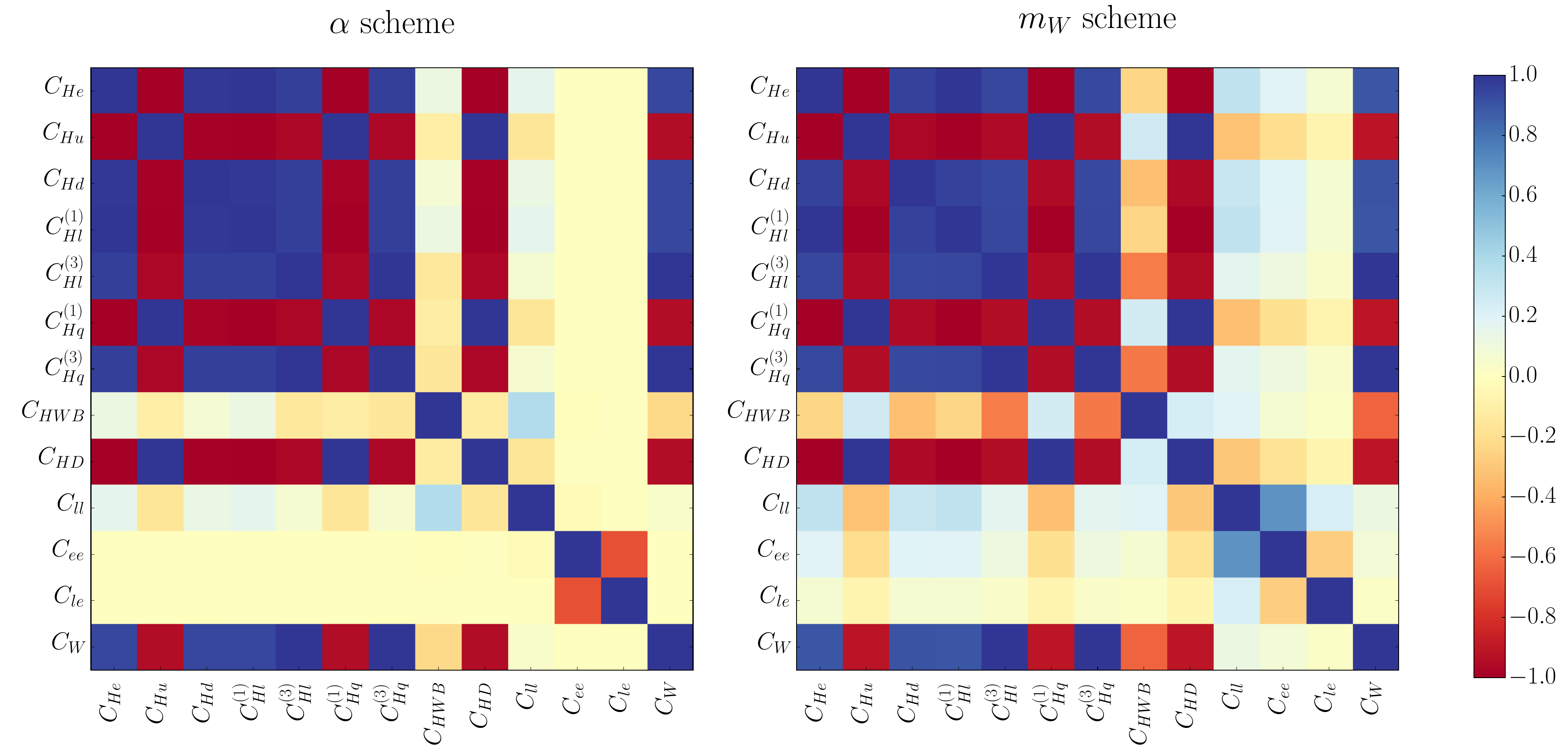}
 \caption{\label{Fig:correlation_matrices}Color map of the correlation matrix among the Wilson coefficients \cite{Berthier:2016tkq,Brivio:2017bnu},
 obtained assuming zero SMEFT error, for the $\{\hat{\alpha}, \hat{m}_Z,\hat{G}_F\}$ input scheme (left)
 and for the $\{\hat{m}_W, \hat{m}_Z,\hat{G}_F\}$ input scheme (right). The fit space is highly correlated, as can be understood
 to be due to the facts that the Lagrangian parameters are unphysical in general, and there is a reparameterization invariance present in LEPI data.}
\end{figure}
In a SMEFT analysis of EWPD a
model is mapped to the SMEFT Wilson coefficients in a tree or loop level matching calculation and
model independent global fit results are used to constrain the Wilson coefficients in a global $\chi^2$ fit.

Initial works pioneering this approach are Refs.~\cite{Grinstein:1991cd,Han:2004az}.
The analysis of Ref.~\cite{Han:2004az} identified
unconstrained directions in the EWPD set and correctly found a highly correlated Wilson coefficient space
in the SMEFT. Recent analyses that do not break these correlations by
assumption (or a chosen marginalization procedure) still find that the EWPD Wilson coefficient space is highly correlated, see
Refs.~\cite{Han:2004az,Cacciapaglia:2006pk,Buchalla:2013wpa,Berthier:2015oma,Berthier:2015gja,Falkowski:2014tna,Berthier:2016tkq,Brivio:2017bnu}.
In determining the constraints on the Wilson coefficients of the SMEFT,
one chooses an input parameter set, and predicts the EWPD PO.
In the $Z,W$ pole results of Refs.~\cite{Berthier:2015oma,Berthier:2015gja} the mapping is
\bea
\{\hat{m}_Z,\hat{m}_h,\hat{m}_t,\hat{G}_F,\haew,\hat{\alpha}_s,\Delta \hat{\alpha} \}
\rightarrow \{m_W,\sigma_h^0,\Gamma_Z,R_\ell^0,R_b^0,R_c^0,A_{FB}^\ell,A_{FB}^c,A_{FB}^b\},
\eea
through $\mathcal{L}_{SMEFT}$. Here the hat superscript indicates an input parameter.
In more detail, a SMEFT fit procedure is as follows:  A set of observables
is denoted $O_i$, $\bar{O}^{LO}_i$, $\hat{O}_i$ for the SM prediction, the SMEFT prediction to first order in the $C^{(6)}$,
and the experimental value of the extracted PO respectively. The measured value
$\hat{O}_i$ is assumed to be a Gaussian variable centered about $\bar{O}_i$ and
the likelihood function ($L(C)$) and $\chi^2$ are defined as
\bea
L(C) = \frac{1}{\sqrt{(2 \pi)^n |V|}} \text{exp} \left(-\frac{1}{2} \left( \hat{O} - \bar{O}^{LO}\right)^T V^{-1} \left( \hat{O} - \bar{O}^{LO}\right)\right),
\quad \chi^2 = - 2 \text{Log}[L(C)],
\eea
where $V_{ij} = \Delta^{exp}_i \rho^{exp}_{ij} \Delta^{exp}_j + \Delta^{th}_i \rho^{th}_{ij} \Delta^{th}_j$
is the covariance matrix with determinant $|V|$.
$\rho^{exp}$/$\rho^{th}$ are the experimental/theoretical correlation matrices
and $\Delta^{exp}$/$\Delta^{th}$ the experimental/theoretical error of the observable $O_i$.
This approach is an approximation, with neglected
effects introducing a theoretical error \cite{Berthier:2015oma,Berthier:2015gja,David:2015waa}.
The theoretical error $\Delta_i^{th}$ for an observable $O_i$ is defined as
$\Delta^{th}_i = \sqrt{\Delta_{i,SM}^2 + \left(\Delta_{i,SMEFT} \times O_i\right)^2}$,
where $\Delta_{i,SM}$, $\Delta_{i,SMEFT}$ correspond to the absolute SM theoretical, and the multiplicative SMEFT theory error.
The $\chi^2$ is
\bea\label{thechisquared}
\chi^2_{C_i^6} = \chi^2_{C_i^6, min} + \left(C_i^6 - C_{i,min}^6 \right)^{T} \mathcal{I} \left(C_i^6 - C_{i,min}^6\right),
\eea
where $C_{i,min}^6$ corresponds to the Wilson coefficients vector minimizing the $\chi^2$ and $ \mathcal{I} $ is the Fisher information matrix.
Recent results \cite{Brivio:2017bnu} using this methodology are shown in Fig.~\ref{Fig:correlation_matrices}.
The effect of modifying the input parameters: $\{\haew,\Delta \hat{\alpha} \} \rightarrow \hat{m}_W$
was been recently examined \cite{Brivio:2017bnu}, which does not change this conclusion.
The Fisher matrices of the SMEFT fit space allow the construction of the SMEFT $\chi^2$ function.
These matrices were developed in a fit using 177 observables \cite{Berthier:2015oma,Berthier:2015gja,Berthier:2016tkq,Brivio:2017bnu} and are available
upon request of the authors of Ref.~\cite{Brivio:2017bnu}.

\subsubsection{Loop corrections to \titlemath{$Z$}{Z} decay}\label{subsec:loops}
The SMEFT and HEFT formalisms allows one to combine data sets into
a global constraint picture in a consistent fashion and these EFTs also allow the systematic determination
of perturbative corrections. One loop calculations in these theories are
absolutely required to gain the full constraining power of the most precisely measured observables, such as the LEPI
PO.  This is not surprising; far more startling is that the complete one loop results of this
form remain undetermined decades after the LEPI data set was reported!
In the case of $\mathcal{O}(y_t^2,\lambda)$ corrections to $\{\Gamma_Z, \Gamma_{Z \rightarrow \bar{\psi \psi}},\Gamma_Z^{had},R_\ell^0,R_b^0\}$, about
thirty loops were recently determined in Ref.~\cite{Hartmann:2016pil} mapping the input parameters to these observables,
while retaining the $\hat{m}_t,\hat{m}_h$ mass scales in the calculation.
The renormalization of the $\mathcal{L}_6$ operators in the Warsaw basis \cite{Grojean:2013kd,Jenkins:2013zja,Jenkins:2013wua,Alonso:2013hga} is
used in this result, which simultaneously provides a check of the terms $\propto y_t^2,\lambda$ that appear in
these observables \cite{Hartmann:2016pil}. The results define a perturbative expansion for the LEPI PO
\bea
\bar{O}_i = \bar{O}^{LO}_i(C_i^6) + \frac{1}{16 \, \pi^2}\left(F_1[C_j^6]
+ F_2[\lambda,C_k^6]\log{\frac{\mu^2}{\hat{m}_h^2}}
+ F_3[y_t^2,C_l^6]\log{\frac{\mu^2}{\hat{m}_t^2}}\right) + \cdots
\eea
The LO results depend on ten Wilson coefficients in the Warsaw basis, defining the $\{C_i\}$, and ${\rm dim}(\{C_j\})\neq {\rm dim}(\{C_k\}) \neq {\rm dim}(\{C_l\}) >{\rm dim}(\{C_i\})$ holds in general,
where $\dim(\{C\})$ here denotes the dimension of a set of coefficients $\{C\}$.
At one loop, considering $\mathcal{O}(y_t^2,\lambda)$ corrections the new SMEFT parameters that appear are \cite{Hartmann:2016pil}
\bea
\{C_{qq}^{(1)}, C_{qq}^{(3)}, C_{qu}^{(1)},C_{uu}, C_{qd}^{(1)},C_{ud}^{(1)},C_{\ell q}^{(1)},C_{\ell q}^{(3)},
C_{\ell u},C_{qe},C_{eu},C_{Hu}, C_{HB}+ C_{HW},C_{uB},C_{uW},C_{uH} \}. \nonumber \\
\eea
It has been shown in Ref.~\cite{Hartmann:2016pil} in this manner that the number of parameters exceeds the number of
precise LEPI PO measurements when one loop corrections are calculated. The LEPI PO are very important to project
into the SMEFT, as for a few observables $\Delta^{exp}_j \sim 0.1 \%$.
When
\bea
\frac{1}{16 \, \pi^2}\left(F_1[C_j^6]
+ F_2[\lambda,C_k^6]\log{\frac{\mu^2}{\hat{m}_h^2}}
+ F_3[y_t^2,C_l^6]\log{\frac{\mu^2}{\hat{m}_t^2}}\right) + \cdots \gtrsim \Delta^{exp}_j \hat{O}_i,
\eea
it is clear that these corrections can have a significant effect on the interpretation of the LEPI PO for the exact same reason.
If this is the case depends on the values of the UV dependent Wilson coefficients and the global constraint picture, which is unknown.
However, we note that recent global fit results indicate directly that some of the four fermion operators that
feed in at one loop are very weakly constrained by lower energy data
\cite{Berthier:2015gja,Berthier:2016tkq,Falkowski:2017pss,Falkowski:2015krw}.
For the near $Z$ pole observables, one can fix $\mu = \hat{m}_Z$, but the new weakly constrained parameters are still present.
Although EFT techniques can sum all of the logs that appear relating various scales,
the extraction and prediction of the LEPI PO is a complex multi-scale problem
with the scales
\bea
0 \ll \hat{m}_\mu^2 \ll \hat{m}_Z^2 < \hat{m}_h^2 < \hat{m}_t^2.
\eea
The required calculations to sum all the logs are not available to date.\footnote{Ideally these results would have been determined
in the decades {\it before} the turn on of LHC.}
These results already establish that LEPI PO data does not constrain the SMEFT parameters appearing at tree level
to the per-mille level in a model independent fashion. This is very good news for hopes of the indirect
techniques discussed in this review discovering evidence for physics beyond the SM at LHC.

Using the determined SMEFT constraints that result from EWPD to study LHC data, one must also run the
determined constraints on $C_{i,j,k,l}(m_Z)$ to the LHC measurement scales.
This also acts to reduce the power of constraints when mapping between the data sets
by renormalization group equation (RGE) running. It is simply not advisable to
set  $C_i^6(\mu) = 0$ in LHC analyses to attempt to incorporate EWPD constraints for all these reasons.
The challenge of combining LEPI PO consistently with Higgs data requires further theoretical development of the SMEFT.

\subsubsection{SMEFT reparameterization invariance}\label{subsec:reparam}
A central feature of interpreting LEP data in the SMEFT is the highly correlated Wilson coefficient fit space.
This results from the unphysical nature of Lagrangian parameters and
the fact that several parameters in $\mathcal{L}_6$ appear at the same order in the power counting
of the theory simultaneously (see Section \ref{lagrangianparam}). A further wrinkle is that
unconstrained directions due to LEPI data in this Wilson coefficient space
are manifest in the Warsaw basis but not in other formalisms.
As a result, these unconstrained directions have caused
enormous confusion over the years.\footnote{For discussion on these unconstrained directions
in the Wilson coefficient space, see Refs.~\cite{SanchezColon:1998xg,Kilian:2003xt,Han:2004az,Grojean:2006nn,Brivio:2017bnu}.} This has lead
to some misplaced intuition that the number of SMEFT parameters is too large to do a consistent analysis of the global data
set. It has also lead to claims that
some operator bases are better related to experimental measurements than others. The logical extension of this thinking
has lead to ad-hoc phenomenological
parameterizations being promoted for the LHC experimental program, which are also argued
to be better related to experimental measurements. (See Section \ref{adhoc} for more discussion.)

The physics of these unconstrained directions is
now understood in an operator basis independent manner \cite{Brivio:2017bnu}.
To understand this result the unphysical nature of Lagrangian parameters is an essential feature,
and these flat directions follow from a scaling argument that is a property of $\bar{\psi} \psi \rightarrow \bar{\psi} \psi$
data.

The scaling argument underlying the reparameterization invariance is simple.
A vector boson can always be transformed between canonical and non-canonical form
in its kinetic term by a field redefinition without physical effect,
due to a corresponding correction in the LSZ formula. Such a shift can be canceled by a corresponding shift in the
$V \bar{\psi} \psi$ coupling. The same set of physical
scatterings can then be parameterized by an equivalence class of fields and coupling parameters in the SMEFT
as a result \cite{Brivio:2017bnu}
 \bea\label{WOW}
 \left(V,g \right) \leftrightarrow \left(V' \, (1+ \epsilon), g' \, (1- \epsilon) \right),
 \eea
where  $\epsilon \sim \mathcal{O}(\bar{v}_T^2/\Lambda^2)$.
This is the SMEFT reparameterization invariance identified in Ref.~\cite{Brivio:2017bnu}.
Denoting $\langle \cdots \rangle_{S_R}$ as the class of $\bar{\psi} \psi \rightarrow V \rightarrow  \bar{\psi} \psi$ matrix elements,
the following operator relations follow from the SM EOM given in Section~\ref{SMEOM}
\bea\label{EOMrelations}
\langle y_h \, g_1^2 Q_{HB} \rangle_{S_R} &=& \langle \sum_{\psi} y_k \, g_1^2 \, \overline \psi_\kappa \, \gamma_\beta  \psi_\kappa \, (H^\dagger \, i\overleftrightarrow D_\beta H)
+   2 \, g_1^2 \,Q_{HD}
- \frac{1}{2} g_1 \, g_2 \, Q_{HWB}  \rangle_{S_R}, \\
\langle  \, g_2^2 Q_{HW} \rangle_{S_R} &=& \langle  g_2^2 \,(\overline q \, \tau^I \gamma_\beta  q + \overline l \, \tau^I \gamma_\beta  l ) \, (H^\dagger \, i\overleftrightarrow D_\beta^I H)
- 2 \, g_1 \, g_2 \, y_h \, Q_{HWB}  \rangle_{S_R}.
\eea

Because of the SMEFT reparameterization invariance, a Wilson coefficient multiplying the left hand side of these equations is
not observable in $\bar{\psi} \psi \rightarrow \bar{\psi} \psi$ scatterings.
The invariance of $S$ matrix elements under field configurations equivalent by use of the EOM  means then, that
the corresponding fixed linear combinations of Wilson coefficients that appear on the right-hand sides of these equations
are also not observable in the $S_R$ matrix elements. These combinations are
EOM equivalent to physical effects that cancel out due to the reparameterization invariance.

The $S_R$ class of data is simultaneously invariant under the two independent reparameterizations (defining $w_{B,W}$)
that leave the products $(g_1 B_\mu)$ and $(g_2W^i_\mu)$ unchanged.
The unconstrained directions in the global fit are found to be
\bea\label{empiracalflatalpha}
w_1= \frac{v^2}{\Lambda^2} \left(\frac{C_{Hd}}{3}-2C_{HD}+ C_{He}+\frac{\CHls}{2} -\frac{\CHqs}{6}-\frac{2 \, C_{Hu}}{3} -1.29 (\CHqt+ \CHlt)+ 1.64 C_{HWB}\right), \nn\\
w_2=\frac{v^2}{\Lambda^2} \left(\frac{C_{Hd}}{3}-2C_{HD}+ C_{He}+\frac{\CHls}{2} -\frac{\CHqs}{6}-\frac{2 C_{Hu}}{3} + 2.16 (\CHqt+ \CHlt)- 0.16 C_{HWB}\right). \nn
\eea
These unconstrained directions
can be projected into the vector space defined by $w_{B,W}$ as \cite{Brivio:2017bnu}
$w_1 = -w_B - 2.59 \, w_W$, and  $w_2= -w_B +4.31 \, w_W$.

We stress that it is important to understand that the existence of these flat directions should not
be considered a sign of the SMEFT having too many parameters to interface with the data.
Conversely, a correct interpretation of this physics is that a consistent EFT formalism retaining all parameters can indicate that hidden
structures such as the reparameterization invariance are present.

\begin{figure}[t]
	\centering
	\begin{subfigure}{0.40 \textwidth}
	\includegraphics[width=0.95\textwidth]{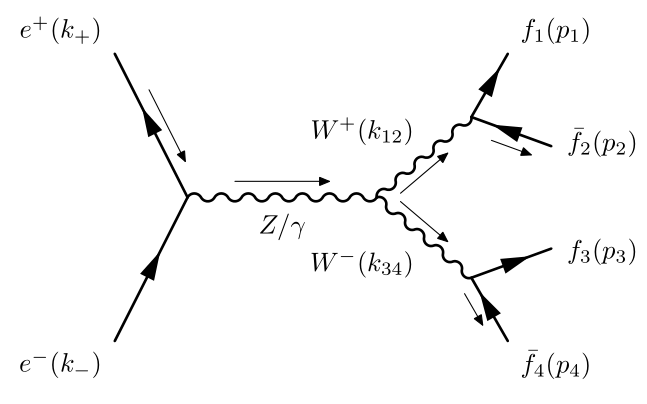}
	\caption{Amplitude $\mathcal{A}_V$}
	\label{fig:CC03_s_channel}
	\end{subfigure}
	\begin{subfigure}{0.40 \textwidth}
	\includegraphics[width=0.95\textwidth]{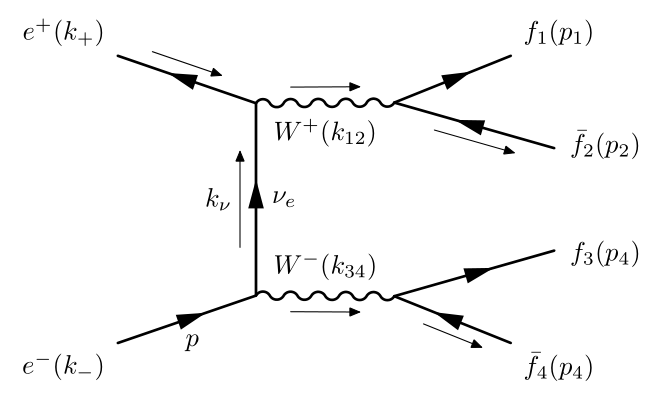}
	\caption{Amplitude $\mathcal{A}_\nu$}
	\label{fig:CC03_t_channel}
	\end{subfigure}
	\caption{The $s$-channel (a) and $t$-channel (b) Feynman diagrams contributing to $e^+ e^- \rightarrow W^+ W^- \rightarrow f_1 \bar{f}_2 f_3 \bar{f}_4$ (from Ref.~\cite{Berthier:2016tkq}). }
	\label{diagrams_WWprod}
\end{figure}

It is required to include data from $\bar{\psi} \psi \rightarrow \bar{\psi} \psi \bar{\psi} \psi$ scattering to lift the flat directions
\cite{Han:2004az}. This is understood to be the case when considering Fig.~\ref{diagrams_WWprod} (a) that contributes to $\bar{\psi} \psi \rightarrow \bar{\psi} \psi \bar{\psi} \psi$
scattering which is not invariant under Eq.~\ref{WOW}.
Fig.~\ref{diagrams_WWprod} (a) contains a TGC vertex, which is the reason the
reparameterization invariance is broken at an operator level. We emphasize that there is a distinction between the scaling argument in
Eq.~\ref{WOW} that applies to $S$ matrix elements in a basis independent manner
and the presence (or not) of an operator contributing to an anomalous TGC vertex. The latter depends
upon the operator basis chosen and unphysical field redefinitions.

A recent approach of using mass eigenstate (unitary gauge) coupling parameters to characterize deviations from the Standard Model
makes the presence of these unconstrained directions even harder to uncover in data analyses. The reason is that EOM relations key to understanding
the reparameterization invariance do not have a (manifest) equivalent in the parameterization chosen,
although the fact that there remain un-probed aspects of the $Z$ boson phenomenology in $\bar{\psi} \psi \rightarrow \bar{\psi} \psi$
scatterings is directly acknowledged in Refs.~\cite{Gupta:2014rxa,Falkowski:2014tna}. Defining correlations
for mass eigenstate parameters in a form that manifestly preserves the consequences of the
reparameterization invariance remains an unsolved problem, and assuming no correlations between these parameters
can bias results by breaking the reparameterization invariance.

As $\bar{\psi} \psi \rightarrow \bar{\psi} \psi \bar{\psi} \psi$  occurs through narrow $W^{\pm}$ states, the requirement
to utilize the narrowness of the $W^\pm$ boson in a consistent theoretical approach is now front and center.
\subsection{LEPII pseudo-observables and interpretation.}
The LEPII data and analysis summary is reported in Ref.~\cite{LEP-2}. A major goal of the
LEPII run was to exploit the sensitivity of scattering
$\sigma_{2\psi}^{4\psi} \equiv \sigma(e^+ e^- \rightarrow W^+ W^- \rightarrow f_1 \bar{f}_2 f_3 \bar{f}_4)$
to $\rm CP$ even Triple Gauge Couplings (TGCs) vertices to test the non-abelian structure of the SM.
For this reason, measurements of $\sigma_{2\psi}^{4\psi}$ were stepped up to
a running energy of $\sqrt{s} = 206.5 \, {\rm GeV}$ through the $W^\pm$ pair production threshold.

The SM prediction of this process was developed in Refs.~\cite{Tsai:1965hq,Flambaum:1974wp,Gaemers:1978hg,Bilchak:1984ur,Gunion:1985mc,Hagiwara:1986vm,Beenakker:1994vn}.
The direct calculation of the process $\sigma_{2\psi}^{4\psi}$, and related differential distributions, is
sufficiently complex that it benefits from using spinor-helicity results developed in Refs.~\cite{Hagiwara:1986vm,Hagiwara:1993ck}.
See the Appendix for the results broken into Helicity eigenstates in the SMEFT. To define radiative corrections to this process in the SM,
the narrowness of the $W^{\pm}$ bosons is exploited in a double pole expansion \cite{Beenakker:1994vn,Beenakker:1996kt,Bardin:1997gc,Denner:2000bj,Denner:2002cg}.
Functionally the program RacoonWW is used \cite{Denner:2002cg} which utilizes this expansion.

Due to the reparameterization invariance present in $e^+ e^- \rightarrow f_1 \bar{f}_2$ scattering, this
data set is critical to lifting the flat directions in the Wilson coefficient space, but this was not the motivation
for extending the LEPI data set to measure $\sigma_{2\psi}^{4\psi}$. Only retrospectively
was it realized how this extension of the LEP data is critical to globally constrain the SMEFT Wilson coefficient space
in a consistent SMEFT analysis \cite{Han:2004az}. Unfortunately, the pseudo-observables
reported for LEPII were subject to less cross checks on SM-like assumptions used in extracting
the PO, and are less directly connected to $S$ matrix elements. The LEPI PO set is focused on the observable cross sections given in Table \ref{EWtable-1}
while LEPII PO generally refer to the reported values of some Lagrangian parameters
treated as extracted pseudo-observable quantities.\footnote{In this latter case
a label of pseudo-nonobservable is perhaps more appropriate, as the distinction between Lagrangian
parameters and $S$ matrix elements has been obscured. We use this tongue-in-cheek nomenclature in what follows to emphasize this point.
Note that the neutral gauge boson parameters $h_i^{Z,\gamma},f_i^{Z,\gamma}$ also are in this class of
pseudo-nonobservable.} This distinction is
important for the LHC program, where proposals exist that are closely related to the LEPII
approach to PO as opposed to the LEPI PO. For this reason, we review what has been understood
about the LEPII case using the SMEFT in recent years.

\subsubsection{TGC pseudo-nonobservables and effective pseudo-observables}
Several studies of $\sigma_{2\psi}^{4\psi}$ in an EFT context
developed the formalism used in LEPII results \cite{Gaemers:1978hg,DeRujula:1991ufe,Hagiwara:1986vm,Hagiwara:1993ck}.
The most general $C,P$ conserving TGCs is introduced as \cite{Hagiwara:1986vm,Hagiwara:1993ck}
\bea\label{def_tgc}
\frac{- \mathcal{L}_{TGC}}{g_{VWW}}=i \bar g_{1}^{V} \left( W_{\mu \nu}^{+} W^{- \mu}- W_{\mu \nu}^{-} W^{+ \mu}\right)V^{\nu} + i \bar \kappa_{V}W^{+}_{\mu}W^-_{\nu}V^{\mu \nu} + i \frac{\bar\lambda_{V}}{\hat m^2_W}V^{\mu \nu} W^{+ \rho}_{\nu}W^{-}_{\rho \mu},
\eea
where $V=\{Z,A\}$, $W^{\pm}_{\mu \nu}=\partial_{\mu} W^{\pm}_{\nu} - \partial_{\nu}W^{\pm}_{\mu}$ and similarly
$V_{\mu \nu} = \partial_{\mu} V_{\nu} - \partial_{\nu} V_{\mu}$. At tree level in the SM
\bea
g_{AWW}= e,\qquad g_{ZWW}= e \cot \theta,\qquad g_1^V=\kappa_V=1,\qquad \lambda_V=0.
\eea
As it stands, this parameterization of SM deviations is not based on a linearly realized
operator formalism \cite{DeRujula:1991ufe,Burgess:1992va}.
It can be considered to be an example
of a non-linearly realized $\rm SU_L(2) \times U_Y(1)$ theory; in modern EFT
Language this requires an embedding in the HEFT. A complete HEFT description contains more interaction terms,
see Ref.~\cite{Brivio:2013pma} for more discussion. This ad-hoc construction
can also be embedded into the SMEFT by relating it to gauge invariant operators as shown in
Ref.~\cite{Gounaris:1996rz}. An up-to-date embedding of this form in the Warsaw basis is given in the Appendix \ref{TGC}.
This embedding is straightforward as this parameterization does not have any gauge dependent defining conditions.

Traditional interpretations of LEPII pseudo-nonobservables,
reported as effective bounds on the TGC parameters shifting the SM predictions $\delta g_1^V, \delta \kappa_V,  \delta \lambda_V$ are problematic.
In many studies, including Refs.~\cite{Han:2004az,Pomarol:2013zra}, the constraints on $\delta g_1^V,  \delta \lambda_V$ reported by
the LEPII experiments are used as observables. These parameters
cannot be treated directly as physical observables to constrain the SMEFT parameter space consistently \cite{Trott:2014dma}.\footnote{
See Section \ref{preliminarymeasurements} for a general discussion on the observable/Lagrangian parameter distinction.}
Losing this distinction in SMEFT studies of $\sigma_{2\psi}^{4\psi}$ related observables leads to spurious results. The reason is
that constraints on the
non-physical TGC parameters are generally developed
using Monte-Carlo tools where vertex corrections of the massive gauge boson couplings are assumed to be ``SM-like''
for all fermion species. This assumption when interpreted as an operator basis independent constraint on the SMEFT,
implies a ``SM-like'' TGC vertex is also required in the SMEFT \cite{Trott:2014dma}
due to the EOM relations linking the relevant Lagrangian parameters.\footnote{Ref.~\cite{Wells:2015uba}
pointed out that the conclusion does not hold in the case of universal theories. On the other hand, see Section \ref{UVassumeaway}
for a discussion on universal theories.}
Attempts to then develop ``effective TGC parameters" were subsequently reported in
Ref.~\cite{Falkowski:2014tna}.\footnote{Ref.~\cite{Falkowski:2014tna} does not specify
the treatment of the expansion used due to the narrowness of the SM states to define the observables. It seems apparent
that a narrow width approximation is implicitly used to define the $\sigma_{2\psi}^{4\psi}$ observable.
This result is then combined with SM predictions defined using the double pole expansion results of Ref.~\cite{LEP-2}.
The inconsistencies so introduced in such a procedure are not suppressed by $\Gamma_W/m_W$.} Ref.~\cite{Berthier:2016tkq} showed that to define a PO
set of  ``effective TGC parameters" for the SMEFT also requires
that the shift to the $W^{\pm}$ mass ($\delta m_W^2$) and width ($\delta \Gamma_W^2$)
must be taken into account in addition to the gauge boson vertex corrections, to accommodate using a double pole expansion to
define the observables. Taking these corrections into account gives
\bea
(\delta g_1^V)^{eff} &=& \delta g_1^V - 2\, \delta D^W(m_W^2) - \delta D^V(m_V^2), \\
(\delta \kappa_V)^{eff} &=& \delta \kappa_V - 2\, \delta D^W(m_W^2)  - \delta D^V(m_V^2).
\eea
We have approximated the dependence in the residue in the effective TGC as $s_{ij} = m_W^2$, $s = m_V^2$,
consistent with a double pole expansion. $\delta D^W$, $\delta D^V$ are not relatively
suppressed by an extra factor of $\Gamma_W/m_W$ or $\Gamma_V/m_V$ as might be expected
to result from the narrow width expansions of the SM gauge bosons. Results
of this form  were reported in Ref.~\cite{Berthier:2016tkq}, as detailed in the
Appendix.
\begin{center}
\begin{table}[t!]
\begin{tabular}{|c|c|c|c|c|c|c|}
\hline
Parameter & Cons. & Ref. & Cons.& Ref. & Cons.& Ref \\ \hline
$g_1^Z$ &  $0.984^{+0.018}_{-0.020}$ & \cite{LEP-2} & $0.975^{+0.033}_{-0.030}$ & \cite{Abdallah:2010zj} & $0.95^{+0.05}_{-0.07}$ & \cite{Falkowski:2014tna} \\
$\kappa_\gamma$ & $0.982 \pm 0.042$ & \cite{LEP-2}& $ 1.024^{+ 0.077}_{-0.081}$ & \cite{Abdallah:2010zj}& $1.05^{+0.04}_{-0.04}$ & \cite{Falkowski:2014tna} \\
$\lambda_\gamma$ & $- 0.022 \pm 0.019$ & \cite{LEP-2} & $0.002 \pm 0.035$ & \cite{Abdallah:2010zj} & $0.00 \pm 0.07$ & \cite{Falkowski:2014tna} \\
\hline
\end{tabular}
\caption{Results for the LEPII pseudo-nonobservables produced while varying the TGC Lagrangian parameters
one at a time under the simultaneous assumption of a ``SM-like'' coupling
of the massive gauge bosons to fermions.
\label{TGCtable}}
\end{table}
\end{center}

\subsubsection{LEP II bounds}
The LEP experiments during the LEPII run extracted limits on the effective parameters
in Eq.~\ref{def_tgc}, both in the individual experiments and in combination. This was a significant
focus of experimental efforts. The focus of the theoretical
community leading up to the LEPII data reporting was to move beyond the naive narrow width
approximation in $\sigma_{2\psi}^{4\psi}$ and to
define SM radiative corrections to hypothesis test the SM at LEPII.
A good summary of the theoretical issues that were priorities going into LEPII
is reported in  Ref.~\cite{Altarelli:300671}. A self-consistent SMEFT approach was not
a theoretical priority as the SM had not yet been validated in a test of the non-abelian
nature of the massive SM gauge boson interactions, and no Higgs-like scalar was known
to be found experimentally.
The situation has now changed due to LHC and LEP results.

LEPII data has been reexamined in recent years to develop a consistent SMEFT interpretation in
Refs.~\cite{Han:2004az,Corbett:2012dm,Corbett:2012ja,Corbett:2013pja,
Pomarol:2013zra,Falkowski:2014tna,Falkowski:2016cxu,Falkowski:2015jaa,Butter:2016cvz,Englert:2014uua,Ellis:2014jta,Englert:2015hrx,Berthier:2016tkq}.
The conclusions found are generally consistent
but the detailed numerical results are subject to significant uncertainties. This is illustrated in Table \ref{TGCtable}
where the quoted results for bounds on the parameters $\delta g_1^V,\delta \kappa_V,\delta \lambda_V$
that have been produced from the experiments, and an external group, are listed.
These LEPII bounds were reported
by varying one parameter at a time while assuming the other TGC parameters vanish in
Ref.~\cite{LEP-2,Abdallah:2010zj}.\footnote{Ref~\cite{Falkowski:2014tna} does not specify
this is the procedure it follows, but we assume this is the practice as no correlation matrix
is reported for the results quoted.} All of these results are produced under the assumption of a ``SM-like'' coupling
of the massive gauge bosons to all fermions in the Monte-Carlo modeling, despite the
fact that imposing such a constraint in a basis independent manner in the general
SMEFT renders this approach functionally redundant \cite{Trott:2014dma}.

Comparing the quoted results in Ref,~\cite{Falkowski:2014tna}
shows the bounds on anomalous TGC parameters are sensitive to
the inclusion of quartic terms in the likelihood.
Expanding a general $\chi^2$ function as defined in Eq.~\ref{thechisquared} in the correction to the observables
one obtains \cite{Berthier:2015gja}
\bea
+ 2\, \sum \limits_{i=1}^{n} \sum \limits_{k,l=1}^{q} \sum  \frac{1}{\Delta_i^2} \,  \, \left[\zeta_{i,k,l} \, C_{k}^{6} \, C_{l}^{6}\right] \left( \hat{O} - O\right)_i + 2 \sum \limits_{i=1}^{n} \sum \limits_{k=1}^{r} \frac{1}{\Delta_i^2}  \gamma_{i,k} C_{k}^{8}
 \left( \hat{O} - O\right)_i ,
\eea
when neglecting correlations between the different observables. These effects are numerically suppressed
relative to the terms
\bea
\sim \sum \limits_{i=1}^{n} \sum \limits_{k=1}^{q_i} \sum \limits_{l=1}^{q_i} \frac{C_{i,k}^{6} \, C_{i,l}^{6}}{(\Delta_i)^2}.
\eea
This is due to the fact that when $\left( \hat{O} - O\right)_i \sim \Delta_i$ a relative suppression of $C_{k}^{8}$
terms by $\Delta_i$ is numerically present.\footnote{This does not
correspond to a power counting suppression as there is no evidence of beyond the SM (BSM) physics.}
Including $C_{i,k}^{6} \, C_{i,l}^{6}$ effects and neglecting $C_{k}^{8}$ corrections in the
$\chi^2$ function is most justified if $\Delta_i << 1$. In the case of TGC parameters treated as
observables, the results in Table \ref{TGCtable} have $\Delta_i \sim 1-10 \%$ at one sigma, while
$\sigma_{2\psi}^{4\psi}$ based results, that only use actual observables, have $\Delta_i \sim 10-50 \%$ \cite{Berthier:2016tkq}.
Retaining $C_{i,k}^{6} \, C_{i,l}^{6}$ terms while neglecting $C_{k}^{8}$ corrections
relies on other numerical effects not overwhelming this relative numerical enhancement in either case.
Unfortunately, it is known that
\begin{itemize}
\item{The number of operators dramatically increases order by order in the SMEFT expansion.
The increase is exponential \cite{Henning:2015alf}, leading to the expectation of a large multiplicity
of $C_{k}^{8}$ parameters compared to the number of $C_{i,k}^{6}$ parameters.}
\item{There are numerical suppressions of the linear interference terms due
to $C_{i,l}^{6}$ following from the Helicity arguments of Refs.~\cite{Dixon:1996wi,Mangano:1990by,Dixon:1993xd,Azatov:2016sqh,Falkowski:2016cxu},
introducing more numerical sensitivity to $C_{k}^{8}$ corrections.}
\end{itemize}
$\sigma_{2\psi}^{4\psi}$ results projected into the SMEFT retaining $C_{i,k}^{6} \, C_{i,l}^{6}$ terms
in the likelihood are subject to substantial theoretical uncertainties for all these reasons. The approach of
Ref.~\cite{Berthier:2015gja,Berthier:2016tkq} is to assign and vary a theoretical error due to neglected higher order
terms in the SMEFT when using LEPII data, and to only use the total and differential $\sigma_{2\psi}^{4\psi}$
results with identified final states, avoiding the use of reported LEPII pseudo-nonobservables. It was noted in
Ref.~\cite{Berthier:2015gja} that when numerical behavior indicates that the neglect or inclusion of quartic
terms have a small effect on the likelihood, without simultaneously
changing the theory error in the fit, can lead to the wrong conclusion on the sensitivity of the fit to higher order effects.
The substantial theoretical uncertainties present when projecting LEPII results into the SMEFT are
acknowledged in Ref.~\cite{Falkowski:2015jaa}. This work also argued that
a likelihood that combines Higgs data and TGC data has numerical behavior that indicates that these
higher order terms have a small effect on the likelihood.

\subsection{LEP SMEFT summary}
The SMEFT interpretation of LEP data demonstrates the challenge of consistently combining the
data sets in this EFT can be overcome. This requires a careful separation of IR and UV assumptions
and a consistent SMEFT analysis.

Inconsistent treatments of the LEPII results treat Lagrangian parameters
as directly observable, even though many of the corresponding vertices are off-shell, and
formalisms used have been subject to gauge dependence. LEPII quantities are extracted with other SMEFT parameters being set to zero
in reported results. This introduces non-intuitive consequences in the SMEFT, and inconsistencies
due to the EOM relations between $\mathcal{L}_6$ operators.
A consistent approach to LEPII results can be developed using the double pole expansion
to exploit the narrowness of the SM massive gauge bosons, and only using the experimental $\sigma_{2\psi}^{4\psi}$
total and differential observables.

 LEPI PO and interpretations are on a much firmer footing.
They are more directly related to measured quantities
and extractions involved cross checks of assumptions of a QED like radiation field with $S$ matrix techniques.
As a direct result, the LEPI PO are not subject to the degree of misinterpretation that has plagued LEPII
interpretations.

The numerical differences between results developed using inconsistent methodology and
more consistent SMEFT interpretations is small, as can be seen comparing results in
Refs.~\cite{Han:2004az,Corbett:2012dm,Corbett:2012ja,Corbett:2013pja,
Pomarol:2013zra,Falkowski:2014tna,Falkowski:2016cxu,Falkowski:2015jaa,Butter:2016cvz,Englert:2014uua,Ellis:2014jta,Englert:2015hrx,Berthier:2016tkq}.
The experimental uncertainties
at LEPII are significant, and no evidence of physics beyond the SM emerged from
the LEP data sets. This fact does not validate, and should not encourage, using inconsistent results and
methods to interpret LHC data. The inconsistencies that can be introduced in SMEFT studies
illustrated with LEP data here can tragically obscure the meaning of a real deviation being discovered
using EFT techniques, if the inconsistency is numerically larger than the experimental errors.
In the presence of unknown UV dependent Wilson coefficients, numerically estimating the size of the inconsistencies
introduced is a severe challenge. This point holds for LHC studies of the Higgs-like boson using EFT methods.

\section{The Higgs-like scalar}\label{Higgsfun}
The experimental determination of the couplings of the newly discovered $J^P = 0^+$ scalar
is essential. It is important to study these couplings in a consistent framework
and to upgrade this approach to a full EFT interpretation. The
SMEFT and HEFT approach are now fully defined at leading order and available to be used,
but transitioning from the currently used formalism known as the "$\kappa$ formalism" to these consistent field theory frameworks is still a work
in progress for the LHC experiments. The challenges
to performing this task in the LHC collider environment are profound, but it is important
to emphasize that these challenges
can be overcome in a consistent program of applying EFT methods to collider studies, including the discovered scalar's
couplings, benefiting from the lessons learned interpreting LEP data.
In this section, we review the current dominant
paradigm for reporting constraints on Higgs properties, known as the $\kappa$
formalism, due to the notation of Ref.~\cite{LHCHiggsCrossSectionWorkingGroup:2012nn}.

\subsection{The \titlemath{$\kappa$}{kappa} formalism}\label{sec:kappas}
\begin{figure}[t]\centering
 \includegraphics[width=.9\textwidth]{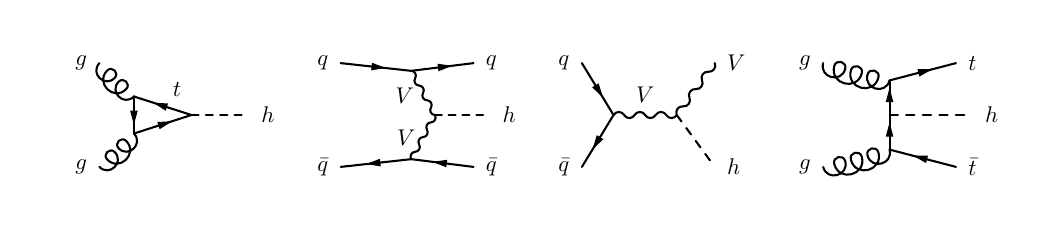}
 \includegraphics[width=.7\textwidth]{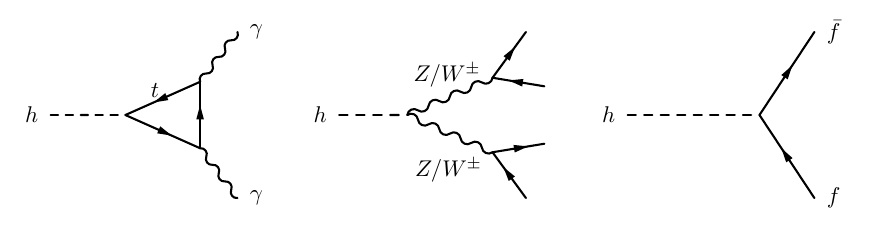}
 \caption{Higgs production (top) and decay modes (bottom) that are rescaled in the $\kappa$
 formalism.}\label{Higgs-prod-decay}
\end{figure}

The $\kappa$ formalism is not an EFT approach to Higgs data as it was set up in Ref.~\cite{LHCHiggsCrossSectionWorkingGroup:2012nn}, but is the
fusion of two approaches. The idea to reweigh the SM couplings and extract and limit deviations in the partial
and total widths of a discovered scalar was laid out in
Ref.~\cite{Zeppenfeld:2000td,Duhrssen:2003tba,Duhrssen:2004cv,Lafaye:2009vr}.
This approach is an ad-hoc rescaling of couplings in the SM without a field theory embedding.
See Fig.~\ref{Higgs-prod-decay} for Feynman diagrams of some of the production and decay modes rescaled.
Some of the rescaled couplings appear in loop diagrams,
which mediate the leading production $gg \rightarrow h$ and decay mode $h \rightarrow \gamma \gamma$
used to probe the properties of this state. That these modes appear first at the one loop level in the SM
is due to the fact that the SM is renormalizable. Introducing such ad-hoc rescalings into
the SM parameters is not a small perturbation for such one loop processes, as the counterterm structure is changed and the theory being used
is no longer the SM. The $\kappa$ formalism can thus only make sense when it is embedded into a consistent
field theory framework that can be renormalized. Refs.~\cite{Zeppenfeld:2000td,Duhrssen:2003tba,Duhrssen:2004cv,Lafaye:2009vr}
did not perform this embedding, although this challenge was understood to be present.

A rescaling approach to the parameters of the SM can be interpreted in principle in the SMEFT or the HEFT.
Systematic field theory embeddings of a rescaling of the Higgs-like scalar couplings,
to interpret the emergence of a signal for this state, appeared in the influential Refs.~\cite{Espinosa:2012ir,Carmi:2012yp,Azatov:2012bz}.\footnote{These works
themselves were also influenced by Ref.~\cite{Hagiwara:1993qt,Low:2009di,Espinosa:2010vn,Barger:2003rs,Manohar:2006gz,GonzalezGarcia:1999fq,Giudice:2007fh,Grinstein:2007iv,Contino:2010mh}.}
Ref.~\cite{LHCHiggsCrossSectionWorkingGroup:2012nn} closely follows in its proposed methodology these works, as it directly notes
in its introduction. The $\kappa$ formalism made a series of assumptions in its detailed implementation
that are not IR assumptions consistent with either field theory embedding. As a result, in both the HEFT and the SMEFT,
a direct relation between the $\kappa$ approach and a general EFT framework is not present.
In both cases, when further UV assumptions are made a mapping can be performed. For example, Refs.~\cite{Espinosa:2012ir,Azatov:2012bz}
are formulated with a general HEFT approach in mind, while Ref.~\cite{Carmi:2012yp} is constructed
in a linear SMEFT formalism in unitary gauge. The $\kappa$ formalism can accommodate large corrections
to the properties of the Higgs-like scalar, and relations due to $\rm SU_L(2) \times U_Y(1)$ linear $\mathcal{L}_6$
operators are not directly imposed. The HEFT also contains parameters whose behavior is analogous
to that of the $\kappa$'s, for example $a_C$ plays the same role as $\kappa_V$.
For these reasons, the rescaling approach, of which the $\kappa$ formalism is a particular example,
has been widely considered to be a restricted version
of the HEFT.\footnote{Works based on this understanding include
Refs.~\cite{Espinosa:2012ir,Azatov:2012bz,Espinosa:2012ir,Espinosa:2012vu,Espinosa:2012im,Azatov:2012rd,Azatov:2012qz,Azatov:2013ura,Farina:2012ea,Farina:2012xp}.
Perhaps the most comprehensive discussion of this embedding is given in Refs.~\cite{Buchalla:2015wfa,Buchalla:2015qju}.}

The $\kappa$ formalism assumes the existence of a CP even scalar resonance with $\hat{m}_h \sim 125$~GeV, whose couplings
have the same Lorentz structure as those of the SM Higgs boson and whose width is $\Gamma \sim \Gamma^{SM}_h$, i.e. narrower
than the energy resolution of the LHC experiments.
The assumption of a narrow resonance even in the presence of SM perturbations is used in Ref.~\cite{LHCHiggsCrossSectionWorkingGroup:2012nn}
to factorize the total event rate into a production cross section and a partial decay width.
Although the exact expansion used for the narrow width is not specified in Ref.~\cite{LHCHiggsCrossSectionWorkingGroup:2012nn},
a naive narrow width assumption is functionally present (and explicitly mentioned in Ref.~\cite{Heinemeyer:2013tqa}
in this context).
A set of scale factors $\kappa_i$ are defined, such that each Higgs production cross section and decay channel is
formally rescaled by a corresponding $\kappa_i^2$. For instance, in the case of the process
$gg\to H \to \gamma\gamma$ one has the parameterization
\begin{equation}
 \sigma(gg\to H)\cdot {\rm BR}(H \to \gamma\gamma) =  \frac{\kappa_g^2 \,\kappa_\gamma^2}{\kappa_H^2}\;\sigma(gg\to H)_{\rm SM} \cdot {\rm BR}(H \to \gamma\gamma)_{\rm SM}\,,
\end{equation}
where $\kappa_H$ rescales the Higgs total width. In the limit $\kappa_i\equiv 1$ the SM is recovered,
while values $\kappa_i\neq 1$ indicate deviations from the SM. A list of relevant $\kappa$ factors is defined as~\cite{LHCHiggsCrossSectionWorkingGroup:2012nn,Heinemeyer:2013tqa}:
\begin{equation}\label{kappa_def}
\begin{aligned}
 \frac{\sigma_{WH}}{\sigma_{WH}^{\rm SM}} &=\kappa^2_{W}\quad
& \frac{\sigma_{ZH}}{\sigma_{ZH}^{\rm SM}} &=\kappa^2_{Z} \quad
& \frac{\sigma_{VBF}}{\sigma_{VBF}^{\rm SM}} &=\kappa^2_{VBF} \quad
& \frac{\sigma_{ggH}}{\sigma_{ggH}^{\rm SM}} &= \kappa^2_g\quad
& \frac{\sigma_{ttH}}{\sigma_{ttH}^{\rm SM}} &=\kappa^2_{t}
\\
\frac{\Gamma_{WW^*}}{\Gamma_{WW^*}^{\rm SM}} &= \kappa^2_{W}
&\frac{\Gamma_{ZZ^*}}{\Gamma_{ZZ^*}^{\rm SM}} &= \kappa^2_{Z}
&\frac{\Gamma_{\gamma\gamma}}{\Gamma_{\gamma\gamma}^{\rm SM}} &= \kappa^2_{\gamma}
&\frac{\Gamma_{Z\gamma}}{\Gamma_{Z\gamma}^{\rm SM}} &= \kappa^2_{Z\gamma}
&\frac{\Gamma_{ff}}{\Gamma_{ff}^{\rm SM}} &= \kappa^2_{f} \,.
\end{aligned}
\end{equation}
The $\kappa$'s are in a sense pseudo-observables, as they are
defined as a rescaling of SM observables in many cases, or inferred quantities while
using the narrowness of the SM gauge bosons to factorize a scattering amplitude.
The assumption of SM like soft radiation effects is present, and no detailed procedure for a correction factor is introduced
to remove these corrections. This pseudo-observable understanding
is not developed in great detail in the specific $\kappa$ proposal, and projecting constraints on these parameters onto the
Higgs couplings requires further information and assumptions.\footnote{More explicit PO interpretations of these parameters have since been advanced
in Refs.~\cite{Gonzalez-Alonso:2014eva,Gonzalez-Alonso:2015bha,Ghezzi:2015vva,Bordone:2015nqa,Greljo:2015sla,deFlorian:2016spz}.}
It deserves to be emphasized that many of these limitations and subtleties are well understood and very clearly stated
in Ref.~\cite{LHCHiggsCrossSectionWorkingGroup:2012nn}.

An example of an assumption adopted in a $\kappa$ fit is the case where
the scaling factors for $VH$ production and $H\to~VV^*$ decay have been defined to be the same. This approximation holds only for tree level computations\footnote{In this section, any reference to perturbative orders is implicitly referred to the EW sector. Radiative QCD corrections can be factorized due to the narrow SM widths, with some exception, in the $\kappa$'s definitions (see Eqn~\ref{kappa_def}).} and under the assumption that deviations in these observables can be interpreted in terms of an anomalous $hZZ$ coupling only, with corrections to e.g. $Zqq$ vertices being
neglected.\footnote{This procedure is beset with inconsistencies that are discussed in Section \ref{functionalredundant}.}
Within these assumptions, the prescriptions of the $\kappa$ formalism are equivalent to the use of the phenomenological
ad-hoc Lagrangian
\begin{equation}
\begin{aligned}
 \mathcal{L}_\kappa =&-\sum_\psi \kappa_\psi \frac{\sqrt2 M_\psi}{\hat{v}}\bar{\psi}\psi h + \kappa_Z \frac{M_Z^2}{\hat{v}} Z_\mu Z^\mu h+ \kappa_W \frac{2M_W^2}{\hat{v}} W^+_\mu W^{-\mu} h,  \\
 &+ \kappa_{g,c} \frac{g_3^2}{16\pi^2 \hat{v}} G_{\mu\nu} G^{\mu\nu} h
 + \kappa_{\gamma,c} \frac{e^2}{16\pi^2 \hat{v}} F_{\mu\nu} F^{\mu\nu} h
 + \kappa_{Z\gamma,c} \frac{e^2}{16\pi^2 c_{\hat{\theta}} \hat{v}} Z_{\mu\nu} F^{\mu\nu} h,
\end{aligned}\label{kappas}
\end{equation}
where $h$ denotes the physical Higgs boson and $F_{\mu\nu}$ is the photon field strength.
For this reason, the $\kappa$'s are often referred to as ``coupling modifiers''. Those in the second line of
Eq.~\ref{kappas} have been defined with an additional ``c'' to underline that they are associated to effective contact interactions.
To match the $\kappa$'s definitions above (with in particular $\kappa_{i,c}=\kappa_i$ for $i=g,\gamma,Z\gamma$), the Lagrangian $\mathcal{L}_\kappa$ must be strictly used for tree-level computations only.
We stress that this parameterization of Lagrangian terms should not be interpreted to give a naive physical meaning
to the $\kappa_i$. The distinction between observables and Lagrangian parameters discussed in Section \ref{preliminarymeasurements} applies,
and formulating the $\kappa_i$ in terms of mass eigenstate shifts does not change this fact.

For processes that are produced at one-loop already in the SM, it is possible to employ the Lagrangian above at NLO,
with the caveat that in this case the parameters $\kappa_{g,c}$, $\kappa_{\gamma,c}$ and $\kappa_{Z\gamma,c}$
appearing in Eq.~\ref{kappas} do not coincide with the $\kappa$'s defined in Eq.~\ref{kappa_def} and in particular
their SM value is $\kappa_{i,c}=0$. Consider the one loop SM process $\sigma(gg \rightarrow h)$, the leading contribution is generated radiatively
with a top or bottom quark running in the loop.
In this case, the amplitude can be formally split separating the various contributions as
\begin{equation}
 \mathcal{A}_{ggH} = \kappa_t \mathcal{A}^{t}_{ggH} + \kappa_b \mathcal{A}^b_{ggH} + \kappa_{g,c},
\end{equation}
where $\mathcal{A}_{ggH}^{t(b)}$ represents the contribution from the top (bottom) loop.

Although the most general approach is that of retaining $\kappa_{g,c}$ as an independent parameter that can capture contributions from BSM diagrams, it is possible to consider the approximation $\kappa_{g,c}=0$. This corresponds to the implicit assumption that the $ggH$ vertex does not receive direct new physics contributions from e.g. a heavy state running in the loop.
As a consequence the factor $\kappa_g$ (see Eq.~\ref{kappa_def}) can be expressed as a function of $\kappa_t$ and $\kappa_b$ according to
\begin{equation}\label{L_kappa}
 \kappa_g^2(\kappa_t,\kappa_b) =\frac{ \kappa_t^2 \,\sigma^{tt}_{ggH} +\kappa_b^2\, \sigma^{bb}_{ggH} +\kappa_t \kappa_b\, \sigma^{tb}_{ggH} }{\sigma^{tt}_{ggH}+\sigma^{bb}_{ggH}+\sigma^{tb}_{ggH}}\,.
\end{equation}
Here $\sigma^{tt(bb)}_{ggH}$ denotes the contribution  of the top (bottom) loop to the $ggH$ cross section, while $\sigma^{tb}_{ggH}$ stands for the interference term.
Similar considerations hold for the $\gamma\gamma$ and $Z\gamma$ decays, for the tree-level case of vector boson fusion ($VBF$) production, where the SM diagrams are modified with $\kappa_{Z,W}$,
and for the total $\Gamma_h$, whose corresponding modifier $\kappa_H$ receives contributions from $\kappa_{f,W,Z,\gamma,g}$ in addition to a BSM component that can be denoted $\kappa_{H,{\rm BSM}}$.

\begin{figure}[t]\centering
 \includegraphics[width=.9\textwidth]{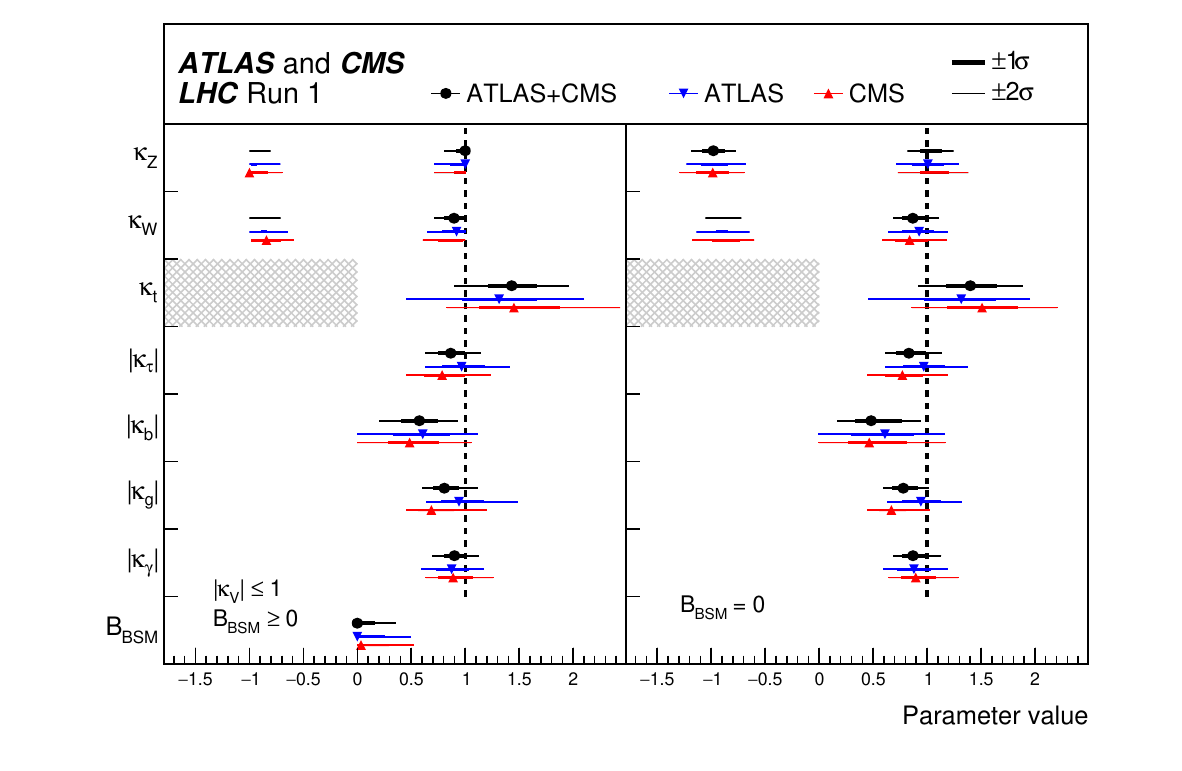}
 \caption{Fit results from Ref~\cite{Khachatryan:2016vau} obtained including $\kappa_{g,c}$, $\kappa_{\gamma,c}$ as free parameters. In the left panel the BSM Higgs width  $\kappa_{H,{\rm BSM}}={\rm B_{BSM}}$ has also been treated as independent, while requiring $|\kappa_V|\leq1$ for $V=Z,W$. The right panel assumes instead  $\kappa_{H,{\rm BSM}}=0$. The hatched area for $\kappa_t$ has been forbidden in the fit, to break the degeneracy due to the absence of sensitivity to the sign of this parameter in the data used.}\label{exp_kappa_1}
\end{figure}

\begin{figure}[t]
\floatbox[{\capbeside\thisfloatsetup{capbesideposition={right,top},capbesidewidth=6cm}}]{figure}[\FBwidth]
{\caption{Fit results from Ref.~\cite{Khachatryan:2016vau} obtained imposing $\kappa_{\gamma,c}=\kappa_{g,c}=\kappa_{H,{\rm BSM}}=0$, i.e. assuming the absence of BSM contributions to the $hGG$ and $h\gamma\gamma$ interactions, as well as in the Higgs width, beyond those stemming from modified Higgs couplings. The hatched area for $\kappa_t$ has also been forbidden.}\label{exp_kappa_2}}
{\includegraphics[width=7cm]{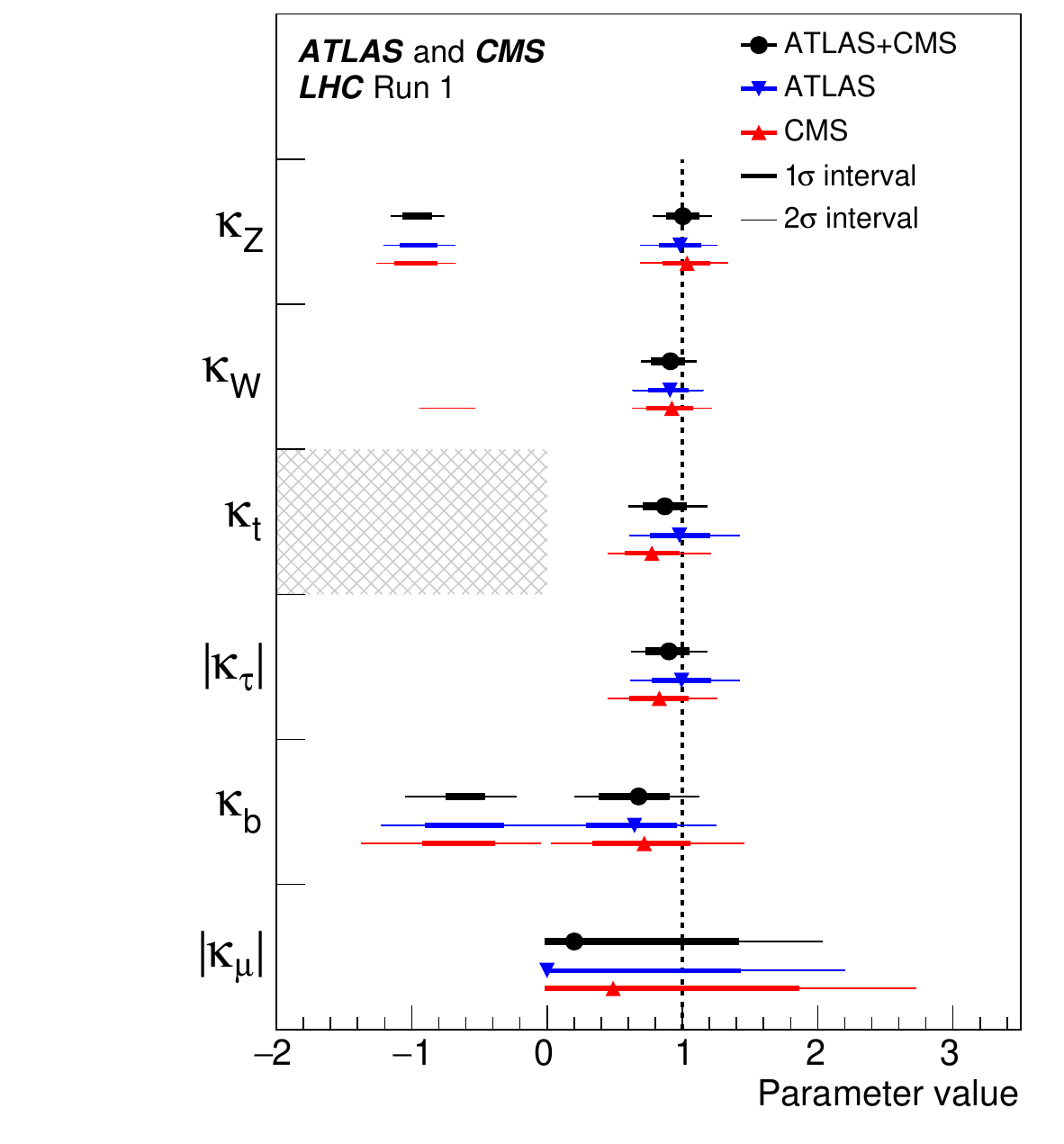}}
\end{figure}

The experimental collaborations have provided limits on the $\kappa$ parameters,
that are extracted from a global fit to Higgs production and decay measurements~\cite{Khachatryan:2014jba,Aad:2015pla,Aad:2015gba,Khachatryan:2016vau}.
Both benchmarks described above in the ggH example have been considered: Figure~\ref{exp_kappa_1} shows the most recent constraints~\cite{Khachatryan:2016vau} obtained including $\kappa_{\gamma,c},\,\kappa_{g,c},\,\kappa_{H,{\rm BSM}}$ (the latter is denoted $\rm B_{BSM}$ in the plot) as free parameters, while Figure~\ref{exp_kappa_2} shows the results for $\kappa_{\gamma,c}=\kappa_{g,c}=\kappa_{H,{\rm BSM}}=0$.
In the first case (Figure~\ref{exp_kappa_1}) the system is under-constrained, and therefore one additional condition is required to constrain all the parameters. Two alternative choices were considered: either imposing $|\kappa_V|=|\kappa_Z|,|\kappa_W|\leq 1$ while allowing $\kappa_{H,{\rm BSM}}\neq 0$ (left panel), or fixing $\kappa_{H,{\rm BSM}}= 0$ (right panel).

The results generically show that $\kappa_t$ is sensitive to which of the two scenarios is assumed,
as expected considering that it gives the dominant contribution to the radiative decay and
production channels. The other parameters do not show a dramatic variation among the different
setups. It also worth noting that, as anticipated in Section~\ref{HEFTsec} the current
uncertainty in the determination of the Higgs couplings is  10 -- 20\% on
average.\footnote{Note that the degeneracy of the $\kappa_t$ sign can be lifted using the method discussed in Refs.~\cite{Biswas:2012bd,Farina:2012xp,Degrande:2012gr}.}

Notably, as the experimental accuracy drops below the 10\%, EW radiative corrections become significant.
As a consequence, it is not appropriate to use the $\kappa$ framework to project
directly the signal-strength measurements into constraints on the Higgs couplings except in some
limited applications.

\begin{figure}[h!]\centering
 \includegraphics[width=.445\textwidth]{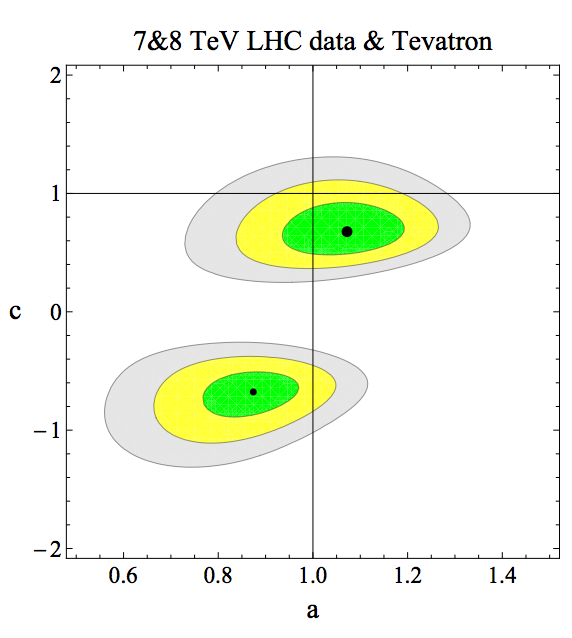}
 \includegraphics[width=.495\textwidth]{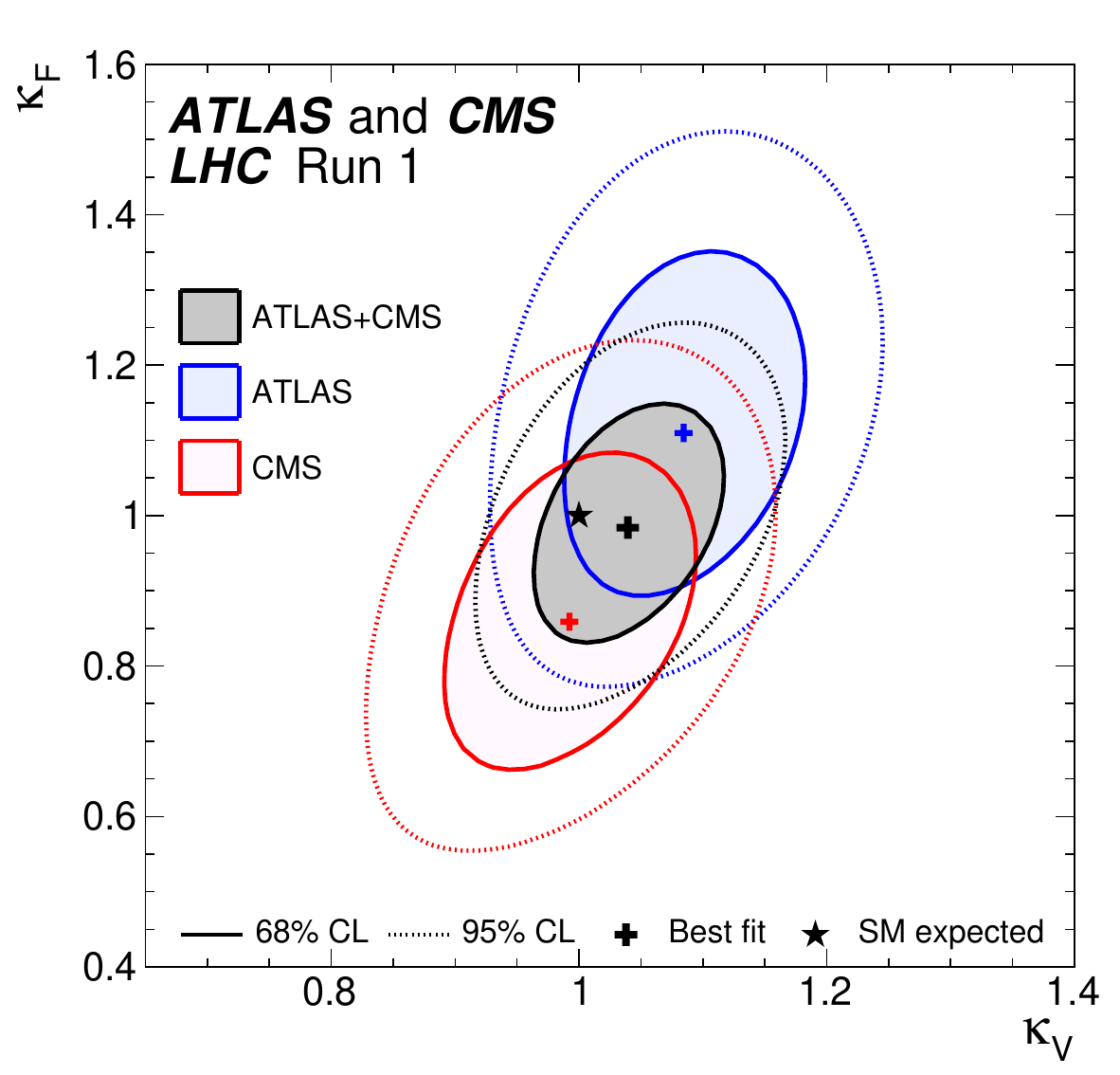}
\caption{Hypothesis test of the discovered scalar as a Higgs-like boson with a universal rescaling
parameter set for the fermions and vector bosons $(\kappa_F,\kappa_V)$ (right) Ref.~\cite{Khachatryan:2016vau}
and using the alternate notation $(c,a)$ (left) around the time of discovery in Ref.~\cite{Espinosa:2012im}.
Note the different scales of the plots.}\label{hypotest}
\end{figure}

The $\kappa$-formalism was constructed as a first probe of the Higgs boson's properties and constitutes
a reasonable framework for the interpretation of the  Higgs dataset collected so far at the LHC. The key strength of this
approach is not that it is an EFT, but that it allows a series of hypothesis tests addressing the question on the consistency
of the properties of the discovered scalar with the SM Higgs. Perhaps the most elegant of these tests
is the two dimensional test with only a universal $(\kappa_F,\kappa_V)$.  A comparison of results of this form
produced at the time of discovery in 2012 in Ref.~\cite{Espinosa:2012im} in Fig.\ref{hypotest} (left) and the combined ATLAS+CMS results
in  Fig.\ref{hypotest} (right) demonstrates the degree to which the Run I data set increasingly supported the hypothesis
that the discovered scalar is the Higgs boson.

\begin{figure}[t]\centering
\includegraphics[width=.495\textwidth]{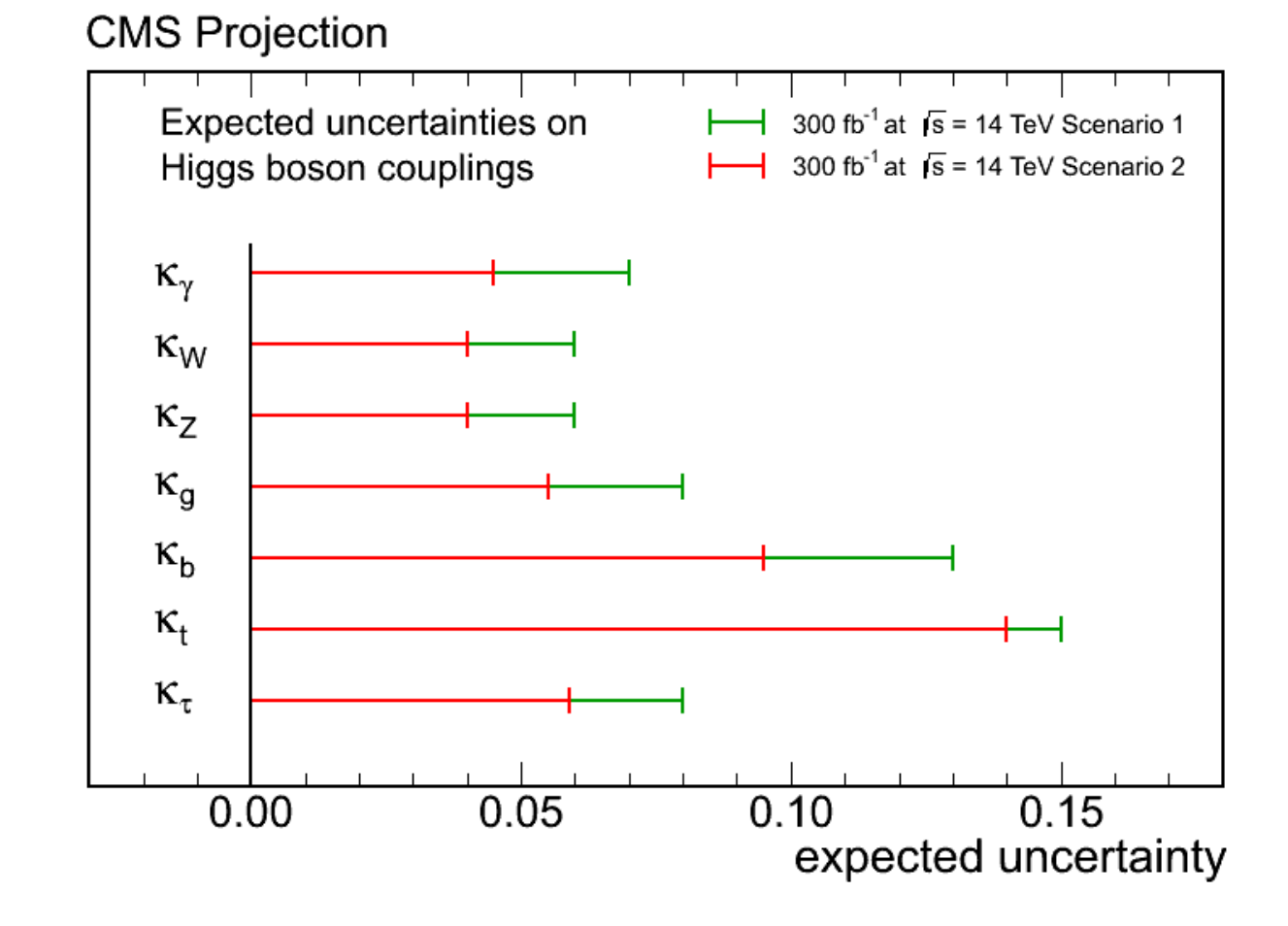}
\includegraphics[width=.495\textwidth]{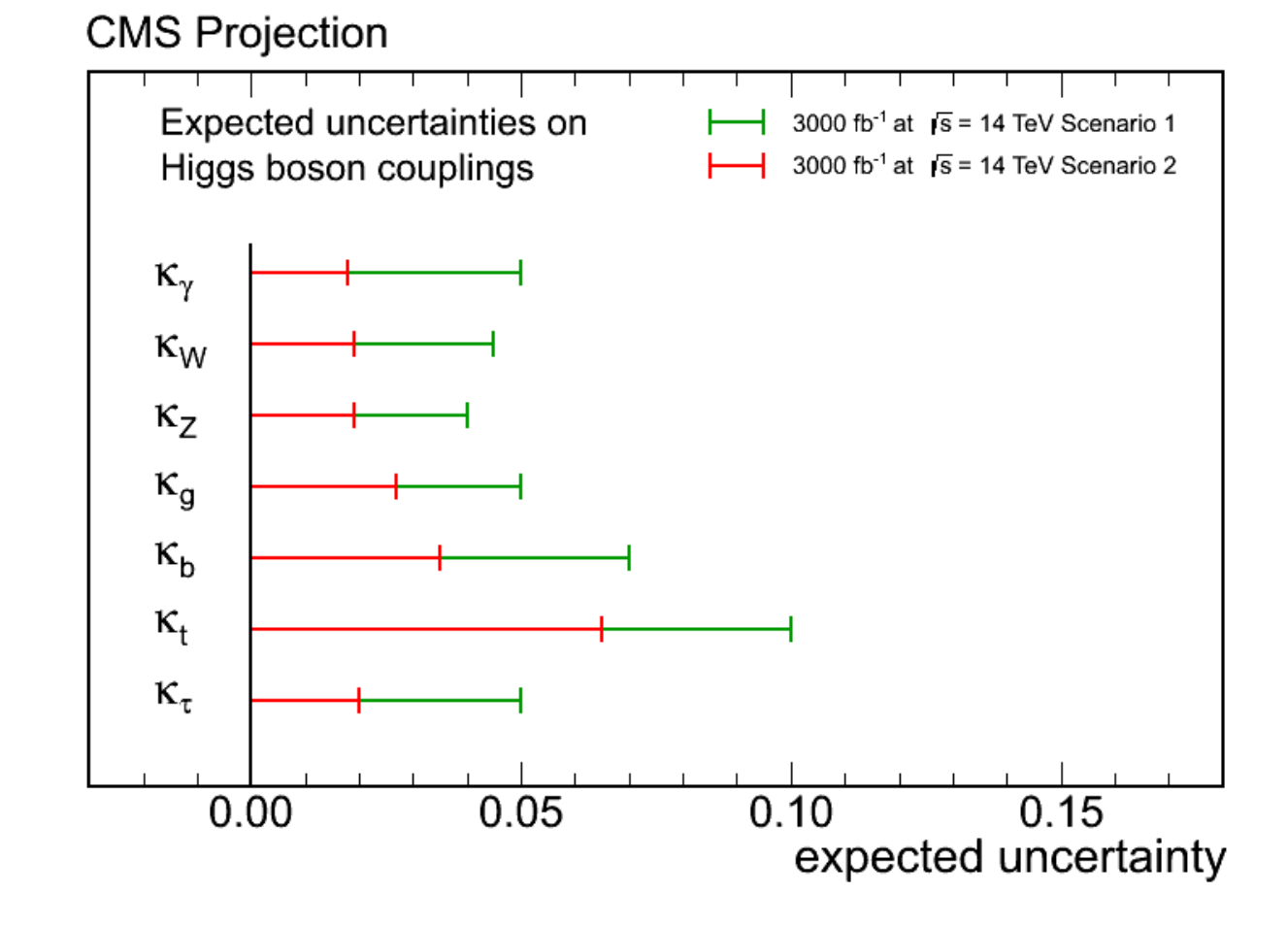}
\caption{Prospects for the measurement of the $\kappa$ parameters at the CMS detector~\cite{CMS-NOTE-2013-002} at $\sqrt{s}=14$~TeV and
with an integrated luminosity of 300~fb$^{-1}$ (left) and 3000~fb$^{-1}$ (right). The red and green lines
are described in the text.}\label{kappa_prospects}
\end{figure}

The $\kappa$ formalism does not constitute a suitable tool for a consistent analysis of the Higgs properties
going forward, as the experimental data improves.
Figure~\ref{kappa_prospects} shows the projections for these measurements at the CMS experiment at 14~TeV and
for integrated luminosities of 300~fb$^{-1}$ (left panel) and 3000~fb$^{-1}$ (right panel)~\cite{CMS-NOTE-2013-002}.
Similar results are expected at ATLAS~\cite{ATL-PHYS-PUB-2014-016}. The plot shows that the
sensitivity will approximately reach the 5 -- 10 \% level, with a significant dependence
on the scaling of experimental errors assumed.
The green lines correspond to a quite conservative case (Scenario~1) in which all
the systematics are assumed to be the same as in the 2012 performance.
The red lines (Scenario~2), instead, are derived rescaling the theoretical uncertainties
by a factor 1/2 and the other systematics by the square root of the luminosity.
Although the projections are extremely uncertain, and subject to a number of unverified assumptions
it is clear that the properties of the Higgs-like scalar will be increasingly resolved experimentally
in Run II and beyond. The $\kappa$ formalism is not the right tool for this era of increasing experimental
precision. Some of its main limitations are
\begin{itemize}
\item The  $\kappa$ formalism is not an EFT as formulated in Ref.~\cite{LHCHiggsCrossSectionWorkingGroup:2012nn}.
As the $\kappa$ formalism can only be related to EFT constructions with a set of further UV assumptions,
it is not guaranteed that it captures a consistent IR limit of an underlying new physics sector. If $\kappa$ fits show deviations
from the SM constructing a consistent inverse map to the UV sector is not guaranteed to be possible as a result.
\item The $\kappa$ formalism is not systematically improvable with perturbative corrections. This is
a fatal flaw  that introduces a multitude of difficulties. These difficulties
always appear in any ad-hoc construction.
Any replacement formalism must be able to systematically determine perturbative corrections
without assuming the SM to address this central flaw. The only known way to accomplish this is with a
well defined effective field theory embedding. As the accuracy of the data descends below
the $\sim 10 \%$ range, higher order calculations simply become indispensable in important channels
sensitive to Higgs properties.
\item The rescalings of parameters that are off-shell vertices, in particular $h \rightarrow V \, V^\star$ is
ambiguous as the off-shell massive gauge boson is not an external state and has no precise definition without a field theory embedding.
\item The $\kappa$ formalism  cannot be consistently used to interface Higgs data with LEP data, due to the different
scales involved in the measurements. Again this requires a field theory embedding to relate the Wilson coefficient
constraints. Similarly, the  $\kappa$ formalism is not a useful tool to interface with even lower energy measurements.
\item The formulation is intrinsically non-gauge invariant, and at best an example of a non-linear realization
of the gauge symmetry of the SM, i.e. a restricted version of the HEFT. The couplings of the scalar to fermions and gauge bosons
are left arbitrary. It is also the case that amplitudes computed with the Lagrangian given in Eq.~\ref{kappas} generically
lead to non-unitary S-matrix elements. This is not a concern if the $\kappa$  formalism is embedded in an EFT
extension of the SM, as such a theory need only be unitary to the cut off scale.
As the $\kappa$ formalism is not so embedded in an EFT without further assumptions,
its lack of unitarity at high energies renders it obviously inconsistent, and unsuitable for
studies of differential distributions at high energies in particular.
\item  The construction of the
 $\kappa$ formalism is not a truly general set of deviations from the SM, but a biased construction
 informed by experimental constraints in a fairly haphazard fashion. For example, having two independent
parameters $\kappa_Z\neq \kappa_W$ introduces a hard breaking of the custodial symmetry, but this is
avoided in many $\kappa$ fits due to experimental constraints (see the discussion in Ref.~\cite{Farina:2012ea}).
On the other hand, these constraints are not consistently determined
in the  $\kappa$ formalism itself, due to its inability to relate LEP and LHC data which requires a field theory embedding.
\item The $\kappa$ formalism includes only couplings with standard Lorentz structures in a renormalizable field theory.
As such, it can only capture deviations in total production/decay rates, and this is another reason
it cannot be consistently used for the analysis of kinematic distributions.
\end{itemize}

Despite all of these flaws, and the clear need to go beyond the $\kappa$ framework, we wish to emphasize that
the $\kappa$ framework and Run I results reported in it were a profound and important achievement.
The theoretical framework to interface with LHC data in a consistent EFT extension of the
SM was simply not available in Run I. As such, the $\kappa$ framework, despite all its flaws, was a
sensible and insightful choice to project the raw experimental data into a useful and informative form.
It is clear that the use of the $\kappa$ framework  to hypothesis test the SM was an informative
application. This data reporting formalism should be maintained into Run II and beyond despite all its limitations
for this application. Nevertheless, it is time to go beyond the $\kappa$ framework. The HEFT and the SMEFT
have now been developed to a sufficient degree that they can be systematically used for this task going forward.

\subsection{Relation of the \titlemath{$\kappa$}{kappa} formalism to SMEFT and HEFT}\label{sec:kappas_efts}
In order to overcome the $\kappa$ framework limitations listed above, it is necessary to switch
from the $\kappa$-parameterization to one given in terms of Wilson coefficients
of a non-redundant EFT basis (see e.g.\cite{David:2015waa,Passarino:2016pzb}). Once such a basis has been chosen,
it is possible to identify a one way mapping between the $\{\kappa_i\}$ and $\{C_i\}$.\footnote{
This does not turn the  $\kappa$'s into a basis, as no gauge invariant field redefinitions result in the mapping.}
In general, each $\kappa_i$ can be decomposed as
\begin{equation}
 \kappa_i^2 = 1 + \Delta \kappa_i,
\end{equation}
with $\Delta\kappa_i$ a linear combination of EFT parameters, whose numerical coefficients are computed
calculating the relevant $\sigma_i$ or $\Gamma_i$ at a given order in the EFT.
We stress that translating the $\kappa$ framework into an EFT form does not just
represent a bare reparameterization, but actually improves
the theoretical consistency of the description. The EFT embedding makes manifest the presence of correlations
among different observables that are required by gauge invariance or other imposed symmetries and structures
such as the SMEFT reparameterization invariance due to the EOM.
The generic result, once again, is a correlated fit space of unphysical Lagrangian parameters, as in
the LEP case represented in Fig.~\ref{Fig:correlation_matrices}.

To illustrate how the procedure is carried out in the SMEFT, consider for instance the decay $h\to\bar bb$. A partial
NLO calculation of this process in the SMEFT has been presented in Refs.~\cite{Gauld:2015lmb,Gauld:2016kuu}. For illustrative purposes,
here we report only the tree-level result computed in the Warsaw basis, which gives
\begin{equation}\label{kappa_b_smeft}
\begin{aligned}
  \kappa_b^2 &=\frac{\mathcal{A}^2(h\to\bar bb)_{\rm SMEFT}}{\mathcal{A}^2(h\to\bar bb)_{\rm SM}} =
 1+ \Delta\kappa_b\,,\\
 \Delta\kappa_b &= 2 \, \bar{v}_T^2\left(C_{H\square}-\frac{C_{HD}}{4}-C_{Hl}^{(3)}+\frac{C_{ll}'}{2}-\frac{C_{dH}}{[Y_d]_{33}}\right)\,.
 \end{aligned}
\end{equation}
The terms $C_{H\square},C_{HD}$ come from normalizing the Higgs' kinetic term to the canonical form, while $C_{Hl}^{(3)},C_{ll}'$ appear due to the shift between the true vev $\bar{v}_T$ and the value inferred from the measurement of $G_F$. Finally, $C_{dH}$ represents the only direct $d=6$ contribution, which is due to the operator $\mathcal{Q}_{dH}$, that perturbs the Yukawa coupling.
See Appendix \ref{shiftsgalore} for details on shift parameters.

Another important example is that of $h\to \gamma\gamma$. In this case it is necessary to carry out the computation at one-loop,
which gives
\begin{equation}
 \kappa_\gamma^2 = \frac{\mathcal{A}^2(h\to\gamma\gamma)_{\rm SMEFT}}{\mathcal{A}^2(h\to\gamma\gamma)_{\rm SM} }= 1 +\Delta\kappa_\gamma^{\rm LO} + \Delta\kappa_\gamma^{\rm NLO}
\end{equation}
where $\Delta\kappa_\gamma^{\rm LO}$ and $\Delta\kappa_\gamma^{\rm NLO}$ include the contributions computed respectively at tree-level and at one-loop in the SMEFT,
both normalized to the SM amplitude calculated at a specific order in perturbation theory. Here we normalize by the one loop SM amplitude.
The tree-level term is easily derived in the Warsaw basis. The SMEFT contributions from CP even operators at LO reads~\cite{Manohar:2006gz}
\begin{equation}\label{kappa_gamma_smeft}
\Delta\kappa_\gamma^{\rm LO} = -\frac{16\pi^2}{e^2 \mathcal{I}^{\gamma}} v_T^2\left(C_{HWB} \, \st \, c_{\hat{\theta}} -C_{HW}\st^2-C_{HB} \, c^2_{\hat{\theta}}\right),
\end{equation}
where $\mathcal{I}^{\gamma}$ encodes the adimensional SM amplitude, whose expression can be found
in Ref.~\cite{Bergstrom:1985hp}. It is interesting to notice that $\Delta\kappa_\gamma^{\rm LO}$ carries
an enhancement of $8\pi^2/e^2$ compared e.g. to $\Delta\kappa_b$. This could give $\mathcal{O}(1)$
deviations for $C_i \sim 1$ and $\Lambda$ as low as about 4~TeV (barring cancellations among the coefficients),
which simply reflects the fact that processes that are radiatively suppressed in the SM are \emph{a priori}
more sensitive to the presence of new physics as the SMEFT has a multi-pole expansion in general.
This was emphasized long ago in Ref.~\cite{Manohar:2006gz}.
The one-loop term $\Delta\kappa_\gamma^{\rm NLO}$ has a much more complex structure
and is not so easy to derive. It contains a large number of $\mathcal{L}_6$ Wilson coefficients, feeding in
relatively suppressed by $16 \pi^2$ compared to the LO results. Many of these Wilson coefficients
do not appear at tree level in the SMEFT, see the results in Refs.~\cite{Hartmann:2015aia,Ghezzi:2015vva,Hartmann:2015oia}.
Note also that an explicit
matching of the $\kappa$'s into the Warsaw basis, extended to all the Higgs two-body decay channels, has been partially
given in Ref.~\cite{Ghezzi:2015vva}.

The mapping procedure illustrated above for the SMEFT case can be done with the HEFT. In this case,
the $\kappa$'s are mapped to combinations of parameters belonging both to the leading and next-to-leading
order Lagrangian. Consider the tree-level expression of $\Delta\kappa_b$. Using the basis of Ref.~\cite{Brivio:2016fzo} one has
\begin{equation}
 \Delta\kappa_b = \frac{32\pi^2v_T^2}{\Lambda^2}(r_2^l-r_5^l)+\frac{[Y_d^{(1)}]_{33}}{[Y_d]_{33}}\,.
\end{equation}
The coefficients $r^l_{2,5}$ are associated to four-fermion operators that enter through $\delta v_T \sim \delta G_F$ (analog to $C_{ll}'$ in \ref{kappa_b_smeft}) and belong to the NLO Lagrangian\footnote{Compared to Eq.~\ref{kappa_b_smeft}, here there is no equivalent of the terms $C_{H\square}$ and $C_{HD}$ because there is no operator in the basis of Ref.~\cite{Brivio:2016fzo} that modifies the Higgs kinetic term. The operator corresponding to $\mathcal{Q}_{Hl}^{(3)}$ was not retained either.} $\Delta\mathcal{L}$.
In the last term, instead, $Y_d^{(1)}$ belongs to $\mathcal{L}_0$: it  is the Yukawa matrix appearing
in the second term of the expansion of the functional $\mathcal{Y}_Q(h)$, defined in Eq.~\ref{Lag0}.
The impact of this term is analogous to that of the operator $\mathcal{Q}_{dH}$ in the SMEFT,
with the difference that in the HEFT case anomalous Higgs couplings appear already at LO due to the singlet nature of the $h$ field.
This is true for the interaction terms with $d\leq 4$, while it is still
necessary to include $\Delta\mathcal{L}$ terms to match $H\gamma\gamma$, $HGG$, $HZ\gamma$ couplings.
Namely, $\kappa_{W,Z}$ and $\kappa_f$ receive leading contributions respectively from $a_C$ and $Y_f^{(1)}$,
while $\kappa_{\gamma}$, $\kappa_{g}$, $\kappa_{Z\gamma}$ are mapped to a combination of
higher order Wilson coefficients.
For instance,\footnote{This result can be compared with Eq.~\ref{kappa_gamma_smeft}.
There is a close correspondence between the coefficients $C_{HWB}\to a_1$, $C_{HB}\to a_B$,
$C_{HW}\to a_W$, while the term $a_{12}$ in the HEFT comes
from the custodial breaking operator $\Tr ( \T W_{\mu\nu} )^2$
that has an equivalent only at $d = 8$ in the SMEFT.}
\begin{equation}
 \Delta\kappa_\gamma^{\rm LO} =
 \frac{8\pi^2}{e^2 \mathcal{I}^{\gamma}}\left( -4\st \, c_{\hat{\theta}} \tilde{a}_1 + c_{\hat{\theta}}^2\tilde{a}_B + \st^2(\tilde{a}_W-4\tilde{a}_{12})\right)\,,
\end{equation}
where the shorthand notation $\tilde{a}_i = C_ia_i$ stands for the product of $C_i$ with the coefficient of the linear term in the function $\mathcal{F}_i(h)=1+2a_i h/v+\dots$

Restricted versions of the SMEFT and the HEFT can be used in this manner to develop one way mappings to
the $\kappa$'s. This can be done as no defining conditions in the $\kappa$ approach are fundamentally
gauge dependent.  As the assumptions of the $\kappa$ formalism itself starts to fail
at the experimental precision where this mapping becomes of interest, this task is not a
high priority. Directly developing the corresponding results in the SMEFT and HEFT to interface with past data and future
LHC results at leading, and next to leading order, is ongoing in a manner that is essentially bypassing the $\kappa$
formalism.

\section{SMEFT developments in the top sector}\label{topsection}

Being the heaviest known particle, and the one with the largest Yukawa coupling, the top quark is possibly the SM state that is closest to new physics sectors. In particular, it represents a sensitive probe of new physics driving the EW symmetry breaking (EWSB): for instance, its mass plays a fundamental role in determining the RG evolution and stability of the Higgs potential in the UV. At the same time, the top is typically expected to exhibit the largest mixing with exotic states in scenarios with non-linear EWSB sectors, such as composite Higgs models or models with warped extra dimensions.

Studying the properties and couplings of the top quark can thus give a unique insight into new physics, which is complementary to that offered by the Higgs boson. Top physics also benefits from a significantly larger dataset compared to Higgs physics, as top quarks are abundantly produced at high energy hadron colliders such as Tevatron and LHC. This has motivated several analyses of the top sector based on the SMEFT approach.

Early studies explored the possibility of constraining its interactions at $e^+e^-$ colliders~\cite{
Ladinsky:1992vv,Chang:1993fu,Grzadkowski:1995te,Grzadkowski:1996kn,Grzadkowski:1996pc,Whisnant:1997qu,Yang:1997iv,Baek:1997ib,Brzezinski:1997av,Grzadkowski:1997cj,Bartl:1998ja,Grzadkowski:1998bh,Grzadkowski:1999iq,Boos:2001sj} (for recent analyses at future lepton colliders see e.g. \cite{AguilarSaavedra:2012vh,Rontsch:2015una,Amjad:2015mma,Cao:2015qta,Vos:2016til,Englert:2017dev,Bernreuther:2017cyi}) and in $\gamma\gamma$ collisions~\cite{Choi:1995kp,Baek:1997ib,Poulose:1997xk,Boos:2001sj,Grzadkowski:2003tf,Grzadkowski:2004iw,Grzadkowski:2005ye}, which constitute a particularly suitable environment to probe CP violating couplings.
Here we give an overview of SMEFT studies of the top sector, focusing on the processes relevant for top physics at the Tevatron and LHC.

In the Warsaw basis there are 28 operators that directly involve the top quark
at $\mathcal{L}^{(6)}$ in unitary gauge (see Table \ref{op59} for the operators definitions)\footnote{Prior to the construction of the Warsaw basis, a systematic parameterization of $d=6$ effects in the top sector was proposed in~\cite{AguilarSaavedra:2008zc,AguilarSaavedra:2009mx}.}:
\begin{equation}\label{operators_a}
\begin{aligned}
 &\Q_{uH},\, \Q_{Hu},\, \Q_{Hq}^{(1),(3)},\, \Q_{Hud},\,
 \Q_{uW},\, \Q_{uB},\, \Q_{uG},\, \Q_{dW},\,\\
& \Q_{qq}^{(1),(3)},\, \Q_{lq}^{(1),(3)},
 \Q_{uu},\, \Q_{ud}^{(1),(8)},\, \Q_{eu},\,
 \Q_{lu},\, \Q_{qe},\, \Q_{qu}^{(1),(8)},\, \Q_{qd}^{(1),(8)},\,\\
 &\Q_{ledq},\, \Q_{quqd}^{(1),(8)},\, \Q_{lequ}^{(1),(3)}.
 \end{aligned}
\end{equation}
In addition to these, other operators can be relevant for a global analysis of the top sector, either because they enter the top couplings due to input parameter definitions (see Appendix \ref{shiftsgalore}) or because they modify other interactions entering top production processes. The first class includes
\begin{equation}\label{operators_b}
 \Q_{H\square},\, \Q_{HD},\, \Q_{HWB},\, \Q_{ll},\, \Q_{Hl}^{(3)},
\end{equation}
while the operators fulfilling the latter condition are
\begin{equation}\label{operators_c}
 \Q_{G},\, \Q_{\tilde G},\, \Q_{HG},\, \Q_{H\tilde G}.
\end{equation}
Considering a general flavor scenario and retaining only the index contractions that select a top quark, the overall number of independent parameters is 1179 (622 absolute values + 557 complex phases). The picture is remarkably simplified in the approximation in which the quarks of the first two generations respect a $\rm U(2)^3$ symmetry (a $\rm U(2)$ for each field $q$, $u$, $d$) and, simultaneously, flavor universality is assumed in the lepton sector. In this case, that represents the customary set of assumptions for global top analyses, the number of independent parameters is reduced to 85, corresponding to 68 absolute values and 17 phases\footnote{The number of independent parameters is further reduced in the presence of a full $\rm U(3)^5$ flavor symmetry, that gives 46 independent quantities (38 absolute values + 8 phases). However, this option is rarely considered in top analyses, as they implicitly assume that new physics effects may impact top physics more significantly compared to processes involving the first two generations.}.

This number is still too large for a complete analysis to be performed, but a rich variety of studies has been carried out in the literature, focusing on specific subsets of the parameter space.
In particular, the operators in Eq.~\ref{operators_b} are usually neglected in top physics analyses, under the assumption that they can be constrained in other classes of measurements. An exception are fits to EWPD, where $C_{WB}$ and $C_{HD}$ are typically retained. These studies allow to constrain top operators of classes 6 and 7, via loop contributions to the gauge bosons self energies~\cite{Greiner:2011tt,Zhang:2012cd} and four fermion operators arising either at tree level or at 1-loop due to RG mixing~\cite{deBlas:2015aea}.
CP violating contributions are also negligible for a large number of top measurements at the LHC, because the spin-averaged SM amplitudes for these processes are dominantly CP-conserving, implying that the interference term $\mathcal{A}_{SM}\mathcal{A}^*_{d=6}$ is typically suppressed. Bounds on the CP-odd parameters are rather inferred from measurements of top polarizations and $t$-$\bar t$ spin correlations~\cite{Kane:1991bg,Atwood:1992vj,Atwood:2000tu,AguilarSaavedra:2006fy,Han:2009ra,Zhang:2010dr,Biswal:2012dr,Bernreuther:2013aga} or from lower energy experiments, such as EDM measurements~\cite{CorderoCid:2007uc,Kamenik:2011dk,Brod:2013cka,Gorbahn:2014sha,Cirigliano:2016njn,Cirigliano:2016nyn} and $B$ meson decays~\cite{Burdman:1999fw,Grzadkowski:2008mf,Drobnak:2011aa}.

The first global fit to top results from Tevatron and LHC has been presented by the TopFitter collaboration in~\cite{Buckley:2015nca,Buckley:2015lku} (see also \cite{Moore:2016jsc,Russell:2017cut}), where 12 independent combinations of 14 Wilson coefficients were constrained using both differential and inclusive measurements.

\begin{figure}[t]
\hspace*{-1cm}
 \includegraphics[width=1.1\textwidth]{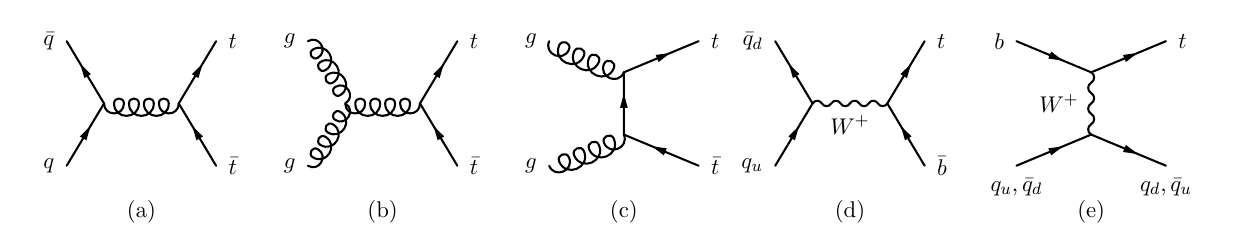}
 \caption{Main diagrams for top pair and single top production in $pp$ collisions in the SM. }\label{diagrams:top_processes}
\end{figure}

The relevant processes for top physics measurements at the LHC are the following:
\begin{enumerate}[label=(\roman*)]
 \item Top pair production $p p \to t \bar t$.

 In the SM, this process is dominated by QCD contributions: both $q\bar q$ and $gg$ initiated diagrams contribute, as shown in Figure~\ref{diagrams:top_processes}~(a)-(c).
 In the SMEFT, diagram (a) is corrected only by contributions proportional to $(C_{uG})_{33}$, modifying the $G t\bar t$ coupling. This is the coefficient with the largest impact on top pair production. Diagrams obtained inserting $(C_{uG})_{ii}$ with $i=\{1,2\}$ in the initial $G q\bar q$ vertex are negligible because their interference with the SM amplitude is proportional to $m_u$ or $m_d$, see Section \ref{EFTcolliderbasics}.
 In addition, $q\bar q$ initiated top pair production receives tree-level SMEFT contributions from the 6 four-quark operators $\Q_{qq}^{(1)}$, $\Q_{qq}^{(3)}$, $\Q_{uu}$, $\Q_{ud}^{(8)}$, $\Q_{qu}^{(8)}$, $\Q_{qd}^{(8)}$~\cite{Zhang:2010dr,AguilarSaavedra:2010zi}, whose impact can be expressed in terms of only four combinations of Wilson coefficients~\cite{Zhang:2010dr}:
\begin{equation}\label{Cs_tt}
 \begin{aligned}
 C_u^1 &= (C_{qq}^{(1)})_{1331}	+(C_{qq}^{(3)})_{1331}+(C_{uu})_{1331},\quad	&
 C_d^1 &= 4 (C_{qq}^{(3)})_{1133} + (C_{ud}^{(8)})_{3311},\\
 C_u^2 &= (C_{qu}^{(8)})_{1133} + (C_{qu}^{(8)})_{3311},		&
 C_d^2 &= (C_{qu}^{(8)})_{1133} + (C_{qd}^{(8)})_{3311}.
 \end{aligned}
\end{equation}
 Diagrams induced by $\Q_{qu}^{(1)}$, $\Q_{qd}^{(1)}$, having color-singlet contractions in the fermion currents, do not interfere with the QCD SM diagrams, but only with the EW production. Their contributions to the total cross section are roughly an order of magnitude smaller that those of the six four-quark operators listed above\footnote{
 Note that the interference of $\Q_{qq}^{(1),(3)}$ and $\Q_{uu}$ contributions with SM QCD production is non-vanishing, as particular linear combinations of these operators are equivalent to the color octet contractions $(\bar q \g_\mu T^A  q)^2$, $(\bar u \g_\mu T^A  u)^2$ via Fierz transformations and the $\rm SU(3)$ completeness relation $T^A_{BC}T^A_{DE}=2 \d_{BE}\d_{DC}-2/3 \d_{BC}\d_{DE}$.
 Among the 3 remaining operators admitting interactions with two top quarks, $\Q_{ud}^{(1)}$ gives contributions that only interfere with the EW SM diagrams. $Q_{quqd}^{(1),(8)}$ contain currents with a $(\bar LR)$ chiral structure, so they give only diagrams whose interference with the SM is suppressed by the mass of the initial state quarks.}.

 For the $gg$ initiated channel, that dominates at high energy, the SM diagrams in Fig.~\ref{diagrams:top_processes}~(b) and (c) can be dressed with $d=6$ contributions from $\Q_{uG}$ in the $Gt\bar t$ vertices.
 This operator contributes, in addition, through a $GGt\bar t$ four-point interaction.
 $\Q_{uG}$ is the operator with the most significant impact on $t\bar t$ production, and it has been extensively studied in the literature, see e.g.~\cite{Atwood:1994vm,Hioki:2009hm,Choudhury:2009wd,Hioki:2013hva,Aguilar-Saavedra:2014iga,Franzosi:2015osa,Barducci:2017ddn}.
 Diagram (b) can also be corrected with an insertion of $\Q_G$ (or $\Q_{\tilde G}$) in the $GGG$ vertex. Due to the helicity structure of the external gluons, this term interferes only with the diagram in Figure~\ref{diagrams:top_processes}~(c) proportionally to $m_t^2$, and thus yields a smaller correction compared to $\Q_{uG}$. The operators $\Q_{HG},\,\Q_{H\tilde G}$ can also contribute inducing a tree $gg\to h\to t \bar t$ diagram. However, this term is suppressed by the Higgs propagator being always largely off-shell.

 \vskip 1em

 The main observables in $ pp \to t\bar t$ is the total cross section, that is currently measured at the LHC with an uncertainty $\lesssim 5\%$~\cite{Aad:2014kva,Khachatryan:2016yzq,Aaboud:2016pbd,Khachatryan:2016mqs,Khachatryan:2016kzg,Sirunyan:2017uhy}.
 Differential cross sections are also important for constraining the relevant Wilson coefficients. In particular, the analysis of the $m_{t\bar t}$ spectrum is useful to target and disentangle four-fermion operators~\cite{Degrande:2010kt}.

 Further, charge asymmetries received much attention in the past, mainly due to a discrepancy with the SM expectation registered by the CDF experiment at Tevatron~\cite{Aaltonen:2011kc}, in the forward-backward asymmetry
\begin{equation}
 A_{FB} = \frac{N(\Delta y > 0)- N(\Delta y < 0)}{N(\Delta y > 0) + N(\Delta y < 0)},\quad
 \Delta y = y_t-y_{\bar t},
\end{equation}
where $y_f$ is the rapidity of the fermion $f$ in the laboratory frame.
The excess was subsequently reduced in analyses with higher statistics, and the most recent combination of the Tevatron measurements finds agreement with the SM expectation at NNLO QCD + NLO EW within $1.6\sigma$~\cite{Aaltonen:2017efp}. Besides refined experimental techniques, an important role in reducing the anomaly has been played by higher order calculations in the SM~\cite{Kidonakis:2011zn,Hollik:2011ps,Kuhn:2011ri,Manohar:2012rs,Czakon:2014xsa,Kidonakis:2015ona,Czakon:2016ckf,Czakon:2017lgo} showing that $A_{FB}$ receives large radiative corrections. \\
At the LHC, the forward-backward asymmetry is washed out by the huge symmetric $gg\to t \bar t$ contributions and by the fact that the initial $pp$ state is forward-backward symmetric.
A better observable for LHC measurements is the central charge asymmetry $A_C$, which is correlated with $A_{FB}$ and defined as~\cite{Diener:2009ee}
 \begin{equation}
 A_{C} = \frac{N(\Delta |y| > 0)- N(\Delta |y| < 0)}{N(\Delta |y| > 0) + N(\Delta |y| < 0)}.
\end{equation}
The $A_{C}$ asymmetry has been measured by the ATLAS and CMS collaborations in $pp$ collisions at 7
and~8~TeV
(see~\cite{Sirunyan:2017lvd} and references therein), finding agreement with the SM expectation.
Interestingly, the two asymmetries $A_{FB}$ and $A_C$ are different linear functions of the parameters in Eq.~\ref{Cs_tt}. Therefore the four combinations of Wilson coefficients can be constrained combining Tevatron measurements of $A_{FB}$ and LHC measurements of $A_C$~\cite{Rosello:2015sck}.

The first systematic EFT studies of $t\bar t$ production were carried out in Refs.~\cite{AguilarSaavedra:2010zi,Zhang:2010dr,Degrande:2010kt}, including a phenomenological analysis considering all the observables mentioned above. The role of angular observables in the top decay products has been explored, for instance, in~\cite{Bernreuther:2013aga,Bernreuther:2015yna}.

Higher order results are also available: the SMEFT amplitude has been computed at NLO QCD including the contribution of a real $(C_{uG})_{33}$~\cite{Franzosi:2015osa}; effects induced via RG mixing were also considered within the class of four-fermion operators~\cite{Jung:2014kxa}.

 \item Single top production $ p p \to q t$ ($q\neq t$), $pp \to t W$.

 Single top production is usually classified into $s$-channel, $t$-channel and $tW$ production.

 Focusing on the $ p p \to q t$ modes, the $s$-channel is characterized by a $b$ in the final state (Fig.~\ref{diagrams:top_processes} (d)) while the $t$-channel has typically a light quark in the final state (Fig.~\ref{diagrams:top_processes} (e)).
 The SM diagrams receive SMEFT corrections from operators entering the $Wtb$ vertex~\cite{AguilarSaavedra:2008gt}, namely $\Q_{uW}$, $\Q_{Hq}^{(3)}$ and the operators in Eq.~\ref{operators_b} (all but $\Q_{H\square}$, that only corrects the top Yukawa coupling), or modifying the $W$ coupling to a light quark pair. Among the latter, only $\Q_{Hq}^{(3)}$ and the terms in Eq.~\ref{operators_b} give relevant contributions, while the interference terms for $\Q_{uW}$, $\Q_{dW}$, $\Q_{Hud}$ are always suppressed by the mass of one of the quarks entering the vertex.
 Among four-fermion operators, $\Q_{qq}^{(1),(3)}$ have the largest impact: due to the chiral structure of the SM diagrams, the interference terms of invariants with at least one right-handed fermion are proportional to the mass of one of the external quarks~\cite{Bach:2014zca}.

 The $pp\to tW$ mode is mostly produced from the $gb$ partonic initial state and and is sensitive to $(C_{uG})_{33}$ in addition to the coefficients of the operators affecting $pp\to q t$~\cite{Ayazi:2013cba,Rindani:2015vya}.

 \vskip 1em
 The total cross section for mono-top production is dominated by the $t$-channel contribution, that has been measured at the LHC  with a precision of $\sim 10\%$~\cite{Chatrchyan:2012ep,Aad:2014fwa,Khachatryan:2014iya,Aaboud:2017pdi,Aaboud:2016ymp,Sirunyan:2016cdg}. The cross section for the $s$-channel is $\sim 20$~times smaller at the LHC, due to the presence of an anti-quark in the initial state \cite{Kidonakis:2010tc}.
 ATLAS and CMS have reported evidence for this process with LHC-Run I data, although with a quite low significance of $3.2$ and $2.5\,\sigma$ respectively~\cite{Aad:2015upn,Khachatryan:2016ewo}
 (see~\cite{Giammanco:2017xyn} for a recent experimental review).

The SM cross section of $pp \to tW$ is also sizable at the LHC (although about 3--4 times smaller than the $t$-channel cross section). Evidence for this process at the LHC has been found already at $\sqrt s=7$~TeV~\cite{Chatrchyan:2014tua,Aad:2015eto}. Recently, the ATLAS and CMS collaborations reported measurements of the total and differential rates at $\sqrt s=13$~TeV~\cite{Aaboud:2016lpj,CMS-PAS-TOP-17-018,Aaboud:2017qyi}, with a maximum precision of $\sim 10\%$ reached by CMS using the full $35.9$~fb$^{-1}$ dataset collected in 2016.
Despite the accessible cross sections, measurements of $tW$ production are challenging at the LHC, mainly because the process interferes with $t\bar t$ production beyond LO in QCD. Disentangling the two signals is a quite complex task~\cite{Kersevan:2006fq,Frixione:2008yi,White:2009yt,Jezo:2016ujg}.

Measurements of the top polarizations in mono-top processes can play an important role in this context, helping to constrain anomalous $Wtb$ interactions~\cite{Aguilar-Saavedra:2014eqa} and four-fermion operators~\cite{Aguilar-Saavedra:2017nik}.

Comprehensive EFT analyses of mono-top can be found in~\cite{Cao:2007ea,AguilarSaavedra:2010zi,Zhang:2010dr}.
NLO QCD corrections to the SMEFT amplitude have been computed for all three channels~\cite{Zhang:2016omx} finding, in particular, that they have a non-trivial impact on differential distributions.

\item Top pair production in association with a neutral gauge boson $pp \to t\bar t V$, $V=\{Z,\gamma\}$.

 The production of a top in association with a neutral gauge boson $Z/\gamma$ can be both $q\bar q$ and $gg$-initiated: the relevant SM diagrams have the same structure as those in Figure~\ref{diagrams:top_processes}~(a)-(c), with a gauge boson emitted from any of the fermion lines.
 In addition to the set of operators that modify $t\bar t$ production, these channels give a unique access to the $Ztt$ and $\g tt$ couplings~\cite{Rontsch:2014cca,Rontsch:2015una}, that are corrected by $\Q_{Hq}^{(1),(3)}$, $\Q_{Hu}$, $\Q_{uW}$, $\Q_{uB}$, $\Q_{dW}$, $\Q_{dB}$ and the operators in Eq.~\ref{operators_b} (except $\Q_{H\square}$).
 Four-fermion operators also contribute to the $q\bar q$-initiated channel. The corresponding corrections behave as in the $pp\to t\bar t$ case.

 The total cross section for both $t\bar t Z$, $t \bar t \gamma$ have been measured at the LHC with an accuracy of $\sim15-20\%$~\cite{Aad:2015uwa,Aad:2015eua,Aaboud:2017era,Sirunyan:2017iyh,Sirunyan:2017uzs}. Their constraining power is quite low at the moment~\cite{Buckley:2015lku}, but shall become significant with higher statistics.
 It has been pointed out in Ref.~\cite{Schulze:2016qas} that cross section ratios of $t\bar t V$/$t\bar t$ are also convenient observables, that allow to isolate and constrain the top dipole moments induced by $(C_{uW})_{33}$, $(C_{dW})_{33}$, $(C_{uB})_{33}$.

 The SMEFT contributions from class 6 and 7 operators to the total cross section and differential distributions of $pp\to t\bar t Z/\gamma$ have been computed at NLO QCD accuracy~\cite{Rontsch:2014cca,Rontsch:2015una,Bylund:2016phk}.

 \item Top pair production in association with a Higgs boson $ pp \to t \bar t h$.

 In the SM $pp\to t\bar t h$ takes place mainly through diagrams analogous to Figs.~\ref{diagrams:top_processes}~(a)-(c), with a Higgs radiated from one of the top propagators. SMEFT corrections to these diagrams are therefore analogous to those for $pp\to t\bar t$, with the addition of possible insertions of the coefficients $C_{H\square}$, $C_{HD}$ and $(C_{uH})_{33}$ in the $tth$ vertex.
 The operator $\Q_{HG}$ ($\Q_{H\tilde G}$) also contributes, inducing new diagrams in which the Higgs is radiated from a gluon line rather than from a top, and a $gg$-initiated diagram containing a $GGGh$ four-point interaction.

 This process represents therefore an interesting bridge between top and Higgs global analyses, providing information complementary to that extracted from Higgs production and decay~\cite{Degrande:2012gr,Bramante:2014gda,Maltoni:2016yxb}.
 In particular, the interplay of the $t\bar t h$ channel with $h$, $hh$ and $h+j$ production at the LHC has been explored with the inclusion of NLO QCD corrections to contributions from $(C_{uH})_{33}$, $(C_{uG})_{33}$, $C_{HG}$~\cite{Maltoni:2016yxb}.
 The analysis of angular observables of the decay products of the tops is also interesting in this context, as it would allow to probe the imaginary part of $(C_{uH})_{33}$, testing the CP nature of the top Yukawa vertex~\cite{Demartin:2014fia,Khatibi:2014bsa,Buckley:2015vsa,Mileo:2016mxg}.

 Due to the low cross section and the presence of large irreducible backgrounds, significant constraints from $pp\to t\bar th$ are likely to be extracted only at the high luminosity phase of the LHC. At present, the ATLAS and CMS collaborations have reported the observation of this process at $\sqrt{s}=13$~TeV. The uncertainties on the measured cross sections are of the order of 15--20\%
 ~\cite{Sirunyan:2018hoz,Aaboud:2018urx}.

 \item Top decays.

 In the SM, top quarks decay nearly 100\% of the time to $b W$. As such, a study of the properties of the top decay products can probe the $Wtb$ interaction to a good accuracy. In the Warsaw basis, this vertex receives tree-level corrections from
 $\Q_{Hq}^{(3)}$ and the operators in Eq.~\ref{operators_b} (except $\Q_{H\square}$), that preserve the Lorentz structure of the SM interaction\footnote{The operators in Eq.~\ref{operators_b} enter the vertex due to the parameter shifts determined with the choice of the input quantities. In particular, $(C_{ll})_{1221}$ and $(C_{Hl}^{(3)})_{11,22}$ are present due to $\hat G_F$ being chosen as an input. On the other hand $C_{HD}$ and $C_{HWB}$ enter when choosing the set $\{\haew,\hat m_Z,\hat G_F\}$ as inputs but do not contribute if $\haew$ is replaced with $\hat {m}_{W}$.}, from $\Q_{Hud}$ that introduces a right-handed coupling, and from $(C_{uW})_{33}$ and $(C_{dW})_{33}$, with a dipole contraction.
 The presence of $\Q_{Hq}^{(3)}$, $\Q_{ll}$, $\Q_{Hl}^{(3)}$, $\Q_{HD}$ or $\Q_{HWB}$ therefore determines a rescaling of the total decay rate, while $\Q_{Hud}$, $\Q_{uW}$, $\Q_{dW}$ impact the kinematic properties of the decay products.
 The four fermion operators $\Q_{qq}^{(3)}$, $\Q_{lq}^{(3)}$ also contribute to this decay for the hadronic and leptonic final states of the $W$ respectively~\cite{AguilarSaavedra:2010zi}.

 The total width of the top has been measured both at the Tevatron~\cite{Abazov:2012vd,Aaltonen:2013kna} and at the LHC~\cite{Khachatryan:2014nda,CMS-PAS-TOP-16-019,Aaboud:2017uqq} with quite large uncertainties.
 More precise measurements (with a 3--4\% accuracy) are available for the helicity fractions of the $W$ boson~\cite{Aaboud:2016hsq,Khachatryan:2016fky}, that are among the most promising observables. They are predicted to the permille accuracy in the SM~\cite{Czarnecki:2010gb} and they are modified only by\footnote{There are in principle corrections proportional to $(C_{Hud,uW,dW})_{ii}$, $i=1,2$ and $(C_{eW})_{jj}$ that enter the $W$ decay vertex. These, however, interfere with the SM amplitude proportionally to the mass of one of the $W$ decay products, which is always negligible.}
 $(C_{Hud})_{33}$, $(C_{uW})_{33}$ and $(C_{dW})_{33}$. Measurements of these quantities (possibly in combination with other angular observables) allow therefore to derive significant constraints on these coefficients, see e.g.~\cite{Chen:2005vr,AguilarSaavedra:2006fy,AguilarSaavedra:2007rs,AguilarSaavedra:2010nx,Bernardo:2014vha,Cao:2015doa,Hioki:2015env}.

 The decay $t\to Wb$ in the presence of anomalous couplings has been computed analytically up to NLO QCD \cite{Drobnak:2010ej,Zhang:2014rja}.
 In particular, Ref.~\cite{Zhang:2014rja} also explored the impact of four-fermion operators.

 Finally, upper limits on the observation of exotic top decays through FCNC can be used to set bounds on the flavor off-diagonal entries of some Wilson coefficients. Relevant processes in this sense include $t\to u_i\, X$, with $u_i=\{u,c\}$ and $X=\{Z,\gamma,g, h\}$~\cite{CorderoCid:2004vi,Ferreira:2005dr,Coimbra:2008qp,Chen:2013qta,Greljo:2014dka}. A full list of the operators contributing (up to NLO QCD accuracy) can be found in~\cite{Zhang:2014rja}.
 Contributions from class 6 and 7 operators to $t\to u_i \, Z/\gamma$ have been computed to NLO QCD~\cite{Drobnak:2010wh,Drobnak:2010by,Zhang:2010bm}. More recently, the calculation has been extended to the $t\to u_i\,h$ channel~\cite{Zhang:2013xya,Zhang:2014rja} and with the inclusion of parton shower effects~\cite{Degrande:2014tta}.
 Ref.~\cite{Durieux:2014xla} included these observables in a global analysis of flavor-changing top couplings.
 Operators giving flavor-changing charged currents and flavor-changing four-fermion interactions were considered in Ref.~\cite{Aguilar:2015vsa}, that also performed a global analysis including constraints from single top production and $B$ and $Z$ decays (see also~\cite{AguilarSaavedra:2010zi}).

 In a complementary approach, Ref.~\cite{Davidson:2015zza}, considered lepton flavor violating top decays $t\to q \ell^+ \ell^{\prime-}$, $\ell\neq \ell^\prime$ induced by four-fermion interactions, showing that their measurement at the LHC can give bounds comparable to those obtained from flavor physics at HERA.

 \item Other processes.

 Other processes can be relevant for constraining the couplings of the top at the LHC.

 For instance, $pp \to t\bar t t\bar t$ and $pp \to t\bar t b\bar b$ would give access to the (3333) flavor contraction of the operators with four quarks, that do not affect significantly any other process above~\cite{Kumar:2009vs,Degrande:2010kt,Zhang:2017mls}.

 Single top production in association with a neutral gauge boson $pp \to t\,Z/\gamma$
 is also of interest, as it can give relevant constraints on FCNC top interactions, complementary to those from decays~\cite{delAguila:1999ac,delAguila:1999kfp,Ferreira:2008cj,AguilarSaavedra:2010rx,Agram:2013koa,Durieux:2014xla,Aguilar-Saavedra:2017vka}.

 Single top production in association with a Higgs boson $pp\to t h$ can instead help setting constraints on $(C_{uH})_{33}$, as this process is sensitive to the sign (an therefore the complex phase) of the top Yukawa coupling
 \cite{Biswas:2012bd,Farina:2012xp,Englert:2014pja,Chang:2014rfa,Kobakhidze:2014gqa,Demartin:2015uha}, and on $(C_{uH})_{3i}$, as to the presence of flavor-changing $twh$ couplings~\cite{AguilarSaavedra:2000aj,Wang:2012gp,Greljo:2014dka}.

\end{enumerate}

\section{Progress and challenges for LHC pseudo-observables, HEFT and SMEFT}\label{stateoftheart}
The key problem of the $\kappa$ formalism is that it is not a systematically improvable framework. It does not have well
defined perturbative corrections, and its results cannot be combined in a predictive fashion with
different data sets. In short, the $\kappa$ framework is not an EFT. To go beyond the $\kappa$ framework
there are currently two main approaches being developed for LHC applications, the pseudo-observables (PO) approach,
and the systematic development of EFT extensions to the SM. Both of these approaches
face significant challenges and are underdeveloped. In the previous sections we have reviewed the extensive
development of EFT extensions to the SM studying LEPI, LEPII and top physics. In this section we summarize the
overall state of affairs partway through LHC Run II.

\subsection{Challenges for pseudo-observables}
Recently the paradigm of pseudo-observables (PO) has been reinvigorated at the LHC.
Initial work in this direction was reported in Ref.~\cite{Passarino:2010qk},
and further precursor studies \cite{Isidori:2013cla,Isidori:2013cga} have been developed into
a theoretical paradigm in Refs.~\cite{Gonzalez-Alonso:2014eva,Gonzalez-Alonso:2015bha,Bordone:2015nqa,Greljo:2015sla}.
These developments are a welcome advance over the $\kappa$ formalism. They are theoretically grounded
in formal expansions \cite{Passarino:2010qk} around the poles of the narrow SM states, factorizing observables
into gauge invariant sub-blocks (see Section \ref{POsectionintro}). This transitions a $\kappa$ approach to a firmer
theoretical footing. The evolution
of the $\kappa$ approach to the results of Ref.~\cite{Gonzalez-Alonso:2014eva} for inclusive Higgs decays is rather direct.
A key strength of the PO approach is that it is defined as a gauge invariant decomposition around the
physical poles in the process that is disconnected from an underlying assumed Lagrangian.
This approach directly exploits the narrowness of the unstable massive states known to exist,
and can be mapped to both the HEFT and the SMEFT in principle as the underlying Lagrangian field theory
is not fixed. This approach is also based on the correct understanding of the distinction between
observable $S$ matrix elements and unphysical Lagrangian parameters (see Section \ref{preliminarymeasurements}).
For this reason, a PO decomposition, at least in
inclusive Higgs decays, can form a sensible bridge (at tree level) to the underlying EFTs in data reporting.
PO decompositions also have some predictive power. By exploiting
crossing symmetry, relations between different classes of observables can be determined in the PO approach,
such as $h \rightarrow V \mathcal{F}$ and $\mathcal{F} \rightarrow h \, V$ \cite{Isidori:2013cla,Isidori:2013cga}
or between $h \rightarrow f_1 \, f_2 \, f_3 \, f_4$ and $f_1 \, f_2 \rightarrow h \, f_3 \, f_4$ \cite{Greljo:2015sla}.

The challenges to the successful development of a model independent PO program at LHC are also clear.
Decomposing all observable amplitudes into a PO set is challenging at the LHC compared
to LEP for a simple reason, the LEP initial state was well defined, while the LHC initial state is an overlap of various
partonic processes convoluted with parton distribution functions. This challenge can be avoided if a narrow width expansion is employed
to factorize up an observable, and this issue is not present for characterizing inclusive Higgs decays, where PO approaches have been
well developed \cite{Gonzalez-Alonso:2014eva} and are clearly applicable.

A further important issue is related to the core model independent strength of the pseudo-observable approach, namely
the lack of a fixed embedding in an explicit EFT Lagrangian that extends the SM. As a direct result, some of the limitations
of the $\kappa$ approach remain. Without mapping to a particular field theory, relations between observables
(unrelated by crossing symmetry) are absent, and radiative emission is ill-defined in general. The idea to accommodate
soft radiation has been to use universal radiator functions, mimicking the approach at LEP, to dress amplitudes. However, unlike at LEP
there is currently no feasible proposal to check the SM-like radiator functions assumption in the LHC environment.
Explicit assumptions of no effects of physics beyond the SM in radiative emissions dressing
the PO have been invoked, but these are UV assumptions, not IR assumptions, and they have no known interpretation
as a precise condition on UV dynamics. It is possible these challenges can be overcome to enable a precision
pseudo-observable program at LHC that extends beyond characterizing inclusive Higgs decays. Assuming
away these pressing issues with UV assumptions reduces the model independence of the PO approach,
and should be avoided if at all possible.

\subsection{Challenges for SMEFT/HEFT}
Any formalism that seeks to improve upon the  $\kappa$ framework
must address its core defects comprehensively, consistently and without invoking UV
assumptions if it seeks to maintain model independence. The
new framework must be able to capture the IR limit of physics beyond the SM, without assuming that the physics
beyond the SM is already known, and allow an inverse map to the underlying theory
if deviations are discovered at LHC, or in
future facilities where LHC data is also used as legacy information.
It is fortunate that EFT is constructed and defined to
exactly meet these demands, that have essentially resulted from the LHC data set not indicating non-SM
resonances around the EW scale. The powerful constraints of
\begin{itemize}
\item{Lorentz invariance and the global symmetry constraints due to the Higgsing of
$\rm SU_L(2) \times U_Y(1) \rightarrow U_{em}(1)$ in the case of the SMEFT,}
\item{local analytic operators extending the SM due to the assumption of a degree of decoupling
$\bar{v}_T \ll \Lambda$, (see Section \ref{basics}),}
\end{itemize}
leads to a predictive and well defined extension of the SM,
that can be systematically constrained experimentally.
The EFT approach allows a well defined characterization of higher order perturbative and non-perturbative
neglected effects defining the {\it approximate theoretical precision} in an analysis,
as is a fundamental part of the EFT  description.
Such an approximate precision can be characterized as a theoretical error in global data analysis
that is varied to represent various cases of the size of the neglected higher order terms.
This error can be continually and appropriately reduced with further development of this
theoretical paradigm. This approach stops one from overinterpreting the data set and being too aggressive on the
constraints found in the EFT framework, considering the limited theoretical precision of the EFT description.
The HEFT has less constraints due to the presence of a singlet scalar in the spectrum,
but still carries powerful constraints due to a local analytic operator expansion
and can accommodate assumed global symmetries as {\it IR assumptions} reducing its complexity.

A core challenge to SMEFT/HEFT is -- which EFT should be used?
The existence of two self-consistent constructions must be
seriously considered. It is not appropriate to casually dismiss
the HEFT construction just because the Higgs-like scalar has converged on the
properties of the SM Higgs to date. The key distinction between
the HEFT and the SMEFT is an IR assumption about the states in the spectrum in the presence
of $\rm SU_L(2) \times U_Y(1) \rightarrow U_{em}(1)$.
Considering the viewpoint laid out in Section \ref{precision-motivation},
on the Higgs potential being an effective parameterization of the true
dynamical mechanism underlying $\rm SU_L(2) \times U_Y(1) \rightarrow U_{em}(1)$, this EFT choice essentially corresponds
to the assumption of one low energy parameterization of such physics
being preferred over the other. $\rm SU_L(2) \times U_Y(1) \rightarrow U_{em}(1)$ can occur due to
weakly coupled or strongly coupled dynamics in the UV sector.
The core problem with making this choice outright is that no good understanding of the low energy limit
of all possible strongly interacting sectors exists, due to the difficulties in
calculating non-perturbative physics. On the other hand, the fact that the
HEFT and the SMEFT seem to be functionally indistinguishable on SM pole processes requiring studies
of tails of distributions to seek out differences (see Section \ref{tailspoles})
indicates that a dedicated pole constraint program formulated in one of these theories can be mapped
to the alternate EFT construction directly.

The efficient way to develop the constraint picture for each EFT is fairly clear
and is undergoing a rapid development in the community.
First, a consolidation/data mining phase that distills past LEPI/LEPII/Tevatron/LHC Run I
and lower energy results into constraints on the Wilson coefficients is developed. The first priority for
this effort is to map the constraints on pole processes (scattering events where $p^2 \sim m^2$ for a SM intermediate state)
to the SMEFT. This restriction to single pole resonantly enhanced processes allows the narrow width
factorization of the dependence on Wilson coefficients in $\mathcal{L}_6$ into those that
are resonantly enhanced, and those suppressed by an additional factor of $\Gamma/m$.
As the data set of such pole processes is limited, it is appropriate to invoke flavour
symmetries when pursuing such bounds, at least initially.

This initial analysis phase is extended with data from the tails of distributions and low energy
observables. The much larger data sets present in these cases allow the flavour symmetries to be
relaxed. Tails of distributions are an important source of information on Wilson coefficients,
but it cannot be avoided that when the EFT expansion is breaking down, predictivity is lost.
This can be the case in the tails of distributions. It is also clear that when examining
tails of distributions the choice between the HEFT and the SMEFT as a field theory approach
is more pressing. Furthermore, in either EFT, the number of parameters grows dramatically
as the IR effect of class 8 $\psi^4$ operators being further suppressed is absent.
These challenges can all be overcome without invoking UV assumptions so long as an appropriate theoretical
error is assigned to EFT studies in such tails of distributions.

A related question is how PDF uncertainties impact the extraction of constraints from LHC measurements of high-energy tails.
PDF uncertainties in the large-$x$ region strongly depend on the partons considered, being largest for gluons and antiquarks. At the LHC with $\sqrt{s}=13$~TeV,
for a partonic c.o.m. energy of $\sqrt{\hat s}\sim2-3$~TeV, they are typically in the ranges (-30\%,+10\%) and (-15\%, +10\%) for $gg$ and $q\bar q$-initiated processes respectively, but they can grow up to 100\% at $\sqrt{\hat s}\sim5$~TeV.\footnote{Additionally, the $gg$, $qg$ and $q\bar q$ PDF luminosities are sensitive to some flavor assumptions and parameterization choices that lead to energy-dependent discrepancies of up to 20--40\% between predictions of different groups. See Ref.~\cite{Gao:2017yyd} for a recent detailed discussion.}
Precision measurements for this class of observables are then particularly challenging and generally possible only for $E\lesssim2$~TeV.
Very different is the situation for dominantly $qq$ or $qg$-initiated processes, that typically carry PDF uncertainties of the order 10-20\% or smaller on the entire spectrum~\cite{Gao:2017yyd}. Around $\sqrt{\hat s}\sim3$~TeV, EFT effects can exceed the PDF uncertainty band for $\Lambda\lesssim10$~TeV, thus allowing the extraction of constraints within the region of validity of the EFT expansion.

A more accurate comparison of PDF uncertainties with possible BSM signals can be done only on a case by case basis, as both quantities are strongly process-dependent.
The reduction of PDF uncertainties at large-$x$, that is crucial for EFT analysis,
will be possible in the near future, thanks to the inclusion of high-energy LHC data in PDF global fits (see e.g.~\cite{Czakon:2016olj} for the impact of top quark pair production measurements).

To combine  data sets measured at disparate energies, it is required to develop the SMEFT
and HEFT to the order of one loop calculations for the most precise observables. It is also clear that the
challenge of statistical estimates of constraints in multi-dimensional Wilson coefficient spaces
are underdeveloped. One loop results are
not available in almost every process of phenomenological interest, although it has been shown they
can have a remarkable impact on standard interpretations of precise measurements such as the LEP EWPD PO
\cite{Passarino:2016owu,Ghezzi:2015vva,Hartmann:2016pil}.
It is unclear if a systematic one loop SMEFT and HEFT paradigm can be developed in time to have maximum impact on
LHC studies.

A non-physics challenge to the HEFT and the SMEFT is the literature is manifestly conflicted. There is little agreement on EFT conventions,
basis choice (the definition of a basis), the meaning of power counting, the degree of constraint on Wilson coefficients,
the possibility of doing model independent EFT studies or not,
and other issues. The great interest in developing EFT extensions of the SM, in response to the
discovery of a Higgs-like scalar and no other resonances in the LHC data set,
has resulted in a significant disarray in the rapidly advancing literature. This makes
this fascinating and important area of research incomprehensible when comparing various parts of the literature, and
effectively unapproachable for
the next generation of students. We hope this review will have a positive
and clarifying impact by removing some of these barriers and resolving some of these issues. We
hope it will encourage students
to work in this area.\footnote{To this end, in the Appendix we have also included some results
on LO corrections to a number of processes
with a unified and common notation, in the Warsaw basis, for the SMEFT.}
There is an enormous amount of important work to do to develop and use the SMEFT and the HEFT to gain the
most out of the unprecedented LHC data set, that is soon to arrive.
\acknowledgments
MT and IB thank the Villum Foundation, NBIA and the Discovery Centre at Copenhagen University for support.
MT thanks the many experts in EFT that he has learned from over the years, and acknowledges
informative and revealing conversations on this subject with Cliff Burgess, Ben Grinstein, Gino Isidori,
Mike Luke, Aneesh Manohar, Giampiero Passarino, Maxim Pospelov and Mark Wise.
MT also thanks past collaborators and members of the Higgs Cross Section Working Group, in particular
Andr\'e Mendes and Michael D\"uhrssen, for discussions and motivation. IB acknowledges instructive conversations with
past collaborators and thanks in particular Bel\'en Gavela and Luca Merlo for many teachings and formative discussions.
We thank Poul Damgaard, Andreas Helset, Yun Jiang, Andr\'e Mendes, Subodh Patil, Duccio Pappadopulo, William Shepherd, Michael Spira,
Anagha Vasudeven and Jordy de Vries for comments on the review.
We welcome comments on the review and encourage feedback and corrections.
\appendix

\section{Cross sections/decay widths of selected processes in the SMEFT}\label{shiftsgalore}
In this Appendix we report the analytic expression of the cross sections for selected EW processes in the SMEFT
including leading order shifts due to $\mathcal{L}_6$.
All the observables are computed at tree-level in the Warsaw basis with the operators defined as in
Ref.~\cite{Grzadkowski:2010es}. We choose the set $\haew, \hat m_Z, \hat G_F,\hat m_h\}$ as
input parameters\footnote{The subscript in $\haew$ will be dropped in the following.} and adopt the notation
and assumptions of Ref.~\cite{Berthier:2015oma}. Many of these results were already reported in the literature in various
bases. A unified presentation with common notational conventions is lacking,
so we have included this summary of known results here, and simultaneously extended the known literature.

We adopt the {\it IR assumption} that
light quark masses are neglected both in the SM predictions and in the SMEFT corrections.
Unless otherwise specified, the SMEFT Lagrangian is assumed to respect an approximate $\rm U(3)^5$ flavour symmetry
as a further {\it IR assumption}, which is only violated through insertions of the Yukawa couplings.
In this scenario, the Wilson coefficients of operators containing chirality-flipping fermion currents
have the\footnote{At lowest order in the linear MFV expansion \cite{Chivukula:1987py,Hall:1990ac,DAmbrosio:2002vsn,Feldmann:2008ja,Kagan:2009bn,Feldmann:2009dc}.} flavour structure
\begin{equation}
\begin{aligned}
C_{\substack{f H\\ rs}}, C_{\substack{f W\\ rs}}, C_{\substack{f B\\ rs}}, C_{\substack{f G\\ rs}}  &\propto \left[Y^\dagger_f\right]_{rs},\\
\\
C_{\substack{H ud \\ rs}}  \propto \left[Y_u \, Y_d^\dag \right]_{rs},\qquad
C_{\substack{ledq\\ rs}} \propto \left[Y^\dag_e Y_d\right]_{rs}, \qquad
C_{\substack{quqd\\ rs}}^{(1),(8)} &\propto \left[Y^\dag_u Y^\dag_d\right]_{rs},
\qquad
C_{\substack{lequ\\ rs}}^{(1),(3)} \propto \left[Y^\dag_e Y^\dag_u\right]_{rs}.
\end{aligned}
\end{equation}
The terms that give very suppressed contributions to interactions involving light fermions proportional to the
light quark masses are neglected here.
Furthermore, the Wilson coefficients have been redefined so as to absorb the factor $\Lambda^{-2}$: we use the dimensionful parameters $C_i' = C_i / \Lambda^2$ with the prime implicitly dropped.
We use the indices $p,\,r,\,s\dots$ for the flavour space and $I,\,J,\,K\dots$ for $\rm SU_L(2)$ consistent with the notational conventions
in Section \ref{SMsection}.

\subsection{Core shifts in the \titlemath{$\{\hat\a,\hat G_F,\hat m_Z\}$}{\{alpha, GF, mZ\}} input scheme}\label{subsec_shifts}
The SM Lagrangian is written in terms of several internal parameters whose values can be determined via the measurement of a few input quantities.
For the EW sector it is customary to adopt the inputs set $\{\hat{\a}, \hat G_F, \hat m_Z,\hat m_h, \cdots\}$, where $\hat\a$ is
the electromagnetic structure constant extracted from Thomson scattering, $\hat G_F$ is the Fermi
constant extracted from muon decay, and $\hat m_Z$, $\hat m_h$ are respectively the measured
masses of the $Z$ and Higgs bosons.\footnote{Arguably a transition to the $\{\hat{m}_W, \hat G_F, \hat m_Z,\hat m_h, \cdots\}$ scheme
is favoured for several reasons as discussed in Ref.~\cite{Brivio:2017bnu}. We do not fight the tide
of historical convention here but note that an $\hat m_W$ input scheme is available to be used in the SMEFTsim package \cite{Brivio:2017btx}.} The numerical values of some of the inputs are reported in Table~\ref{tab.input_values}.
\begin{table}[t]
\begin{tabular}{ccc}
 \hline
Parameter & Input Value & Ref.  \\ \hline
$\hat{\a}$ & $1/137.035999139(31) $ &  \cite{Olive:2016xmw,Mohr:2012tt} \\
$\hat{G}_F$ & $1.1663787(6) \times 10^{-5} $ GeV$^{-2}$&  \cite{Mohr:2012tt,Olive:2016xmw} \\
$\hat{m}_Z$ & $91.1876  \pm 0.0021$ GeV& \cite{Z-Pole,Olive:2016xmw,Mohr:2012tt} \\
$\hat{m}_h$ & $125.09 \pm 0.22 $ GeV& \cite{Olive:2016xmw} \\
\hline
\caption{Current best fit values of the core input parameters. }\label{tab.input_values}
\end{tabular}
\end{table}
The other relevant Lagrangian parameters are fixed by the following definitions:
\begin{equation}
 \begin{aligned}
 \hst^2 &= \frac{1}{2}\left[1-\sqrt{1-\frac{4\pi\hat{\a}}{\sqrt2\hat G_F \hat m_Z^2}}\right],\quad
 &
 \hat e &=\sqrt{4\pi\hat{\a}},\\
 \hat g_1 &= \frac{\hat{e}}{\hct},
 &
 \hat g_2 &= \frac{\hat{e}}{\hst},\\
 \hat v_T &= \frac{1}{2^{1/4}\sqrt{\hat G_F}},
 &
 \hat m_W^2 &= \hat m_Z^2\hct^2.
 \end{aligned}
\end{equation}
When applied in the SMEFT, this procedure introduces a mismatch between the quantities determined from the input measurements
and the parameters defined in the canonically normalized Lagrangian consistent with an on-shell EFT construction.
Denoting the former quantities with a hat and the latter with a bar, a generic parameter $\kappa$ receives a shift from its SM value given by
\begin{equation}
 \d \kappa = \bar\kappa - \hat\kappa\,.
\end{equation}
 In the SM limit ($C_i\to 0$) hatted and bar quantities coincide. It is convenient to define the quantities
\begin{align}
 \d G_F &= \frac{1}{\sqrt{2} \,  \hat{G}_F} \left(\sqrt{2} \, C^{(3)}_{\substack{Hl}} - \frac{C_{\substack{ll}}'}{\sqrt{2}}\right), \\
 \d m_Z^2 &= \frac{1}{2 \, \sqrt{2}} \, \frac{\hat{m}_Z^2}{\hat{G}_F} C_{HD} + \frac{2^{1/4} \sqrt{\pi \hat{\alpha}} \, \hat{m}_Z}{\hat{G}_F^{3/2}} C_{HWB},\\
 \d m_h^2 &=  \frac{\hat m_h^2}{\sqrt2\hat G_F}\left(-\frac{3C_H }{2\lambda}+2 C_{H\square}-\frac{C_{HD}}{2}\right), \\
\d  m_W^2 &= \hat m_W^2 \left(\sqrt2 \d G_F + 2 \frac{\d g_2}{\hat g_2}\right), \\
&=  -\frac{\hsdt \bar v_T^2}{4\hcdt}\left( \frac{\hct}{\hst} C_{HD} +\frac{\hst}{\hct}(4\CHlt-2C_{ll}') +4C_{HWB}\right),
 \end{align}
Using this notation, related results are\footnote{To define the SMEFT in $R^\xi$ gauge the approach used here to define the diagonalization of
the mass eigenstate fields is advantageous, see Refs.~\cite{Hartmann:2015oia,Dedes:2017zog}.}
\begin{align}
 \d v_T^2 =\bar v_T^2 - \hat v_T^2&= \frac{\d G_F}{\hat G_F}\\
 \d g_1 =\bar g_1-\hat g_1&=\frac{\hat g_1}{2\hcdt}\left[\hst^2\left(\sqrt2 \d G_F+\frac{\d m_Z^2}{\hat m_Z^2}\right)+\hct^2\,\hsdt\bar v_T^2 C_{HWB}\right]\nonumber\\
 &=\frac{\hat g_1}{\hcdt}\frac{\bar v_T^2}{4}\left[\hst^2\left(C_{HD}+4\CHlt-2C_{ll}'\right)+2\hsdt C_{HWB}\right],
\end{align}
\begin{align}
 \d g_2 =\bar g_2 - \hat g_2&=-\frac{\hat g_2}{2\hcdt}\left[
 \hct\left(\sqrt2 \d G_F+\frac{\d m_Z^2}{\hat m_Z^2}\right)+\hst^2\,\hsdt\bar v_T^2 C_{HWB}\right]\nonumber\\
 &=-\frac{\hat g_2}{\hcdt}\frac{\bar v_T^2}{4}\left[\hct^2\left(C_{HD}+4\CHlt-2C_{ll}'\right)+2\hsdt C_{HWB}\right]\\
 \d \st^2 = s_{\bar \theta}^2- \hst^2 &= 2\hct^2\hst^2\left(\frac{\d g_1}{\hat g_1}-\frac{\d g_2}{\hat g_2}\right) +\bar v_T^2  \frac{\hsdt\hcdt}{2} C_{HWB}\nonumber\\
 &=\frac{\hsdt}{8\hcdt\sqrt{2}\hat G_F}\left[\hsdt\left(C_{HD}+4\CHlt-2C_{ll}'\right)+4 C_{HWB}\right]
\end{align}

 \subsubsection*{Couplings of the gauge bosons to fermions}
The photon couplings do not receive corrections in this input parameter set, due to the fact that $\a$ is an input and, at the same time, the $U_{\rm em}(1)$ gauge symmetry is preserved in the SM.
The relevant Lagrangian term is
\begin{equation}
 \mathcal{L}_{A,\rm{eff}} = -2\sqrt{\pi\hat \a}\,Q_\psi \, J^{\psi, em}_\mu\, A^\mu\,,
\end{equation}
where $J^{\psi,em}_\nu$ is the electromagnetic current with the fermion $\psi=\{\ell,u,d\}$.
On the other hand, the $Z$ and $W$ couplings to fermions are modified. Using the notation of Refs~\cite{Berthier:2015oma, Berthier:2015gja}, the former can be parameterized as
\begin{equation}\label{def_Zcurrent}
 \mathcal{L}_{Z,{\rm eff}}  =  \hat g_{Z}  \,   \left(J_\mu^{Z \ell} Z^\mu + J_\mu^{Z \nu} Z^\mu + J_\mu^{Z u} Z^\mu +  J_\mu^{Z d} Z^\mu \right),
\end{equation}
where $\hat g_{Z} = -\hat g_2/\hct= - \, 2 \, 2^{1/4} \, \sqrt{\hat{G}_F} \, \hat{m}_Z$, $(J_\mu^{Z \psi})^{pr} = \bar{\psi}_p \, \gamma_\mu \left[(\bar{g}^{\psi}_V)_{pr}- (\bar{g}^{\psi}_A)_{pr} \, \gamma_5 \right] \psi_r$ for $\psi = \{u,d,\ell,\nu \}$.
The couplings' normalization is such that
$$\gsm{\psi}{V} = T_3/2 - Q_\psi  s_{\hat{\theta}}^2,\qquad \gsm{\psi}{A} = T_3/2$$ with $T_3 = \pm 1/2$
and $Q_\psi = \{-1,2/3,-1/3 \}$ for $\psi = \{\ell,u,d\}$.
The couplings deviate from the SM expressions as
\begin{equation}
\delta (g^{\psi}_{V,A})_{pr} = (\bar{g}^{\psi}_{V,A})_{pr} - (\gsm{\psi}{V,A})_{pr},
\end{equation}
with $F[C_1,C_2,C_3 + \cdots]_{pr} = (C_{\substack{1 \\ pr}} + C_{\substack{2 \\ pr}} + C_{\substack{3 \\ pr}} + \cdots)/(4 \sqrt{2} \hat{G}_F)$
\begin{align}
\delta (g^{\ell}_V)_{pr}&=\delta \bar{g}_Z \, (\gsm{\ell}{V})_{pr} - F[C_{\substack{H e}},C^{(1)}_{\substack{H \ell}},C^{(3)}_{\substack{H \ell}}]_{pr} + \delta s_\theta^2\, \delta_{pr}, \\
%%%
\delta(g^{\ell}_A)_{pr}&=\delta \bar{g}_Z \, (\gsm{\ell}{A})_{pr} +
F[C_{\substack{H e}},-C^{(1)}_{\substack{H \ell}},-C^{(3)}_{\substack{H \ell}}]_{pr},  \\
%%%
\delta(g^{\nu}_{A/V})_{pr}&=\delta \bar{g}_Z \, (\gsm{\nu}{A/V})_{pr} -
F[C^{(1)}_{\substack{H \ell}},-C^{(3)}_{\substack{H \ell}}]_{pr},
\\
% %%%
\delta (g^{u}_V)_{pr}&=\delta \bar{g}_Z \, (\gsm{u}{V})_{pr}  +
F[-C_{\substack{H q}},C^{(3)}_{\substack{H q}},-C_{\substack{H u}}]_{pr} - \frac{2}{3} \delta \hst^2 \, \delta_{pr},\\
%%%
\delta(g^{u}_A)_{pr}&=\delta \bar{g}_Z \, (\gsm{u}{A})_{pr}
-F[C_{\substack{H q}}^{(1)},-C^{(3)}_{\substack{H q}},-C_{\substack{H u}}]_{pr},  \\
%%%
\delta (g^{d}_V)_{pr}&=\delta \bar{g}_Z \,(\gsm{d}{V})_{pr}
-F[C_{\substack{H q}}^{(1)},C^{(3)}_{\substack{H q}},C_{\substack{H d}}]_{pr} +  \frac{1}{3} \delta \hst^2 \, \delta_{pr}, \\
%%%
\delta(g^{d}_A)_{pr}&=\delta \bar{g}_Z \,(\gsm{d}{A})_{pr}
+F[-C_{\substack{H q}}^{(1)},-C^{(3)}_{\substack{H q}},C_{\substack{H d}}]_{pr},
\end{align}
and
\begin{equation}
 \delta \bar{g}_Z =- \frac{\delta G_F}{\sqrt{2}} - \frac{\delta m_Z^2}{2\hat{m}_Z^2} + \frac{s_{\hat{\theta}} \, c_{\hat{\theta}}}{\sqrt{2} \hat{G}_F} \, C_{HWB}.
\end{equation}
For the charged currents
\begin{equation}\label{def_Wcurrent}
\mathcal{L}_{W,eff} = - \frac{ \sqrt{2 \,\pi \, \hat{\a}}}{\hst} \left[(J_{\mu}^{W_\pm, \ell})_{pr} W_\pm^\mu + (J_{\mu}^{W_\pm, q})_{pr} W_\pm^\mu\right],
\end{equation}
with
\begin{align}
(J_\mu^{W_{+}, \ell})_{pr} &=   \, \bar{\nu}_p \, \gamma^\mu \,\left(\bar{g}^{W_{+},\ell}_V - \bar{g}^{W_{+},\ell}_A \gamma_5 \right)\, \ell_r,\\
(J_\mu^{W_{-}, \ell})_{pr} &= \, \bar{\ell}_p \, \gamma^\mu \, \left(\bar{g}^{W_{-},\ell}_V - \bar{g}^{W_{-},\ell}_A \gamma_5 \, \right) \nu_r,
\end{align}
and analogously for quarks.
In the SM
\begin{equation}
 (\bar g_{V}^{W_\pm,\ell})^{SM}_{pr}= (\bar g_A^{W_\pm,\ell})^{SM}_{pr} = \frac{(U_{PMNS}^\dagger)_{pr}}{2},
 \qquad\qquad
  (\bar g_V^{W_\pm,q})^{SM}_{pr}=(\bar g_A^{W_\pm,q})^{SM}_{pr} = \frac{V_{pr}}{2},
\end{equation}
where $V$ is the CKM matrix.
In the SMEFT $\bar g_{V/A}^{W_\pm,\psi} = (\bar g_{V/A}^{W_\pm,\psi})^{SM} + \d \bar g_{V/A}^{W_\pm,\psi}$ where, for the flavour diagonal component:
\begin{align}
\delta(g^{W_{\pm},\ell}_V)_{rr} = \delta(g^{W_{\pm},\ell}_A)_{rr}  &= \frac{1}{2\sqrt{2} \hat{G}_F} \left(C^{(3)}_{\substack{H \ell \\ rr}} + \frac{1}{2} \frac{\hct}{\hst} \, C_{HWB} \right)
- \frac{1}{4} \frac{\delta s_\theta^2}{s^2_{\hat{\theta}}},\\
\delta(g^{W_{\pm},q}_V)_{rr} = \delta(g^{W_{\pm},q}_A)_{rr}  &=  \frac{1}{2\sqrt{2} \hat{G}_F} \left(C^{(3)}_{\substack{H q \\ rr}} + \frac{1}{2} \frac{\hct}{\hst} \, C_{HWB} \right)
- \frac{1}{4} \frac{\delta s_\theta^2}{s^2_{\hat{\theta}}}.\label{delta_gW_quarks}
\end{align}
the off diagonal components are the obvious generalization of this result.
\subsubsection*{Triple Gauge Couplings}\label{TGC}
Going from the SM to the SMEFT, the TGC couplings get redefined by a subset of $\mathcal{L}_6$ operators, so that $\bar{g}_1^V = g_1^V+ \delta g_1^V$, $\bar{\kappa}_V= \kappa_V + \delta \kappa_V$, $\bar{\lambda}_V = \lambda_V+\delta \lambda_V$ with
\begin{align}
\d g_1^{A}  &= 0,  &
\d g_1^Z &=  \frac{1}{ 2 \sqrt{2}\hat{G}_F}\left(\frac{\hst}{\hct}+\frac{\hct}{\hst}  \right) C_{HWB} -
\frac{1}{2}\d\st^2\left(\frac{1}{\hst^2}+\frac{1}{\hct^2}\right), \\
\d \kappa_A &=   \frac{1}{ \sqrt{2}\hat{G}_F}\frac{\hct}{\hst} C_{HWB},
& \d\kappa_Z &=\frac{1}{ 2 \sqrt{2}\hat{G}_F}\left(- \frac{\hst}{\hct}+\frac{\hct}{\hst}  \right)C_{HWB} -  \frac{1}{2}\d\st^2\left(\frac{1}{\hst^2}+\frac{1}{\hct^2}\right),  \\
\d \lambda_{A} &=  6 \hst  \frac{\hat{m}^2_W}{g_{AWW}}C_W,  &
\d\lambda_{Z} &=  6 \hct  \frac{\hat{m}^2_W}{g_{ZWW}}C_W.
\end{align}
Notice that three relations between parameters in this input scheme (at the level of $\mathcal{L}_6$ \cite{Hagiwara:1993ck})
hold in the SMEFT: $\delta \kappa_Z = \delta g_1^Z - t_{\that}^2 \delta \kappa_A$, $\delta \lambda_A = \delta \lambda_Z$ and $\delta g_1^A = 0$.
\subsubsection*{Fermion masses and Yukawa couplings}
If the assumption of massless fermions is relaxed, the measured masses of quarks and leptons can be incorporated in the set of input parameters and they allow to determine the Yukawa couplings through the definition
\begin{equation}\label{Yhat}
 \hat Y_f = 2^{3/4}\hat m_f \sqrt{\hat G_F}\,.
\end{equation}
In the SM the coupling of the Higgs boson to fermions is then $g_{h\bar ff}^{SM} =\hat  Y_f/\sqrt2$.
In the SMEFT it is shifted as $\bar g_{h\bar ff} = g_{h\bar ff}^{SM} + \d g_{h\bar ff}$ where~\cite{Alonso:2013hga}
\begin{equation}
 \d g_{h\bar ff} =\frac{\hat Y_f}{\sqrt2}\left[\bar v_T^2\left(C_{H\square}-\frac{C_{HD}}{4}\right) -\frac{\d G_F}{\sqrt2} \right]-\frac{\bar v_T^2}{\sqrt2}C^*_{fH}\,.
\end{equation}

\subsection{Generic \titlemath{$2\to2$}{2->2} scattering processes via gauge boson exchange}
\subsubsection*{\boldmath Scattering $\ell^{+} \ell^{-} \to \bar{f} f\,,\quad f=\{\ell'\neq \ell,u,c,b,d,s\} $}
The general $s$ channel differential cross section $d\sigma(\ell^{+} \ell^{-} \to f \, \bar{f})/d\ct$, valid on and off resonance scattering, has been computed in the SMEFT in  Ref.~\cite{Berthier:2015oma}. The result includes the contributions from $Z$ and $\gamma$ exchange, the effect of $\psi^4$ operators and the interference of all of these terms, up to leading order in the interference of the $\psi^4$ operators with the SM amplitude.
Initial and final state radiation (including possible $\alpha_s$ corrections to final state fermions) have been neglected, together with fermion masses. Consistently with the other results reported here, a $\rm U(3)^5$ flavour symmetry is assumed, which allows to neglect interference effects with operators of the form $LRRL, LRLR$, that are proportional to SM Yukawas. Finally, the initial $e^+,e^-$ are taken to be unpolarized. The final expression, in Feynman gauge, reads
\footnote{Here we correct a factor of $1/\pi$ in the first line compared to Ref.~\cite{Berthier:2015oma}, which has a typo.}
 \begin{eqnarray}
\label{generaldifferential}
\frac{1}{N_c}\frac{d\sigma}{d \ct} &=&  \frac{\hat{G}_F^2 \hat{m}_Z^4}{\pi}
\, \bar{\chi}(s) \left[\left(|\bar{g}^{\ell}_V|^2 + |\bar{g}^{\ell}_A|^2\right) \, \left(|\bar{g}^{f}_V|^2 + |\bar{g}^{f}_A|^2\right) \left(1+ c_\theta^2 \right)
- 8 \, {\rm Re}\left[\bar{g}^{\ell}_A  \bar{g}^{\ell,\star}_V \right] \, {\rm Re}\left[\bar{g}^{f}_A \bar{g}^{f, \star}_V \right] c_\theta \right],  \nonumber \\
&+& \frac{|\hat{\alpha}|^2 \, |Q_\ell|^2 \, |Q_f|^2 \, \pi}{2 \, s} \left(1+ c_\theta^2 \right) + \frac{\hat{G}_F \hat{m}_Z^2 Q_\ell \, Q_f}{\sqrt{2}} \,
\left[\alpha^\star \frac{\bar{g}^{\ell}_V \, \bar{g}^{f}_V \left(1+ c_\theta^2 \right) + 2 \, c_\theta \, \bar{g}^{\ell}_A \, \bar{g}^{f}_A}{s - \bar{m}_Z^2 + i \,  \bar{w}(s)} + {\rm h.c.} \right],  \nonumber \\
&+& \frac{Q_\ell \, Q_f}{32} \, \left[\alpha^\star \, C_{LL,RR}^{\ell, f} \, (1+ c_\theta)^2  + {\rm h.c.}  \right] +
\frac{Q_\ell \, Q_f}{32} \, \left[\alpha^\star \, C_{LR}^{\ell, f} \, (1- c_\theta)^2  + {\rm h.c.}  \right], \nonumber \\
&+&  \left(\frac{\hat{G}_F \hat{m}_Z^2}{16 \, \sqrt{2} \, \pi} \right)\left[ \left(\frac{s }{s - \bar{m}_Z^2 + i \, \bar{w}(s)} \right)
C_{LL,RR,LR}^{\ell, f,\star} (\bar{g}^{\ell}_V \pm \bar{g}^{\ell}_A)(\bar{g}^{f}_V \pm \bar{g}^{f}_A) \left(1+ c_\theta^2 \right) + {\rm h.c.} \right], \nonumber \\
&+&  \left(\frac{\hat{G}_F \hat{m}_Z^2}{16 \, \sqrt{2} \, \pi} \right) \left[ \left(\frac{s}{s - \bar{m}_Z^2 + i \, \bar{w}(s)}\right)
 C_{LL,RR,LR}^{\ell, f,\star} \, (\bar{g}^{\ell}_A \pm \bar{g}^{\ell}_V)(\bar{g}^{f}_A \pm \bar{g}^{f}_V) \, 2 \, c_\theta + {\rm h.c.}\right].
 \end{eqnarray}
where in the last two lines the couplings combinations with signs (++, $--$, +$-$) are associated to the $LL$, $RR$ and $LR$ operators respectively. We have also defined
\begin{equation}
 \bar{\chi}(s) = \frac{s}{(s - \bar{m}_Z^2)^2 + | \bar{w}(s)|^2}\,,
\end{equation}
where $\bar{w}(s)$ represents the Breit-Wigner distribution \cite{Breit:1936zzb}, that can be either expressed as an $s$ dependent width ($\bar{w}(s) = s \, \bar{\Gamma}_Z /\bar{m}_Z$, which is the approach used at LEP) or alternatively using directly the real part of the complex pole $\bar{w}(s) =\bar{\Gamma}_Z \,  \bar{m}_Z$.
The parameter $\ct$ is the cosine of the angle between the incoming $\ell^-$ and the outgoing $\bar{f}$, and $s = (p_{\ell^+} + p_{\ell^-})^2$. $N_C$ is the dimension of the $\rm SU(3)$ group
of the produced fermion $f$.

Flavour indices on the $\psi^4$ operator Wilson coefficients and the effective gauge couplings  have been suppressed.
Reintroducing them, one has $C^\star \to C^\star_{\ell \, \ell \, f \, f}, C^\star_{\ell \, f \, f \, \ell}, C^\star_{f \, \ell \, \ell \, f}$ for $C^\star_{LL,RR}$.
For the $LR$ operators $C^\star \to C^\star_{\ell \, \ell \, f \, f}$ is as in the previous chirality cases, while the cases $ C^\star_{\ell \, f \, f \, \ell},C^\star_{f \, \ell \, \ell \, f}$ vanish.

Expanding linearly in the Wilson coefficient, the differential cross section is shifted compared to the SM prediction by\footnote{
Here we correct a $1/\pi$ compared to the result in the first three lines that is derived
from Ref.~\cite{Berthier:2015oma}.}
\begin{equation}
 \begin{aligned}
\label{generaldifferentialshift}
&\frac{1}{N_c} \d \left(\frac{d\sigma}{d \ct}\right)=\\
& \frac{\hat{G}_F^2 \hat{m}_Z^4}{\pi}
  \chi(s) \left[2  {\rm Re}\left[(\gsm{\ell}{V})^*   \d g^{\ell}_V +(\gsm{\ell}{A})^*   \d g^{\ell}_A\right]   \left(|\gsm{f}{V}|^2 + |\gsm{f}{A}|^2\right) \left(1+ \ct^2 \right)
+  \left(\ell \leftrightarrow f \right) \right], \\
&- \frac{8 \hat{G}_F^2 \hat{m}_Z^4}{\pi}   \chi(s) \left[
 {\rm Re}\left[\d g^{\ell}_A  (\gsm{\ell}{V})^* + (\gsm{\ell}{A})^*  \d g^{\ell,\star}_V \right]   {\rm Re}\left[\gsm{f}{A} (\gsm{f}{V})^* \right] \ct +  \left(\ell \leftrightarrow f \right) \right], \\
 &+ \frac{\hat{G}_F^2 \hat{m}_Z^4}{\pi}
  \d \chi(s) \bigg[\left(|\gsm{\ell}{V}|^2 + |\gsm{\ell}{A}|^2\right)   \left(|\gsm{f}{V}|^2 + |\gsm{f}{A}|^2\right) \left(1+ \ct^2 \right)\\
&\hspace*{3cm}- 8   {\rm Re}\left[\gsm{\ell}{A}  (\gsm{\ell}{V})^* \right]   {\rm Re}\left[\gsm{f}{A} (\gsm{f}{V})^*\right] \ct \bigg], \\
 &+ \frac{\hat{G}_F \hat{m}_Z^2 Q_\ell   Q_f}{\sqrt{2}}
\left[\alpha^\star  \chi_2(s)   \frac{(\d g^{\ell}_V   \gsm{f}{V}+ \gsm{\ell}{V}   \d g^{f}_V) \left(1+ \ct^2 \right) + 2   \ct
\left(\d g^{\ell}_A   \gsm{f}{A} +  \gsm{\ell}{A}   \d g^{f}_A\right) }{s} + {\rm h.c.} \right], \\
&+ \frac{\hat{G}_F \hat{m}_Z^2 Q_\ell   Q_f}{\sqrt{2}}
\left[\alpha^\star  \d \chi_2(s)   \frac{\gsm{\ell}{V}   \gsm{f}{V} \left(1+ \ct^2 \right) + 2   \ct
\gsm{\ell}{A}   \gsm{f}{A}}{s} + {\rm h.c.} \right], \nn
&+ \frac{Q_\ell   Q_f}{32}   \left[\alpha^\star   C_{LL,RR}^{\ell, f}   (1+ \ct)^2  + {\rm h.c.}  \right] +
\frac{Q_\ell   Q_f}{32}   \left[\alpha^\star   C_{LR}^{\ell, f}   (1- \ct)^2  + {\rm h.c.}  \right], \\
&+  \left(\frac{\hat{G}_F \hat{m}_Z^2}{16   \sqrt{2}   \pi} \right)\left[ \chi_2(s)
C_{LL,RR,LR}^{\ell, f,\star} (\gsm{\ell}{V} \pm \gsm{\ell}{A})(\gsm{f}{V} \pm \gsm{f}{A}) \left(1+ \ct^2 \right) + {\rm h.c.} \right], \\
&+  \left(\frac{\hat{G}_F \hat{m}_Z^2}{16   \sqrt{2}   \pi} \right) \left[  \chi_2(s) C_{LL,RR,LR}^{\ell, f,\star}   (\gsm{\ell}{A} \pm \gsm{\ell}{V})(g^{f,SM}_A \pm \gsm{f}{V})   2   \ct + {\rm h.c.}\right].
 \end{aligned}
\end{equation}
Here
\begin{align}
\chi(s) &= |\Xi(s)|^2/s, \quad     &\d \chi(s) &= \frac{1}{s} \left[\Xi(s) \,  \d \Xi^\star(s) + \d \Xi(s) \, \Xi^\star(s)\right],  \\
\chi_2(s) &=\Xi(s), \quad  & \d \chi_2(s) &= \d \, \Xi(s),
\end{align}
with
\begin{align}
\Xi(s) &= \frac{s}{s -  \hat{m}_Z^2 + i (w(s))_{SM} }, \\
\d \Xi(s) &= \frac{s}{[s -  \hat{m}_Z^2 + i (w(s))_{SM}]^2} \left[ - i \d w(s) \right].
\end{align}
The shift in the Breit-Wigner distribution depends on the specific form assumed for $w(s)$:
\begin{align}
 \bar{w}(s) &= s \frac{\bar{\Gamma}_Z}{\bar{m}_Z}& \to&
 &
\d w(s) &= s  \frac{\d \Gamma_Z}{{\hat{m}_Z}}, \\
\bar{w}(s) &=  \bar{\Gamma}_Z \bar{m}_Z& \to&
&
\d w(s) &=  \hat{m}_Z \,\d \Gamma_Z.
\end{align}
The expression for the $Z$ width correction $\d\Gamma_Z$ is given in Eq.~\ref{d_Gamma_Z}.

\subsubsection*{\boldmath Scattering $\bar f\, f\to \bar f\,f$}
Here we consider the particular case where the initial and final states fermion $f$ are identical. Two kinematic channels, $s$ and $t$, are present in this process. Adopting the same set of approximations and assumptions
as above, the differential cross section for Bhabha scattering ($e^+\ e^- \to e^+\ e^-$) in the SMEFT is given by~\cite{Berthier:2015oma}
\begin{equation}
\begin{aligned}
 \frac{d \sigma}{d c_\theta} &= \frac{2 \, \hat{G}_F^2 \hat{m}_Z^{4}}{\pi s}\left[
( |\bar{g}^{\ell}_V|^2 + |\bar{g}^{\ell}_A|^2 )^2 \left(\frac{u^2 + s^2}{(t-\bar{m}_Z^2)^2} +  \frac{\bar{\chi}(s)}{s} \left(u^2 + t^2 \right) +  2 \, \bar{\chi}(s) \frac{u^2 (1 - \bar{m}_Z^2/s)}{t - \bar{m}_Z^2} \right), \right. \\
&\, \left. - 4 \, {\rm Re}\left[\bar{g}^{\ell *}_V \bar{g}^{\ell}_A\right]^2 \left(\frac{s^2 - u^2}{(t-\bar{m}_Z^2)^2} +  \frac{\bar{\chi}(s)}{s} \left(u^2 - t^2 \right) - 2 \, \bar{\chi}(s) \frac{u^2 (1 - \bar{m}_Z^2/s)}{t - \bar{m}_Z^2} \right) \right], \\
&+ \frac{ \sqrt{2} \, \hat{G}_F \hat{m}_Z^2}{s} \left[\hat{\alpha}^* \frac{(\bar{g}^{\ell}_V)^2 (u^2 + t^2) + (\bar{g}^{\ell}_A)^2 (u^2-t^2)}{s \left(s - \bar{m}_Z^2 + i \bar{w}(s) \right)} + \hat{\alpha}^* \frac{(\bar{g}^{\ell}_V)^2 (u^2 + s^2) + (\bar{g}^{\ell}_A)^2 (u^2-s^2) }{t \left(t - \bar{m}_Z^2 \right)} + h.c. \right], \nn
 &+ \frac{\sqrt{2} \, \hat{G}_F \hat{m}_Z^2 \, u^2}{s}\left[\frac{\hat{\alpha}^*}{t}
 \frac{(\bar{g}^{\ell}_V)^2 + (\bar{g}^{\ell}_A)^2}{\left(s - \bar{m}_Z^2 + i \bar{w}(s) \right)} +
  \frac{\hat{\alpha}}{s} \frac{(\bar{g}^{\ell,\star}_V)^2 + (\bar{g}^{\ell,\star}_A)^2}{\left(t - \bar{m}_Z^2 \right)}  \right],  \\
 &+  \frac{2 \, \pi \, \hat{\alpha}^{2}}{s}\left[\frac{u^2 + s^2}{t^2}+ \frac{u^2+t^2}{s^2} + \frac{2 u^2}{ts}\right]
 + \frac{\hat{\alpha}}{4 s}\left[ 2\left(\frac{u^2}{s} + \frac{u^2}{t} \right)C_{LL,RR}^\star + \left(\frac{t^2}{s}+ \frac{s^2}{t} \right)C^\star_{LR} + h.c \right], \\
 &+ \frac{\hat{G}_F \hat{m}_Z^2}{4 \sqrt{2} \pi s} \left[\frac{4 u^2 \left(\bar{g}^{\ell}_A \pm \bar{g}^{\ell}_V\right)^2 C^\star_{LL,RR}  + 2 t^2\left((\bar{g}^{\ell}_V)^2 - (\bar{g}^{\ell}_A)^2\right)C^\star_{LR}}{s-\bar{m}_Z^2 +i w(s)} + h.c \right],  \\
  &+ \frac{\hat{G}_F \hat{m}_Z^2}{4 \sqrt{2} \pi s} \left[\frac{4 u^2 \left(\bar{g}^{\ell}_A \pm \bar{g}^{\ell}_V\right)^2 C^\star_{LL,RR}  + 2 s^2\left((\bar{g}^{\ell}_V)^2 - (\bar{g}^{\ell}_A)^2\right)C^\star_{LR}}{t-\bar{m}_Z^2} + h.c \right].
\end{aligned}
\end{equation}
The shift from the SM result is~\cite{Berthier:2015gja}
\begin{equation}
 \begin{aligned}
\delta \left(\frac{d \sigma_{e^+e^- \rightarrow e^+ e^-}}{d \, {\rm cos}({\theta})}\right) =&\frac{2 \, \hat{G}_F^2 }{\pi s}
 \left[ \frac{u^2 \, F_{3}^{+} + s^2 \,F_{3}^{-}}{P(t)^2} + \frac{u^2 \, F_{3}^{-}+ t^2 \, F_{3}^{+}}{P(s)^2} + \frac{2\, u^2 \, F_{3}^{+}}{P(s)P(t)}\right],  \\
 &+ \frac{2\sqrt{2} \hat{G}_F \hat{\alpha}}{s}\left[\frac{u^2 F_7^{+} + t^2 F_7^{-}}{s P(s)} + \frac{u^2 F_7^{+} + s^2 F_7^{-} }{t P(t)} +\frac{u^2 F_7^{+}}{t P(s)} + \frac{u^2 F_7^{+}}{s P(t)}\right],  \\
 &+\frac{2 \hat{G}_F}{\pi s} \left[ F_{4} u^2 \left( \frac{1}{P(s)}+ \frac{1}{P(t)} \right) + F_{5} \left(\frac{t^2}{P(s)} + \frac{s^2}{P(t)}\right) \right],  \\
&+ \frac{\hat{\alpha}}{2 s} \left[ 2 \left(\frac{u^2}{s}+\frac{u^2}{t}\right)C_{LL/RR} + \left(\frac{t^2}{s} +\frac{s^2}{t}\right)C_{LR}\right].
 \end{aligned}
\end{equation}
where $P(x) = x/\hat{m}_Z^2 - 1$.
The factors $F_i$ are defined as follows\footnote{Note that in the $U(3)^5$ limit and for $F=\ell$, the operator $\Q_{ll}$ admits two independent flavor contractions and both contribute to $C_{LL}$, so that $C_{LL} = (C_{ll} + C_{ll}')$. This also applies to $\Q_{qq}^{(1)}$, $\Q_{qq}^{(3)}$, contributing to $C_{LL}$ for $F=q$.}:
\begin{equation}
 \begin{aligned}
F_3^{\pm}&= 4 (N_{VA}^{\ell})^3 G_{VA}^{\ell} \delta G_{VAAV}^{\ell} \pm 8(N_{VA}^{\ell})^2 \delta G_{VVAA}^{\ell}, &
F_4 &= \frac{(\gsm{\ell}{V} \pm \gsm{\ell}{A})^2}{\sqrt{2}}  C_{LL/RR}, \\
F_5 &=  \frac{(\gsm{\ell}{V})^2-(\gsm{\ell}{A})^2}{2 \sqrt{2}}C_{LR}, &
F_7^{\pm} &= 2 \gsm{\ell}{V} \delta g_{V}^{\ell} \pm 2 \gsm{\ell}{A} \delta g_A^{\ell},
\end{aligned}
\end{equation}
where
\begin{align}
N_{VA}^{\ell}&= \gsm{\ell}{V} \gsm{\ell}{A},&
G_{VA}^{\ell}&=\frac{(\gsm{\ell}{V})^2 + (\gsm{\ell}{A})^2}{(\gsm{\ell}{V}\gsm{\ell}{A})^2}, &
\delta G_{ijkl}^{\ell} &= \frac{\delta g_i^{\ell}}{ \gsm{\ell}{j}}+\frac{\delta g_k^{\ell}}{ \gsm{\ell}{l}}.
\end{align}

\subsection{Electroweak observables near the Z pole}
Analytic expressions for the electroweak precision observables in the SMEFT can be extracted from the general parameterization of $2\to2$ scattering given in the previous section. This section summarizes the results, using the notation of Ref~\cite{Berthier:2015oma}.
\subsubsection*{Partial and total Z widths}
In the SMEFT, at tree level, one has
\begin{align}
 \bar{\Gamma} \left(Z \to f \bar{f} \right) &= \frac{ \, \sqrt{2} \, \hat{G}_F \hat{m}_Z^3 \, N_c}{3 \pi} \left( |\bar{g}^{f}_V|^2 + |\bar{g}^{f}_A|^2 \right), \\
 \bar{\Gamma} \left(Z \to {\rm Had} \right) &= 2 \, \bar{\Gamma} \left(Z \to u \bar{u} \right)+ 3  \, \bar{\Gamma} \left(Z \to d \bar{d} \right).
 \end{align}
These expressions can be written down separating the SM contribution from the SMEFT correction:
\begin{equation}
 \bar{\Gamma} \left(Z \to f \bar{f} \right)= \Gamma_{Z \to f \bar{f}}^{SM} + \delta \Gamma_{Z \to f \bar{f}}
\end{equation}
for each fermion $f$. The same kind of relation holds for the total width $\bar{\Gamma}_{Z}$.
Specifically, the shifts in each channel read:
\begin{align}
\delta \Gamma_{Z \to \ell \bar{\ell}}=& \frac{\sqrt{2}   \hat{G}_F \hat{m}_Z^3}{6 \pi}   \left[ -\d g^{\ell}_A + \left(-1 + 4 \hst^2 \right) \d g^{\ell}_V \right] + \delta\Gamma_{Z \to \bar{\ell}   \ell, \psi^4}, \\
\d \Gamma_{Z \to \nu \bar{\nu}}=& \frac{\sqrt{2}   \hat{G}_F \hat{m}_Z^3}{6 \pi}   \left[ \d g^{\nu}_A +  \d g^{\nu}_V \right] + \d \Gamma_{Z \to \nu \bar{\nu},\psi^4} , \\
\d \Gamma_{Z \to {\rm Had}}=& 2   \d \Gamma_{Z \bar{u} u} + 3   \d \Gamma_{Z \bar{d} d}, \\
&= \frac{   \sqrt{2}   \hat{G}_F \hat{m}_Z^3}{\pi} \left[ \d g^{u}_A - \frac{1}{3} \left(- 3 + 8 s_{\that}^2 \right) \d g^{u}_V - \frac{3}{2} \d g ^{d}_A + \frac{1}{2}\left(- 3 + 4 s_{\that}^2 \right) \d g^{d}_V  \right]+ \d \Gamma_{Z \to {\rm Had}, \psi^4}, \\
\d \Gamma_{Z} =& 3\d \Gamma_{Z \to \ell \bar{\ell}} + 3 \d \Gamma_{Z \to \nu \bar{\nu}} +\d \Gamma_{Z\to {\rm Had}}, \\
=&  \frac{   \sqrt{2}   \hat{G}_F \hat{m}_Z^3}{  2  \pi} \left[ \d g^{\nu}_A +  \d g^{\nu}_V -\d g^{\ell}_A + \left(-1 + 4 \hst^2 \right) \d g^{\ell}_V \right. \nonumber \\
& \left. + 2 \d g^{u}_A - \frac{2}{3} \left(- 3 + 8 s_{\that}^2 \right) \d g^{u}_V - 3 \d g ^{d}_A + \left(- 3 + 4 s_{\that}^2 \right) \d g^{d}_V \right] \nonumber \\
&  +\d \Gamma_{Z \to {\rm Had}, \psi^4} + 3 \d \Gamma_{Z \to \ell \bar{\ell}, \psi^4} + 3 \d \Gamma_{Z \to \nu \bar{\nu},\psi^4}.
\label{d_Gamma_Z}
\end{align}
The corrections due to four-fermion operators have been generically denoted by $\d\Gamma_{Z\to f \bar f, \psi^4}$ and can be
derived directly from Eq.~\ref{generaldifferential}. Their expressions are not given here as they are suppressed
by $\bar{v}_T \, \Gamma_Z/m_Z^2$ beyond the power counting suppression, see Ref~\cite{Berthier:2015oma} for details.

The ratios of decay rates are defined in the SM as $R^0_{f}=\Gamma_{Z\to {\rm Had}}/\Gamma_{Z \to \bar{f} f}$ where $f$ can be a charged lepton $\ell$ or
a neutrino. These are shifted by  $\bar{R}^0_{f}= R^0_{f} + \d R^0_{f}$ with
\begin{equation}
\d R^0_{f}=\frac{1}{(\Gamma_{Z \to f \bar{f}}^{SM})^2} \bigg[ \d \Gamma_{Z \to {\rm Had}} \Gamma_{Z \to f \bar{f}}^{SM} - \d \Gamma_{Z \to f \bar{f}}  \Gamma_{Z \to {\rm Had}}^{SM} \bigg].
\end{equation}
When $f$ is an identified quark, the ratio  $R^0_{q}$ is defined as the inverse of the lepton case.
\subsubsection*{Forward-backward asymmetries}
The forward backward asymmetry for the scattering $\ell^+\ell^-\to f\bar f$ is defined as
\begin{equation}
A_{FB} = \frac{\sigma_F - \sigma_B}{\sigma_F + \sigma_B},
\end{equation}
where $\sigma_F$ is defined by $\theta \in \left[0,\pi/2\right]$ and $\sigma_B$ by $\theta \in \left[\pi/2, \pi \right]$ with $\theta$ the angle between the incoming $\ell^-$ and the outgoing $\bar{f}$. In the SM:
\begin{equation}
A_{FB}^{0, f}= \frac{3}{4} A_{\ell} A_{f},\quad\text{ with }
\quad A_{\ell}= 2 \frac{g^{\ell}_V g^{\ell}_A}{ (g^{\ell}_V)^2 + (g^{\ell}_A)^2},
\quad A_{f}= 2 \frac{g^{f}_V g^{f}_A}{ (g^{f}_V)^2 + (g^{f}_A)^2}.
\end{equation}
In the SMEFT $\bar{A}_{f}$ can be written as
\begin{equation}
\bar{A}_f = \frac{2 \bar{r}_f}{1 + \bar{r}_f^2}\,\,,\qquad \bar{r}_f = \frac{\bar{g}^{f}_V}{\bar{g}^{f}_A}\,
\end{equation}
and it is shifted due to modifications of the $Z$ couplings as
$\bar{A}_f = (A_{f})^{SM}+\d A_{f}$ with
\begin{equation}
\d A_{f} = (A_{f})^{SM} \left( 1 - \frac{2 (r_{f}^2)^{SM}}{1+ (r_{f}^2)^{SM}}\right) \d r_f
\end{equation}
and
\begin{equation}
 r_f = (r_{f})^{SM} \left( 1 + \d r_{f} \right)\,,\qquad  \d r_{f} =  \frac{\d g^{f}_V}{\gsm{f}{V}} - \frac{\d g^{f}_A}{\gsm{f}{A}}\,.
\end{equation}
The corresponding correction to $A_{FB}^{0,f}$ is
\begin{equation}
\d A_{FB}^{0,f} =\frac{3}{4} \left[ \d A_{\ell} \,  (A_{f})^{SM} + (A_{\ell})^{SM} \, \d A_{f}\right].
\end{equation}
Corrections due to four-fermion operators are negligible, as the forward backward asymmetry measurements are direct cross section measurements extracted near the $Z$ pole.

\subsection{Properties of the \titlemath{$W^\pm$}{W} boson}\label{appendix_W_properties}
\subsubsection*{W width}
The partial $W^\pm$ widths in the SMEFT read~\cite{Berthier:2016tkq}
\begin{align}
\bar{\Gamma}_{W \rightarrow \bar f_p f_r} =& \Gamma_{W \rightarrow \bar f_p f_r}^{SM} + \delta \Gamma_{W \rightarrow \bar f_p f_r},\\
\Gamma_{W \rightarrow \bar f_p f_r}^{SM} =& \frac{N_C \, |V^f_{pr}|^2 \sqrt{2} \hat{G}_F \, \hat{m}_W^3}{12 \pi},\\
\delta\Gamma_{W \rightarrow \bar f_p f_r} =& \frac{N_C \, |V^f_{pr}|^2 \sqrt{2} \hat{G}_F \, \hat{m}_W^3}{12 \pi} \left( 4 \delta g_{V/A}^{W, f} + \frac{1}{2} \frac{\delta m_W^2}{\hat{m}_W^2}\right).
\end{align}
As above, $N_C$ depends on the color representation of final state fermions and $V^f$ corresponds to the CKM ($f=q$) or PMNS ($f=\ell$) matrix. In the lepton case, as the neutrino flavour of the decay of a $W^\pm$ boson is not identified, the sum over the neutrino species gives $ \sum_r |V^\ell_{pr}|^2 = 1$. As a result, the total width is $\bar{\Gamma}_W = \Gamma_W^{SM} + \delta \Gamma_W$
with
\begin{align}\label{shift_Wwidth}
\Gamma^{SM}_W &= \frac{3\sqrt{2}\hat G_F \hat{m}_W^3}{4\pi},&
\d\Gamma_W &= \Gamma^{SM}_W\left(\frac{4}{3}\delta g^\ell_W + \frac{8}{3}\delta g^q_W +\frac{\delta m_W^2}{2 \hat{m}_W^2}\right).
\end{align}
Here $\hat{m}_W$ is the standard model value of the $W$-mass at tree level in terms of the input parameters, $\hat{m}_W = c_{\that} \, \hat {m}_Z$.

\subsection{Scattering \titlemath{$\ell^+ \ell^- \to 4 f$}{ll->4f} through \titlemath{$W^\pm$}{W} currents}
The doubly resonant contribution to the $\ell^+ \ell^- \to f_1 \bar f_2 f_3 \bar f_4$ scattering via $W^\pm$ currents was computed in the SMEFT in Ref.~\cite{Berthier:2016tkq}.
There are two relevant diagrams contributing for each fixed final state, as illustrated in Figure~\ref{diagrams_WWprod}.
The total amplitude can be written as $\mathcal{A} = \mathcal{A}_V + \mathcal{A}_\nu$ with
\begin{align}
 \mathcal{A}_V &= \mathcal{A}_{\ell\ell\to WW,V}^{\lambda_{12}\lambda_{23}\lambda_+\lambda_-} D^W(s_{12})D^W(s_{23}) \mathcal{A}_{W^+\to f_1\bar f_2}^{\lambda_{12}}  \mathcal{A}_{W^-\to f_3\bar f_4}^{\lambda_{34}}\,,\\
 \mathcal{A}_\nu &= \mathcal{A}_{\ell\ell\to WW,\nu}^{\lambda_{12}\lambda_{23}\lambda_+\lambda_-} D^W(s_{12})D^W(s_{23}) \mathcal{A}_{W^+\to f_1\bar f_2}^{\lambda_{12}}  \mathcal{A}_{W^-\to f_3\bar f_4}^{\lambda_{34}}\,.
\end{align}
Here $\lambda_\pm$ is the helicity of the initial $\ell^\pm$ and  $\lambda_{12}, \, \lambda_{34} = \{0,+,-,L\}$ are the helicities of the $W^+$ and $W^-$ boson respectively (see Refs~\cite{Hagiwara:1986vm,Hagiwara:1985yu,Berthier:2016tkq} for further details on this spinor helicity formalism). The contribution of the longitudinal helicity ($L$) vanishes in the limit of massless fermions and therefore it can be neglected. Each amplitude has been decomposed as the product of three helicity-dependent sub-amplitudes (for $WW$ production via $V=\{Z,\gamma\}$ or $\nu$ exchange and for the decay of each $W$) and of the $W^\pm$ propagators, parameterized as:
\begin{equation}
 D^W(s_{ij}) = \frac{1}{s_{ij}-\bar m_W^2+i\bar \Gamma_W \bar m_W+i\epsilon}\,,
\end{equation}
with $s_{ij} = s_{12}$ for the $W^+$ and $s_{ij}=s_{34}$ for the $W^-$. In the SMEFT and with the $\{\hat\a,\hat G_F, \hat m_Z\}$ input scheme both the $W$ pole mass and width are shifted compared to the SM prediction.
The propagators the need to be expanded up to linear order in the Wilson coefficients as~\cite{Berthier:2016tkq}
\begin{align}
 D^W(s_{ij}) &= \frac{1}{s_{ij}-\hat m_W^2+i\hat \Gamma_W \hat m_W+i\epsilon}\left[1+\d D^W(s_{ij})\right]\,,\\
 \d D^W(s_{ij}) &= \frac{1}{s_{ij}-\hat m_W^2+i\hat \Gamma_W \hat m_W}\left[\left(1-\frac{i\hat \Gamma_W}{2\hat m_W}\right)\,\d m_W^2-i\hat m_W\d\Gamma_W\right]\,.\label{def_Ds}
\end{align}

The expressions of $\mathcal{A}_{\ell\ell\to WW,V}^{\lambda_{12}\lambda_{23}\lambda_+\lambda_-}$ , $\mathcal{A}_{\ell\ell\to WW,\nu}^{\lambda_{12}\lambda_{23}\lambda_+\lambda_-}$ in the SMEFT for each $\lambda_{ij}$ assignment are listed in Tables~\ref{tab.Wprod_Vpm_amp},~\ref{tab.Wprod_Vmp_amp} and ~\ref{tab.Wprod_nu_amp}. Those for $\mathcal{A}_{W^+\to f_1\bar f_2}^{\lambda_{12}}$, $\mathcal{A}_{W^-\to f_3\bar f_4}^{\lambda_{34}}$ are instead in Table~\ref{tab.Wdecay_amp}.
The tables use the notation
\begin{equation}
 \begin{aligned}
& & D^V(s) &= \frac{1}{s-\hat{m}_V^2+i \hat\Gamma_V \hat m_V+i\epsilon}, \\[2mm]
F_1^Z &= -\hat{g}_{Z,eff} \, g_{ZWW}\, \bar{g}_L^e,\qquad&
F_2^Z &= -\hat{g}_Z\, g_{ZWW} \,\bar{g}_R^e, &
F_1^{\gamma} = \bar{F}_2^{\gamma}&=\sqrt{4\pi \hat{\alpha}}\, g_{AWW},&
 \end{aligned}
\end{equation}
for the $V$ couplings and propagators, and $\lambda^{1/2}(x,y,z)$ is the square root of the K\"all\'en function
\begin{equation}\label{def_lambda_function}
 \lambda(x,y,z) = \left[x^2+y^2+z^2-2 xy-2xz -2 yz\right]\,.
\end{equation}
The quantities $\bar g_1^V, \bar\kappa_V$, $\bar\lambda_V$ are the triple gauge couplings in the parameterization of Eq.~\ref{def_tgc}.
The phase space is parameterized as follows: $\theta$ is the angle between the momenta of the incoming $e^-$ and the outgoing $W^+$ in the center-of-mass frame; $\tilde\phi_{12(34)}$ and $\tilde\theta_{12(34)}$ are the azimuthal and polar angle of the momentum of the final state fermion $f_{1(3)}$ in the rest frame of the $W^{+(-)}$ boson.

The total spin averaged cross section is
\begin{align}
  \bar{\sigma}(s) &=  \int \frac{\sum |\mathcal{A}|^2}{8 s}  \frac{d s_{12} d s_{34}}{\left(2\pi\right)^2}\left[ \frac{\bar{\beta}_{12}}{8 \pi} \frac{d \cos \tilde{\theta}_{12}}{2}\frac{d \tilde{\phi}_{12}}{2 \pi}\right]\left[ \frac{\bar{\beta}_{34}}{8 \pi} \frac{d \cos \tilde{\theta}_{34}}{2}\frac{d \tilde{\phi}_{34}}{2 \pi}\right] \left[ \frac{\bar{\beta}}{8 \pi} \frac{d \cos {\theta}}{2}\frac{d {\phi}}{2 \pi}\right], \label{eq:cross_section_expression}
\end{align}
where, for $\ell=e$
\begin{align}
 \sum |\mathcal{A}|^2 &= | D^W(s_{12})  D^W(s_{34})|^2 \sum_{\lambda_{12},\lambda_{12}'}  \sum_{\lambda_{34},\lambda_{34}'} \left( \mathcal{A}^{\lambda_{12}}_{W^+\to f_1\bar f_2} \right)\left( \mathcal{A}^{\lambda_{12}'}_{W^+\to f_1 \bar f_2}\right)^*  \left( \mathcal{A}^{\lambda_{34}}_{W^-\to f_3 \bar f_4}\right)  \left(\mathcal{A}^{\lambda_{34}'}_{W^-\to f_3 \bar f_4}\right)^* \nonumber \\
 &\times  \sum \limits_{\lambda_{+}}  \sum \limits_{\lambda_{-}} \left( \mathcal{A}_{ee \rightarrow WW}^{\lambda_{12}\lambda_{34},\lambda_+,\lambda_-}\right)  \left( \mathcal{A}_{ee \rightarrow WW}^{\lambda_{12}' \lambda_{34}',\lambda_+,\lambda_-}\right)^*,
\end{align}
and the $WW$ production amplitudes contain both the $V$ and $\nu$ exchange contributions.

The $\beta$-factors are
\begin{equation}
 \bar\beta =\sqrt{1-\frac{2(s_{12}+s_{34})}{s}+\frac{(s_{12}-s_{34})^2}{s^2}},\qquad \bar\beta_{ij}=1
\end{equation}
and the integration over the parameters can be performed numerically in the regions
\begin{equation}
\begin{aligned}
 \tilde \phi_{ij} &\in [0,2\pi], \qquad&
 s_{34}&\in[0,(\sqrt s-\sqrt s_{12})^2],\\
 \cos\theta, \cos\tilde\theta_{ij} &\in [-1,1],&
 s_{12} &\in[0,s].
\end{aligned}
\end{equation}

\begin{table}[t] % W prod V +-
\centering
\tabcolsep 8pt
\begin{tabular}{|c|c|c|}
\hline
$\lambda_{12}$ &$\lambda_{34}$& $\mathcal{A}_{e^+ e^- \rightarrow W^+ W^-,V}^{\lambda_{12}\lambda_{34}+-}$   \\
\hline \hline
$0$&$0$&$- F_1^V\left(\bar{g}_1^V\left(s_{12}+s_{34}\right)+\bar{\kappa}_V s\right)\lambda^{1/2}\sin \theta D^{V}\left(s\right) \, /\, (2\sqrt{s_{12}}\sqrt{s_{34}})$  \\

$+$&$+$&$\quad F_1^V \left(2 \bar{g}_1^V \bar{m}_W^2 +\bar{\lambda}_V s\right)\lambda^{1/2}\sin\theta D^{V}\left(s\right)\,/\,(2\bar{m}_W^2)$\\
$-$&$-$&$\quad F_1^V \left(2 \bar{g}_1^V \bar{m}_W^2 +\bar{\lambda}_V s\right)\lambda^{1/2}\sin\theta D^{V}\left(s\right) \,/\,(2\bar{m}_W^2)$\\
$0$&$-$&$-\sqrt{s}\lambda^{1/2}\left(F_1^V\cos \theta - F_1^V\right) \left(\bar{g}_1^V \bar{m}_W^2 +\bar{\kappa}_V \bar{m}_W^2 +\bar{\lambda}_V s_{12}\right)D^{V}\left(s\right)\,/\,(2\sqrt{2}\sqrt{s_{12}}\bar{m}_W^2)$ \\
$0$&$+$&$\quad \sqrt{s}\lambda^{1/2}\left(F_1^V\cos \theta + F_1^V  \right) \left(\bar{g}_1^V \bar{m}_W^2 +\bar{\kappa}_V \bar{m}_W^2 +\bar{\lambda}_V s_{12}\right)D^{V}\left(s\right)\,/\,(2\sqrt{2}\sqrt{s_{12}}\bar{m}_W^2)$  \\
$+$&$0$&$\quad \sqrt{s}\lambda^{1/2}\left(F_1^V\cos \theta - F_1^V \right) \left(\bar{g}_1^V \bar{m}_W^2 +\bar{\kappa}_V \bar{m}_W^2 +\bar{\lambda}_V s_{34}\right)D^{V}\left(s\right)\,/\,(2\sqrt{2}\sqrt{s_{34}}\bar{m}_W^2)$ \\
$-$&$0$&$-\sqrt{s}\lambda^{1/2}\left(F_1^V\cos \theta + F_1^V  \right) \left(\bar{g}_1^V \bar{m}_W^2 +\bar{\kappa}_V \bar{m}_W^2 +\bar{\lambda}_V s_{34}\right)D^{V}\left(s\right)\,/\,(2\sqrt{2}\sqrt{s_{34}}\bar{m}_W^2)$  \\
$+$&$-$&$0$\\
$-$&$+$&$0$\\
\hline
\end{tabular}
\caption{ \label{tab.Wprod_Vpm_amp}The W-production matrix elements $\mathcal{A}_{e^+ e^- \rightarrow W^+ W^-,V}^{\lambda_{12}\lambda_{34}+-}$
for $\lambda_{12},\lambda_{34}=\{0,+,-\}$, $V=\{Z,\gamma\}$ and $\lambda^{1/2} = \lambda^{1/2}\left(s,s_{12},s_{34}\right)$. }
\end{table}
~
\begin{table}[t] % W prod V -+
\centering
\tabcolsep 8pt
\begin{tabular}{|c|c|c|}
\hline
$\lambda_{12}$ &$\lambda_{34}$& $\mathcal{A}_{e^+ e^- \rightarrow W^+ W^-,V}^{\lambda_{12}\lambda_{34}-+}$   \\
\hline \hline
$0$&$0$&$-  F_2^V\left(\bar{g}_1^V\left(s_{12}+s_{34}\right)+\bar{\kappa}_V s\right)\lambda^{1/2}\sin \theta D^{V}\left(s\right)\,/\,(2\sqrt{s_{12}}\sqrt{s_{34}})$  \\
$+$&$+$&$\quad F_2^V \left(2 \bar{g}_1^V \bar{m}_W^2 +\bar{\lambda}_V s\right)\lambda^{1/2}\sin\theta D^{V}\left(s\right)\,/\,(2\bar{m}_W^2)$\\
$-$&$-$&$\quad F_2^V \left(2 \bar{g}_1^V \bar{m}_W^2 +\bar{\lambda}_V s\right)\lambda^{1/2}\sin\theta D^{V}\left(s\right)\,/\,(2\bar{m}_W^2)$\\
$0$&$-$&$-\sqrt{s}\lambda^{1/2}\left(F_2^V\cos \theta  +F_2^V \right) \left(\bar{g}_1^V \bar{m}_W^2 +\bar{\kappa}_V \bar{m}_W^2 +\bar{\lambda}_V s_{12}\right)D^{V}\left(s\right)\,/\,(2\sqrt{2}\sqrt{s_{12}}\bar{m}_W^2)$ \\
$0$&$+$&$\quad\sqrt{s}\lambda^{1/2}\left(F_2^V\cos \theta  -F_2^V \right) \left(\bar{g}_1^V \bar{m}_W^2 +\bar{\kappa}_V \bar{m}_W^2 +\bar{\lambda}_V s_{12}\right)D^{V}\left(s\right)\,/\,(2\sqrt{2}\sqrt{s_{12}}\bar{m}_W^2)$  \\
$+$&$0$&$\quad\sqrt{s}\lambda^{1/2}\left(F_2^V\cos \theta+F_2^V \right) \left(\bar{g}_1^V \bar{m}_W^2 +\bar{\kappa}_V \bar{m}_W^2 +\bar{\lambda}_V s_{34}\right)D^{V}\left(s\right)\,/\,(2\sqrt{2}\sqrt{s_{34}}\bar{m}_W^2)$ \\
$-$&$0$&$-\sqrt{s}\lambda^{1/2}\left(F_2^V\cos \theta  -F_2^V \right) \left(\bar{g}_1^V \bar{m}_W^2 +\bar{\kappa}_V \bar{m}_W^2 +\bar{\lambda}_V s_{34}\right)D^{V}\left(s\right)\,/\,(2\sqrt{2}\sqrt{s_{34}}\bar{m}_W^2)$  \\
$+$&$-$&$0$\\
$-$&$+$&$0$\\
\hline
\end{tabular}
\caption{ \label{tab.Wprod_Vmp_amp}The W-production matrix elements
$\mathcal{A}_{e^+ e^- \rightarrow W^+ W^-,V}^{\lambda_{12}\lambda_{34}-+}$  for
$\lambda_{12},\lambda_{34}=\{0,+,-\}$, $V=\{Z,\gamma\}$ and $\lambda^{1/2} = \lambda^{1/2}\left(s,s_{12},s_{34}\right)$. }
\end{table}

\begin{table}[t] % Wprod nu
\tiny
\centering
\tabcolsep 8pt
\begin{tabular}{|c|c|c|}
\hline
$\lambda_{12}$ &$\lambda_{34}$& $\left(\mathcal{A}_{e^+ e^- \rightarrow W^+ W^-,\nu}^{\lambda_{12}\lambda_{34}+-}\right)/\left(2\pi\hat{\alpha} \left(\bar{g}_V^{W_\pm,\ell}\right)^2\right)$   \\
\hline \hline
$0$&$0$&$\frac{2 \sin\theta}{s_{\that}^2  \sqrt{s_{12}}\sqrt{s_{34}} \lambda^{1/2} }\left(\left(s^2-\left(s_{12}-s_{34}\right)^2\right) - \dfrac{8 s s_{12} s_{34}}{s - s_{12} - s_{34} + \lambda^{1/2} \cos \theta}\right)$  \\
$+$&$+$&$\quad- \frac{4 \sin\theta}{s_{\that}^2  \lambda^{1/2}}\left(s  + \dfrac{-s\left(s_{12}+ s_{34}\right)-\left(s_{12}-s_{34}\right)\left(-s_{12}+s_{34} + \lambda^{1/2} \right)}{s - s_{12} - s_{34} + \lambda^{1/2} \cos \theta}\right)$ \\
$-$&$-$&$- \frac{4 \sin\theta}{s_{\that}^2  \lambda^{1/2}}\left(s  + \dfrac{-s\left(s_{12}+ s_{34}\right)+\left(s_{12}-s_{34}\right)\left(s_{12}-s_{34} + \lambda^{1/2} \right)}{s - s_{12} - s_{34} + \lambda^{1/2} \cos \theta}\right)$ \\
$0$&$-$&$-\frac{4 \left(1-\cos \theta\right) \sqrt{s}}{s_{\that}^2 \sqrt{2}\sqrt{s_{12}}\lambda^{1/2}}\left(\left(s+s_{12}-s_{34}\right) - \dfrac{2s_{12} \left(s- s_{12}+ s_{34} - \lambda^{1/2} \right)}{s - s_{12} - s_{34} + \lambda^{1/2} \cos \theta}\right)$\\
$0$&$+$&$-\frac{4 \left(1+\cos \theta\right)\sqrt{s}}{s_{\that}^2 \sqrt{2}\sqrt{s_{12}}\lambda^{1/2}}\left(\left(s+s_{12}-s_{34}\right) - \dfrac{2s_{12} \left(s- s_{12}+ s_{34} + \lambda^{1/2} \right)}{s - s_{12} - s_{34} + \lambda^{1/2} \cos \theta}\right)$\\
$+$&$0$&$\quad\frac{4 \left(1-\cos \theta\right)\sqrt{s}}{s_{\that}^2 \sqrt{2}\sqrt{s_{34}}\lambda^{1/2}}\left(\left(s-s_{12}+s_{34}\right) - \dfrac{2s_{34} \left(s+ s_{12}- s_{34} - \lambda^{1/2} \right)}{s - s_{12} - s_{34} + \lambda^{1/2} \cos \theta}\right)$\\
$-$&$0$&$\quad\frac{4  \left(1+\cos \theta\right)\sqrt{s}}{s_{\that}^2 \sqrt{2}\sqrt{s_{34}}\lambda^{1/2}}\left(\left(s-s_{12}+s_{34}\right) - \dfrac{2s_{34} \left(s+ s_{12}- s_{34} + \lambda^{1/2} \right)}{s - s_{12} - s_{34} + \lambda^{1/2} \cos \theta}\right)$\\
$+$&$-$&$- \frac{4}{s_{\that}^2}\dfrac{s\sin\theta\left(1-\cos\theta\right)}{s - s_{12} - s_{34} + \lambda^{1/2} \cos \theta}$\\
$-$&$+$&$\quad\frac{4}{s_{\that}^2}\dfrac{s\sin\theta\left(1+\cos\theta\right)}{s - s_{12} - s_{34} + \lambda^{1/2} \cos \theta}$\\
\hline
\end{tabular}
\caption{ \label{tab.Wprod_nu_amp} The $W^\pm$ pair production matrix elements  $\mathcal{A}_{e^+ e^- \rightarrow W^+ W^-,\nu}^{\lambda_{12}\lambda_{34}+-}$ for helicities $\lambda_{12},\lambda_{34}=\{0,+,-\}$
and we have used the notation $\lambda^{1/2} = \lambda^{1/2}\left(s,s_{12},s_{34}\right)$.}
\end{table}

\begin{table}[t] % W decay
\renewcommand{\arraystretch}{1.5}
\begin{tabular}{|c|c|}
\hline
$\lambda_{12}$ & $\mathcal{M}_{W^{+} \rightarrow f_1 \bar{f}_2}^{\lambda_{12}}/C \sqrt{2\pi \hat{\alpha}}$   \\
\hline \hline
$0$&$ \frac{- 2 \bar{g}_V^{W_+,f_1}}{s_{\that}}\sqrt{s_{12}}\sin \tilde{\theta}_{12}$ \\
$+$&$ \frac{\bar{g}_V^{W_+,f_1}}{s_{\that}}\sqrt{s_{12}}\sqrt{2}\left(1-\cos \tilde{\theta}_{12}\right) \, e^{i \tilde{\phi}_{12}}$\\
$-$&$ \frac{\bar{g}_V^{W_+,f_1}}{s_{\that}}\sqrt{s_{12}}\sqrt{2}\left(1+\cos \tilde{\theta}_{12}\right) \, e^{-i \tilde{\phi}_{12}}$\\
\hline
\end{tabular}
~~
\begin{tabular}{|c|c|}
\hline
$\lambda_{34}$ & $\mathcal{M}_{W^{-} \rightarrow f_3 \bar{f}_4}^{\lambda_{34}}/C'\sqrt{2\pi \hat{\alpha}}$   \\
\hline \hline
$0$&$ \frac{2 \, \bar{g}_V^{W_-,f_3}}{s_{\that}}\sqrt{s_{34}}\sin \tilde{\theta}_{34}$ \\
$+$&$ \frac{- \bar{g}_V^{W_-,f_3}}{s_{\that}}\sqrt{s_{34}}\sqrt{2}\left(1-\cos \tilde{\theta}_{34}\right) \, e^{-i \tilde{\phi}_{34}}$\\
$-$&$ \frac{-\bar{g}_V^{W_-,f_3}}{s_{\that}}\sqrt{s_{34}}\sqrt{2}\left(1+\cos \tilde{\theta}_{34}\right) \, e^{+i \tilde{\phi}_{34}} $\\
\hline
\end{tabular}
\caption{Decay amplitudes $\mathcal{A}_{W^\pm\to f_i\bar f_j}^{\lambda_{ij}}$. $C,C'=\{1,\sqrt3\}$ for final state leptons and quarks respectively.}\label{tab.Wdecay_amp}
\end{table}

\subsubsection*{Universality in $\beta$ decays}
It is possible to place bounds on combinations of four fermion operators
and $W^\pm$ vertex corrections by comparing the extraction of $G_F$ from $\mu^- \to e^- + \bar{\nu}_e + \nu_\mu$ decays to the value determined in semileptonic $\beta$ decays~\cite{Cirigliano:2009wk}.
Assuming $\rm U(3)^5$ universality in the SMEFT, this represents a constraint on the unitarity of the CKM matrix and it translates into a bound on the following  combination of operators
\begin{equation}
\d |V_{CKM}|^2 = \frac{\sqrt{2}}{\hat{G}_F} \left(- C_{l q}^{(3)}  +C_{l l}' + C_{Hq}^{(3)} - C_{H l}^{(3)}\right).
\end{equation}

\subsection{Higgs production and decay}
In the following we list the expressions of the partonic cross sections for the main Higgs production processes
and the partial width for the relevant Higgs decay channels, computed at tree level in the SMEFT.

\subsubsection*{\boldmath $gg\to h$ and $h\to gg$}
The dominant Higgs production mechanism at the LHC is via gluon fusion. In the SM this process is generated at one loop,
as the SM is renormalizable. The leading contribution comes from a top quark loop and it can be computed in the $m_t\to\infty$ approximation\footnote{In this approximation, the gluon-fusion cross section is now known at N3LO in QCD corrections~\cite{Anastasiou:2015ema}.}, where the contact interaction $h G^A_{\mu\nu}G^{A\mu\nu}$ is present. In the SMEFT, this coupling receives a contribution from the operator $\Q_G$. In addition, the operator $\Q_{\tilde G}$ introduces a CP violating Lorentz structure that does not interfere with the SM amplitude.

The leading SM contribution (at tree-level in the EFT obtained integrating out the top quark) gives the
partonic cross section~\cite{Wilczek:1977zn,Dawson:1990zj,Djouadi:1991tka}
\begin{equation}
 \s^{SM}(gg\to h) = \frac{G_F\a_s^2}{32\sqrt2 \pi}\abs{I^g}^2
\end{equation}
where $I^g$ is a Feynman integral that accounts for the top-quark loop contribution and the result is understood in a distribution sense
multiplying a suppressed delta function. Including QCD corrections up to NLO\footnote{The normalization is such that $I^g=1/3$ if QCD corrections are omitted and in the limit $m_h/m_t\to 0$.}
~\cite{Dawson:1990zj,Djouadi:1991tka,Bergstrom:1985hp,Manohar:2006gz}:
\begin{equation}
 I^g = \left(1+\frac{11}{4}\frac{\a_s}{\pi}\right) \int^1_0 dx\int^{1-x}_0 dy\,\frac{1-4 x y}{1- (m_h^2/m_t^2) x y } \, \simeq0.375.
\end{equation}
Compared to the SM, the total cross section in the SMEFT is rescaled by~\cite{Manohar:2006gz,Alonso:2013hga}
\begin{equation}
\frac{ \sigma(gg \to h)}{\sigma^{SM}(gg \to h)}
\simeq \abs{1+ \frac{16 \pi^2 \bar v_T^2}{\bar g_3^2 I^g}C_{HG}}^2+\abs{ \frac{16 \pi^2 \bar v_T^2}{\bar g_3^2 I^g} C_{\tilde{HG}}}^2\,.
\end{equation}
The decay $h\to gg$ (which is not observable at the LHC) proceeds through the same diagrams if initial and final state gluon emission
is neglected and  therefore (for this limited case) it is modified by the same relative correction:
\begin{equation}
 \frac{ \Gamma(h \to gg )}{\Gamma^{SM}(h \to gg)} \simeq \frac{ \sigma(gg \to h)}{\sigma^{SM}(gg \to h)}.
\end{equation}

\subsubsection*{\boldmath $hV$ associated production}
The amplitude for $Vh$ associated production can be decomposed as~\cite{Isidori:2013cga,Buchalla:2013mpa}\footnote{See also Ref.~\cite{Degrande:2016dqg}.}
\begin{equation}
 \mathcal{A}_{hV} = \frac{iN_V g_V^2 \hat m_V}{q^2-\hat m_V^2+i\hat\Gamma_V \hat m_V}\,J^{V,\psi}_\nu\,\epsilon^*_\mu\, T_V^{\mu\nu}
\end{equation}
where $g_V = \{\bar g_2, \bar g_2/\hct\}$ and $N_V=\{1/\sqrt2,1\}$ for $V=\{W,Z\}$ respectively and $\epsilon^*_\mu$ denotes
the polarization vector of the $V$ boson.\footnote{The massive vector boson is not an external state and does not appear
in the Hilbert space of the SMEFT. Here we are considering the approximate experimental reconstruction of the massive vector boson $V$
with kinematics being dominated by the approximate on-shell region of phase space.}
 The fermionic currents are defined as in Eqs.~\ref{def_Zcurrent},~\ref{def_Wcurrent}. The tensor $T_V^{\mu\nu}$ can be decomposed
in the sum of four independent Lorentz structures and form factors~\cite{Isidori:2013cga}
\begin{equation}\label{Tmunu_decomposition}
 T_V^{\mu\nu} = f_1^V(q^2) g^{\mu\nu}+f_2^V(q^2)q^\mu q^\nu+f_3^V(q^2)(p\cdot q \, g^{\mu\nu}-q^\mu p^\nu) + f_4^V(q^2)\epsilon^{\mu\nu\rho\s} p_{\rho}q_\s\,,
\end{equation}
where $q$ denotes the four-momentum of the fermion pair ($q^2=s$) and $p$ is the four-momentum of the outgoing $V$ boson.
In the SM, at tree level and in unitary gauge: $f_1^{V,SM}(q^2)=-m_V^2 f_2^{V,SM}(q^2)\equiv 1$, {$f_{3,4}^{V,SM}(q^2)\equiv 0$}.

\begin{figure}[t]
 \includegraphics[width=\textwidth]{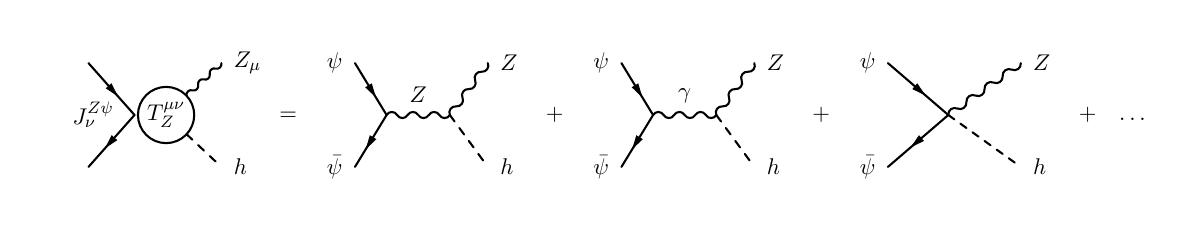}
 \caption{Diagrams for $hZ$ associated production in the SMEFT.  The dots stand for diagrams
 in which the Higgs is coupled to the fermion current, that are suppressed by small
 Yukawa couplings in the $\rm U(3)^5$ flavour symmetric limit and can be neglected. In the case of $hW$ production there are only two diagrams: with a $W$ in $s$-channel and the 4-point interaction.  }\label{diagrams_hZ}
\end{figure}

In the SMEFT, the amplitude receives corrections that can be decomposed as follows
\begin{equation}
 \d\mathcal{A}_{hV} = \frac{iN_V g_V^2 \hat m_V}{q^2-\hat m_V^2 + i\hat\Gamma_V \hat m_V}\,\epsilon^*_\mu\,\left[
 \d J_\nu^{V,\psi} (T_{V}^{\mu\nu})^{SM} + (J_\nu^{V,\psi})^{SM}\d T_V^{\mu\nu}
 \right] + \d\mathcal{A}_{hV},
\end{equation}
where the first term contains corrections to the SM diagram,  while $\d \mathcal{A}_{hV}$ stands for the corrections from extra diagrams that are present only in the SMEFT case.
In particular $ \d J_\nu^{V,\psi}$ contains the shift in the fermion couplings to the $V$ boson that can be inferred from the results in Section~\ref{subsec_shifts} and
$$\d T_V^{\mu\nu}=\d f_1^V(q^2) g^{\mu\nu}+\d f_2^V(q^2)q^\mu q^\nu+\d f_3^V(q^2)(p\cdot q \, g^{\mu\nu}-q^\mu p^\nu) + \d f_4^V(q^2)\epsilon^{\mu\nu\rho\s} p_{\rho}q_\s\,$$
accounts for modifications to the $VVh$ interaction and the $V$ propagator (in the $W$ case). For $hW$ production, $\d \mathcal{A}_{hW}$ corresponds to a contribution from the 4-point interaction $udhW$, while for $hZ$ production $\d \mathcal{A}_{hZ}$ contains both the contribution from the 4-point interaction $\bar\psi\psi hZ$ and that stemming from the diagram with
photon exchange in the $s$-channel (see Figure~\ref{diagrams_hZ}).

In the SMEFT the corrections $\d f_i^V(q^2)$ and the amplitude shifts $\d\mathcal{A}_{hV}$ can be expressed as linear functions in the Wilson coefficients.
For the case $V=Z$, with the Warsaw basis and in the $\rm U(3)^5$ limit \cite{Buchalla:2013mpa}\footnote{Here we correct factors of 2
compared to results quoted in Ref.~\cite{Isidori:2013cga} for only a subset of SMEFT operators.}
\begin{align}
 \d f_1^Z(q^2)=& \d D^Z(q^2)+\bar v_T^2\left(C_{H\square} +\frac{C_{HD}}{4}-\CHlt+\frac{C_{ll}'}{2}\right)\,,\label{f1Z_AP}\\
 \d f_2^Z(q^2)=& -\frac{1}{\hat m_Z^2}\d f_1^Z(q^2)\,,\\
 \d f_3^Z(q^2)=& \frac{2\bar v_T^2}{\hat m_Z^2}\left[\hct^2C_{HW}+\hst^2 C_{HB}+\hst\hct C_{HWB}\right]\,,\\
 \d f_4^Z(q^2)=& -\frac{2\bar v_T^2}{\hat m_Z^2}\left(\hct^2C_{H\tilde W}+\hst^2 C_{H \tilde B}+\hst\hct C_{H\tilde WB}\right)\,,\\
 \d\mathcal{A}_{hZ} =& \,
   \frac{2i\bar e\bar v_T}{q^2}Q_\psi\,J_\nu^{\psi,em}\,\epsilon^*_\mu\bigg[
 \left(\hsdt(C_{HW}-C_{HB})-\hcdt C_{HWB}\right)(p\cdot q \,g^{\mu\nu}-q^\mu p^\nu)
 +\nonumber\\
 &\qquad\qquad\qquad+\left(\hsdt(C_{H\tilde B}-C_{H\tilde W})+\hcdt C_{H\tilde WB}\right)\epsilon^{\mu\nu\rho\s}p_\rho q_\sigma
 \bigg]
 +\nonumber\\
 &+ 2 i\hat m_Z\,\epsilon^*_\mu\,\,\, \bar\psi_s\g^\mu\left(C_{\substack{H\psi\\ sr}}P_R + (\CHqs\pm\CHqt)_{sr}P_L\right)\psi_r.\label{hZprod_4point_diagram}
\end{align}
In~\ref{f1Z_AP}, $\d D^Z(q^2)$ denotes the correction due to the modified $Z$-width in the propagator: choosing the Breit-Wigner distribution to be $\bar\Gamma_Z \bar m_Z$, it is given by
\begin{equation}
 \d D^Z(q^2)= \frac{-i\hat m_Z\d\Gamma_Z}{q^2-\hat m_Z^2+i\hat \Gamma_Z \hat m_Z}\,.
\end{equation}
 The first two lines of \ref{hZprod_4point_diagram} contain the contribution from the $s$-channel photon exchange and $J^{em,\psi}_\nu =  \bar\psi\g_\nu\psi$ is the electromagnetic current. The last line accounts for the 4-point interaction $\bar \psi\psi hZ$ with $\psi$ a quark (the expression for a lepton current is analogous). Here  $P_{L,R}=(1\mp\g_5)/2$ are the left and right chirality projectors and $s,r$ are flavour indices. The left-handed couplings is $\sim (\CHqs-\CHqt)$ for $\psi=u$ and $ (\CHqs+\CHqt)$ for $\psi=d$.

For the charged current case $V=W^+$:
\begin{align}
 \d f_1^W(q^2)=&\, \d D^W(q^2) +\bar v_T^2\left(  C_{H\square} -\left(2+\frac{1}{\hcdt}\right)\frac{C_{HD}}{4}+\frac{C_{ll}'-2\CHlt}{2\hcdt}-\frac{\hsdt }{\hcdt}C_{HWB}\right)\,,\\
 \d f_2^W(q^2)=& -\frac{1}{\hat m_Z^2}\d f_1^W(q^2)\,,\\
 \d f_3^W(q^2)=& \frac{2\bar v_T^2}{\hat m_W^2} C_{HW}\,,\\
 \d f_4^W(q^2)=& -\frac{2\bar v_T^2}{\hat m_W^2} C_{H\tilde W}\,,\\\label{hWprod_4point_diagram}
 \d\mathcal{A}_{hW} =& \,
 -2\sqrt2i \,\hat m_W\,\epsilon^*_\mu\,\,\, \bar u_{L,a}\g^\mu (\CHqt\, V_{CKM})_{sr}\, d_{L,r}.
\end{align}
Here $\d D^W(s)$ is the correction to the $W$ propagator due to the shift in the $W$ pole mass and width defined in~\ref{def_Ds}.

The partonic cross sections are completely determined by the amplitude structure given above. Their analytic expressions are simple in the case in which only the $V=\{W,Z\}$ mediated diagrams are retained. This approximation is justified as these diagrams are assumed to largely dominate the process in the kinematic region selected for the experimental measurement.
With this simplification, the SM partonic cross sections at fixed $q^2$ are~\cite{Ellis:1975ap}
\begin{align}
\sigma ( \bar{\psi} \psi \to Z h)^{\rm SM}  &=
\frac{2 \pi \alpha^2 [(\gsm{\psi}{V})^2 + (\gsm{\psi}{A})^2]}{3  N_c  \hst^4 \hct^4}
\, \, \frac{|\vec{p}_h|}{\sqrt{q^2}} \, \frac{|\vec{p}_h|^2 + 3 \hat m_Z^2}{(q^2- \hat m_Z^2+i\hat \Gamma_Z \hat m_Z)^2}~,   \\
\sigma ( \bar{\psi}_s \psi_r \to W h)^{\rm SM} &=
 \frac{\pi\alpha^2 |V_{rs}|^2}{18 \hst^4}
 \, \, \frac{|\vec{p}_h|}{\sqrt{q^2}} \, \frac{|\vec{p}_h|^2 + 3 \hat m_W^2}{(q^2- \hat m_W^2+i\hat \Gamma_W \hat m_W)^2}~,
\end{align}
where $|\vec{p}_h| = [\lambda(m_h^2,m_V^2,q^2)/(4 q^2)]^{1/2}$
is the center of mass momentum of the Higgs-like boson and $V_{rs}$ denotes CKM matrix elements. The function $\lambda(x,y,z)$ was defined in~\ref{def_lambda_function}.
In the SMEFT these expressions are modified according to\footnote{Partial results for this expression were reported in
Ref.~\cite{Isidori:2013cga}.}
\begin{equation}
 \begin{aligned}
\frac{\sigma^{\rm BSM} (\psi \, \bar{\psi} \to V h)}{\sigma^{\rm SM} (\psi \, \bar{\psi} \to V h)} & =
\left| f^V_1 (q^2) \right|^2
+  3 ~{\rm Re} \left[f^V_1(q^2)  f^{V*}_3(q^2)\right]  \frac{ \hat m_V^2 (q^2 + \hat m_V^2 -\hat m_h^2)}{ |\vec{p}_h|^2 +3 \hat m_V^2 }  \\
& + \frac{\hat m_V^2 q^2}{ |\vec{p}_h|^2 +3 \hat m_V^2 } \left[ \left| f^V_3(q^2) \right|^2  (3 \hat m_V^2 + 2  |\vec{p}_h|^2) + 2  |\vec{p}_h|^2  \left| f^V_4(q^2)\right|^2 \right]
\\
&
+\frac{2}{(g_{V,SM}^{V,\psi})^{2} + (g_{A,SM}^{V,\psi})^{2}}\left[ g_{V,SM}^{V,\psi}\, \d g_{V}^{V,\psi} + g_{A,SM}^{V,\psi}\, \d g_{A}^{V,\psi}\right],
\label{eq:Rcross}
 \end{aligned}
 \end{equation}
where $f_i^V(q^2)=f_i^{V,SM}(q^2)+\d f_i^V(q^2)$ and the notation $g_{\chi,SM}^{V,\psi}$ denotes the SM coupling of the fermion $\psi$ with chiral structure $\chi=\{V,A\}$ to the $V$ boson.

\subsubsection*{VBF production}
\begin{figure}[t]
 \includegraphics[width=\textwidth]{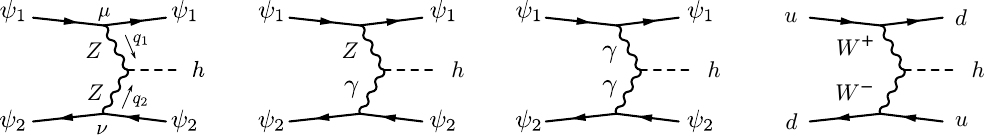}
 \caption{Diagrams contributing to VBF Higgs production in the SMEFT.}\label{diagrams_VBF}
\end{figure}

Adopting a formalism similar to that employed for $Vh$ associated production, the generic amplitude for Higgs production via $VV=\{ZZ,W^+W^-\}$ fusion can be written as
\begin{equation}
 \mathcal{A}_{VBF,VV} = \frac{iN_V^2g_V^3 \hat m_V}{(q_1^2-\hat m_V^2+i\hat \Gamma_V \hat m_V)(q_2^2-\hat m_V^2+i\hat \Gamma_V \hat m_V)}J^{V,\psi_1}_\mu J^{V,\psi_2}_\nu T^{\mu\nu}_V,
\end{equation}
where the currents and momenta are labeled as in Figure~\ref{diagrams_VBF} and, in the limit of massless fermions, the tensor $T_V^{\mu\nu}$ can be decomposed into three Lorentz structures\footnote{An alternative, equivalent decomposition was used in~\cite{Greljo:2015sla}.}:
\begin{equation}
 T_V^{\mu\nu} = f_1^V(q_1^2,q_2^2) g^{\mu\nu}+f_3^V(q_1^2,q_2^2)(q_1\cdot q_2 \, g^{\mu\nu}-q_1^\nu q_2^\mu) + f_4^V(q_1^2,q_2^2)\epsilon^{\mu\nu\rho\s} q_{1\rho}q_{2\s}\,.
\end{equation}
For the $ZZ$ fusion case $f_1^{Z,SM}(q_1^2,q_2^2) \equiv 1$, $f_{3,4}^{Z,SM}(q_1^2,q_2^2)\equiv0$, while in the SMEFT
\begin{align}
 f_1^Z(q_1^2,q_2^2)=& 1+\d D^Z(q_1^2)+\d D^Z(q_2^2)+\bar v_T^2\left(C_{H\square} +\frac{C_{HD}}{4}-\CHlt+\frac{C_{ll}'}{2}\right)\,,\\
 f_3^Z(q_1^2,q_2^2)=& -\frac{2\bar v_T^2}{\hat m_Z^2}\left[\hct^2C_{HW}+\hst^2 C_{HB}+\hst\hct C_{HWB}\right]\,,\\
 f_4^Z(q_1^2,q_2^2)=& \frac{2\bar v_T^2}{\hat m_Z^2}\left(\hct^2C_{H\tilde W}+\hst^2 C_{H \tilde B}+\hst\hct C_{H\tilde WB}\right)\,.
 \end{align}
In the SMEFT the neutral current process receives additional contributions from diagrams with $Z\g$ and $\g\g$ fusion, whose amplitudes are
\begin{align}
 \mathcal{A}_{VBF,Z\gamma} =& \,
   2i\bar e g_Z\bar v_T\,\left(\frac{Q_{\psi_2} J_\mu^{Z,\psi_1}J_\nu^{\psi_2,em}}{(q_1^2-\hat m_Z^2+i\hat \Gamma_Z \hat m_Z)q_2^2}
   + \frac{Q_{\psi_1}J_\mu^{\psi_1,em}J_\nu^{V,\psi_2,}}{(q_2^2-\hat m_Z^2+i\hat \Gamma_Z \hat m_Z)q_1^2}\right) \times \nonumber\\
   &
   \bigg[
 \left(\hsdt(C_{HB}-C_{HW})+\hcdt C_{HWB}\right)(q_1\cdot q_2 \,g^{\mu\nu}-q_2^\mu q_1^\nu)
 +\nonumber\\
 &\qquad+\left(\hsdt(C_{H\tilde W}-C_{H\tilde B})-\hcdt C_{H\tilde WB}\right)\epsilon^{\mu\nu\rho\s}q_{1\rho} q_{2\sigma}
 \bigg]\,,\\
 \mathcal{A}_{VBF,\gamma\gamma} =& \,
   \frac{4i\bar e^2\bar v_T}{q_1^2q_2^2}Q_{\psi_1}Q_{\psi_2}\,J_\mu^{\psi_1,em}J_\nu^{\psi_2,em}\,\bigg[
 \left(\hst^2C_{HW}+\hct^2C_{HB}-\hct\hst C_{HWB}\right)(q_1\cdot q_2 \,g^{\mu\nu}-q_2^\mu q_1^\nu)
 +\nonumber\\
 &\qquad+\left(\hst^2 C_{H\tilde W}+\hct^2C_{H\tilde B}-\hct\hst C_{H\tilde WB}\right)\epsilon^{\mu\nu\rho\s}q_{1\rho} q_{2\sigma}
 \bigg]\,.
\end{align}
The total shift in the neutral-current VBF production can then be written as
\begin{equation}
 \begin{aligned}
\d\mathcal{A}_{VBF,{\rm n.c.}} =&
\frac{i\bar g_2^3 \hat m_Z}{\hct^3(q_1^2-\hat m_Z^2+i\hat \Gamma_Z \hat m_Z)(q_2^2-\hat m_Z^2+i\hat \Gamma_Z \hat m_Z)}\cdot\\
&\left[
\d J_\mu^{Z,\psi_1} (J_\nu^{Z,\psi_2}T_{Z}^{\mu\nu})^{SM} + \d J_\nu^{Z,\psi_2} (J_\mu^{Z,\psi_1}T_{Z}^{\mu\nu})^{SM}+ (J_\mu^{Z,\psi_1}J_\nu^{Z,\psi_2})^{SM}\d T_Z^{\mu\nu}
 \right]+\\
 &+\mathcal{A}_{VBF,Z\gamma}+\mathcal{A}_{VBF,\gamma\gamma}\,.
 \end{aligned}
\end{equation}
For the charged current case  $f_1^{W,SM}(q_1^2,q_2^2) \equiv 1$, $f_{3,4}^{W,SM}(q_1^2,q_2^2)\equiv0$ and in the SMEFT
\begin{align}
f_1^W(q_1^2,q_2^2)=&\, 1+\d D^W(q_1^2) +\d D^W(q_2^2)+\nonumber\\
&\quad+\bar v_T^2\left(  C_{H\square} -\left(2+\frac{1}{\hcdt}\right)\frac{C_{HD}}{4}+\frac{C_{ll}'-2\CHlt}{2\hcdt}-\frac{\hsdt }{\hcdt}C_{HWB}\right)\,,\\
 f_3^W(q_1^2,q_2^2)=& -\frac{2\bar v_T^2}{\hat m_W^2} C_{HW}\,,\\
 f_4^W(q_1^2,q_2^2)=& \frac{2\bar v_T^2}{\hat m_W^2} C_{H\tilde W}\,.
\end{align}
with $\d D^W(s)$ defined in Eq.~\ref{def_Ds}.
Since there are no additional contributions from different diagrams, in this case the overall shift in the amplitude, compared to the SM is
\begin{equation}
\begin{aligned}
  \d\mathcal{A}_{VBF,{\rm c.c.}} =
\frac{i\bar g_2^3 m_W}{2(q_1^2-m_W^2)(q_2^2-m_W^2)}\bigg[&
\d J_\mu^{W_+,q} (J_\nu^{W_-,q}T_{W}^{\mu\nu})^{SM} + \d J_\nu^{W_-,q} (J_\mu^{W_+,q}T_{W}^{\mu\nu})^{SM}+\\
&  +(J_\mu^{W_+,q}J_\nu^{W_-,q})^{SM}\d T_W^{\mu\nu}
 \bigg],
 \end{aligned}
\end{equation}
where the expression has been specialized to the quark current case.

\subsubsection*{\boldmath $h\to \bar ff$}
The decay width of the Higgs boson into a fermion pair is given by
\begin{align}
\Gamma(h \to \bar{f} \, f) =  \frac{(\bar g_{hf\bar f})^2 \, \hat m_h \, N_c}{8 \, \pi} \left(1 - 4 \frac{m_f^2}{\hat m_h^2} \right)^{3/2},
\end{align}
where $N_c$ is the number of colors of the final state fermion $f$. The generic coupling is~\cite{Alonso:2013hga}
\begin{align}
\left[\bar g_{hf\bar f}\right]_{rs}
&= \frac{1}{\sqrt2} \left[\hat Y_f\right]_{rs}  \left[ 1+ \bar v_T^2\left(C_{H\square}-\frac{C_{HD}}{4}\right)-\frac{\d G_F}{\sqrt2} \right]   - \frac{\bar v_T^2}{\sqrt{2}} C^*_{\substack{f H \\ rs}} ,
\qquad f=u,d,e
\end{align}
 with $[\hat Y_f]_{rs} = 2^{3/4}\sqrt{\hat G_F} \,[m_f]_{rr} \,\d_{rs}$ as defined in Eq.~\ref{Yhat}.

\subsubsection*{\boldmath $h\to V\bar ff' $}
The decay of the Higgs into an experimentally reconstructed nearly on-shell vector boson $V=\{W^\pm,Z\}$ and a fermion pair $\bar ff'$ proceeds through
the same diagrams that give $Vh$ associated production. The amplitude of this process can then be decomposed
in the same way~\cite{Isidori:2013cla} exploiting the narrow width of the intermediate vector boson.
\begin{equation}
 \mathcal{A}_{h\to V\bar \psi\psi'}= \frac{iN_V g_V^2m_V}{q^2-m_V^2}\, J^{V,\psi}_\nu\, \epsilon^*_\mu\, T^{\mu\nu}_V
\end{equation}
where $q$ is the four-momentum of the fermion pair in the final state and the momentum $p$ appearing
in the decomposition of $T_V^{\mu\nu}$ (Eq.~\ref{Tmunu_decomposition}) is again that of the outgoing $V$ boson.

Because the two processes have the same diagram forms, the relative correction to the partial width is also analogous
to that of~\ref{eq:Rcross}. The contributions from the 4-point contact interactions and from a $\g$-mediated diagram
in $Z f\bar f$ production:
\begin{equation}
 \begin{aligned}
  \frac{d\Gamma(h\to V \psi\bar \psi)/dq^2}{d\Gamma(h\to V \psi\bar \psi)^{SM}/dq^2} =&
\left| f^V_1 (q^2) \right|^2
-  3 ~{\rm Re} \left[f^V_1(q^2)  f^{V*}_3(q^2)\right]  \frac{ m_V^2 (q^2 + m_V^2 - m_h^2)}{ |\vec{p}|^2 +3 m_V^2 }  \\
& + \frac{ m_V^2 q^2}{ |\vec{p}|^2 +3 m_V^2 } \left[ \left| f^V_3(q^2) \right|^2  (3  m_V^2 + 2  |\vec{p}|^2) + 2  |\vec{p}|^2  \left| f^V_4(q^2)\right|^2 \right]
\\
&
+\frac{2}{(g_{V,SM}^{V,\psi})^{2} + (g_{A,SM}^{V,\psi})^{2}}\left[ g_{V,SM}^{V,\psi}\, \d g_{V}^{V,\psi} + g_{A,SM}^{V,\psi}\, \d g_{A}^{V,\psi}\right],
 \end{aligned}
\end{equation}
where $|\vec p|=[\lambda(m_h^2,m_V^2,q^2)/(4q^2)]^{1/2}$ is the momentum of the $V$ boson in the final state and the differential SM width is~\cite{Buchalla:2013mpa}\footnote{
Ref.~\cite{Buchalla:2013mpa} corrects an overall factor of 1/2 compared to the result in Ref.~\cite{Isidori:2013cla}.}
\begin{equation}
 \begin{aligned}
  \frac{d\Gamma(h\to V \psi\bar \psi)^{SM}}{dq^2}=&
  \frac{N_V^2g_V^4\, [(g_{V,SM}^{V,\psi})^2+g_{A,SM}^{V,\psi})^2]}{96\pi^3m_h^3}\,\,\frac{3 m_V^2+ |\vec p|^2}{(q^2-m_V^2)^2}\,(q^2)^{3/2}\,|\vec p|\\
=&   \frac{N_V^2g_V^4\, [(g_{V,SM}^{V,\psi})^2+g_{A,SM}^{V,\psi})^2]}{768\pi^3m_h^3}\,\,\frac{12 q^2 m_V^2+ \lambda(m_h^2,m_V^2,q^2)}{(q^2-m_V^2)^2}\,\,\lambda^{1/2}(m_h^2,m_V^2,q^2).
 \end{aligned}
\end{equation}
The form factors read\footnote{Note the difference in sign in $f_3(q^2)$ and $f_4(q^2)$ w.r.t. the $Vh$ production case, which is due to the fact that here $p$ and $q$ are both outgoing momenta, while in the production $q$ is incoming and $p$ is outgoing. The form factors for $h\to V f\bar f$ decay coincide with those for VBF production in the limit of massless fermions.}
\begin{align}
 f_1^Z(q^2)=&\, 1+\d D^Z(q^2)+\bar v_T^2\left(C_{H\square} +\frac{C_{HD}}{4}-\CHlt+\frac{C_{ll}'}{2}\right)\,,\\
 f_2^Z(q^2)=& -\frac{1}{\hat m_Z^2}\d f_1^Z(q^2)\,,\\
 f_3^Z(q^2)=& -\frac{2\bar v_T^2}{\hat m_Z^2}\left[\hct^2C_{HW}+\hst^2 C_{HB}+\hst\hct C_{HWB}\right]\,,\\
  f_4^Z(q^2)=& \frac{2\bar v_T^2}{\hat m_Z^2}\left(\hct^2C_{H\tilde W}+\hst^2 C_{H \tilde B}+\hst\hct C_{H\tilde WB}\right),
   \end{align}
\begin{align}
 f_1^W(q^2)=&\,1+ \d D^W(q^2) +\bar v_T^2\left(  C_{H\square} -\left(2+\frac{1}{\hcdt}\right)\frac{C_{HD}}{4}+\frac{C_{ll}'-2\CHlt}{2\hcdt}-\frac{\hsdt }{\hcdt}C_{HWB}\right)\,,\\
 f_2^W(q^2)=& -\frac{1}{\hat m_Z^2}\d f_1^W(q^2)\,,\\
 f_3^W(q^2)=& -\frac{2\bar v_T^2}{\hat m_W^2} C_{HW}\,,\\
 f_4^W(q^2)=& \frac{2\bar v_T^2}{\hat m_W^2} C_{H\tilde W}\,.
 \end{align}
The contributions from the diagram $h\to Z \gamma^*\to Z f\bar f$ and from the 4-point contact interactions $\bar\psi\psi hZ$, $\bar\psi\psi hW$ have the same form, with an opposite sign, as those given for $hV$ associated production in Eqs.~\ref{hZprod_4point_diagram}, \ref{hWprod_4point_diagram}.

\subsubsection*{\boldmath $h\to \gamma\gamma$}
At one loop in the SM, the partial width of the Higgs decay into photons is~\cite{Bergstrom:1985hp,Manohar:2006gz}
\begin{equation}
 \Gamma^{SM}(h\to\g\g) = \frac{\sqrt 2\hat G_F \hat\a^2 \hat m_h^3}{16 \pi^3}\abs{I^\g}^2
\end{equation}
where $I^\gamma$ is a Feynman parameter integral including both contributions from the top quark (with leading QCD corrections) and $W$-boson loops. Its analytic expression is given in Ref~\cite{Manohar:2006gz} and numerically it is $I^\gamma \approx -1.65$~\cite{Manohar:2006gz}.
At tree level in the SMEFT, the decay rate is modified as~\cite{Manohar:2006gz,Alonso:2013hga}
\begin{equation}
\frac{ \Gamma(h\to \gamma \gamma) }{\Gamma^{SM}(h \to \gamma \gamma)}
\simeq \abs{1+\frac{8 \pi^2 \bar v_T^2}{I^\gamma}  \mathscr{C}_{\gamma \gamma} }^2+\abs{ \frac{8 \pi^2 \bar v_T^2} {I^\gamma} \widetilde {\mathscr{C}}_{\gamma \gamma} }^2
\end{equation}
where
\begin{align}
\mathscr{C}_{\gamma \gamma} &= \frac{1}{\bar g_2^2}C_{HW}+\frac{1}{\bar g_1^2}C_{HB}-\frac{1}{\bar g_1 \bar g_2}C_{HWB}\\
\tilde{\mathscr{C}}_{\gamma \gamma} &= \frac{1}{\bar g_2^2}C_{H\tilde{W}}+\frac{1}{\bar g_1^2}C_{H\tilde{B}}-\frac{1}{\bar g_1 \bar g_2}C_{H\tilde{W}B}.
\end{align}

\subsubsection*{\boldmath $h\to Z \gamma$}
The one-loop rate for the Higgs decay into $Z\g$ in the SM reads~\cite{Bergstrom:1985hp}
\begin{equation}
 \Gamma^{SM}(h\to Z\g) = \frac{\sqrt2 G_F \hat\a^2 \hat m_h^3}{8\pi^3}\left(1-\frac{\hat m_Z^2}{\hat m_h^2}\right)^3\abs{I^{Z\g}}^2,
\end{equation}
where, including both top and $W$ loop contributions,  $I^{Z\g} \approx -2.87$~\cite{Bergstrom:1985hp,Manohar:2006gz}.
The modification of the decay rate in the SMEFT has the form~\cite{Manohar:2006gz,Alonso:2013hga}
\begin{equation}
\frac{ \Gamma(h\to \gamma Z) }{\Gamma^{SM}(h \to \gamma Z)}
\simeq \abs{1+\frac{8 \pi^2 \bar v_T^2}{I^{Z\g}}\mathscr{C}_{\gamma Z}  }^2+\abs{ \frac{8 \pi^2 \bar v_T^2} {I^{Z\g}} \widetilde {\mathscr{C}}_{\gamma Z}}^2
\end{equation}
with
\begin{align}
\mathscr{C}_{\gamma Z} &= \frac{1}{\bar g_1 \bar g_2}C_{HW}  - \frac{1}{\bar g_1 \bar g_2}C_{HB} -\left(\frac{1}{2 \bar g_1^2}-\frac{1}{2 \bar g_2^2}\right) C_{HWB}\\
\tilde{\mathscr{C}}_{\gamma Z} &= \frac{1}{\bar g_1 \bar g_2}C_{H\tilde W}  - \frac{1}{\bar g_1 \bar g_2}C_{H\tilde B} -\left(\frac{1}{2 \bar g_1^2}-\frac{1}{2 \bar g_2^2}\right) C_{H\tilde WB}.
\end{align}
Note the inverse gauge coupling dependence that follows from the choice to not scale the Wilson coefficients by
a gauge coupling.

\subsection{Top quark properties}
\subsubsection*{Top width}

The width of the top quark can be computed at tree-level in the SMEFT using the narrow width approximation for the $W$ boson:
\begin{equation}
 \bar\Gamma(t\to b \bar f_p f_r) = \bar\Gamma(t\to b W^+) \overline{\rm Br}(W^+\to \bar f_p f_r)\,.
\end{equation}
The narrow width approximation and the SMEFT approximation do not commute. We perform first the narrow width approximation and then the SMEFT expansion, finding, in the notation of Sec.~\ref{appendix_W_properties}
\begin{equation}\label{top_width_smeft}
\begin{aligned}
\bar\Gamma(t\to b \bar f_p f_r) =& \,\Gamma(t \to b W^+)^{SM}\,\frac{\Gamma_{W\to \bar f_p f_r}^{SM}}{ \Gamma_W^{SM}} + \d\Gamma(t \to b W^+)\,\frac{\Gamma_{W\to \bar f_p f_r}^{SM}}{\Gamma_W^{SM}}\\
&
+\,\Gamma(t \to b W^+)^{SM}\,\frac{\d \Gamma_{W\to \bar f_p f_r}}{ \Gamma_W^{SM}}
-\Gamma(t \to b W^+)^{SM}\,\frac{\Gamma_{W\to \bar f_p f_r}^{SM}}{\Gamma_W^{SM}}\,\frac{\d\Gamma_W}{\Gamma_W^{SM}}\,.
\end{aligned}
\end{equation}
Summing over the polarizations of the $W$ boson one has
\begin{align}
 \Gamma(t\to b W^+)^{SM} =& \frac{\bar g_2^2}{64\pi}\frac{\hat m_t}{\hat m_W^2}\lambda^{1/2}(\hat m_t^2,\hat m_b^2,\hat m_W^2)
 |V_{tb}|^2\left(1+x_W^2-2x_b^2-2x_W^4+x_W^2 x_b^2+x_b^4\right),
\\
 \d\Gamma(t\to b W^+) =& \frac{\bar g_2^2}{64\pi}\frac{\hat m_t}{\hat m_W^2}\lambda^{1/2}(\hat m_t^2,\hat m_b^2,\hat m_W^2)\Bigg[ \nonumber\\
 &
 2\Re\left[V_{tb}^*\,\left(\d (g_V^{W,q})_{33}+\d (g_A^{W,q})_{33}\right)\right]\left(1+x_W^2-2x_b^2-2x_W^4+x_W^2 x_b^2+x_b^4\right) \nonumber\\
 & - 12 x_W^2 x_b \,\Re\left[V_{tb}^*\,\left(\d (g_V^{W,q})_{33}-\d (g_A^{W,q})_{33}\right)\right] \\
 &
 +\frac{6 }{\hat G_F}\,x_W (1-x_W^2-x_b^2) \left(\Re\left(V_{tb} C_{\substack{uW\\33}^*}\right)-\Re\left(V_{tb} C_{\substack{dW\\33}}\right)x_b\right)
 \Bigg], \nonumber
\end{align}
where $x_i= \hat m_i/\hat m_t$, $\lambda(x,y,z)$ is defined in Eq.~\ref{def_lambda_function} and $\d g_{V,A}^{W,q}$ are the shifts for the $W$ couplings to quarks defined in Eq.~\ref{delta_gW_quarks}. These results are in agreement with previous calculations, see e.g.~\cite{AguilarSaavedra:2006fy,Zhang:2010dr}. Note that if $\Q_{Hud}$ is included, it also contributes to the latter as
\begin{equation}
 \d(g_V^{W,q})_{rr} = -\frac{\bar v_T^2}{4} C_{\substack{Hud\\rr}}^*+\dots,\qquad\qquad
 \d(g_A^{W,q})_{rr} = \frac{\bar v_T^2}{4} C_{\substack{Hud\\rr}}+\dots,
\end{equation}
making these quantities complex.

\section{Low energy precision measurements in LEFT}
\subsubsection*{\boldmath $\nu$--lepton scattering}\label{nulep}
The scattering process $\nu \, e^\pm \to \nu \, e^\pm$ can be described by the following Effective Lagrangian~\cite{Berthier:2015gja}
\begin{equation}
\label{nueffective}
\mathcal{L}_{\nu e}= - \frac{ \hat{G}_F}{\sqrt{2} } \left[ \bar{e} \gamma^{\mu} \left( (\bar{g}^{\nu e}_V) - (\bar{g}^{\nu e}_A) \gamma^5 \right) e \right] \left[ \bar{\nu} \gamma_\mu \left( 1-  \gamma^5 \right) \nu \right].
\end{equation}
the shifts are then
$\bar{g}^{\nu e}_V = g^{\nu e}_V + \d g^{\nu e}_V$, $\bar{g}^{\nu e}_A = g^{\nu e}_A + \d g^{\nu e}_A$ where
\begin{align}
\d (g^{\nu e}_V)&= 2 \left(\d g^{\ell}_V + 2 \d g^{\ell, W_{\pm}}_V \right)+ 4 \d g^{\nu}_V \left(- \frac{1}{2} + 2 s_{\that}^2\right)   - \frac{1}{2 \sqrt{2} \hat{G}_F} \left(2C_{l l} + 2 C_{ll}' + C_{ l e} \right) -  \frac{\d m_W^2}{m_W^2},  \\
\d (g^{\nu e}_A) &= 2  \left(\d g^{\ell}_A + 2\d g^{\ell, W_{\pm}}_A \right) - 2
 \d g^{\nu}_V  - \frac{1}{2 \sqrt{2} \hat{G}_F} \left(2C_{l l} + 2C_{ll}' - C_{l e} \right)- \frac{\d m_W^2}{m_W^2}.
\end{align}
these shifts add the contributions of $W$ and $Z$ exchange. Depending on the neutrino flavour some terms are absent. The shift that is relevant for $g_{A,V}^{\nu_\mu e}$ does not have a
$\d M_W^2$ or $\d g^{\ell, W_{\pm}}_{V,A}$ contribution, whereas a shift for $g_{A,V}^{\nu_\mu \mu}$ has both contributions.

\subsubsection*{\boldmath $\nu$--Nucleon scattering}
The scattering process $\nu \, N \to \nu \, X$ via $Z$ exchange can be described by the following Effective Lagrangian~\cite{Berthier:2015gja}
\begin{equation}
\mathcal{L}^{NC}_{\nu \,q}=-\frac{\hat{G}_F}{\sqrt{2}} \left[\bar{\nu}\gamma^{\mu}\left(1-\gamma^5 \right)\nu\right] \left[\bar{\epsilon}_L^q \bar{q}\gamma_{\mu}\left(1-\gamma^5\right)q +\bar{\epsilon}_R^q \bar{q}\gamma_{\mu}\left(1+\gamma^5\right)q \right].\label{shiftZ}
\end{equation}
where $q=\{u,d\}$.
At tree level in the SM: $(\epsilon_L^q)^{SM} = \gsm{q}{V} + \gsm{q}{A}$ and $(\epsilon_R^q)_{SM} = \gsm{q}{V} - \gsm{q}{A}$. The redefinition of the Z couplings and the corrections due to $\psi^4$ operators lead to a shift in $\epsilon_L^q$ and $\epsilon_R^q$ of the form
 $\bar{\epsilon}_{L/R}^q = \epsilon_{L/R}^q + \d \epsilon_{L/R}^q$ with
\begin{align}
\d \epsilon_L^u &= -\frac{1}{2 \sqrt{2} \hat{G}_F}\left(C_{l q}^{(1)} + C_{l q}^{(3)}\right) + \d g_V^u + \d g_A^u + 4 \d g_V^{\nu}(\epsilon_L^u)^{SM},  \\
\d \epsilon_L^d &= -\frac{1}{2 \sqrt{2} \hat{G}_F}\left(C_{l q}^{(1)} - C_{l q}^{(3)}\right) + \d g_V^d + \d g_A^d + 4 \d g_V^{\nu}(\epsilon_L^d)^{SM},  \\
\d \epsilon _R^u &=  -\frac{1}{2 \sqrt{2} \hat{G}_F} C_{l u} + \d g_V^u - \d g_A^u + 4 \d g_V^{\nu} (\epsilon_R^u)^{SM}, \\
\d \epsilon_R^d &= - \frac{1}{2 \sqrt{2} \hat{G}_F} C_{l u} + \d g_V^u - \d g_A^u + 4 \d g_V^{\nu} (\epsilon_R^u)^{SM}.
\end{align}
The scattering $\nu \, N \to \ell \, X$ and the inverse process take place through $W$ exchange and can be described by
\begin{equation}
\mathcal{L}=-\frac{\hat{G}_F}{\sqrt{2}} \left[\bar{\ell}\gamma^{\mu}\left(1-\gamma^5 \right)\nu\right] \left[\bar{\Sigma}^{ij}_L \bar{u}_{i}\gamma_{\mu}\left(1-\gamma^5\right)d_{j} \right] +\hc,
\end{equation}
where at tree level in the SM $(\Sigma_L^{ij})_{SM} = V_{CKM}^{ij}$. This coupling receives corrections from redefinitions of the $W$ mass and couplings, so that $\bar{\Sigma}^{ij}_L= (\Sigma_L^{ij})_{SM} + \d \Sigma_L^{ij}$ with
\begin{equation}
 \label{shiftW}
\d \Sigma_L^{ij} =\left[- \frac{\d m_W^2}{\hat m_W^2}  + 2 \, \d g_V^{q,W} + 2 \, \d g_V^{\ell, W}  - \frac{1}{\sqrt{2}\hat{G}_F}C_{l q}^{(3)} \right] V_{CKM}^{ij}.
\end{equation}
The inclusion of a right-handed coupling $\bar{\Sigma}_R^{ij}$ is not forbidden in principle in the SMEFT, it can be generated by the operator $\Q_{\ell e d q}$. This contribution has been neglected here because such a correction is
proportional to the Yukawa matrices in the $\rm U(3)^5$ limit assumed, and therefore vanishes in the limit of massless fermions.

Charged and neutral current process are related by~\cite{LlewellynSmith:1983ie}
\begin{equation}
\frac{d^2 \, \sigma(\nu N \to \nu X)}{d \, x \, d \, y} = g_{L,eff}^2 \, \frac{d^2 \, \sigma(\nu N \to \mu^- X)}{d \, x \, d \, y} + g_{R,eff}^2 \, \frac{d^2 \, \sigma(\bar{\nu} N \to \mu^+ X)}{d \, x \, d \, y}.
\end{equation}
for the scattering variables $x = - q^2 / (2  p_N  \cdot q)$, $y = (p_N \cdot q)/(p_N \, \cdot  p_\nu)$,
where $q^2$ is the momentum transfer and $p_N$, $p_\nu$ are respectively the nucleon and neutrino momenta.
The effective couplings $g_{L/R,eff}$ receive corrections in the SMEFT so that $\bar{g}_{L/R,eff}^2=g_{L/R,eff}^2+ \d g_{L/R,eff}^2$ and they can be expressed in terms of the $\epsilon_{L/R}^q$ parameters as
\begin{align}\label{gLR_eff}
\bar{g}_{L/R,eff}^2 &= \sum_{i,j} \left[\left|\bar{\epsilon}_{L/R}^{u^i} \right|^2 + \left|\bar{\epsilon}_{L/R}^{d^j} \right|^2 \right] \, \left|(\bar{\Sigma}_L^{ij})\right|^{-2}.
\end{align}
Relevant quantities for these processes are the ratios of cross sections
\begin{align}
R^{\nu} &= \frac{\sigma\left( \nu N \to \nu X\right)}{\sigma \left(\nu N \to \ell^- X\right)} = g_{L,eff}^2 + r \,g_{R,eff}^2,&
R^{\bar{\nu}} &= \frac{\sigma\left(\bar{\nu} N \to \bar{\nu} X\right)}{\sigma \left(\bar{\nu} N \to \ell^+ X\right)} = g_{L,eff}^2 + \frac{g_{R,eff}^2}{r},
\end{align}
where the factor $r$ in an ideal experiment with full acceptance (in the absence of sea quarks) is given by $r=1/3$.  SMEFT contributions to the $r$ parameter can be neglected as long as dimension-6 operators have a negligible impact on the parton distributions of nucleons. This condition is plausibly verified, as such corrections are expected to scale as $\Lambda_{\rm QCD}^2/\Lambda^2$.

Finally, the parameter $\kappa$ defined by
\begin{equation}
 \kappa = 1.7897 \, g_{L,eff}^2+ 1.1479 g_{R,eff}^2 - 0.0916 \, h_{L,eff}^2 - 0.0782 \, h_{R,eff}^2
\end{equation}
has been used to report data, e.g. by the CCFR collaboration~\cite{McFarland:1997wx}. Here $g_{L/R,eff}$ are the couplings introduced in Eq.~\ref{gLR_eff} and
\begin{equation}
\bar{h}_{L/R,eff}^2 =\sum_{i,j} \left[\left|\bar{\epsilon}_{L/R}^{u^i} \right|^2 - \left|\bar{\epsilon}_{L/R}^{d^j} \right|^2 \right] \, \left|(\bar{\Sigma}_L^{ij})\right|^{-2}.
\end{equation}

\subsubsection*{Neutrino Trident Production}\label{trident}
Neutrino trident production is the pair production of leptons from the scattering of a neutrino off the Coulomb field of a
nucleus, $\nu \,  N \to \nu  \, N \, \ell^+ \, \ell^-$. The SM calculation of this process is well known,
see Refs.\cite{Belusevic:1987cw,Brown:1973ih,Altmannshofer:2014pba}.
Using the notation of the effective Lagrangian in Eq.\ref{nueffective}, the constraint on the SMEFT is through the ratio of the partonic cross sections
\begin{equation}
\frac{\bar{\sigma}_{SMEFT}}{\sigma_{SM}} = \frac{(\bar{g}^{\nu_e e}_V)^2 + (\bar{g}^{\nu_e e}_A)^2}{(\gsm{\nu_e e}{V})^2 + (\gsm{\nu_e e}{A})^2}.
\end{equation}

\subsubsection*{Atomic Parity Violation}
For Atomic Parity Violation (APV) the standard Effective Lagrangian is given by~\cite{Berthier:2015gja}
\begin{equation}
\mathcal{L}_{eq}= \frac{\hat{G}_F}{\sqrt{2}} \, \left[ \sum_q \bar{g}^{eq}_{AV} \left(\bar{e} \gamma_\mu \gamma^5 e \right) \left( \bar{q} \gamma^\mu q\right) + \bar{g}^{eq}_{VA} \left( \bar{e} \gamma_\mu e \right) \left( \bar{q} \gamma^\mu \gamma^5 q \right) \right].
\end{equation}
In the SM:  $\gsm{e q}{AV} = 8 \,\gsm{q}{V} \, \gsm{\ell}{A} $ and $\gsm{e q}{VA} = 8 \, \gsm{q}{A}\, \gsm{\ell}{V} $ and the relevant couplings are for $q = u,d$. The effective shifts are
\begin{align}
 \d g^{eu}_{AV} =&\frac{1}{2\sqrt{2} \hat{G}_F} \left( - C_{l q}^{(1)} +C_{l q}^{(3)} - C_{l u} + C_{e u} + C_{q e}\right) + 2 \left(1- \frac{8}{3}s_{\that}^2\right) \d g^{\ell}_A  - 2 \d g^{u}_V, \\
\d g^{eu}_{VA} =& \frac{1}{2\sqrt{2} \hat{G}_F} \left( - C_{l q}^{(1)} +C_{l q}^{(3)} + C_{l u} + C_{e u} - C_{q e}\right) +  2 \d g^{u}_A \left(-1 + 4 s_{\that}^2 \right)+ 2 \d g^{\ell}_V, \\
\d g^{ed}_{AV} =& \frac{1}{2\sqrt{2} \hat{G}_F} \left( - C_{l q}^{(1)} -C_{l q}^{(3)} - C_{l d} + C_{e d} + C_{q e}\right) + 2 \left(-1 + \frac{4}{3}s_{\that}^2\right)\d g^{\ell}_A - 2 \d g^{d}_V, \\
\d g^{ed}_{VA} =& \frac{1}{2\sqrt{2} \hat{G}_F} \left( - C_{l q}^{(1)} -  C_{l q}^{(3)} + C_{l d} + C_{e d} - C_{q e}\right) + 2 \d g^{d}_A \left(-1 + 4 s_{\that}^2 \right)
- 2 \d g^{\ell}_V.
\end{align}
It is convenient to define the four combinations
$$\bar{g}^{ep}_{AV/VA} = 2 \bar{g}^{eu}_{AV/VA} + \bar{g}^{ed}_{AV/VA}, \qquad\bar{g}^{en}_{AV} = \bar{g}^{eu}_{AV/VA} + 2 \bar{g}^{ed}_{AV/VA},$$
that are shifted from their SM values by
\bea
\d g^{ep}_{AV/VA} &=& 2 \d g^{eu}_{AV/VA} + \d g^{ed}_{AV/VA}, \\
\d g^{en}_{AV/VA} &=& \d g^{eu}_{AV/VA} + 2 \d g^{ed}_{AV/VA}.
\eea
The weak charge $Q_{W}^{Z,N}$ of an element $X^A_Z$ defined by \cite{Agashe:2014kda,Erler:2013xha,Blunden:2012ty}
\begin{equation}
Q_{W}^{Z,N} = - 2 \left[ Z \left(g^{ep}_{AV} + 0.00005 \right) + N \left( g^{en}_{AV} + 0.00006 \right)\right]\left(1- \frac{\bar{\alpha}}{2 \pi}\right),
\end{equation}
is measured very precisely for thallium~\cite{Vetter:1995vf} and cesium~\cite{Wood:1997zq} and in the SMEFT it is shifted by
\begin{equation}
\d Q_W^{Z,N} = - 2 \left[Z \d g_{AV}^{ep} + N \d g_{AV}^{en} \right]\left(1- \frac{\hat{\alpha}}{2 \pi}\right)
\end{equation}
compared to the SM value.
\subsubsection*{Parity Violating Asymmetry in eDIS}

For inelastic polarized electron scattering $e_{L,R}\,N \to e \, X$ the right-left asymmetry A is defined as  \cite{Agashe:2014kda}:
\begin{equation}
A = \frac{\sigma_R - \sigma_L}{\sigma_R + \sigma_L},
\end{equation}
where
\bea
\frac{A}{Q^2} &=& a_1 + a_2 \frac{1-(1-y)^2}{1+(1-y)^2} \\
a_1 &=& \frac{3\hat{G}_F}{5 \sqrt{2} \pi \hat{\alpha}}\left(g_{AV}^{eu} - \frac{1}{2} g_{AV}^{ed}\right), \\
 a_2 &=& \frac{3\hat{G}_F}{5 \sqrt{2} \pi \hat{\alpha}}\left(g_{VA}^{eu} - \frac{1}{2} g_{VA}^{ed}\right).
\eea
Here $Q^2 \ge 0$ is the momentum transfer and $y$ is the fractional energy transfer in the scattering $y \simeq Q^2/s$.
In the SMEFT $\bar{g}_{AV/VA}^{eq} = g_{AV/VA}^{eq} + \d g_{AV/VA}^{eq}$ so that $a_1$ and $a_2$ receive the corrections~\cite{Berthier:2015gja}
\bea
\d a_1 &=& \frac{3\hat{G}_F}{5 \sqrt{2} \pi \hat{\alpha}}\left(\d g_{AV}^{eu} - \frac{1}{2} \d g_{AV}^{ed}\right), \\
\d a_2 &=& \frac{3\hat{G}_F}{5 \sqrt{2} \pi \hat{\alpha}}\left(\d g_{VA}^{eu} - \frac{1}{2} \d g_{VA}^{ed}\right).
\eea

\subsubsection*{M\o ller scattering}
Parity Violation Asymmetry ($A_{PV}$) in M\o ller scattering can be parameterized with the standard Effective Lagrangian
\begin{equation}
 \label{moller}
\mathcal{L}_{ee}= \frac{\hat{G}_F }{\sqrt{2}} g_{AV}^{ee} \left(\bar{e} \gamma^{\mu} \gamma^5 e \right)\left(\bar{e} \gamma_{\mu} e\right).
\end{equation}
In the SM $g_{AV}^{ee} = 8 \gsm{\ell}{V} \gsm{\ell}{A} = \frac{1}{2} \left(1 - 4 \hst^2\right)$. In the SMEFT we have the correction~\cite{Berthier:2015gja}
\begin{equation}
\d  g_{AV}^{ee} = \frac{1}{ \sqrt{2} \hat{G}_F}\left(-C_{l l}-C_{ll}' + C_{ee}\right)  - 2 \d g_V^{\ell} - 2 \left(1 - 4 \hst^2\right) \d g_A^{\ell},
\end{equation}
so that the parity violating asymmetry $A_{PV}$ is expressed as
\begin{equation}
\frac{A_{PV}}{Q^2} = - 2 g_{AV}^{ee} \frac{\hat{G}_F}{\sqrt{2} \pi \hat{\alpha}} \frac{1-y}{1+y^{4}+(1-y)^4}.
\end{equation}
$Q^2 $ and $y$ are defined as above.

\bibliographystyle{science_arxiv}
\bibliography{bibliography_V2}

\end{document}